\documentclass[onecolumn,authoryear]{els-mrw} 

\usepackage{amsmath,amssymb,amsfonts,amsthm,makeidx,graphicx}
\usepackage{txfonts}
\usepackage{helvet}
\usepackage{epsfig}

\newcommand{\leqsim}{\,\raisebox{-0.6ex}{$\buildrel < \over \sim$}\,}
\newcommand{\geqsim}{\,\raisebox{-0.6ex}{$\buildrel > \over \sim$}\,}

\newcommand{\pom}{{\rm I\!P}}
\newcommand{\reg}{{\rm I\!R}}
\newcommand{\alphapom}{\alpha_{_{\rm I\!P}}}

\newcommand{\xpom}{x_{_{\rm I\!P}}}

\def\ds{\displaystyle}
\def\be{\begin{equation}}
\def\ee{\end{equation}}
\def\bea{\begin{eqnarray}}
\def\eea{\end{eqnarray}}

\def\ds{\displaystyle}

\def\qb{\bar{\rm q}}

\def\qr{{\rm q}}

\def\gr{{\rm g}}

\def\a0{\alpha_0}
\def\as{\alpha_s}

\def\asq{\alpha_s(Q^2)}

\def\pom{{\rm I\! P}}
\def\reg{{\rm I\! R}}
\def\asl{\raise.15ex\hbox{/}\kern-.57em a}
\def\bsl{\raise.15ex\hbox{/}\kern-.57em b}
\def\csl{\raise.15ex\hbox{/}\kern-.57em c}
\def\dsl{\raise.15ex\hbox{/}\kern-.57em d}
\def\psl{\raise.15ex\hbox{/}\kern-.57em p}
\def\ksl{\raise.15ex\hbox{/}\kern-.57em k}
\def\qsl{\raise.15ex\hbox{/}\kern-.57em q}
\def\ppsl{\raise.15ex\hbox{/}\kern-.57em p'}
\def\kpsl{\raise.15ex\hbox{/}\kern-.57em k'}
\def\qpsl{\raise.15ex\hbox{/}\kern-.57em q'}
\def\ptlsl{\raise.15ex\hbox{/}\kern-.57em \partial}

\begin{document}

\chapter{HERA at low $x$ }

\author[1]{A M Cooper-Sarkar}%
\author[2]{P R Newman}%


\address[1]{\orgname{University of Oxford}, \orgdiv{Department of Physics}, \orgaddress{Denys Wilkinson Building, Keble Rd, Oxford OX1 3RH, UK}}
\address[2]{\orgname{University of Birmingham}, \orgdiv{School of Physics \& Astronomy}, \orgaddress{Edgbaston, Birmingham B15 2TT, UK}}

\articletag{Chapter Article tagline: update of previous edition,, reprint..}

\maketitle


\begin{glossary}[Nomenclature]
\begin{tabular}{@{}lp{34pc}@{}}
HERA &High Energy Ring Accelerator\\
DIS & Deep Inelastic Scattering\\
QCD &Quantum ChromoDynamics\\
QED & Quantum ElectroDynamics\\
pQCD & perturbative Quantum ChromoDynamics\\
PDF & Parton Distribution Function\\
DGLAP &Dokshitzer, Gribov, Lipatov, Altarelli, Parisi QCD evolution equations\\
BFKL &Balitsky, Fadin, Kuraev, Lipatov QCD evolution equations\\
(N(N))LO & (Next to (Next to)) Leading Order \\
BK & Balitsky, Kovchegov non-linear QCD evolution equations\\
JIMWLK & Jalilian-marian, Iancu, McLerran, Weigert, Leonidov, Kovner non-linear QCD evolution equations\\
CGC & Colour Glass Condensate\\
DVCS & Deeply Virtual Compton Scattering \\
GPD & Generalised Parton Distribution \\
DDIS & Diffractive Deep Inelastic Scattering \\
LRG & Large Rapidity Gap \\
DPDF & Diffractive Parton Distribution Function \\
\end{tabular}
\end{glossary}

\begin{abstract}[Abstract]
The field of Deep Inelastic Scattering is introduced with particular reference to the High Energy Ring Accelerator HERA.
The formalism of the Quark Parton Model and its extension in perturbative
Quantum Chromodynamics is outlined, defining the concept of a Parton
Distribution Function. The HERA data on inclusive and semi-inclusive processes
are reviewed with particular emphasis on results at low $x$ and low $Q^2$.
Their theoretical imterpretation in terms of conventional QCD DGLAP evolution
and its extension to BFKL evolution in the low-$x$ regime is explored together
with the possibility of the onset of a new non-linear regime which may lead to
gluon saturation. The focus then moves to diffractive processes both
quasi-elastic and diffractive dissociation, where more detailed hints of
saturation are found.

\end{abstract} 

\section{Introduction}
Deep Inelastic Scattering (DIS) is the scattering of point-like particles, leptons, such as electrons,
muons and neutrinos,
from larger particles, hadrons, such as protons, neutrons and nuclei, with the aim of determining the nature of the
constituents of the former and of mapping the kinematic behaviour of these constituents. These constituents are
called partons; quarks, anti-quarks and gluons. The DIS experiments allow us to measure the fractional momentum of
the hadron taken by the struck parton, $x$, and thus to measure the momentum distributions of the partons.
These distributions, for all
flavours of quark and for the gluons, are known as Parton Distribution Functions (PDFs). These distributions evolve
with the 4-momentum involved in the interaction (generally termed the scale of the interaction) as predicted by
Qunarum Chromodynamics (QCD). Over the course of the last 50 years, these measurements first established QCD as the
correct theory of the strong interaction, which binds the partons into the proton and then were used to make
accurate measurements of the PDFs, which are essential input to the predictions of all scattering cross sections
at the Large Hadron Collider.
However, the measurement of the parton distributions and the features of QCD are of interest in their own right. QCD
is a rich theory, its field carrier, the gluon, is self-interacting, unlike the photon of QED and this gives
rise to novel features. Firstly, to the running of the strong coupling, which decreases as scale increases,
such that we may make successful perturbative calculations (pQCD), which have served us well for more than 40 years.
Secondly, to the possibility of dense gluon population in the proton, where we need to go beyond the conventional
successful caluclations of QCD into a new regime. The reality of this new regime was first glimpsed at HERA in the
kinemtic region where the fractional momementum of the struck quark, $x$, is very small -- the low-$x$ region.
This is the subject of the current article. We begin by outlining the role of DIS in the development of QCD. A more detailed account can be found in~\cite{Devenish:2004pb}.

\section{DIS experiments}\label{chap:exp-dat}

HERA was an electron-proton collider which ran at the DESY Laboratory in Hamburg from 1992-2007. Two experiments, ZEUS and H1,
with somewhat different detector designs, measured the data. A total of 500 pb$^{-1}$ of data were taken per experiment and
this was split roughly equally between electron $e^-$ and positron $e^+$ beams of energy $E_e = 27$ GeV. The proton beam was
run at 4 different energies, $E_p= 460, 575, 800$ and $920$ GeV, giving centre of mass energies, $\sqrt{s} = 225,251,300$ and
$318$ GeV, respectively. Early data were run at $E_p = 800$ GeV, the bulk of the data were run at $920$ GeV and two short runs
were made at the lower energies. After the experiments had finished taking data in 2007 and completed most of their
individual experimental analyses, the data from H1 and ZEUS were combined taking full account of their systematic uncertainties, such that the combination represents a 'best of both worlds', with each experiment calibrating the other, reflecting the differing strengths of each. Most of the data presented here uses these combined data, except when a physics point is
best presented using early data.

Deep Inelastic Scattering can be pictured as the scattering of a lepton (in the case of HERA electron or positron) on a proton by exchange of a virtual vector boson.
The commonly used Lorentz invariants will be defined for the basic process 
\[
               l N \rightarrow l' X
\]
where $l,l'$ represent leptons (lepton is taken to include 
antileptons, unless it is necessary to distinguish them), $N$ represents 
a nucleon and $X$ represents the
hadronic final state particles.
\begin{figure}[tp]
\centerline{\psfig{figure=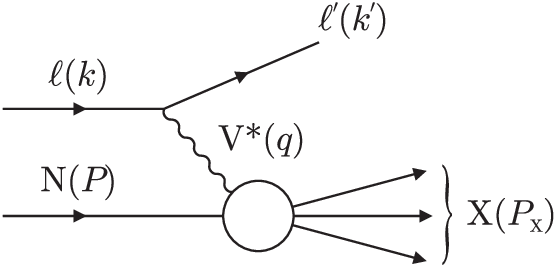,width=0.6\textwidth}}
\label{scatter}
\caption{Schematic diagram of lepton-hadron scattering via Vector-Boson 
Exchange. From~\cite{Devenish:2004pb}}.
\label{fig:lpscat}
\end{figure}
\noindent
The associated four vectors are $k,k'$ for the
incoming and outgoing leptons respectively, and $P$ for the target 
(or incoming)
nucleon. The process is mediated by the exchange of a  virtual vector boson, 
$V^*$($\gamma, W$ or $Z$), see Fig.~\ref{fig:lpscat}. This boson has four 
momentum given by $$q = k - k'$$
and the four vector $P_X$ of the hadronic final state system $X$ is given by
$$P_X = P + q$$
Various Lorentz invariants are useful in the description of the kinematics of 
the process:$$s = ( P + k )^2$$
the centre of mass energy squared for the $lp$ interaction; $$Q^2 = -q^2$$
the (negative of) the invariant mass squared of the virtual exchanged boson;
$$x = Q^2/2P.q$$
the Bjorken $x$ variable, 
which will be  interpreted in the quark-parton model as the fraction 
of the momentum of the incoming nucleon taken by the struck quark, 
and $$y = P.q/P.k$$
which gives a measure of the amount of energy transferred between the 
lepton and the hadron systems. It is also related to the scattering
angle in the lepton-quark centre-of-mass frame.
Note that (ignoring masses) $$Q^2 = s x y$$
so that only two of these quantities are independent. 

\begin{figure}[tbp]
\vspace{-1cm}
\centerline{\psfig{figure=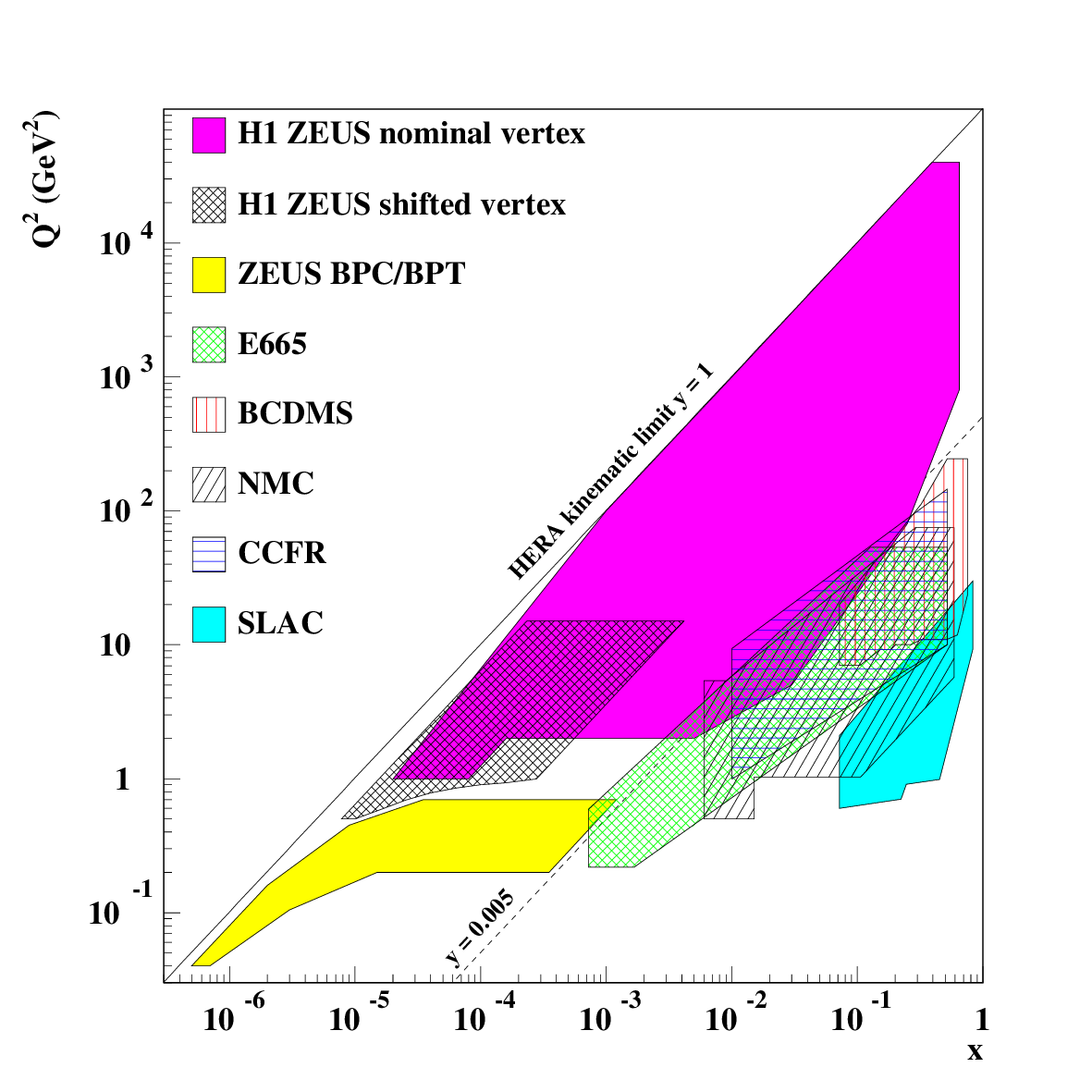,height=.4\textheight}} 
\caption{Regions in the ($x,Q^2$) plane covered by DIS experiments. Both HERA ad earlier fixed target DIS experiants are shown. From~\cite{Devenish:2004pb}.}
\label{fig:kinreg}
\end{figure}

The general form for the differential cross section for $e-p$ scattering by exchange of neutral current (NC) vector bosons
($\gamma^*$ or $Z^*$)
is
\begin{equation}
\frac {d^2\sigma (l^{\pm}N) } {dxdQ^2} =  \frac {2\pi\alpha^2} {Q^4 x}  
\left[Y_+\,F_2^{lN}(x,Q^2) - y^2 \,F_L^{lN}(x,Q^2)
\mp Y_-\, xF_3^{lN}(x,Q^2) \right],
\label{eq:hq2-NCxsec}
\end{equation}
where $Y_\pm=1\pm(1-y)^2$, $F_2, xF_3$ and $F_L$ are the structure functions of
the hadron and the notation $lN$ indicates lepton scattering from a nucleon.
 The term $xF_3$ is only non-negligible if $Z$ exchange is important and this only happens for
 $Q^2$ values, $Q^2 \geqsim 5000$GeV$^2$. Similarly the exchange of charged current (CC) vector bosons ($W^+,W^-$)
 is only non-negligible at such high $Q^2$. Fig.~\ref{fig:kinreg} shows the kinematic reach of HERA. One can see
 that low $x$,
 $x \leqsim 0.01$, is only probed at low $Q^2$, $Q^2 \leqsim 500$ GeV$^2$,  and so we do not consider these cases further and we adopt the simplified form of the differential cross section due to virtual photon exchange.
\begin{equation}
{d^2\sigma(lh)\over dxdQ^2}={2\pi\alpha^2\over xQ^4}
\Bigl[Y_+F_2^{lN}(x,Q^2)-y^2F_L^{lN}(x,Q^2)\Bigr]
\label{eqn:qpm-sig2-xq2}
\end{equation}

\begin{figure}[t]
\centerline{\psfig{figure=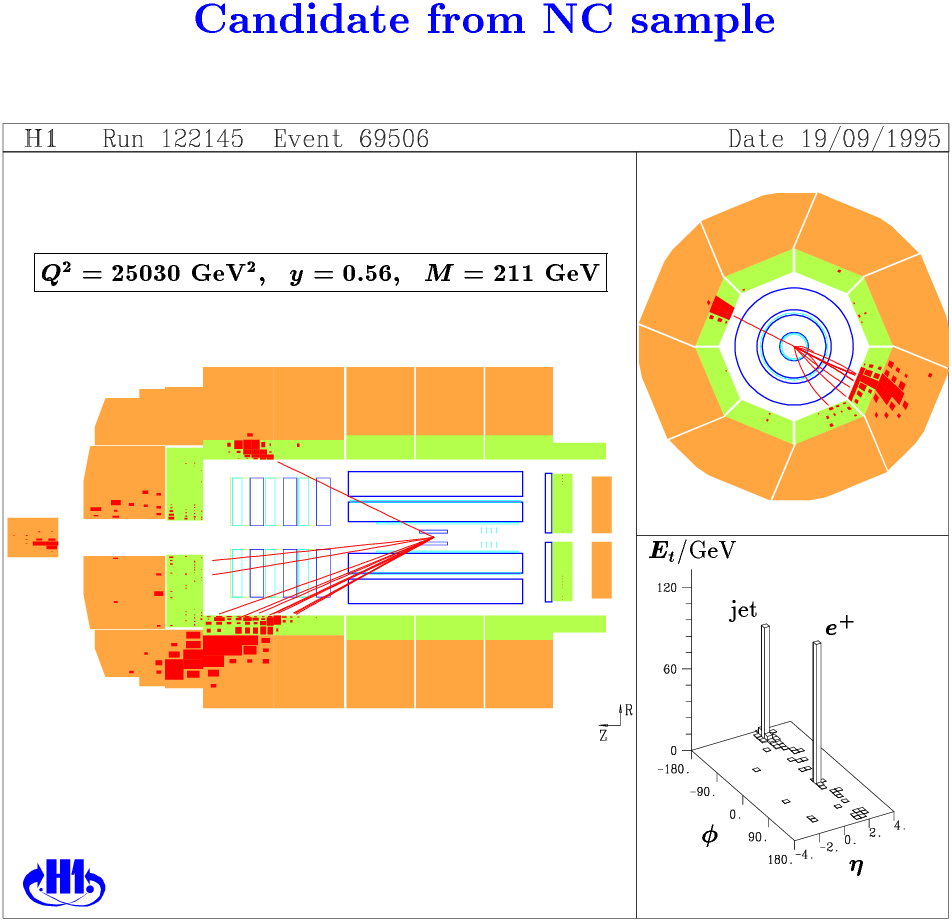,height=.3\textheight}} 
\caption{A DIS NC event in the H1 detector at HERA. From ~\cite{Devenish:2004pb}.}
\label{fig:H1_nc}
\end{figure}

Now let us consider how the variables  $x,Q^2,y$ are determined from quantities
measured in the HERA experiments. 
At HERA the signature of a DIS NC event is one in which the scattered
beam electron is recorded in the EM section of the calorimeter and the
total transverse momentum of the final state is approximately zero. An
example of a high $Q^2$ NC event in the H1 detector is shown in 
Fig.~\ref{fig:H1_nc}. Looking at this diagram we can easily see the scattered electron and the
hadron energy deposits. Both of these can be measured and the kinematic variables can be determined in various different ways.
Only the method using electron variables will
be described here, but other methods involving the hadron variables allow cross
calibration and access to kinematic regions where the resolution on the electron variables
is poor.

We measure the electron kinematic variables as follows.
The proton axis is conventionally taken to define 
the positive $z$ direction from which all angles are measured.
If $E_e$, $E_p$ are the electron and proton beam 
energies, $E^\prime$ and $\theta_e$ the energy and angle of the 
scattered electron to the beam-line, then:
\begin{equation}
s=4E_eE_p~~~~~~~{\rm and}~~~~~
y_e=\bigl(1-{E^\prime\over E_e}\sin^2{\theta_e\over 2}\bigr),
\label{eq:hk1}
\end{equation}
\begin{equation}
Q_e^2 = 4E_eE^\prime\cos^2{\theta_e\over 2} = 
{E^{\prime 2}\sin^2\theta_e\over 1-y_e},
\label{eq:hk2}
\end{equation}
\begin{equation}
x_e = {E_eE^\prime(1-\sin^2{\theta_e\over2})\over E_p(E_e-
E^\prime\sin^2{\theta_e\over 2})}.
\label{eq:hk3}
\end{equation}

Note the main point that the fractional momentum of the struck quark $x$ can actually be
measured simply by measuring the 4-vector of the outgoing electron in an NC scatter.

\clearpage
\subsection{The parton model}

We will now briefly review he interpretation of the structure functions in terms of the parton model and QCD,
details are given in the contribution of D Soper to this encyclopedia.

\begin{figure}[ht]
\centerline{\psfig{figure=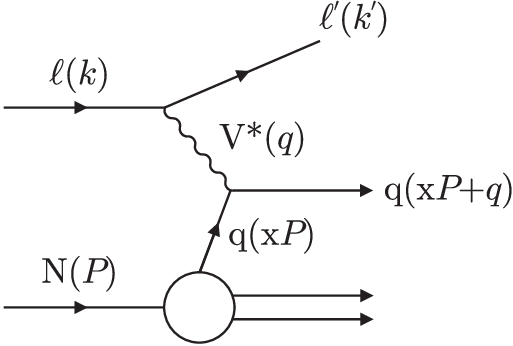,height=0.2\textheight}}
\caption{Schematic diagram of lepton-hadron scattering in the 
quark-parton model. From~\cite{Devenish:2004pb}.}
\label{qpm}
\end{figure}

This parton picture grew out of the attempt by Feynman
to provide a simple physical picture of the scaling that had been predicted 
by Bjorken and observed in the first high energy deep 
inelastic electron scattering experiments at SLAC, where
$F_2$ was observed to be independent of $Q^2$ for $x$ values around
$x \sim 0.25$.
The quark-parton model (QPM) 
states that the nucleon is full of point-like
non-interacting scattering centres known as partons. The lepton-hadron 
reaction cross-section is then approximated by an incoherent sum of
elastic lepton-parton scatters via an exchanged vector boson, 
see Fig.~\ref{qpm}. 
\noindent
Only spin  $1\over 2$ partons (quarks or antiquarks) 
interact directly with the leptonic probe particles. Thus the
electromagnetic scattering of two spin  $1\over 2$ point-like particles is
the underlying fundamental process and this can be calculated in QED.

The result of the calculation is 
\begin{equation}
 \frac{d\sigma}{dy}=\frac{e^2 e^{\prime 2}}{8\pi Q^4}[1 + (1-y)^2] s^{\prime}
\label{eqn:x1} 
\end{equation}
for fermions of charge $e$, the electron and $e^{\prime}$, the quark, with $s^{\prime}$ being the centre of mass energy of the electron-quark scatter.
Since $e^\prime=e_ie$ (where $e_i=2/3, -1/3$) this may be written in the form
\begin{equation}
 \frac{d\sigma}{dy}=\frac{2\pi e_i^2 \alpha^2}{ Q^4}[1 + (1-y)^2] s^{\prime}
\label{eqn:x2}
\end{equation}
Assuming that the quark takes a fraction $x'$ of the proton's 4-momentum then $s^{\prime}= x's$. Considering Fig.~\ref{qpm}, the 4-momentum squared of the outgoing quark is $$(x'P+q)^2= x^{\prime 2}P^2 +2x^{\prime}Pq+q^2 = 0$$
because quarks are assumed massless. But $P^2$ is also $\sim 0$ since the mass of the Nucleon is negligible compared to the scale $Q^2$, and we get $$2x^{\prime}Pq+q^2=2x^{\prime}Pq-Q^2=0$$ so that $$x^{\prime} = Q^2/2Pq =x$$ That is that the measurable kinematic quantity $x$ can be identified with the fractional momentum of the struck quark.

To obtain a cross-section for electron-hadron scattering from this result
a distribution function $q_i(x)$ 
which gives the probability 
that the struck quark $i$ carries a fraction $x$ of the hadron's momentum
$p$ must be defined.  This gives the parton number, or density, distribution. 
It is also useful to define the momentum distribution $xq(x)$
 and this is what is meant by the term
 parton distribution function or PDF.

Thus the double differential cross-section for incoherent scattering
off all possible types of quark with a specific value of $x$ is 
\begin{equation}
 \frac{d^2\sigma}{dxdy}=\frac{2\pi \alpha^2}{ Q^4} [1 + (1-y)^2] s 
\sum_i e_i^2 xq_i(x)
\label{eqn:qpms1}
\end{equation}
or equivalently
\begin{equation}
 \frac{d^2\sigma}{dxdQ^2}=\frac{2\pi \alpha^2}{x Q^4} [1 + (1-y)^2]  
\sum_i e_i^2 xq_i(x)
\label{eqn:qpms}
\end{equation}

Now we compare this QPM prediction to the general form of the lepton hadron cross section in
Eq.~\ref{eqn:qpm-sig2-xq2} and we see that
the parton model predicts
\begin{equation}
        F_2(x,Q^2) = \Sigma_i e_i^2(xq_i(x) + x\bar q_i (x)),
\label{eqn:f2gen1}
\end{equation}
where we have generalised the result to include antiquarks, which are also spin-1/2 point-like particles.
 Thus the parton model predicts Bjorken scaling: $F_2$ depends only on $x$ not on $Q^2$.

The parton model also predicts,
$$F_L^{lh}(x,Q^2) = 0$$ 
as a consequence of scattering from spin ${1\over 2}$ partons, and this was also verified approximately in early data.

These were not the only observations which led to the QPM becoming established. In lepton-hadron scattering we
may use hydrogen or deuterium targets so we may scatter from proton and neutron targets. This will give a
different mixture of $u-$ and $d-$ type quarks. One can also use neutrinos and antineutrinos as as the probe and
these interactions give us different $F_2^{\nu N}$ and $xF_3^{\nu N}$ structure functions with different flavour
dependencies. One can derive that for scattering of isoscalar targets
\[
F_2^{lN}  \approx \frac{5}{18} F_2^{\nu N}
\]
and the early data on charged lepton scattering and the
neutrino scattering structure functions are indeed related by $\sim 5/18$.
The (anti-) neutrino scattering data also allowed us to established that there are antiquarks in the nucleon and to establish
the picture that there are in fact 3 valence-quarks in the nucleon, which give it is flavour properties,
and a 'sea' of quark-antiquark pairs which are flavourless. These data also allowed us
to measure (roughly) the shapes of the valence and sea quark momentum distributions.

However, there were remaining issues.
Firstly, one would expect that the sum over the momentum
quarks and antiquarks in the nucleon a
\begin{equation}
x\Sigma(x) = \Sigma_i (xq_i(x) +\bar xq_i(x))
\end{equation}
integrated over $x$ would be equal to unity
\begin{equation}
\int_0^1 dx x\Sigma(x) = 1 
\end{equation}
This may be checked experimentally, since
\[x\Sigma(x) = F_2^{\nu N} \] and the result
\begin{equation}
\int_0^1 dx F_2^{\nu N} \sim 0.5
\end{equation}
implies that there is more momentum in the nucleon than that carried by the
quarks and antiquarks.
Secondly, the partons are assumed to be essentially free particles in the QPM, but at the
same time are confined inside the hadron.
Thirdly, although Bjorken scaling is approximately satisfied, there is clear 
evidence for scale breaking at large and small $x$.
These problems led  
to the development of the theory of QCD.

\subsection{Development of the QCD improved parton model}

\begin{figure}[htbp]
\vspace*{13pt}
\centerline{\psfig{figure=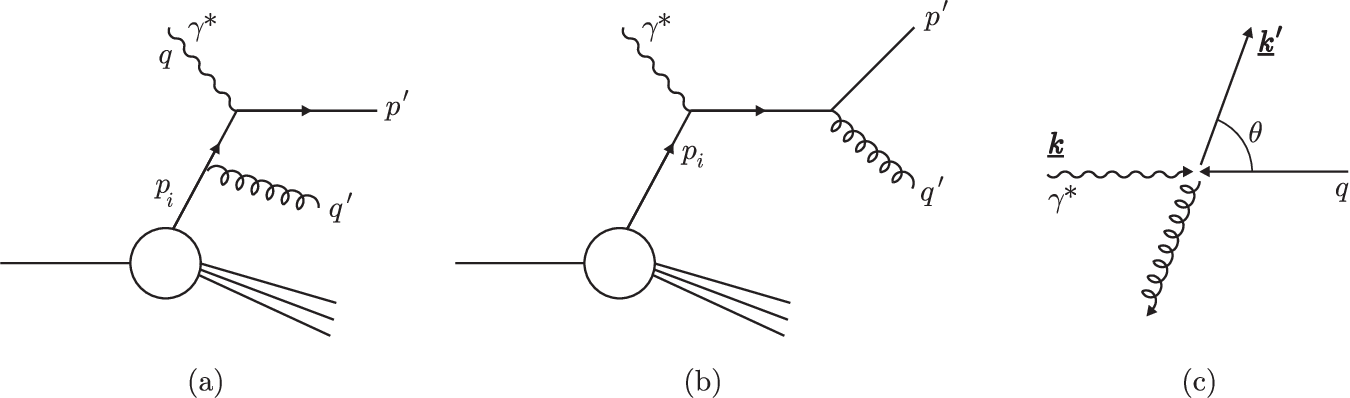,width=1.\textwidth}} 
\caption{Diagrams for gluon radiation from quarks in DIS via QCDC,
together with the parton CM frame. From~\cite{Devenish:2004pb}.}
\label{fig:g_rad}
\end{figure}

\begin{figure}[htbp]
\vspace*{13pt}
\centerline{\psfig{figure=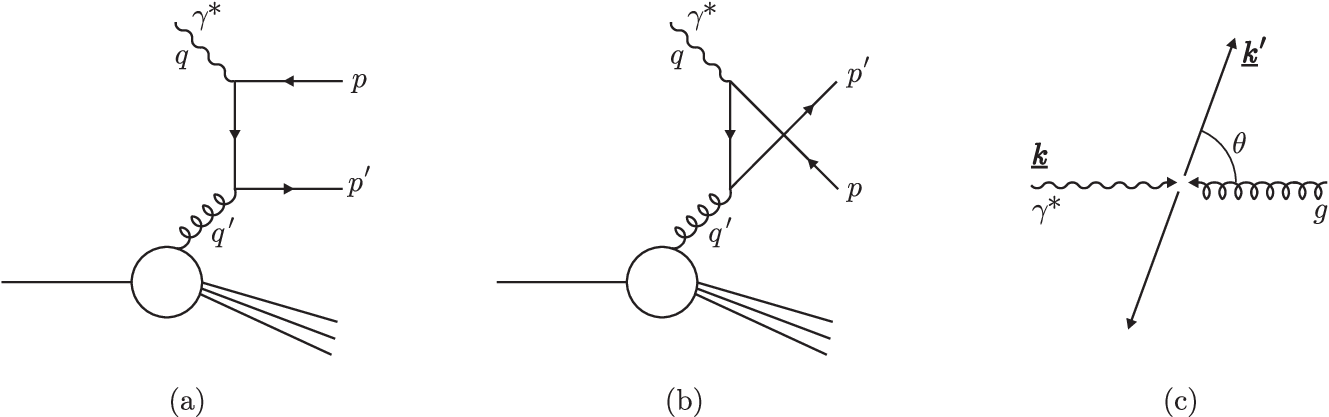,width=1.\textwidth}} 
\caption{The parton-level diagrams for BGF production of $\qr\qb$, 
together with the parton CM frame. From~\cite{Devenish:2004pb}.}
\label{fig:bgf2}
\end{figure}

The simple QPM diagram of Fig.\ref{qpm} is augmented by
the QCD Compton process, as shown in Fig.~\ref{fig:g_rad},
and the BGF process, as shown in Fig.~\ref{fig:bgf2}.
In these diagrams a parton with momentum $p_i= y p$ interacts 
with the $\gamma^*$ of virtuality $Q^2$ and 4-momentum $q$. 
Define the variable 
$z=Q^2/(2p_i.q)=x/y$, the analogue of Bjorken $x$ for the 
$\gamma^*$-parton scattering.
The cross-section for the hard interaction
is calculable in pQCD and the overall $\gamma^*$p cross-section is then
given by a convolution of this cross section with the probability
of finding the parton with momentum fraction $y$ in the proton.
The results for the QCDC and BGF processes each add an extra term to the point-like scatter of Fig~\ref{qpm}. This can be framed as a
redefiniton of the parton density such that it includes the parton splitings and thus gains a scale dependence $q(x) \to q(x,Q^2)$. QCD cannot predict
the shape of the parton distribution at the starting scale $Q^2_0$ (which should be chosen large enough that
perturbative calculations are valid, i.e that the strong coupling $\alpha_s(Q^2_0)$ is small) but it can predict
how it will evolve with $Q^2$.
\begin{equation}
{\partial q_i(x,Q^2)\over \partial\ln Q^2}={\as\over 2\pi}\int_x^1
{dy\over y}\left[q_i(y,Q^2)P_{qq}(x/y)+
g(y,Q^2)P_{qg}(x/y)\right].
\label{eqn:dqdt2}
\end{equation}
In this equation there are two 'splitting functions' $P_{qq}$ and $P_{gq}$ which can be interpreted as the
probability for $\qr\to\qr(z)\gr(1-z)$ splitting (for the QCDC process) and $\gr\to \qr(z)\qb(1-z)$ splitting
(for the BGF process), respectively.

\begin{figure}[tbp]
\vspace*{13pt}
\centerline{\psfig{figure=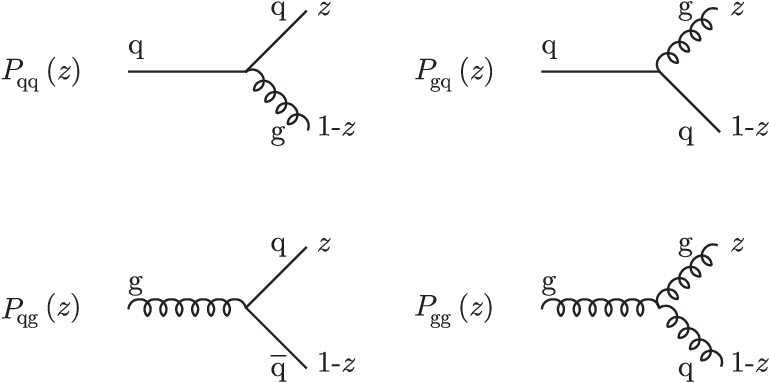,height=.25\textheight}} 
\caption{The DGLAP splitting functions. From~\cite{Devenish:2004pb}.}
\label{fig:spltfcns}
\end{figure}

Just as the quark density evolves, so will the
gluon density 
\begin{equation}
{\partial g(x,Q^2)\over \partial\ln Q^2}={\as\over 2\pi}\int_x^1
{dy\over y}\left[\sum_iq_i(y,Q^2)P_{gq}(x/y)+
g(y,Q^2)P_{gg}(x/y)\right].
\label{eqn:dgdt2}
\end{equation}
where we have introduced two more splitting functions which describe 
the probabilities for $\qr\to \gr(z)\qr(1-z)$ and $\gr\to \gr(z)\gr(1-z)$
splitting. 
The splitting functions are summarised graphically in Fig.~\ref{fig:spltfcns}.

The two equations Eqs~\ref{eqn:dqdt2} and~\ref{eqn:dgdt2} above are known as the DGLAP
evolution equations, ~\cite{Gribov:1972ri,Gribov:1972rt,Lipatov:1974qm,Dokshitzer:1977sg,Altarelli:1977zs}.
At leading order ($\alpha_s^0$) the structure function $F_2$ can be related to the evolved parton
distributions as
\begin{equation}
        F_2(x,Q^2) = \Sigma_i e_i^2(xq_i(x,Q^2) + x\bar q_i (x,Q^2)),
\label{eqn:f2gen2}
\end{equation}
and the structure function $F_L$ is still zero. But there is one more sophistication of this result needed even at leading order.

\begin{figure}[htbp]
\vspace*{13pt}
\centerline{\psfig{figure=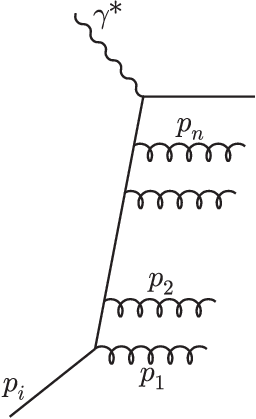,height=.25\textheight}} 
\caption{Multiple gluon emissions from a quark line. From~\cite{Devenish:2004pb}.}
\label{fig:qmanyg}
\end{figure}

The leading order QCD corrections
to the parton densities are all weighted by $\as\ln(Q^2/Q_0^2)$
which could divergeat high $^2$. However,
the QCD coupling $\as$ runs $\asq \propto 1/\ln(Q^2)$. Thus divergences at high $Q^2$ are avoided.
 The final step in setting up the DGLAP equations with a correct
asymptotic behaviour is to replace $\as$ with $\asq$.
The `leading logs' that are now being summed in the DGLAP equations
correspond to diagrams with multiple gluon emission
as shown in Fig.~\ref{fig:qmanyg}, with an important restriction.
To pick out the leading $\as\ln(Q^2/Q_0^2)$ terms requires
that the transverse momenta of the successive steps shown in the diagram
must be ordered in the transverse momentum of the gluons
\begin{equation} 
\mu^2<p^2_{1t}<p^2_{2t}<\dots <p^2_{nt}<Q^2,
\label{eqn:pt-ord}
\end{equation}
This approximation is often known as the leading log 
approximation (LLA).

Beyond leading order the splitting functions are expanded as power series
in $\as(Q^2)$ e.g. $$
P_{qq}(z,\as)=P^{(0)}_{qq}(z)+
{\as\over 2\pi}P^{(1)}_{q_iq_j}(z)+\dots $$
The relationship between the parton distributions and the measurable structure functions becomes more complex
e.g.
\begin{eqnarray}
{F_2(x,Q^2)\over x}&=&\int{d\xi\over \xi}\biggl[\sum_ie^2_iq_i(\xi,Q^2)
  C_q\left({x\over \xi},\as\right)
  +\nonumber \\
&&\bar{e}^2g(\xi,Q^2)C_g\left({x\over \xi},\as\right)\biggr],
\label{eqn:dglap_f2}
\end{eqnarray}
where $\bar{e}^2=\sum_ie^2_i$ and the sums run over all active quark 
and antiquark flavours. $C_q$ an $C_g$ are the coefficient functions, 
which are also be expanded as power series in $\as$ beyond leading order.
The $F_L$ structure function is non-zero, e.g. at NLO
\bea
{F_L(x,Q^2)\over x}&=&{\as\over 2\pi}\int_x^1{d\xi\over \xi}\biggl[
\sum_ie^2_i{8\over 3}\left({x\over \xi}\right)q_i(\xi,Q^2)+
\nonumber \\
&&\bar{e}^24\left({x\over \xi}\right)\left(1-{x\over \xi}\right)
g(\xi,Q^2)\biggr]
\label{eqn:fl_qg}
\eea
At small $x$ the gluon term becomes dominant and controls the behaviour of $F_L$ as well as the evolution of $F_2$.

Complete QCD calculations for the DIS process are now available to Next-to-Next-to-Next to leading order (NNNLO or N3LO). Although the formalism has become more complex, it remains true that given the shapes of the
parton distributions at a given input scale, QCD evolution can predict them at any higher scale and can
thus predict the measurable structure functions for DIS experiments.
But how do we determine the initial distributions?
This has been done by parameterising the
$x$ dependence of the parton density at $Q^2_0$ and determining the
parameters by fitting to the DIS data.
A function of the form
\[
xf(x,Q^2_0)=Ax^\alpha(1-x)^\beta P(x)
\]
is used for each flavour of PDF, where $P(x$) is a polynomials in $x$ or $\sqrt{x}$.
These could be Bernstein or Chebyshev polynomials, or one can
use a neural net. 
Such PDF fits are done by three major global fitting groups: CT,~\cite{Hou:2019efy}, MSHT,~\cite{Bailey:2020ooq}, and NNPDF,~\cite{NNPDF:2021njg}.
This is the topic of the contribution of F Olness to this encyclopedia.

There were some expectations for the values of the exponents $\alpha$ and $\beta$. In
particular there were predictions from Regge theory, see Sec.~\ref{sec:lowq2interp}, for the exponent $\alpha $ which controls the behaviour
at low $x$. It turned out that the data defied our simple expectations. This will be discussed in Sec.~\ref{sec:lowxinterp}.

\clearpage
\section{HERA data}
Here we collect together some relevant results from HERA.
We will begin by showing the measurements of the reduced cross section from the final combination of the H1 and ZEUS
data of both run 1 and run 2 of HERA,~\cite{H1:2015ubc}.
The reduced cross section $$\sigma_r =[F_2(x,Q^2)-(y^2/Y_+)F_L(x,Q^2)]$$
is obviously related
to the differential cross section of Eq.~\ref{eqn:qpm-sig2-xq2}
which is the directly measurabe quantity. For most of the kinematic plane the
kinematic variable $y$ is relatively small and the reduced cross section gives $F_2$.
Recalling, that to leading order in QCD,
$$F_2(x,Q^2) = \Sigma_i e_i^2(xq_i(x,Q^2) + x\bar q_i (x,Q^2)) $$
Thus $F_2$ shows us the behaviour of the quark and antiquark distributions as a function of $Q^2$

\begin{figure}[tbp]
\vspace*{13pt}
\centerline{\psfig{figure=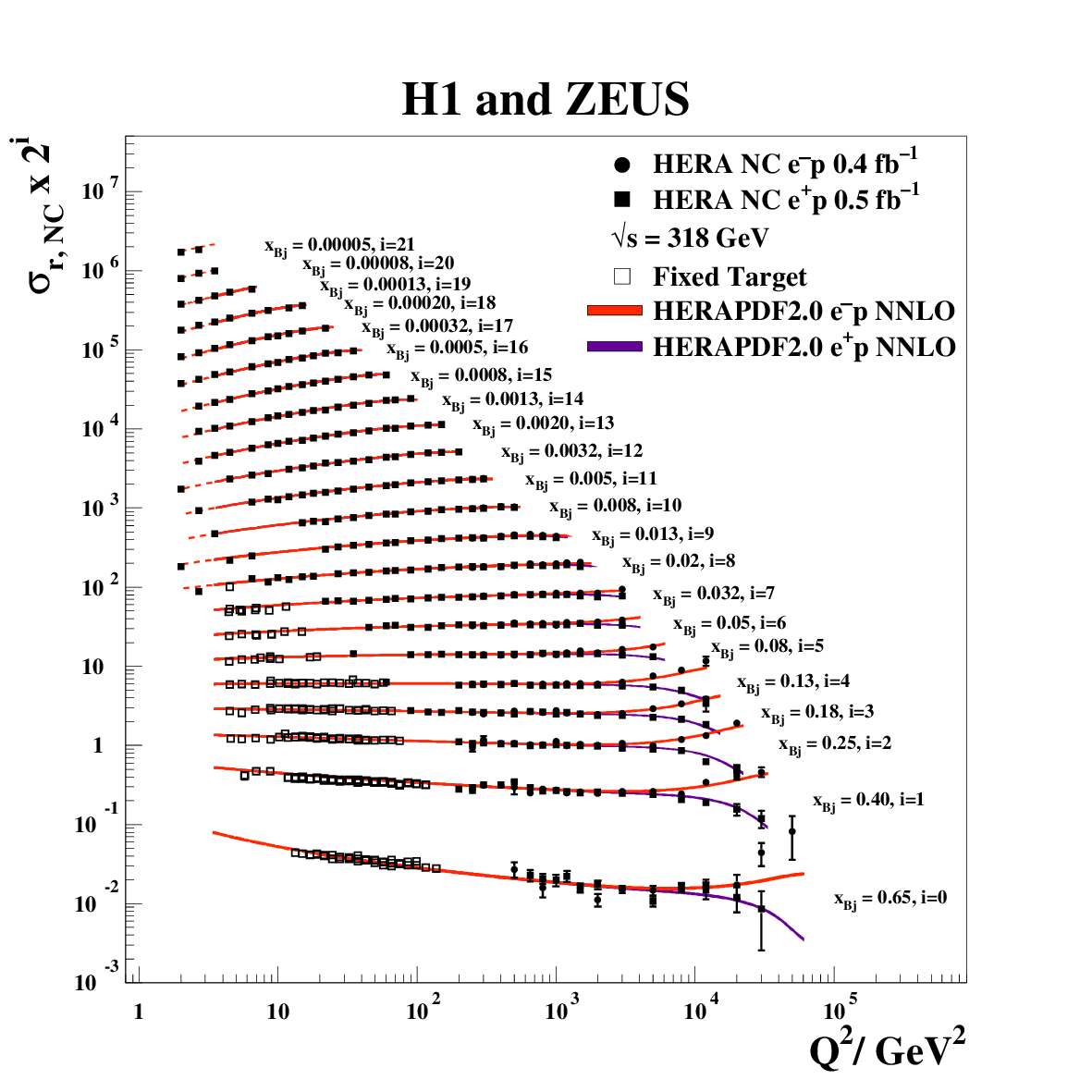,width=0.9\textwidth}} 
\caption{The combined HERA data for the inclusive NC $e^+p$ 
          and $e^-p$ reduced cross sections 
          together with fixed-target data, ~\cite{BCDMS:1989qop,NewMuon:1996fwh}, 
          and the predictions of HERAPDF2.0 NNLO. 
          The bands represent the total uncertainties on 
          the predictions. Dashed lines indicate extrapolation into 
          kinematic regions not included in the fit. From ~\cite{H1:2015ubc}.
}
\label{fig:nnloQ23pt5scal}
\end{figure}

In Fig.~\ref{fig:nnloQ23pt5scal} we show the reduced cross section as a function of $Q^2$ for various $x$ values.
An important feature is the consistency of the data from fixed target experiments and from the HERA collider.

\begin{figure}[tbp]
\vspace{-0.3cm} 
\centerline{
\epsfig{file=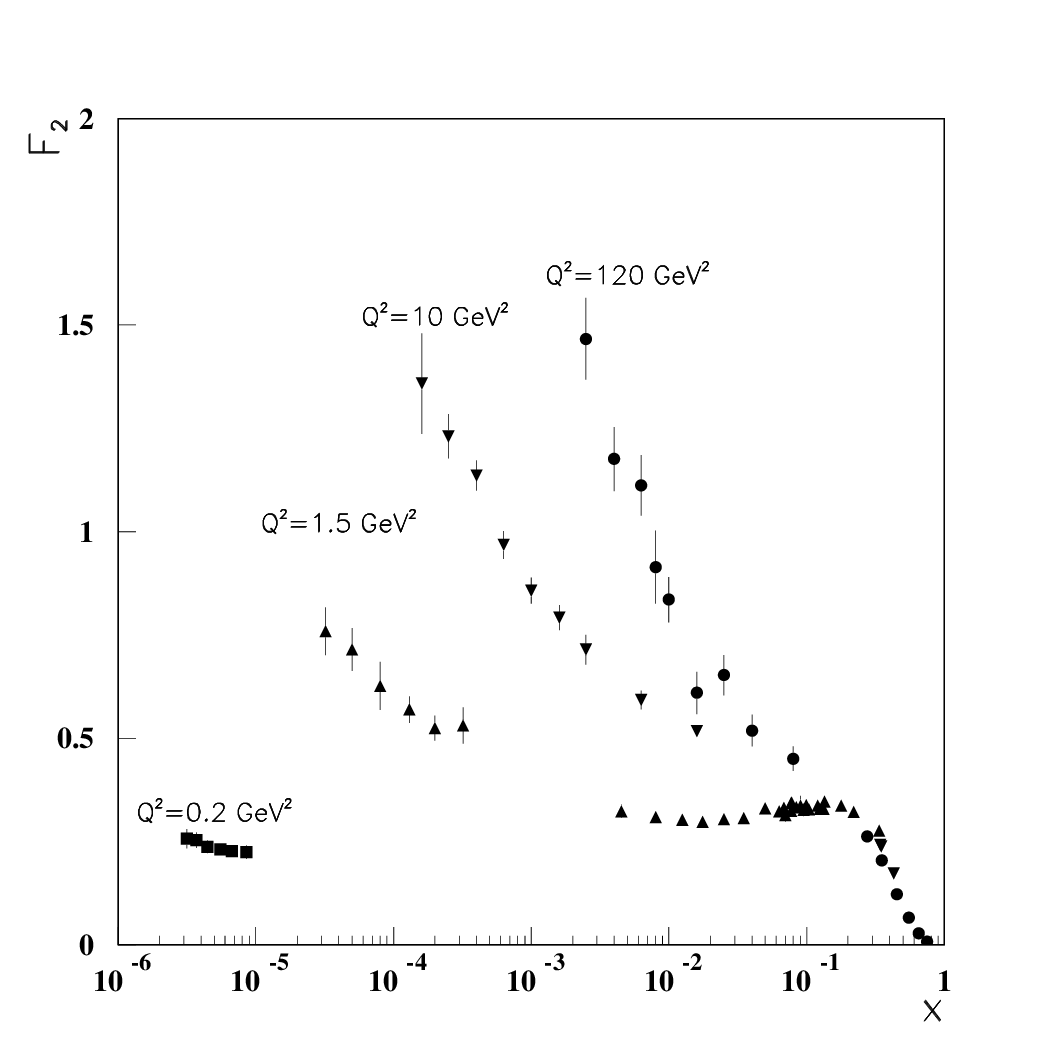 ,width=0.49\textwidth}
\epsfig{file=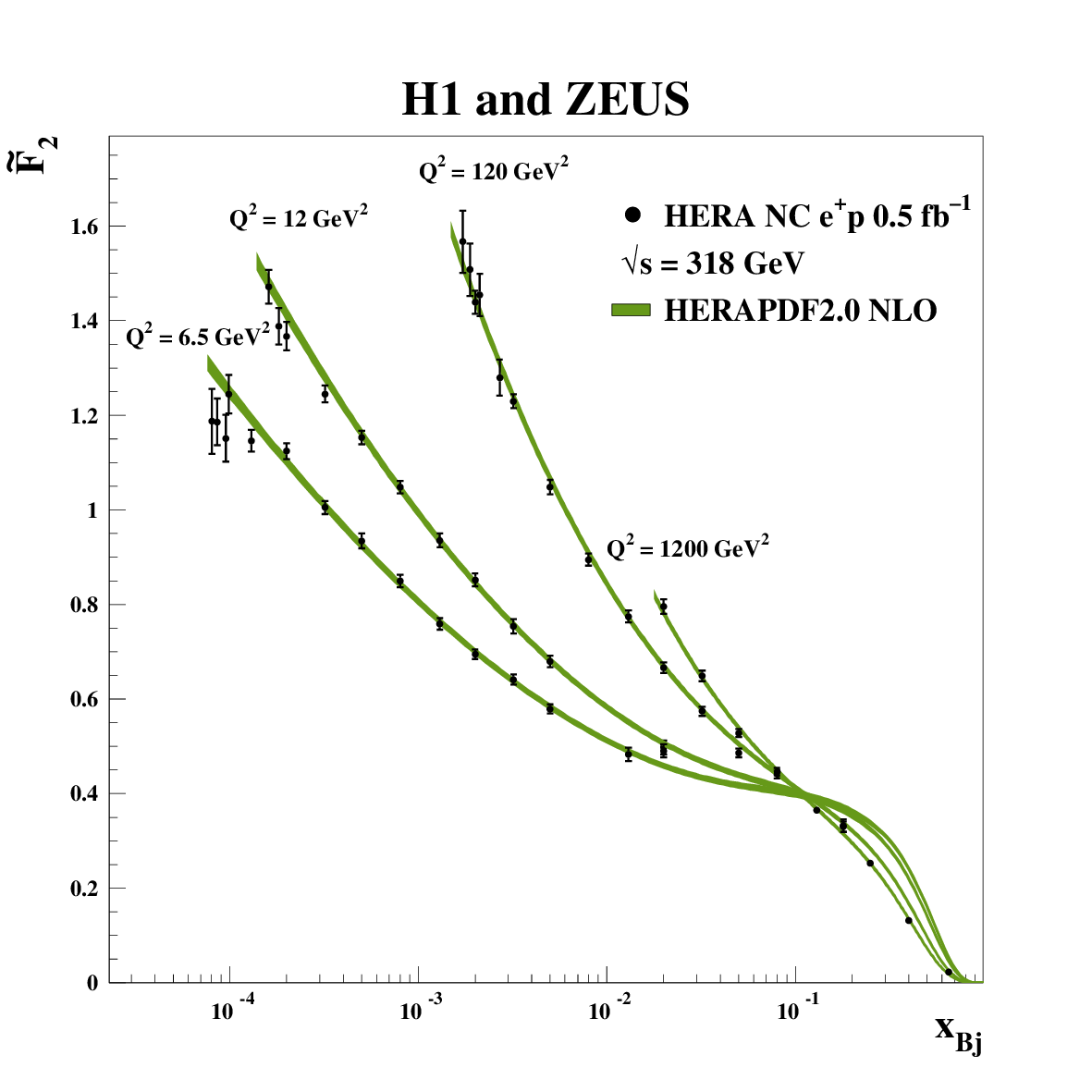 ,width=0.49\textwidth}}
\vspace{0.5cm}
\caption {$F_2$ data showing the strong rise at small $x$. Left: the HERA data
are from the 1994 runs and the data for $Q^2=1.5\,$GeV$^2$ and $x>10^{-3}$ 
(triangles) are from fixed target experiments.
Right:  The structure function $\tilde{F_2}$,
as extracted from the measured reduced cross sections of the final combined HERA data, 
for four values of $Q^2$
together with the predictions of HERAPDF2.0 NLO. 
The bands represent the total uncertainty on the predictions. From ~\cite{H1:2015ubc}.
}
\label{fig:f2}
\end{figure}

To illustrate one of the most striking features of HERA data we plot the data versus $x$ at fixed $Q^2$. This is
done in Fig.~\ref{fig:f2} for both the final HERA data and
for early data, and the quantity plotted is $F_2$ rather than $\sigma_r$.
The figures show that $F_2$ rises steeply as $x$ decreases and that the strength of
the rise increases as $Q^2$ increases. The rise in $x$ is strong, at
$Q^2=120\,$GeV$^2$ $F_2$ increases by almost a factor of 3 as $x$
decreases from $0.08$ to $0.003$. It is worth remarking 
that while a range of possible behaviours had been predicted for 
$F_2$ at small $x$, including a strong rise, when the first preliminary HERA
data showing a rising $F_2$ was presented at the Durham Workshop in 
1993 it caused great excitement. The early fixed target data for 
$Q^2=1.5\,$GeV$^2$ and $x>10^{-3}$ (triangles), shown in
Fig.~\ref{fig:f2} left hand side, explains why --  these data showed a
fairly flat behaviour as $x$ decreased with only the merest hint of a 
possible rise.
At very low $Q^2$, $Q^2\sim 0.2$GeV$^2$ $F_2$ no longer rises steeply. The behaviour of 
$F_2$ for such very 
small $Q^2$ is interesting and will be covered in some detail at 
HERA. This is covered in more detail in section~\ref{sec:lowq2interp}.

In order to extract $F_2$ we must correct $\sigma_r$ for the contribution of $F_L$.
For modern data this is done by using the PDFs extracted from a full QCD fit 
to the reduced cross section data and using the
relationship~\ref{eqn:fl_qg} to obtain $F_L$ from the fitted parton distributions.

\begin{figure}[tbp]
\vspace{-0.3cm} 
\centerline{
\epsfig{file=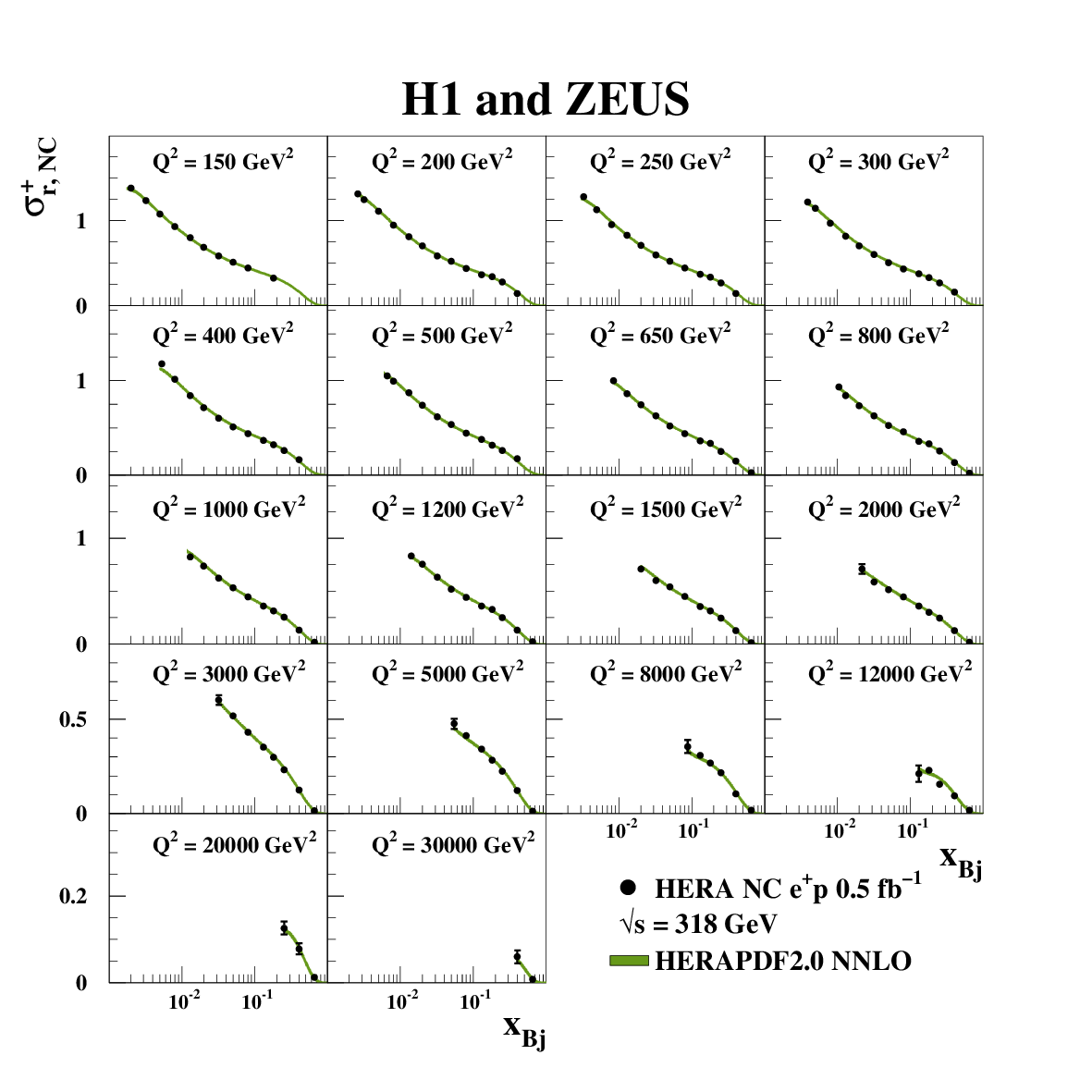   ,width=0.9\textwidth}}
\vspace{0.5cm}
\caption {The combined high-$Q^2$ HERA inclusive NC $e^+p$ 
          reduced cross sections  
          at $\sqrt{s} = 318$\,GeV with overlaid predictions from 
          HERAPDF2.0 NNLO.
          The bands represent the total uncertainties on the predictions. From ~\cite{H1:2015ubc}.
}
\label{fig:nnloQ23pt5ncepc}
\end{figure}
\begin{figure}[tbp]
\vspace{-0.3cm} 
\centerline{
\epsfig{file=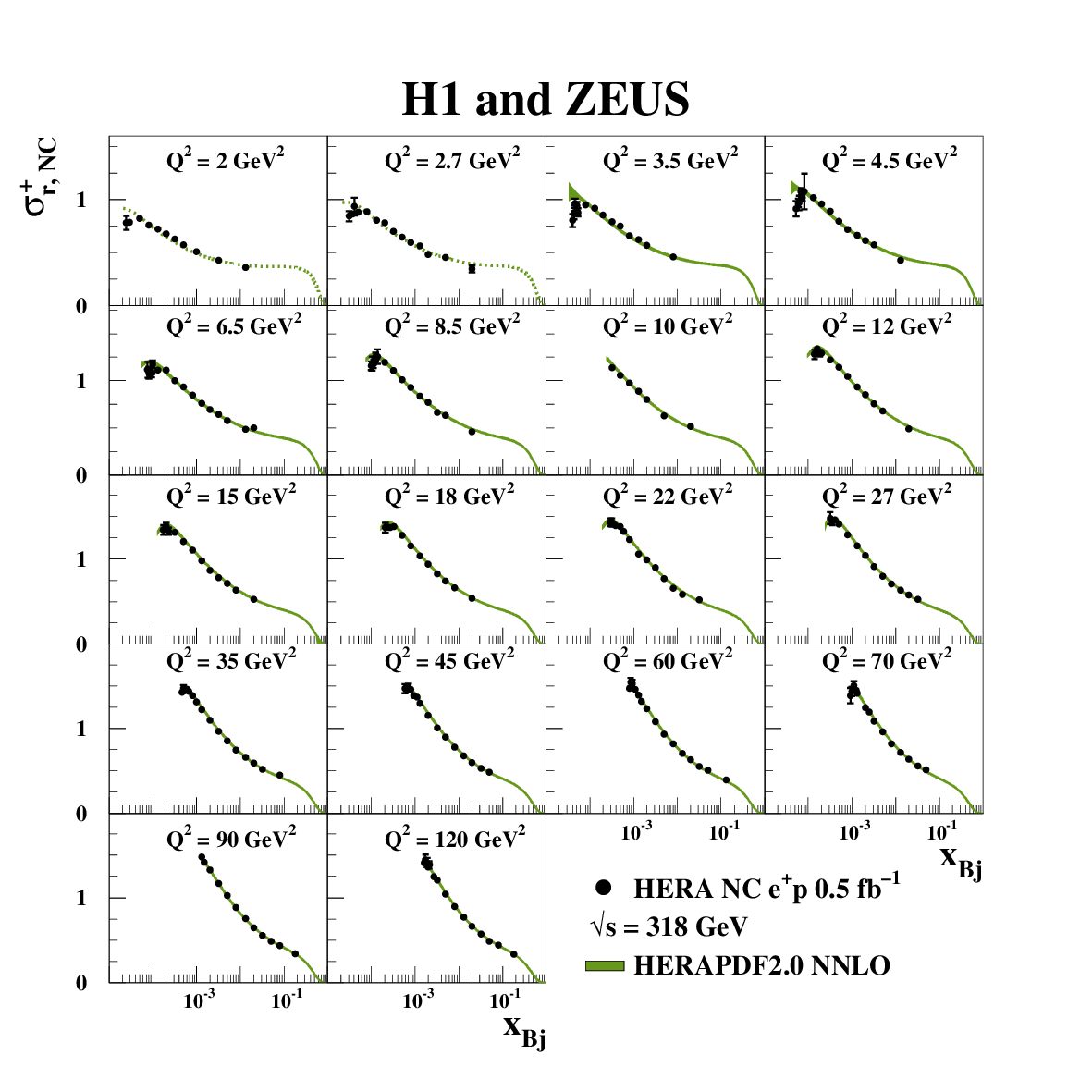   ,width=0.9\textwidth}}
\vspace{0.5cm}
\caption {The combined low-$Q^2$ HERA inclusive NC $e^+p$ 
reduced cross sections  
at $\sqrt{s} = 318$\,GeV with overlaid predictions from HERAPDF2.0 NNLO. 
     The bands represent the total uncertainties on the predictions.   
Dotted lines indicate extrapolation into 
kinematic regions not included in the
fit. From ~\cite{H1:2015ubc}.
}
\label{fig:nnloQ23pt5ncepb}
\end{figure}

\begin{figure}[tbp]
\centerline{
\epsfig{file=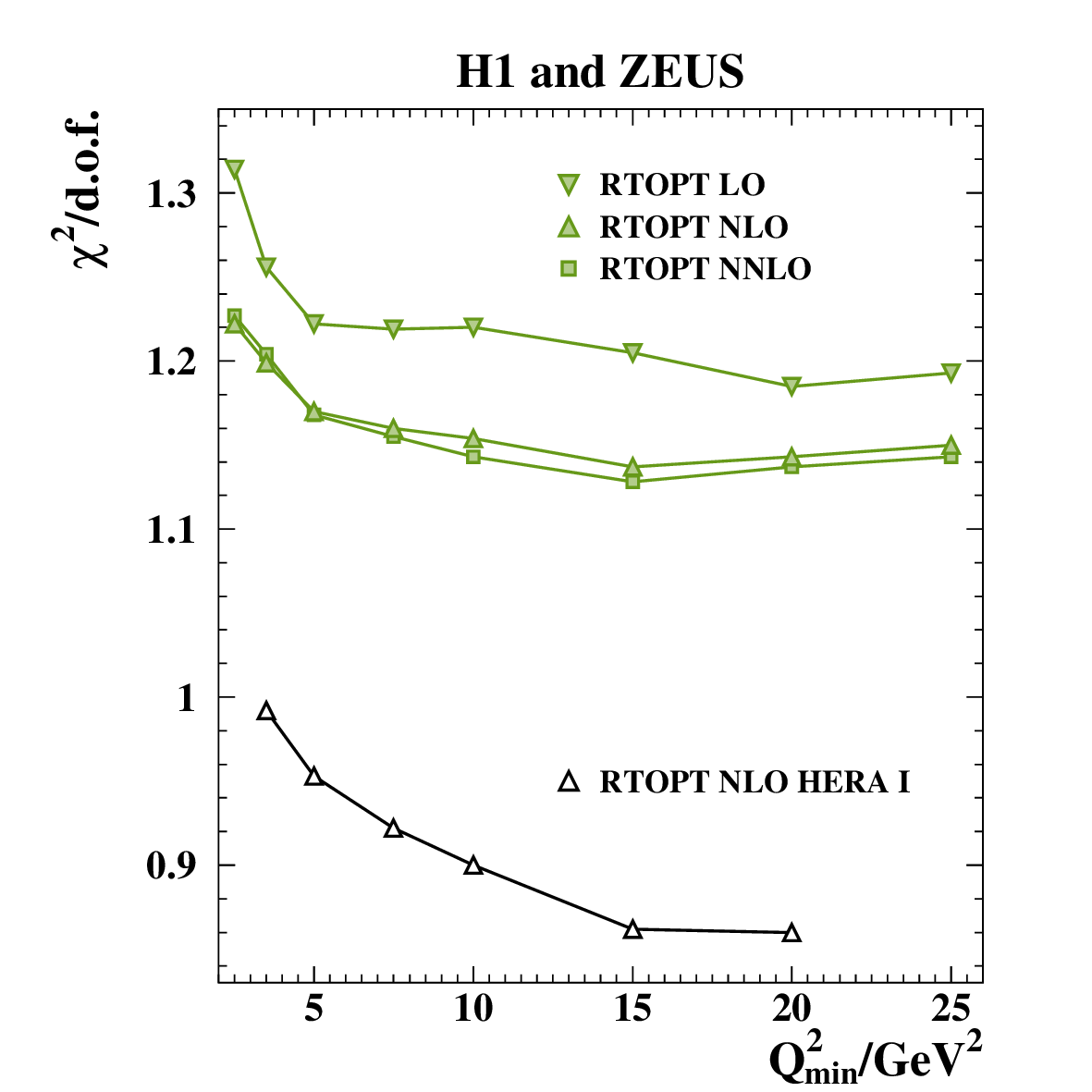 ,width=0.7\textwidth}}
\caption { 
The dependence of $\chi^2/{\rm d.o.f.}$ on $Q^2_{\rm min}$ 
of the LO, NLO and NNLO fits to the HERA combined inclusive data. From ~\cite{H1:2015ubc}.
Also shown are values for an NLO fit to the 
combined HERA\,I data~\cite{H1:2009pze}.
}
\label{fig:chiscan}
\end{figure}

We now show the reduced cross sections as a function of $x$ in different $Q^2$ bins for a very wide
range of $Q^2$ values. Figs~\ref{fig:nnloQ23pt5ncepc} and ~\ref{fig:nnloQ23pt5ncepb} show the reduced cross section as a function of $x$ in
high $Q^2$ and low $Q^2$ bins respectively, with the NNLO QCD fit overlaid. Whereas the high $Q^2$ data are very well fitted, in the low $Q^2$ bins of Fig.~\ref{fig:nnloQ23pt5ncepc}
we can see that the NNLO QCD fit does not desribe the data so well at low $x$ values, where the impact of $F_L$ on
the reduced cross section can be seen clearly as a turn over due to the $-y^2/Y_+. F_L$ term  
(recall that high $y$ occurs at low $x$ for a given $Q^2$ because of the relationship $y=Q^2/sx$).
In fact it seems that the data would like a larger $F_L$ than that predicted by the theory.
The $\chi^2$ scan of the NNLO QCD fit vs the minimum $Q^2$ value of data input to the fit, $Q^2_{\rm min}$, is shown in Fig.~\ref{fig:chiscan}. One can see that, beyond leading order, the data are considerably less well fitted for $Q^2 < 10$GeV$^2$.
This will be discussed further in Section~\ref{sec:lowxinterp} where modifications to the theory at low $x$ and $Q^2$ are discussed.
\clearpage

\subsection{$F_L$}
\label{sec:fl}

$F_L$ is an important
quantity because of its rather direct relation to $xg$ -- the momentum
distribution of the gluon in the proton. One may express the cross-section as
\begin{eqnarray}
{d^2\sigma\over dxdQ^2}&=&{2\pi\alpha^2\over Q^4 x}Y_+
\left(1-{y^2\over Y_+}{F_L\over F_2}\right)F_2(x,Q^2) 
\label{eqn:fl}
\end{eqnarray}
To measure $F_L$ directly at a fixed value of $(x,Q^2)$ requires 
data at two different values of $y$ and hence at two different centre of 
mass energies (because of the constraint $Q^2=sxy$).
Such a measurement was made by both ZEUS and H1 by varying the proton beam energy down to $575$ and $460$ GeV down
from $920$ such that $\sqrt{s}=251$ and $225$ GeV not $318$ GeV.

\begin{figure}[htbp]
\begin{center}
\includegraphics[width=\columnwidth]{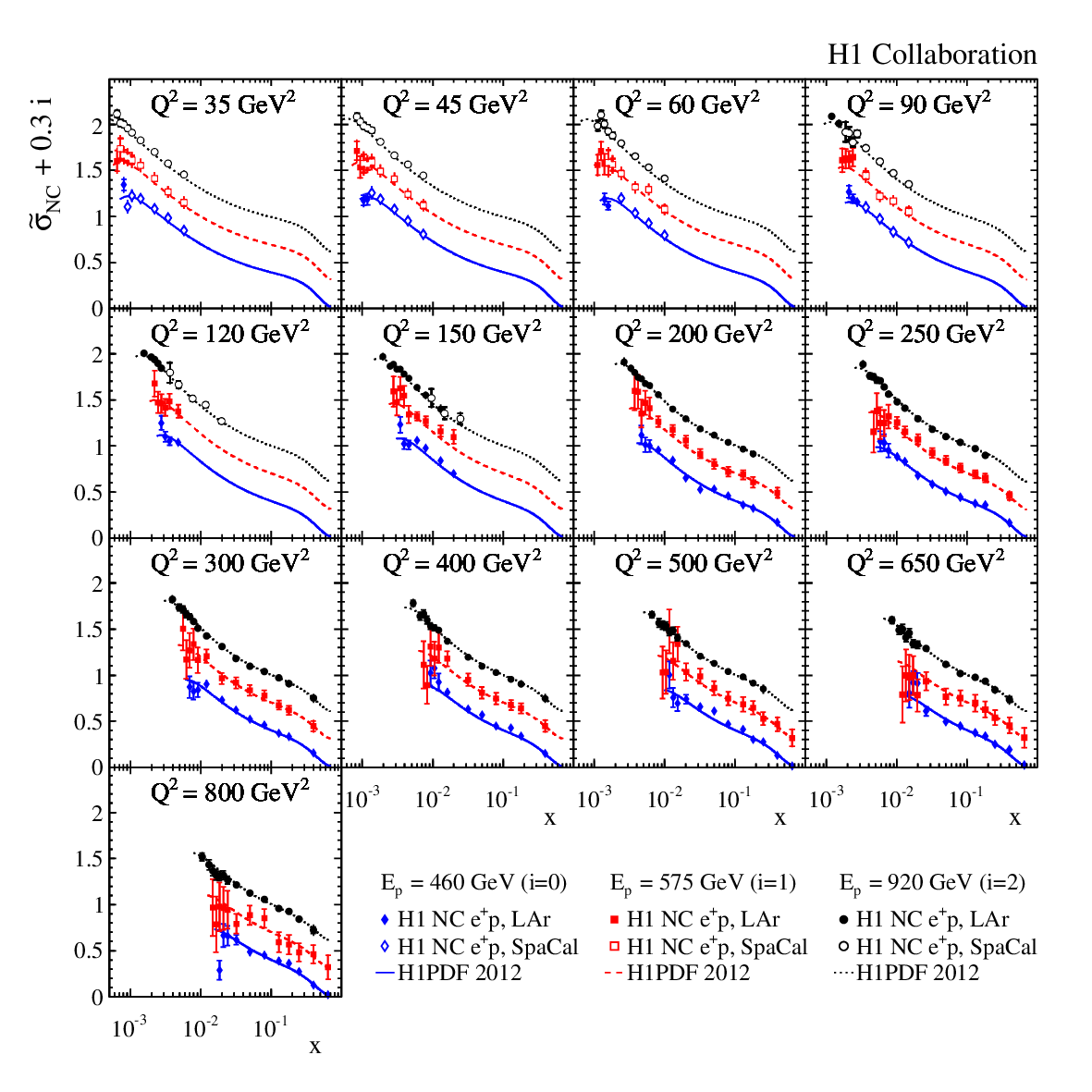}
\end{center}
\caption{ The reduced cross section $\tilde{\sigma}_{NC}(x,Q^2)+0.3i$
  measured at three proton beam energies $E_p=460$~GeV (diamonds, i=0),
  $575$~GeV (squares, i=1) and $920$~GeV (circles, i=2). 
  The inner
  error bars represent the statistical errors, the full error bars include
  the statistical and systematic uncertainties added in quadrature,
  excluding the normalisation uncertainty. The curves represent the
  prediction from an NLO QCD fit. From ~\cite{H1:2013ktq}.}
\label{fig:xsec-xq2}
\end{figure}
The reduced cross section at three different beam energies for H1 data,~\cite{H1:2013ktq}, is shown in Fig.~\ref{fig:xsec-xq2}.

\begin{figure}[htbp]
\begin{center}
\includegraphics[width=\columnwidth]{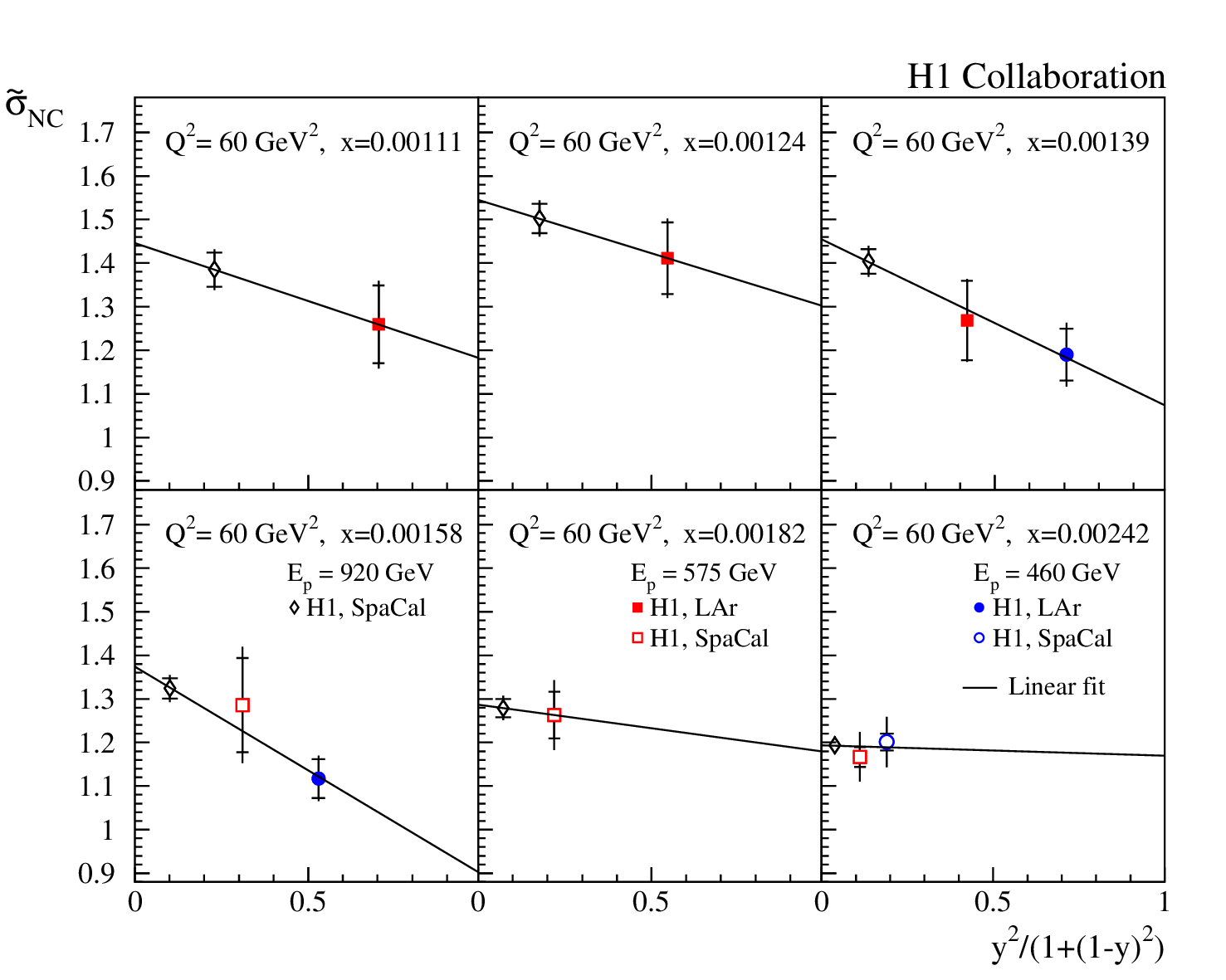}
\end{center}
\caption{\ The reduced cross section $\tilde{\sigma}_{NC}(x,Q^2)$ as
  a function of $y^2/(1 + (1-y)^2)$ for six values of $x$ at $Q^2 =
  60$~GeV$^2$, measured for proton beam energies of $E_p = 920,~575$
  and $460$~GeV. The inner error bars denote the statistical error,
  the outer error bars show statistical and systematic uncertainties
  added in quadrature. The luminosity uncertainty is not included in
  the error bars. From ~\cite{H1:2013ktq}. }
\label{fig:rosenbluth}
\end{figure}

The method to extract $F_L$
is illustrated in Fig.~\ref{fig:rosenbluth}, where the reduced cross section is plotted against $y^2/Y_+$ for different beam energies, in $x,Q^2$ bins. $F_L$ is extracted from the slope of the plot and $F_2$ from its intercept.

\begin{figure}[htbp]
\begin{center}
  \includegraphics[width=0.49\columnwidth]{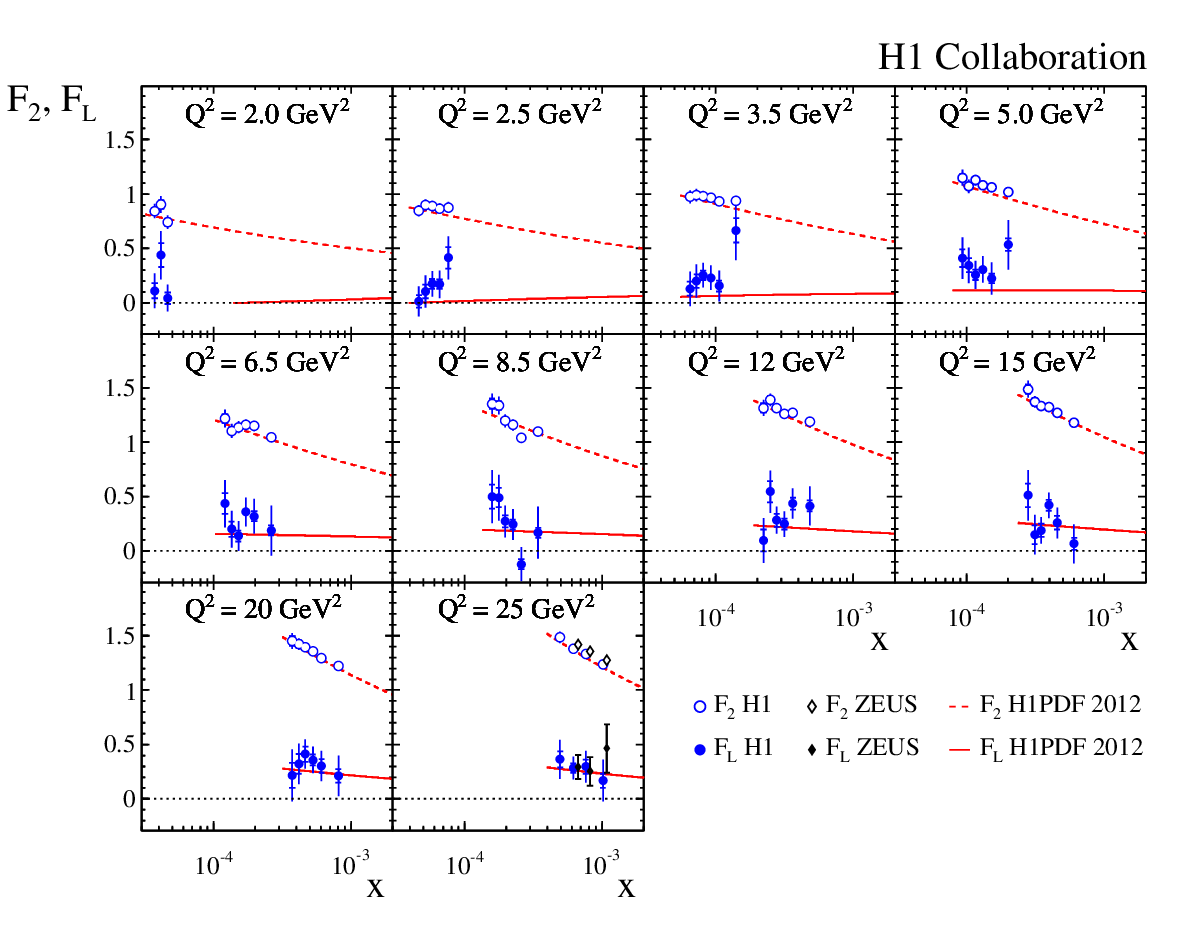}
  \includegraphics[width=0.49\columnwidth]{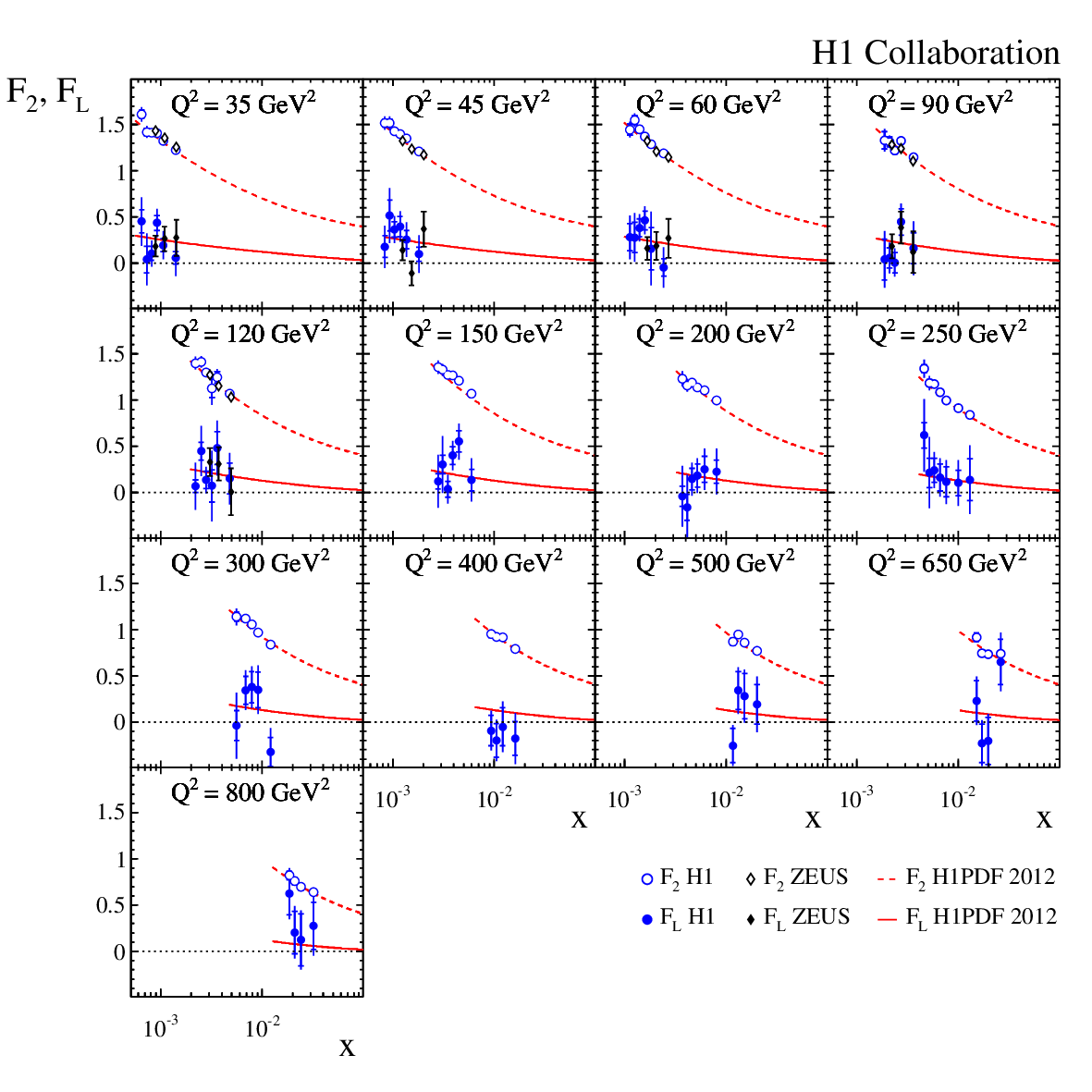}
\end{center}
\caption{The proton structure functions $F_L(x,Q^2)$ (solid
  symbols) and $F_2(x,Q^2)$ (open symbols) measured by H1 (circles)
  and ZEUS (diamonds) in the region: left $2\leq Q^2 \leq 25$~GeV$^2$;
  right $35\leq Q^2 \leq 800$~GeV$^2$
  The inner error bars represent the
  statistical uncertainties, the full error bars include the
  statistical and systematic uncertainties added in quadrature,
  including all correlated and uncorrelated uncertainties. The curves
  represent the prediction from an NLO QCD fit. From ~\cite{H1:2013ktq}.}
\label{fig:FLxq2}
\end{figure}

Figure~\ref{fig:FLxq2} shows the $F_L$ and corresponding $F_2$ values at the same $x,Q^2$.
Superficially the NLO DGLAP QCD fit shown describes the data on $F_l$ reasonably well, but the but errors are large.
When these data are averaged over
$x$ and displayed as a function $Q^2$ one can see that the DGLAP predictions for $F_L$ lie low.
This is shown and discussed in the context of theoretical models in Section~\ref{sec:lowxinterp}.

\clearpage
\subsection{$F_2^{c\bar{c}}$}
\label{sec:f2cc}

The structure functions studied so far are inclusive objects
However, it is possible to study the contribution of heavy flavours to the reduced cross section.
This is particularly
interesting since heavy flavours are dominantly produced through the 
boson-gluon fusion process (BGF), which involves the 
gluon density $xg$ directly. 

The advent of HERA, with a reach to much smaller $x$ values, together 
with detectors capable of measuring details of the DIS
final state gives the prospect of accurate data on $F_2^{c\bar{c}}$
over a wide range of $x$ and $Q^2$. We will concentrate on charm production here because
the data on beuty production from HERA are not very dicsrimnating. Most of
the data on charm production in DIS from HERA has come observation of $D^*$ and $D$
decays. 

To get from the measured $ep\to eD^*X$ cross-section to the inclusive 
charm production cross-section one needs to extrapolate to the
full $D^*$ phase space, to know the probability that the charm quark
will fragment into a $D^*$ meson ($P(c\to D^*)$) and to correct 
for the small fraction of indirect $D$ mesons produced through 
$B$ meson decay or fragmentation. All of these steps involve extra uncertainty
in comparison to the inclusive measurements.
 Finally $F_2^{c\bar{c}}$ is related to the 
 $ep\to ec\bar{c}X$ cross-section by
\begin{equation}
{d^2\sigma(c\bar{c})\over dxdQ^2}={2\pi\alpha^2\over Q^4x}
\left(Y_+F_2^{c\bar{c}}-y^2F_L^{c\bar{c}}\right),
\label{eq:f2cc}
\end{equation}
ZEUS and H1 have combined their final data on charm  and on beauty reduced cross sections, see~\cite{H1:2018flt}.
\begin{figure}[h]
\center
\epsfig{file=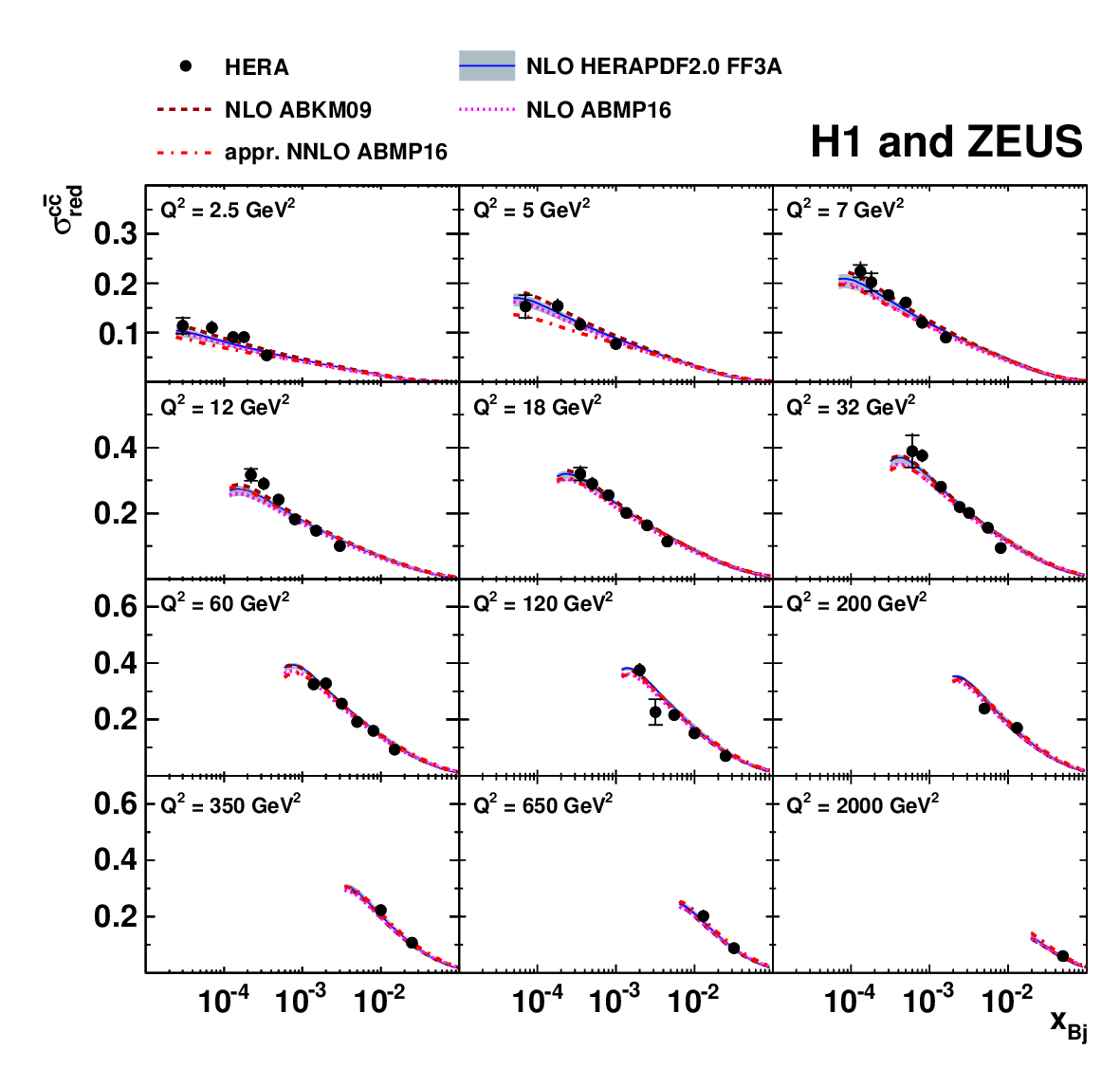 ,width=0.49\textwidth}
\epsfig{file=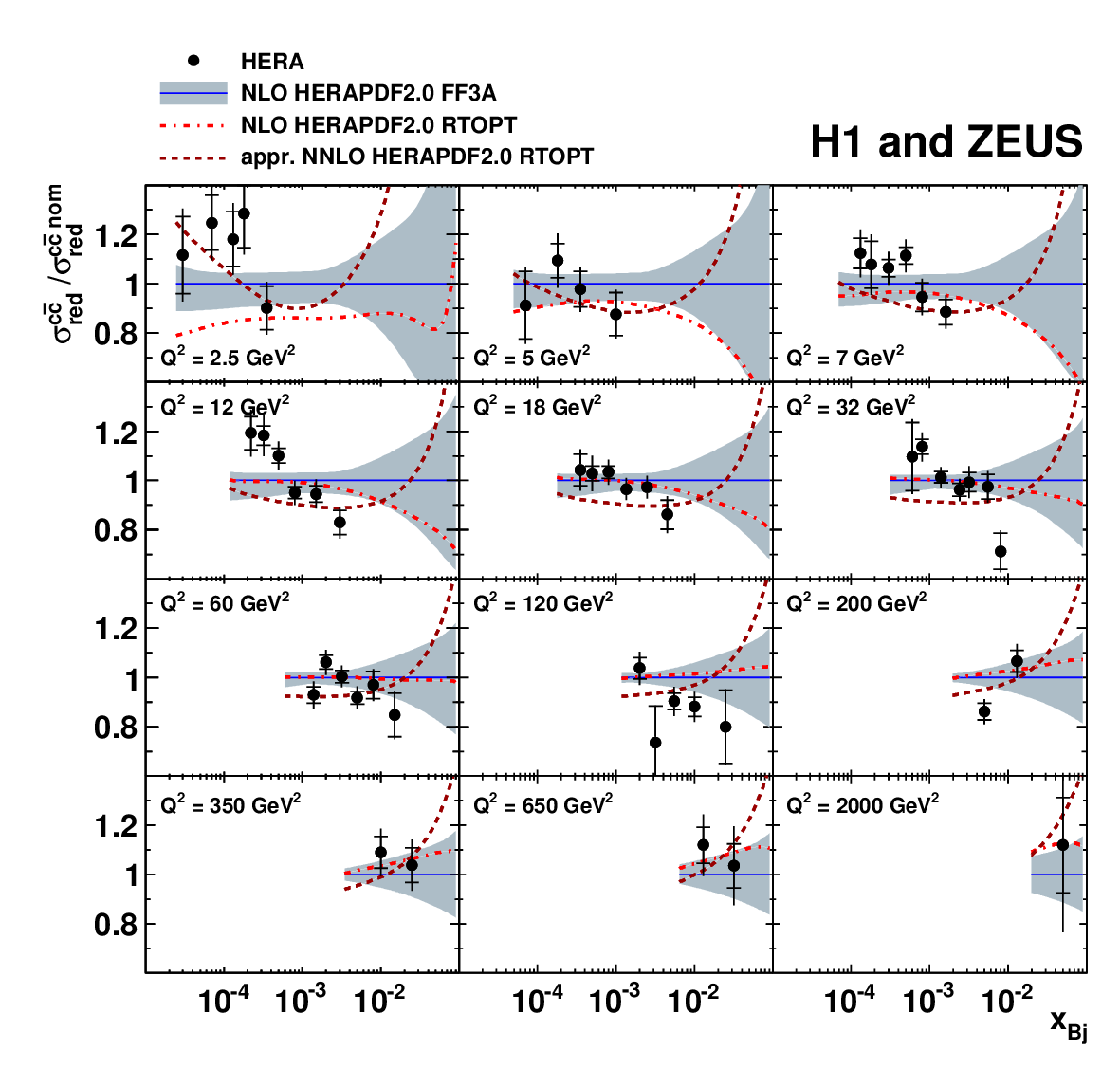 ,width=0.49\textwidth}
\caption{Left: combined reduced charm cross sections (full circles) as a function of $x$ for given values of $Q^2$, 
  compared to the NLO QCD FFNS predictions based on the HERAPDF2.0 FF3A (solid lines), and various other FFN scheme
  calculations
  The shaded bands on the HERAPDF2.0 FF3A predictions show the theory uncertainties obtained by adding PDF,
  scale and charm-quark mass uncertainties in quadrature.
  Right: Ratio of reduced charm cross sections as a function of $x$ for given values of $Q^2$ for the combined
  data~(full circles) and the NLO (dashed-dotted lines) and NNLO (dashed lines) VFNS predictions,
  obtained using the appropriate HERAPDF2.0 PDF sets, with respect to the FFNS NLO predictions,
  $\sigma^{ c\overline{c}\ {\rm nom}}_{\rm red}$, obtained using HERAPDF2.0 FF3A (solid lines with shaded uncertainty bands).
  The uncertainties for the VFNS predictions (not shown) are of similar size to those presented for the FFNS calculation.
  From ~\cite{H1:2018flt}.
 }
\label{fig:data-charm-theory} 
\end{figure}
The plots in Fig.~\ref{fig:data-charm-theory} show the ZEUS and H1 combined reduced cross sections for charm
data (left) and their ratio to various calculations in pertrubative QCD (right). When considering heavy flavour production it
is necessary to think about the treatment of heavy quarks in QCD. There have been two approaches: the Fixed Flavour Number
(FFN) scheme in which only 3 flavours are considered in the proton and heavy flavours are always calculated explicitly by
$g \to q\bar{q}$ splitting, taking into account the mass of the heavy quarks and; ii) the Zero-Mass Variable Flavour Number (ZMVFN) scheme
in which the number of quarks in the proton is considered to change as the scale increases, such that at $Q^2$ values much
above $4mc^2$ the charm quark is considered massless (similarly the bottom quark for $Q^2> 4mb^2$) and all log terms are
resummed just as for the light quarks. In the former case mass terms are treated properly but $ln(Q^2/mc^2)$ diagrams are
not resummed,
in the latter resummations are treated properly but at lower $Q^2$ mass terms are not. There are also General Mass Variable
Flavour Nummber schemes (GMVFN) which combine the features of the ZMVFN and FFN schemes. Looking at Fig.~\ref{fig:data-charm-theory}
it is evident that the theoretical calulations do not match the data, particularly for $Q^2\sim 12 $GeV$^2$. This is true whether a FFN (FF3A) calculation or a GMVFN (RTOPT) calculation is made.

These results will be discussed further in Section~\ref{sec:lowxinterp}, which discusses discrepant bahviour at low $x$.

\clearpage
\subsection{$F_2^p$ at very low $Q^2$}
\label{sec:lowq2}

\begin{figure}[tbp]
\vspace{-0.3cm} 
\centerline{
\epsfig{file=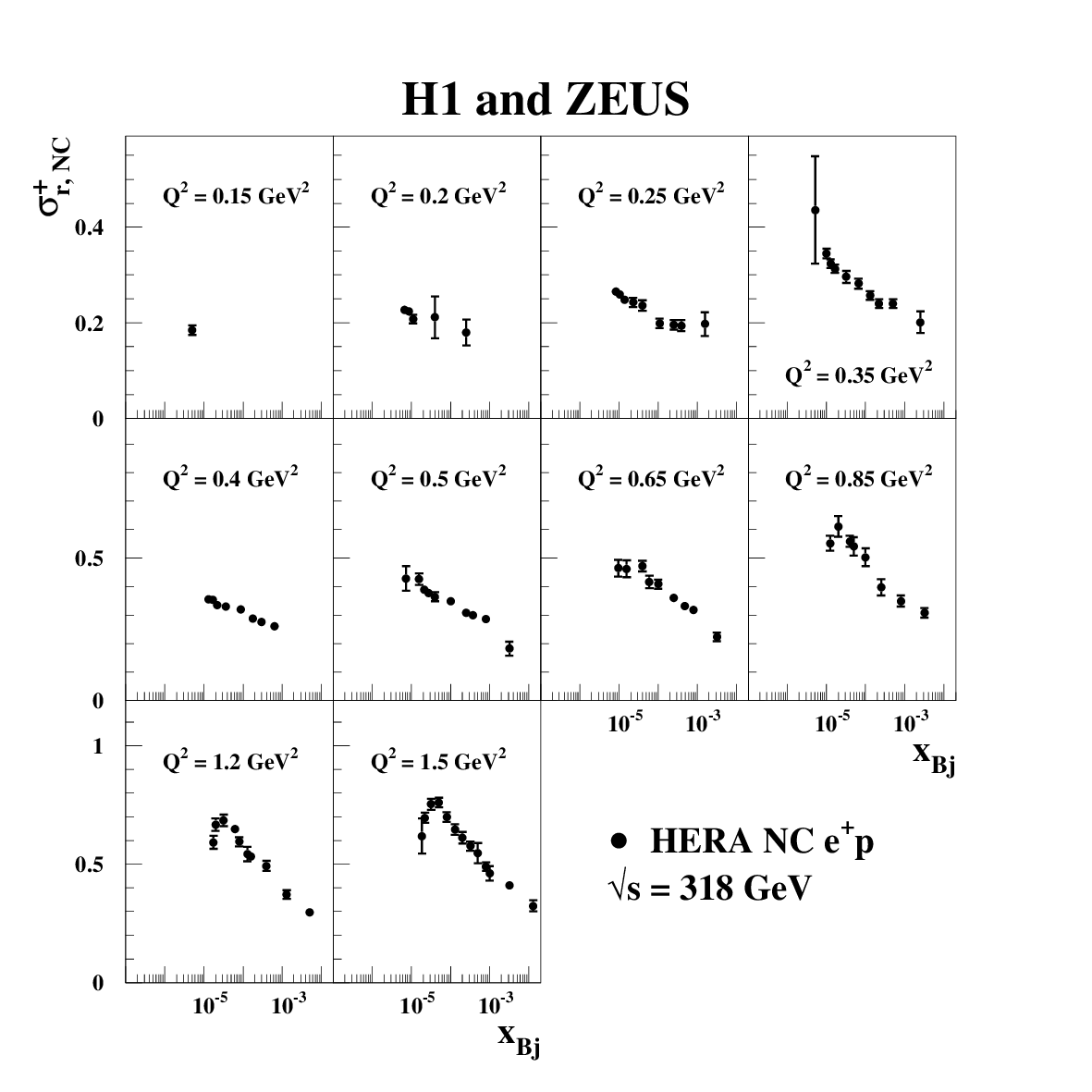   ,width=0.9\textwidth}}
\vspace{0.5cm}
\caption {The combined HERA data for the inclusive NC $e^+p$ 
reduced cross sections at $\sqrt{s} = 318$\,GeV at very low $Q^2$. From ~\cite{H1:2015ubc}.
Error bars represent the total uncertainties.
}
\label{fig:NCvlQ2-920}
\end{figure}

\begin{figure}[tbp]
\vspace{-0.5cm} 
\centerline{
    \includegraphics[width=0.9\textwidth]{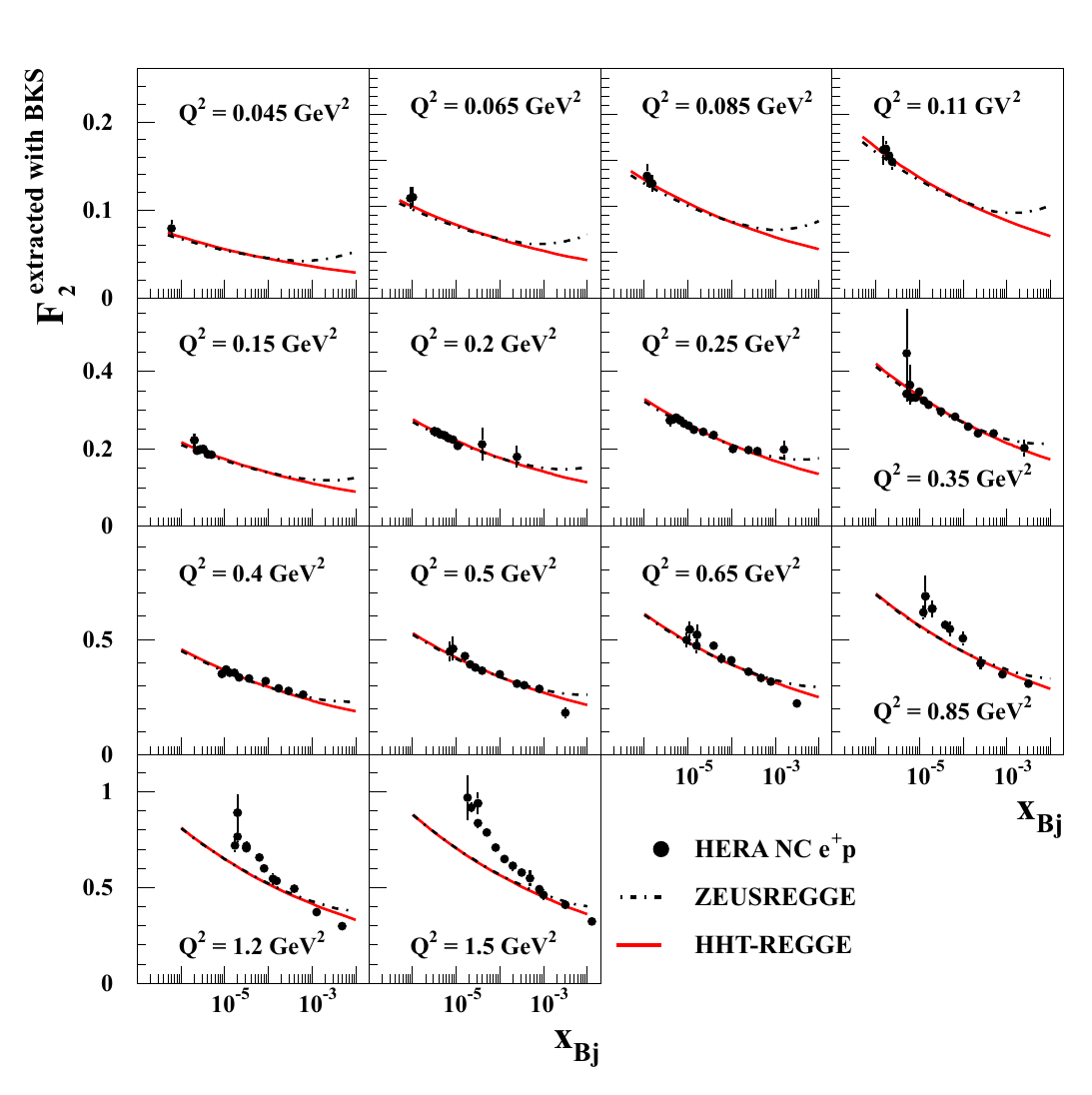}}
\caption {The structure-function $F_2$ as a function of 
          $x_{\rm Bj}$ for low $Q^2$ as extracted from the
          HERA NC $e^+p$ cross sections, 
          together with predictions from the Regge framework
          see Sec~\ref{sec:lowq2interp} for details. From ~\cite{Abt:2017nkc}.
}
  \label{fig:regge-noft}
\end{figure}

A remarkable feature of the proton structure function $F_2$ is its
rapid rise at low $x$ -- as already shown in Fig.~\ref{fig:f2}.
The persistence of this rise down to $Q^2$ values as
small as $0.65\,$GeV$^2$ challenges our understanding of QCD.
For smaller values of $Q^2$, the rise of the structure function at small
$x$ diminishes, as shown in  terms of the reduced cross section in Fig.~\ref{fig:NCvlQ2-920} and terms of the $F_2$
extracted from it in Fig.~\ref{fig:regge-noft}.
These very low $Q^2$ data allow us to examine the transition from the perturbative to non-perturbative regions.

\begin{figure}[tbp]
\vspace*{13pt}
\centerline{\psfig{figure=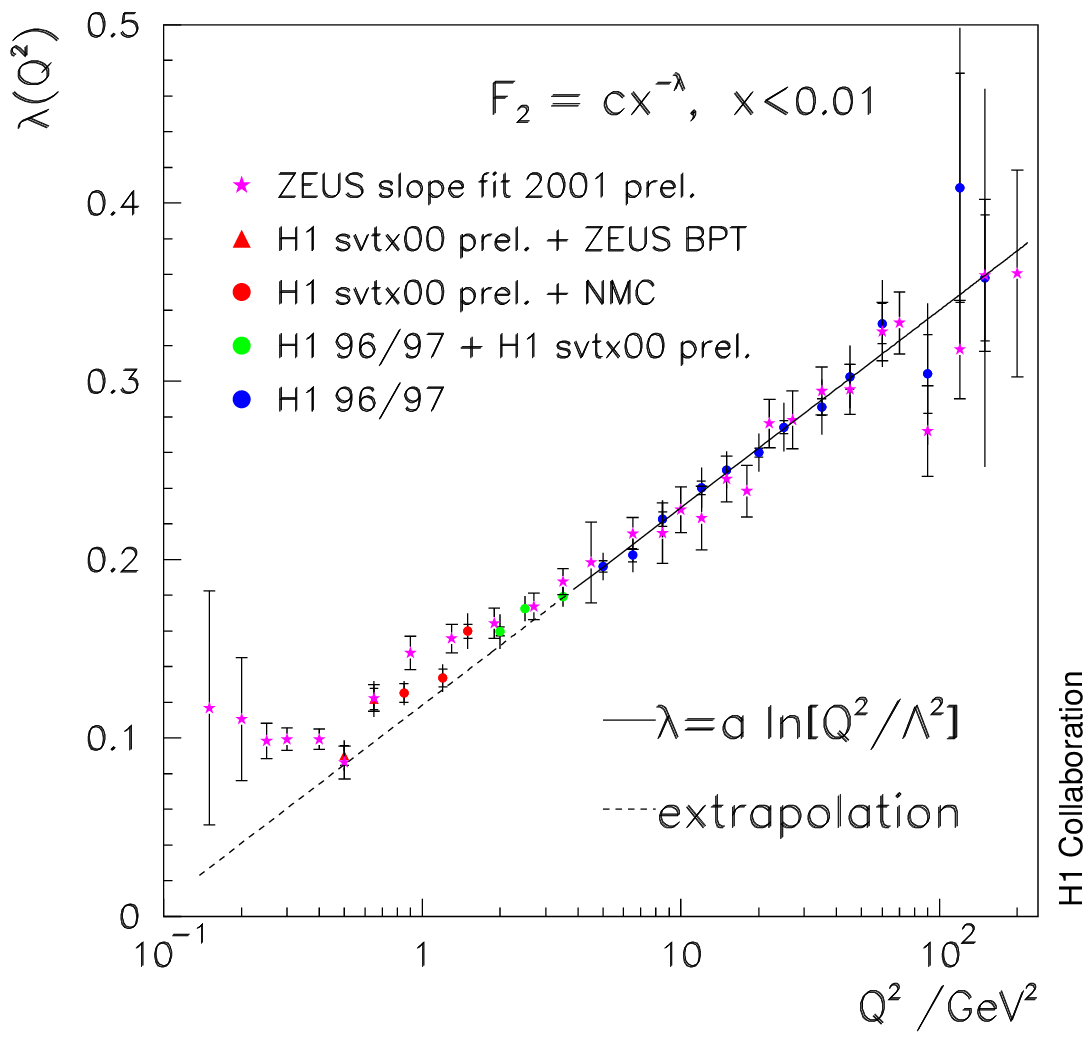,width=.45\textwidth}
           \psfig{figure=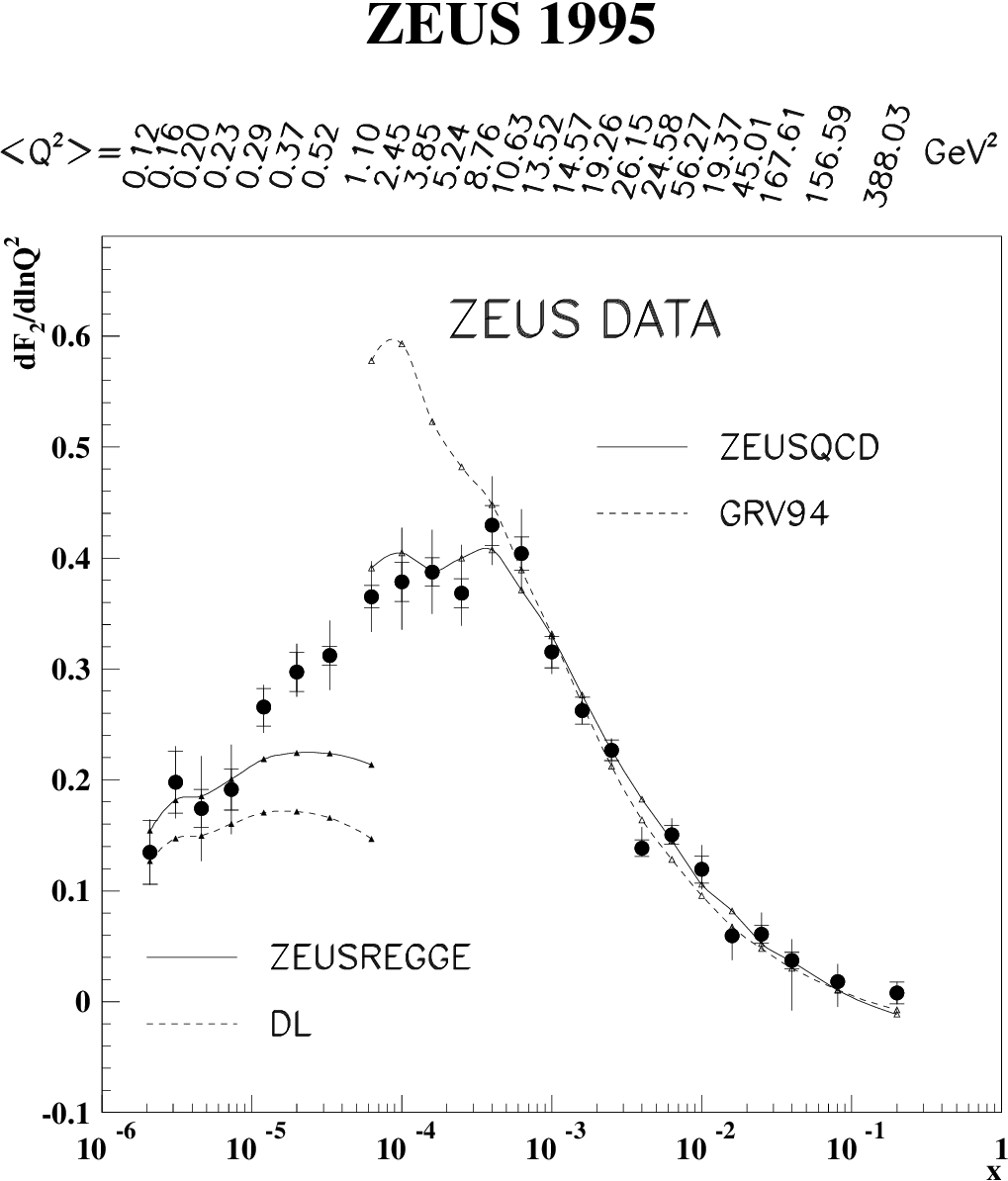,width=.45\textwidth}} 
\caption{Left: the effective slope of $F_2$ at low $x$, 
  $\ds F_2\sim x^{-\lambda(Q^2)}$. From \cite{Gayler:2002ah}. Right:
  The slope  $\ds {\partial F_2\over \partial \ln Q^2}$ as a
function of $x$ from early ZEUS data. From~\cite{ZEUS:1998agx}.}
\label{fig:lambdaeff}
 \end{figure}

The data can be characterised in terms of two slopes,
with respect to $\ln Q^2$ and $\ln(1/x)$ respectively.
The latter is often written as
$\ds \lambda(Q^2) = {\partial F_2\over \partial \ln(1/x)}$ such that
$\ds F_2(x,Q^2)\sim x^{-\lambda(Q^2)}$, and is shown in Fig.~\ref{fig:lambdaeff}.

 This left hand side of this figure shows that there is a region at low $Q^2$, $Q^2\leqsim 0.65$GeV$^2$ for which this slope is flat in accordance with Regge predictions, see Sec.~\ref{sec:lowq2interp}. For higher $Q^2$
 it rises in accordance with the predictions of perturbative QCD, see Sec~\ref{sec:lowxinterp}.

 The slope $\ds {\partial F_2\over \partial \ln Q^2}$ is shown in the right-hand-side of Fig.~\ref{fig:lambdaeff}
 as a function of $x$.
This is another result from HERA that caused a lot 
     of interest and discussion because the turn over at low $x$ was interpreted as evidence for saturation.
     At its simpleast this is the idea that when gluon densities are becoming very large the gluons may recombine
     and the evolution equations must become non-linear.
     We will discuss this further in section~\ref{sec:lowq2interp}.
    However, note that in Fig.~\ref{fig:lambdaeff},
because of the strong correlation between $x$ and $Q^2$ inherent in deep
inelastic kinematics, the slope  $\ds {\partial F_2\over \partial \ln Q^2}$ is not evaluated at a fixed values of
$Q^2$, but rather at the mean values shown along the top of the plot for
each data point. The plot shows that the rate of rise of $F_2$ as a
function of $Q^2$ is starting to abate as $x<10^{-3}$ and $Q^2<2\,$GeV$^2$.
This could be because we are entering the non-perturbative region, rather than evidence for saturation.
The interpretation of the low $Q^2$ data is discussed further in Section~\ref{sec:lowq2interp}.

\clearpage
\section{Low $x$ physics at HERA}
\label{sec:lowxinterp}
Now we will discuss all of the above data in terms of their theoretical interpretations.

\begin{figure}[tbp]
\vspace*{13pt}
\centerline{\psfig{figure=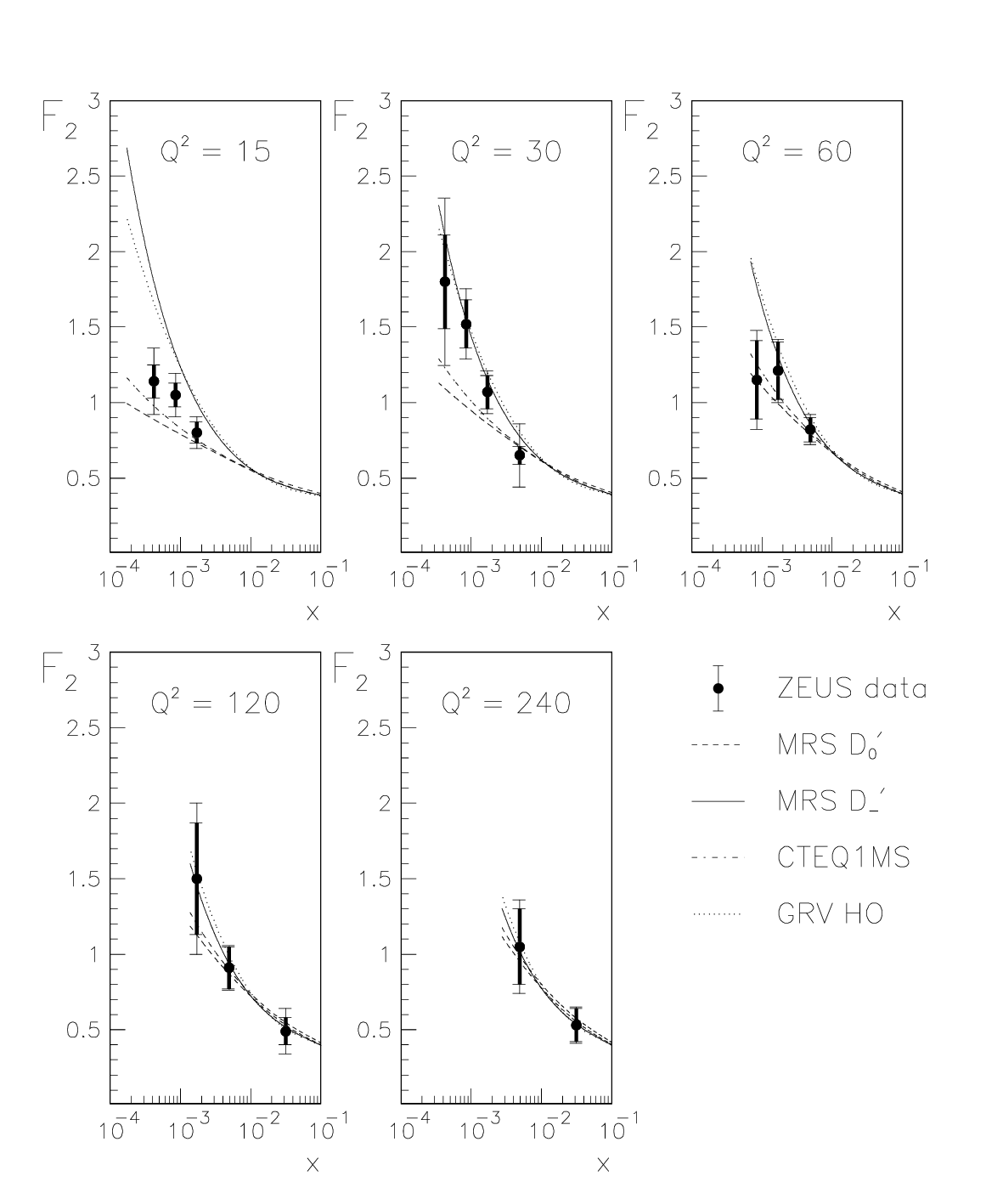,height=.5\textheight}} 
\caption{The first ZEUS $F_2$ data from HERA, published 1993. From~ \cite{ZEUS:1993ppj}. }
\label{fig:f2-z93}
\end{figure}

We have seen the striking rise of $F_2$ at low-$x$ in Fig~\ref{fig:f2}, but to get some feeling for why this caused
some excitement we can look at the very early data in Fig.~\ref{fig:f2-z93}.
  The data are compared to contemporary predictions. The predictions labelled CTEQ1 and MRS$D_0$ are the conventional
  predictions from QCD fits to PDFs based on previous data, which did not probe lower than $x\sim 0.01$. For these
  predictions the input parton distributions to DGLAP evolution had flat gluon and seaquark distributions at a
  starting scale of $Q^2_0\sim 4$GeV$^2$. The curves
  labelled GRV and MRS$D_-$ are also from QCD PDF fits but with unconventional input. In particular the MRS$D_-$ curve
  tries to account for QCD evolution according to the BFKL equation, which will be discussed later, by starting with a steep input gluon distribution and the GRV curve
  is based on an idea to start the QCD evolution at very low $Q^2\sim 0.3$GeV$^2$, where perturbative QCD is not
  expected to work, but a steeprise of $F_2$ can be generated at higher $Q^2$ by evolution from this low scale.
  It was a major suprise that the non-conventional pictures appeared to fit the data better.

  With hindsight it is hard to understand why we were so sure that a flat input was expected.
  Reasons why this was conventional are discussed in Section~\ref{sec:lowq2interp}.
However, rising of $F_2$ at small-$x$ is actually predicted by pQCD, but the predictions
do not give a scale in either $x$ or $Q^2$ at which one might expect
to see such effects.

In the DGLAP formalism at leading order~\footnote{It is simplest to make the explanations at leading order, the physics message remains unchanged at higher orders although the formalism becomes more complex} the gluon splitting 
functions
\begin{equation}
\ P_{gq} \to \frac{4}{3z},\ P_{gg} \to \frac{6}{z}
\label{eq:smallx}
\end{equation}
\noindent
are singular as $z\to 0$. Thus the gluon distribution will
become large as $x \to 0$, and its contribution to the evolution of the
parton distribution becomes dominant. In particular the gluon
will `drive'  the quark singlet distribution, and hence the structure 
function $F_2$, to become large as well, the rise increasing in
steepness as $Q^2$ increases. Quantitatively,
\begin{equation}
 \frac{dg(x,Q^2)}{d \ln Q^2} \simeq \frac{\alpha_s(Q^2)}{2\pi} \int^1_x
\frac{dy}{y} \frac{6}{z} g(y,Q^2)
\label{eq:glev}
\end{equation}
\noindent
may be solved subject to the nature of the boundary function $xg(x,Q^2_0)$.
Inputing a flat gluon at $Q^2_0$, the solution is (from~\cite{DeRujula:1974mnv,Ball:1994du}) 
\begin{equation}
 xg(x,Q^2) \simeq \exp\left(2 \left[\xi (Q^2_0,Q^2)\ln\frac{1}{x}\right]
^{\frac{1}{2}}\right)
\label{eq:DLLAg}
\end{equation}
\noindent
where
\begin{equation}
 \xi(Q^2_0,Q^2) = \int^{Q^2}_{Q^2_0} \frac{dq^2}{q^2} \frac{3\alpha_s(Q^2)}
{\pi}
\label{eq:DLLAxi}
\end{equation}
\noindent
Given a long enough evolution length from $Q^2_0$ to $Q^2$, this will
generate a steeply rising gluon distribution at small $x$, starting from 
a flattish behaviour of $xg(x,Q^2)$ at $Q^2 = Q^2_0$.

Over the $x,Q^2$ range of HERA data this solution mimics a power 
behaviour, $xg(x,Q^2) \sim x^{-\lambda_g}$, with 
\begin{equation}
\lambda_g = \left(\frac{12}{\beta_0}
\frac{ \ln(t/t_0)}{\ln(1/x)}\right)^{\frac{1}{2}}
\end{equation}
\noindent
where $t=\ln(Q^2/\Lambda^2)$, $t_0=\ln(Q_0^2/\Lambda^2)$.
This steep behaviour of the gluon generates a similarly
steep behaviour of $F_2$ at small $x$, $F_2 \sim x^{-\lambda}$, where
$\lambda \sim \lambda_g$.

So why was the observed rise of $F_2$ unexpected?  Because, before the advent
of HERA data, the starting  scale for perturbative evolution had always been 
taken to be $Q^2_0 \geqsim 4\,$GeV$^2$, to be sure that a perturbative 
calculation would be valid. The evolution length from $Q^2$ of 4 to 
$15\,$GeV$^2$ is not large enough to generate a steep slope from a flat input.

There are two assumptions which could be challenged here: i ) the value $Q^2_0 \sim 4\,$GeV$^2$ is not set in stone,
in fact modern PDF analyses start from $Q^2_0 \sim 1.5\,$GeV$^2$ and; ii) the input itself need not necessarily be flat.
First considering i) there was a further surprise when HERA data was extended to lower $Q^2$ and we could see that the rise of $F_2$ continues down to $ Q^2 \sim 1$GeV$^2$, we have seen this in Figs.~\ref{fig:nnloQ23pt5ncepb} and~\ref{fig:regge-noft}.
This would seem to imply that we
need a steep input even at rather low $Q^2$. One may worry that the value of $\alpha_s(Q^2_0)$ is becoming too
high for perturbative QCD to be applicable at such low $Q^2$. However, the inputs are not necessarily flat.
This brings us back to ii) should one consider a non flat input. The motivation for a steeply rising input
came from the BFKL equation,~\cite{Fadin:1975cb,Kuraev:1976ge,Kuraev:1977fs,Balitsky:1978ic}.

At small $x$ terms in
$\ln(1/x)$ are becoming large and the conventional $\ln Q^2$ 
summation of the DGLAP equations does not account for this. It may also be 
necessary to sum $\ln(1/x)$ terms. Such
a summation is performed by the BFKL
equation. To leading order in
$\ln(1/x)$ with fixed $\alpha_s$, this predicts a steep power law behaviour
\begin{equation}
 xg(x,Q^2) \sim f(Q^2)\ x^{-\lambda_g}
\label{eq:BFKLsol}
\end{equation}
\noindent 
where 
\begin{equation}
 \lambda_g = \frac{3\alpha_s}{\pi} 4\ln2 \simeq 0.5
\label{eq:BFKLlam}
\end{equation}
\noindent 
(for $\alpha_s \simeq 0.2$, as appropriate for $Q^2 \sim 4\,$GeV$^2$).
Thus the observation of a steep behaviour in $F_2$ at a relatively low $Q^2$
was seized upon by those who saw this as evidence for BFKL. Higher
statistics data showed that an effective power 
$\lambda_g\approx 0.5$ was actually too steep, but the calculation above is only to leading order
and more data compatible lower slopes of $\lambda_g\approx 0.3$ are predicted by more sophisticated calculations.

\begin{figure}[tbp]
\vspace*{13pt}
\centerline{\psfig{figure=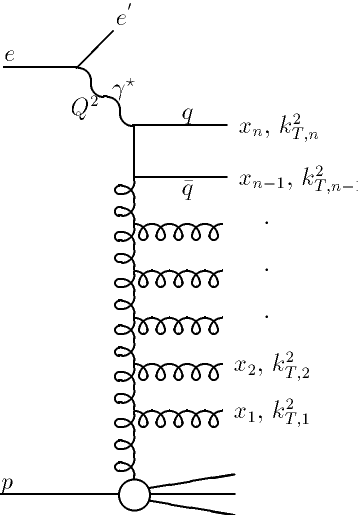,width=0.49\textwidth}
            \psfig{figure=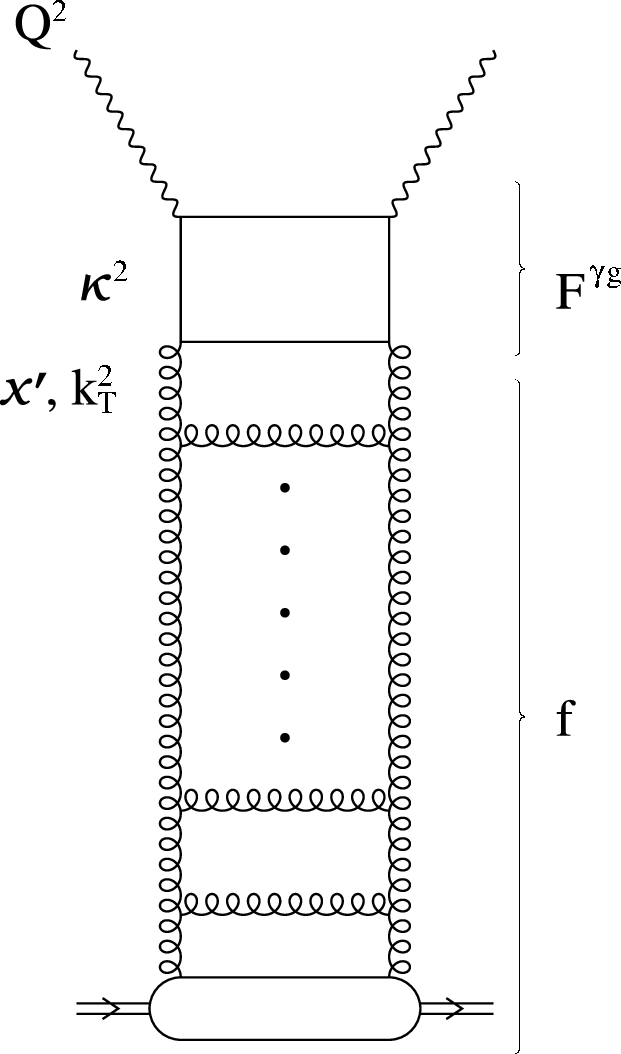,width=.49\textwidth}} 
\caption{Left: gluon emissions in the BGF proces. Right: the BFKL gluon ladder diagram}
\label{fig:ladder}
\end{figure}

The BFKL equation concerns the behaviour of the gluon. Referring back to the gluon emissions from the quark line of
Figure~\ref{fig:qmanyg}, we known that the DGLAP equations also involve gluon emissions from a gluon line, with a
quark-antiquark emission connecting to the $\gamma^*$, as for the BGF process. The corresponding diagram is shown in
Fig.~\ref{fig:ladder}, left-hand-side.
In the DGLAP formalism these emissionsare $p_t$ ordered just as
they were in Eqn~\ref{eqn:pt-ord}. The difference in the BFKL formalism is that these emissions need not be ordered.
Then using the optical theorem to related the total cross section to the imaginary part of the forward eleastic
scattering amplitude, we get the gluon ladder diagram of Fig~\ref{fig:ladder} right-hand-side.
Since the transverse momenta in the BFKL ladder are
not ordered this allows of some physics processes not allowed in DGLAP, for example emission high $p_t$ jets
in the forward direction, see subsection~\ref{sec:forwardjets}

BFKL phenomenlogy has advanced a great deal in more recent years and NLL $ln(1/x)$ resummation
may now be incorporated with NNLO DGLAP $lnQ^2$ resummation in the QCD fits, see subsection~\ref{sec:lnxresum}.
\begin{figure}[tbp]
\vspace*{13pt}
\centerline{\psfig{figure=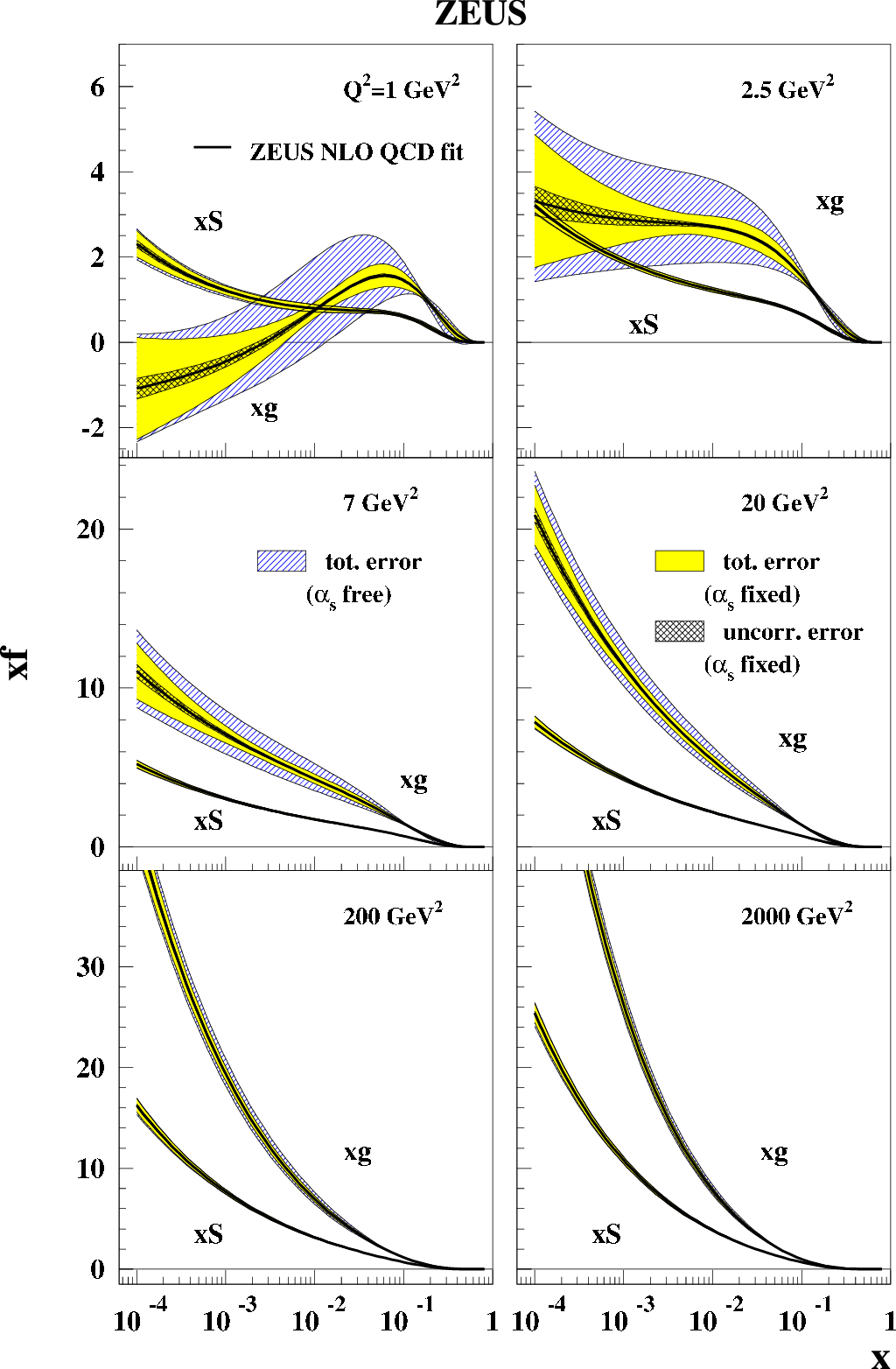,height=.6\textheight}} 
\caption{Gluon and sea quark momentum distributions compared. From a NLO
QCD fit by ZEUS using the full HERA-I data sample. From~\cite{ZEUS:2002xjx}.}
\label{fig:glusea}
\end{figure}

\begin{figure}[tbp]
\vspace*{13pt}
\centerline{\psfig{figure=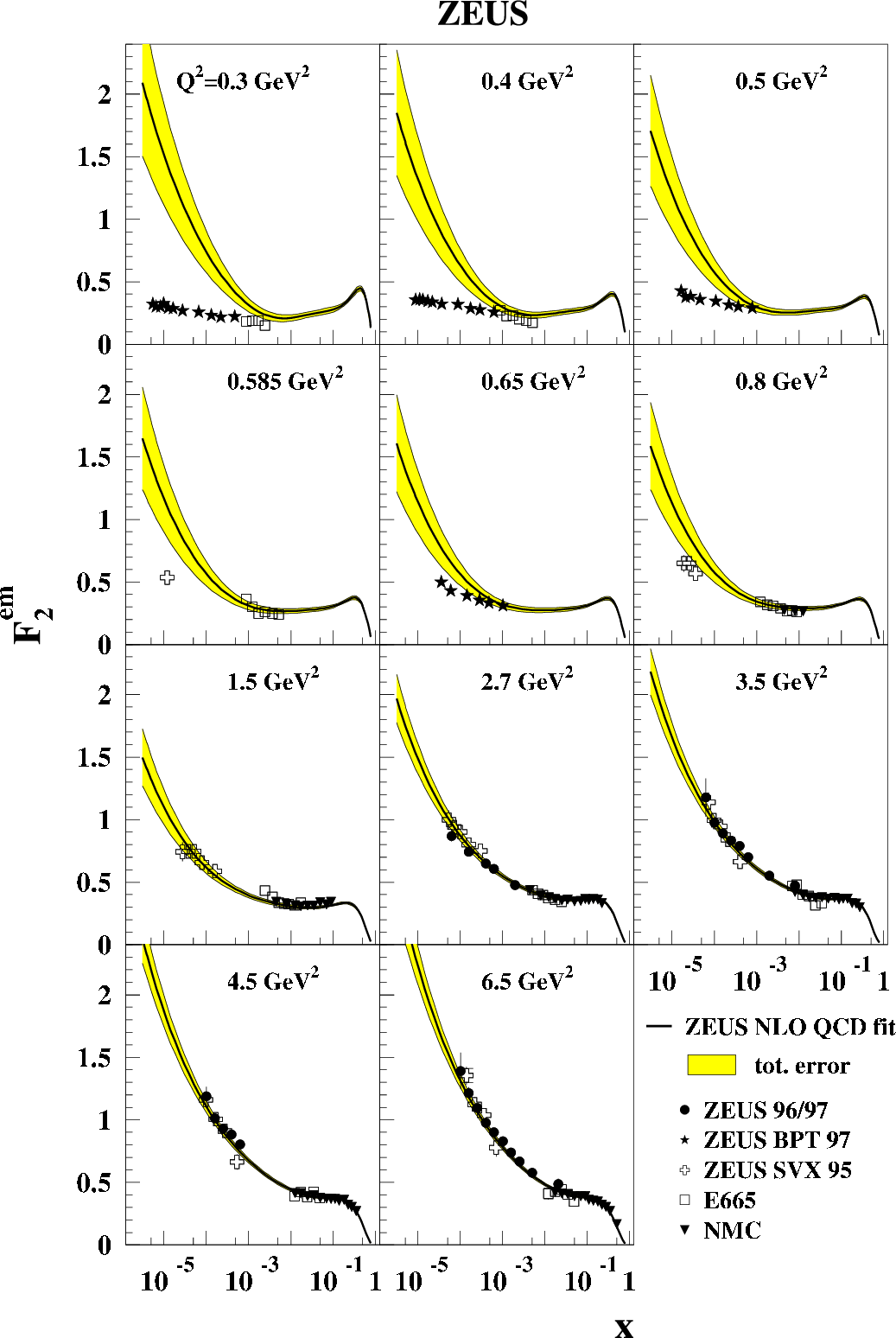,height=.6\textheight}} 
\caption{ZEUS $F_2$ data at low $Q^2$ compared to the ZEUS-S pQCD fit.
From~\cite{ZEUS:2002xjx}.}
\label{fig:lowq2}
\end{figure}

A steep input distribution can be input to pure DGLAP fits, in order to avoid the need for use of the BFKL evolution,
but is this successful?
This begs the question, which input is steep? is it the initial distributions of both the sea-quarks and gluons?
As more HERA data were collected in the early years of HERA running, DGLAP NLO QCD global fits
gave good $\chi^2$ fits to the HERA data,
starting with a conventional flat gluon and sea input, provided that a low 
starting scale $Q^2_0 \sim 2\,$GeV$^2$ was used.
But as we have seen in Fig.~\ref{fig:regge-noft},
the steep slope of $F_2$ extends further down in $Q^2$ to 
$Q^2 \sim 1\,$GeV$^2$. By the end of the HERA-I run in 1999, the precision of the data had increased 
sufficiently that the low-$x$ behaviour of the sea and the gluon distributions
could be fitted separately. In Fig.~\ref{fig:glusea} we show the total sea and gluon PDFs for various
$Q^2$ values down to  $Q^2 = 1$GeV$^2$, below the starting scale for evolution. The data are still reasonably well described
even at this low scale, see Fig.~\ref{fig:lowq2},  but these results led to a new kind of surprise since
it became clear that the flat input which had been used for both the sea and 
the gluon ($\lambda_S = \lambda_g \sim 0$) was actually a compromise between
a sea distribution which remains steep, $\lambda_S \sim 0.2$ down to 
$Q^2 \sim 1~$GeV$^2$, and a gluon distribution which is becoming valence-like, 
$\lambda_g < 0$ at low $Q^2$. Fig.~\ref{fig:glusea} shows the result from an NLO QCD fit made by ZEUS
using the complete HERA-I data sample, \cite{ZEUS:2002xjx}, but it remains true for the full HERA combination, for NNLO analysis and
for all the conventional DGLAP based QCD fits of CT, MSHT and NNPDF. 
This behaviour contradicts the
original argument that the  steep behaviour of the gluon distribution is 
driving the steep behaviour of the sea which is measured in $F_2$.

Fig. ~\ref{fig:lowq2} also shows the 
limits on the kinematic range of applicability of the DGLAP formalism. The fit can no longer describe the data 
for  $Q^2 \leqsim 1$GeV$^2$ 
Furthermore, even though 
the DGLAP formalism works for $Q^2 \geqsim 1~$GeV$^2$, 
doubts remain about its applicability when 
$\lambda_g < \lambda_S$.
Does the data interpretation beyond the DGLAP formalism?
At at low $x$ the form of the DGLAP
equations is such that, at LO, one has  
\[
F_2(x,Q^2)\sim x\Sigma(x,Q^2),~~~{\rm and}~~~
{\partial F_2(x,Q^2)\over \partial \ln Q^2}\sim  P_{qg}\otimes xg(x,Q^2).
\]
$F_2$ and its $Q^2$ evolution are essentially the experimental variables that we are using to determine the sea and the
gluon PDFs respectively.
But the behaviour of the slope $\partial F_2/\partial\ln Q^2$ depends on the splitting function as well as on the
gluon density. If it could be established that 
unconventional behaviour of this slope is attributable to $P_{qg}$, rather than to the gluon PDF then this would
indicate the need for an alternative evolution equation.
This question can be investigated by the measurement of other quantities that depend on the gluon differently than
$\partial F_2(x,Q^2)\over \partial \ln Q^2$,
for example $F_L$ and $F_2^{c\bar{c}}$. We have seen these measurements in Secs~\ref{sec:fl} and~\ref{sec:f2cc}, we discuss the insights that have been gained from $F_2^{c\bar{c}}$ in Sec.~\ref{sec:f2ccinterp}. The role of $F_L$ enters
into the fitting of the reduced cross section of the inclusive data and is discussed in that context in the next section.

\subsection{alternative fits to $\sigma_r$}

As more HERA data were collected it became evident that the DGLAP QCD fits to these data at low $Q^2$, which for
HERA also implies low $x$, are not all that good, see Fig~\ref{fig:chiscan}, whih shows
the $\chi^2$ of the HERAPDF2.0 NNLO, NLO and LO QCD fits to the full HERA combined cross section data as a function of
the minimum $Q^2$ of data entering the fit. It is evident that as $Q^2_{min}$ decreases the $\chi^2$ deteriorates.
Examination of the plots of $\sigma_r$ vs $x$ at low $Q^2$, see Fig.~\ref{fig:nnloQ23pt5ncepb}, showed us that the turn
over of the reduced cross section for low $x$, $x\leqsim 10^{-4}$ is not well described. Recall that this turn over is due to
the contribution of $F_L$ in $\sigma_r=F_2-y^2/Y_+F_L$, so that it seems that the fit wants a larger $F_L$ than that predicted by
the DGLAP formalism. Given that the $F_L$ is strongly related to the gluon (see Eq.~\ref{eqn:fl_qg})
and the gluon of the DGLAP formalism is becoming valence like, ie smaller at low-$x$ and $Q^2$,
this could be interpreted as meaning that the DGLAP formalism needs some modification.
We consider two alternatives: i) the use of NLL $ln(1/x)$ resummation in addition to NNLO DGLAP,~\cite{xFitterDevelopersTeam:2018hym},
and the need for higher twist terms at low $Q^2$,~\cite{Abt:2016vjh}, in subsections~\ref{sec:lnxresum},
\ref{sec:htlowq2} respectively. However, note that both these approaches are still limited to $Q^2 \geqsim 1$GeV$^2$.
It is informative to consider the transition to very low $Q^2$ as we will do in Section~\ref{sec:lowq2interp}.

\subsubsection{$ln(1/x)$ resummation}
\label{sec:lnxresum}
\begin{figure*}[t]
  \begin{centering}
    \includegraphics[width=0.49\textwidth]{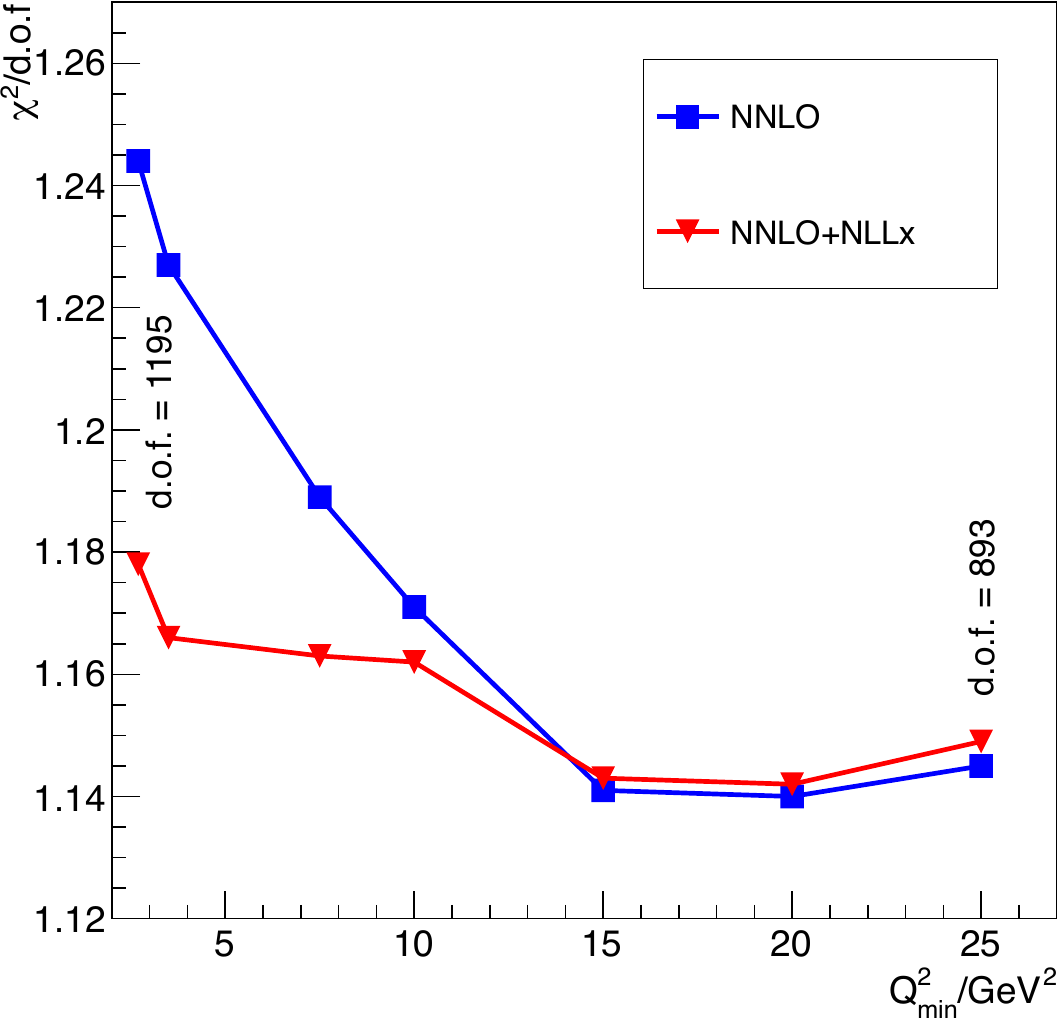}
    \includegraphics[width=0.49\textwidth]{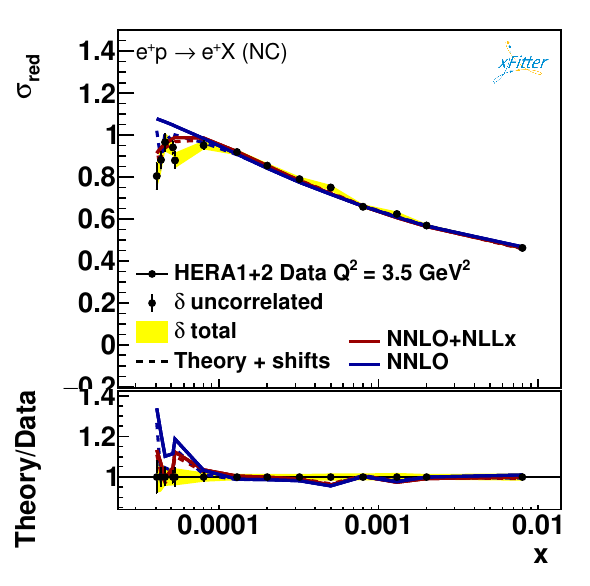}
    \caption{Left: the $\chi^2$/d.o.f. as a function of $Q^2_{\rm min}$ for NNLO vs NNLO+NLLx fits to to
      HERA data. Right: The predictions for the reduced cross section for the fits to HERA data
      (NNLO+NLL$x$) and without (NNLO) $\ln(1/x)$ resummation. From ~\cite{xFitterDevelopersTeam:2018hym}.
      }
    \label{fig:lnxchi2}
  \end{centering}
\end{figure*}

\begin{figure*}[t]
  \begin{centering}
    \includegraphics[width=0.8\textwidth]{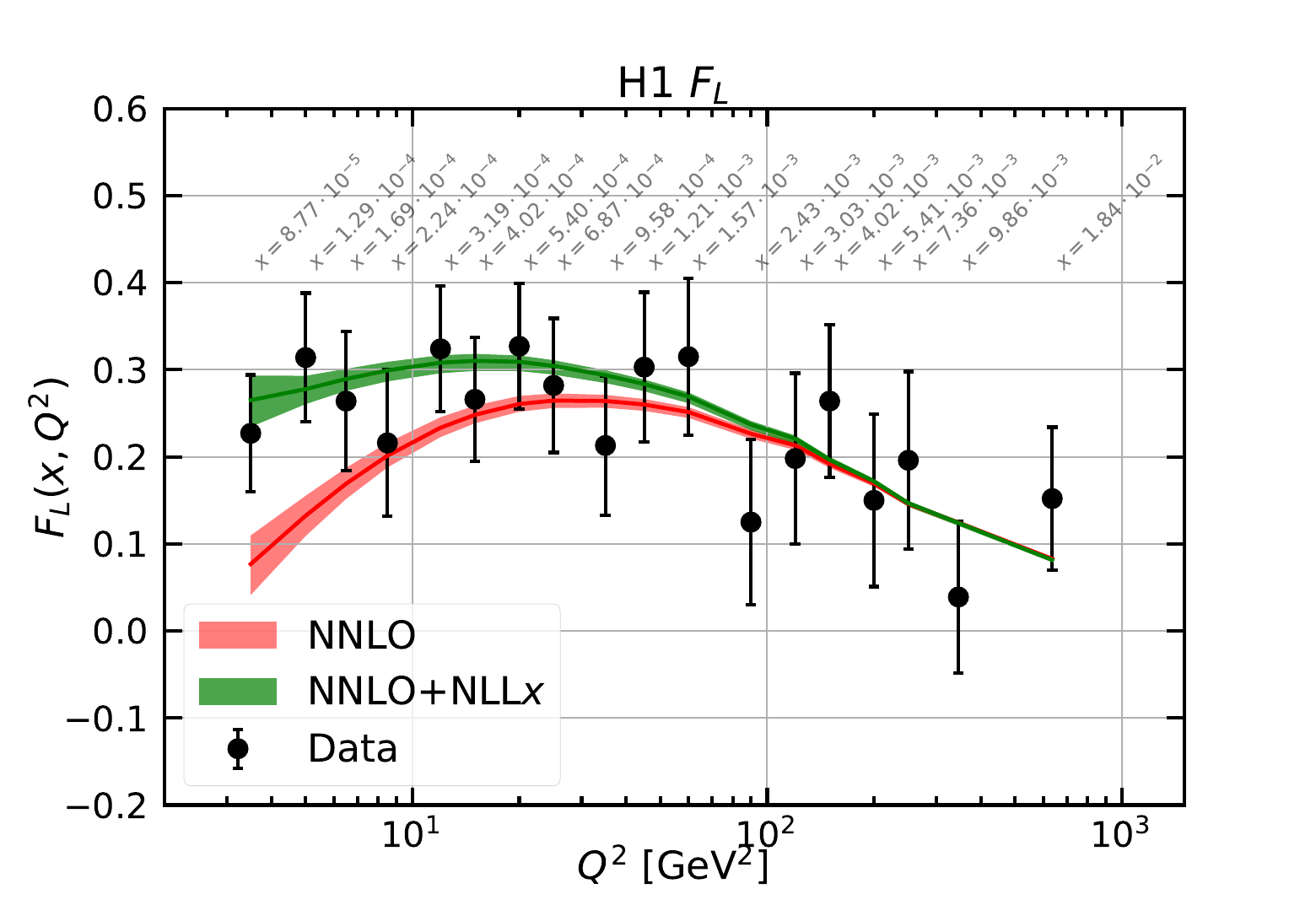}
    \caption{The predictions for $F_L$ for the fits to HERA data
      (NNLO+NLL$x$) and without (NNLO) $\ln(1/x)$ resummation. From ~\cite{xFitterDevelopersTeam:2018hym}.}
    \label{fig:flsigr}
  \end{centering}
\end{figure*}

In recent years the programm HELL has become available,~\cite{Bonvini:2017ogt}, to perform next-to-leading $ln(1/x)$ resummations together with next-to-next-to leading order NNLO $ln(Q^2)$ resummations. This has been used by the NNPDF collaboration to
produce the NNPDF3.1sx PDF,~\cite{Ball:2017otu}, set and by the HERA collaboration,~\cite{xFitterDevelopersTeam:2018hym}.
In Fig~\ref{fig:lnxchi2} (left-hand-side) we show the difference in the $\chi^2$ vs $Q^2_{min}$
plot for NNLO vs NNLO+NLLX treatment using the HERAPDF formalism.
The NNLx treatment leads to a decrease in $\chi^2$ for the fit of $\sim 70$ points.
This is essentially because it
provides a much better description of the turn over of the reduced cross section at small $x$, see Fig.~\ref{fig:lnxchi2}
(right-hand-side),
since it predicts a
somewhat larger $F_L$ than the conventional DGLAP NNLO approach, see Fig~\ref{fig:flsigr}

\begin{figure*}[t]
  \begin{centering}
    \includegraphics[width=0.32\textwidth]{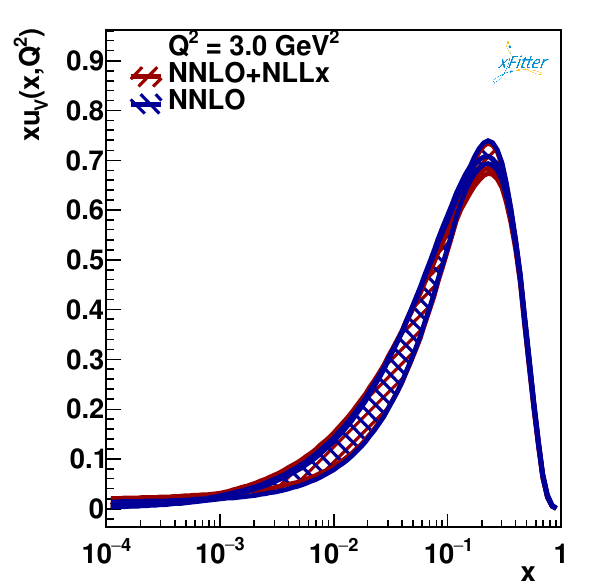}
    \includegraphics[width=0.32\textwidth]{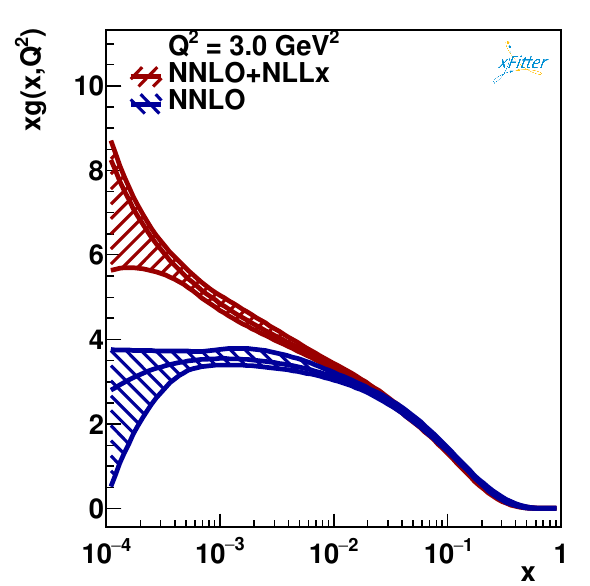}
    \includegraphics[width=0.32\textwidth]{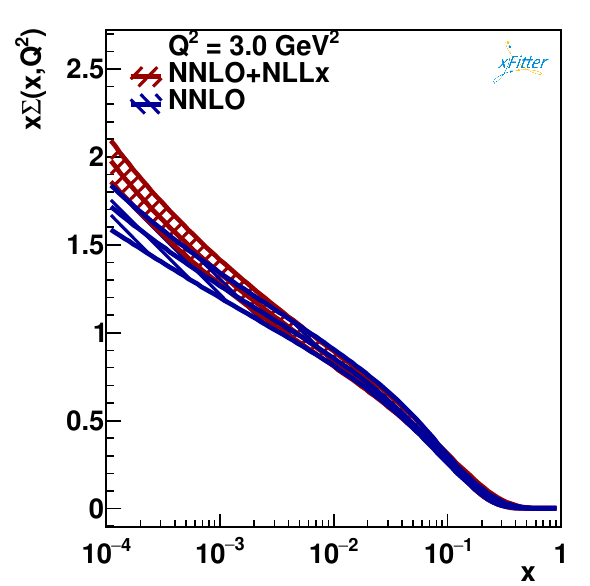}
    \caption{The up valence PDF $xu_v$, the gluon PDF $xg$ and the
      total singlet PDF $x\Sigma$ for the final fits with
      (NNLO+NLL$x$) and without (NNLO) $\ln(1/x)$ resummation. From ~\cite{xFitterDevelopersTeam:2018hym}.}
    \label{fig:finalpdf}
  \end{centering}
\end{figure*}

The PDFs with and without the $ln(1/x)$ resummation are shown in Fig.~\ref{fig:finalpdf}. Here we can see that the
inclusion of this low-$x$ resummation results in a gluon which is unsuppressed at low $x$, contrary to the somewhat
counter-intuitive behaviour of the pure DGLAP fit, for which the gluon density is falling, and it is thus not possible
to state that the gluon is 'driving'
the sea by $g\to q\bar{q}$ splitting. However, once the BFKL-type $ln(1/x)$ resummation is included, we recover the rising gluon density and we may thus also have to consider gluon-gluon recombination and saturation, see Sec~\ref{sec:lowq2interp}

\subsubsection{higher twist}
\label{sec:htlowq2}
\begin{figure}[tbp]
\centerline{
\epsfig{figure=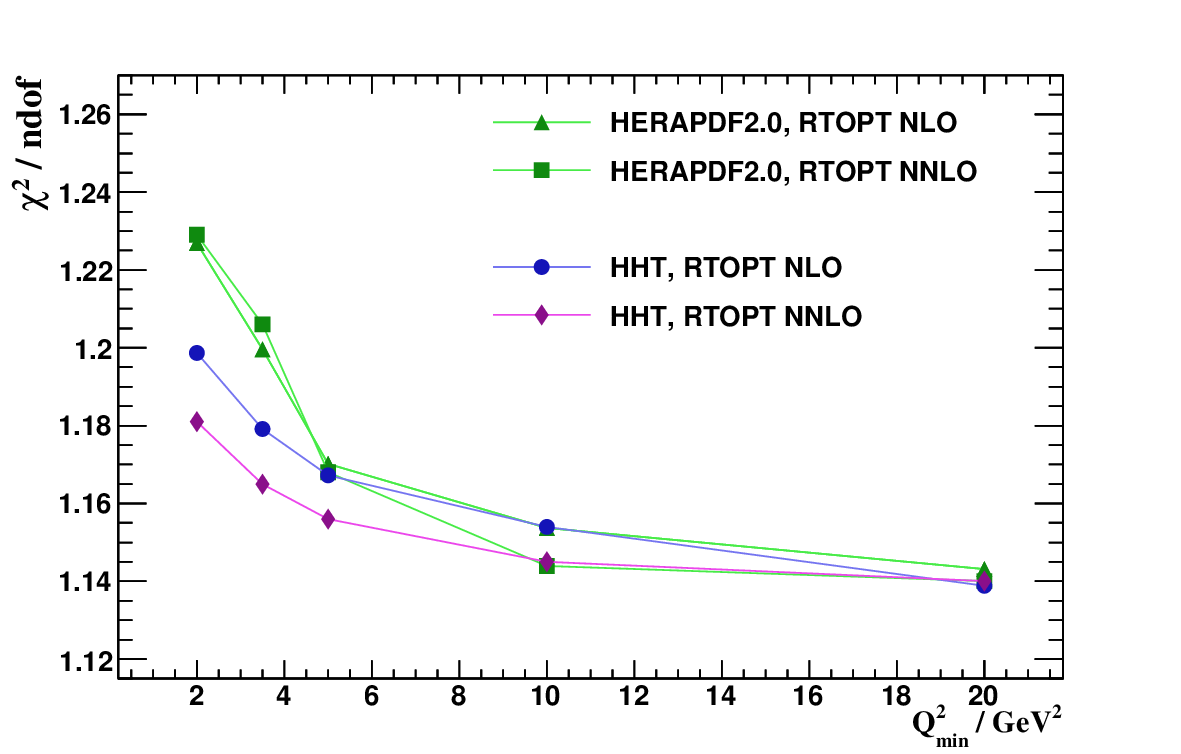 ,width=0.7\textwidth}}
\caption { 
The $\chi^2$/ndof versus $Q^2_{\rm min}$ for HHT and HERAPDF2.0 fits
at NNLO and NLO. From~\cite{Abt:2016vjh}
}
\label{fig:htchiscan}
\end{figure}

There are other approaches to improving the fit to the HERA reduced cross section at low $x$ and $Q^2$. One of the simplest its to introduced higher twist terms into the formalism.
Higher-twist effects have a $1/Q^{2n}$ dependence and are thus important at low $Q^2$. They are usually considered at
high $x$, which is not accessible at low $Q^2$ for HERA data, but it is possible that there could be low $x$
higher twist effects due to gluon-gluon interactions and recombination at high gluon density,~\cite{Bartels:2000hv}. Such effects are
visualised as gluon ladders with recombining gluons.
The leading-twist perturbative QCD 
forms of the structure functions $F_2$ and $F_{\rm L}$ 
are augmented by simple twist-4 terms
\begin{eqnarray}
\label{eq:flht}
F_{\rm 2}^{\rm HT} &= F_{2}^{\rm DGLAP} &(1 + A_2^{\rm HT}/Q^2)~,\\
\label{eq:flht2}
F_{\rm L}^{\rm HT} &= F_{\rm L}^{\rm DGLAP} &(1 + A_{\rm L}^{\rm HT}/Q^2)~,
\end{eqnarray}
where $A_2^{\rm HT}$ and $A_{\rm L}^{\rm HT}$ are free parameters  in the fits. It is found that only the $A_{\rm L}^{\rm HT}$ term is needed to produce fits with a better $\chi^2$, see Figure~\ref{fig:htchiscan}. This is also in accordance with the predictions of ~\cite{Bartels:2000hv}.

\begin{figure}[tbp]
\vspace{-0.3cm} 
\centerline{
  \epsfig{figure=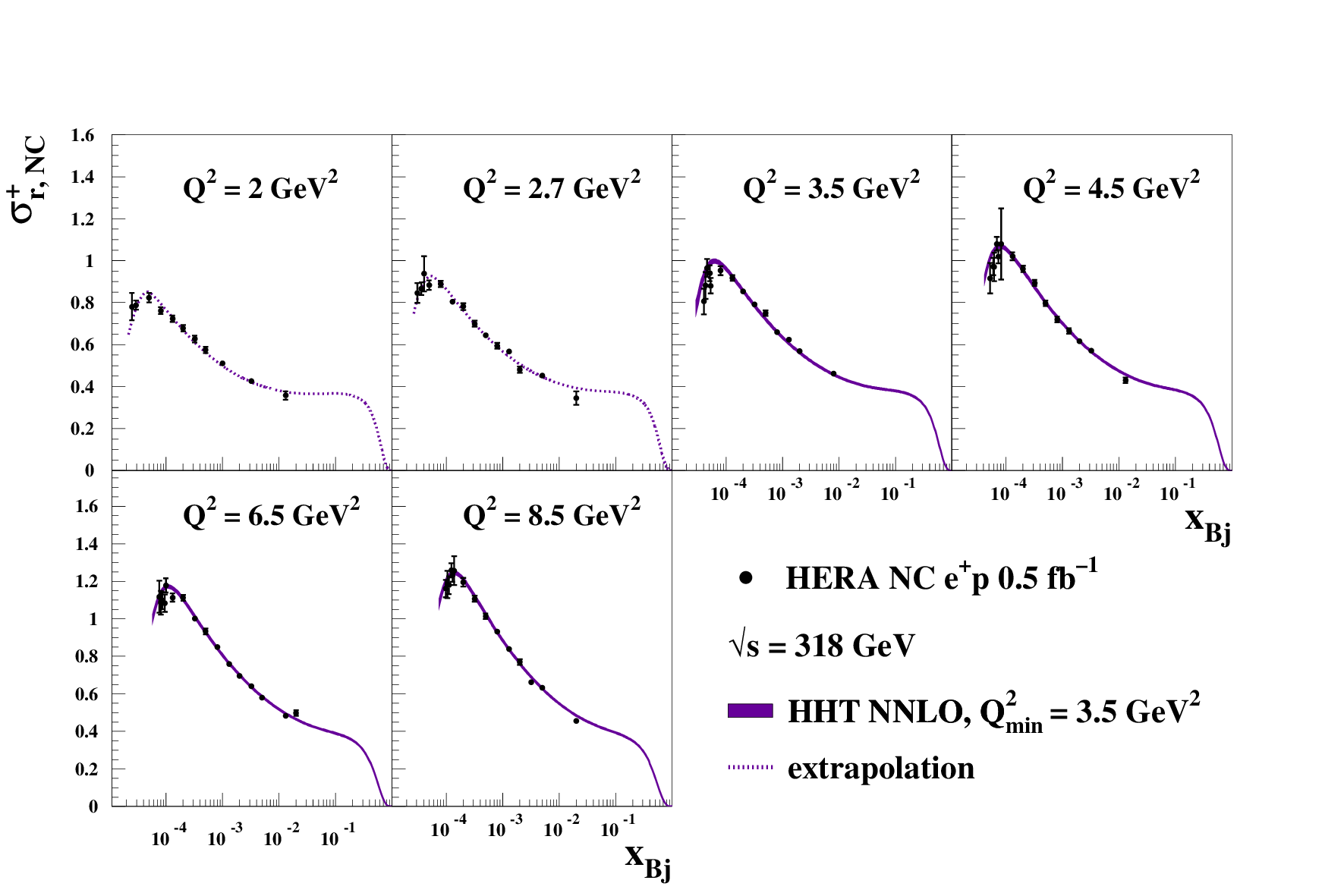,width=0.9\textwidth}}
\centerline{
  \epsfig{figure=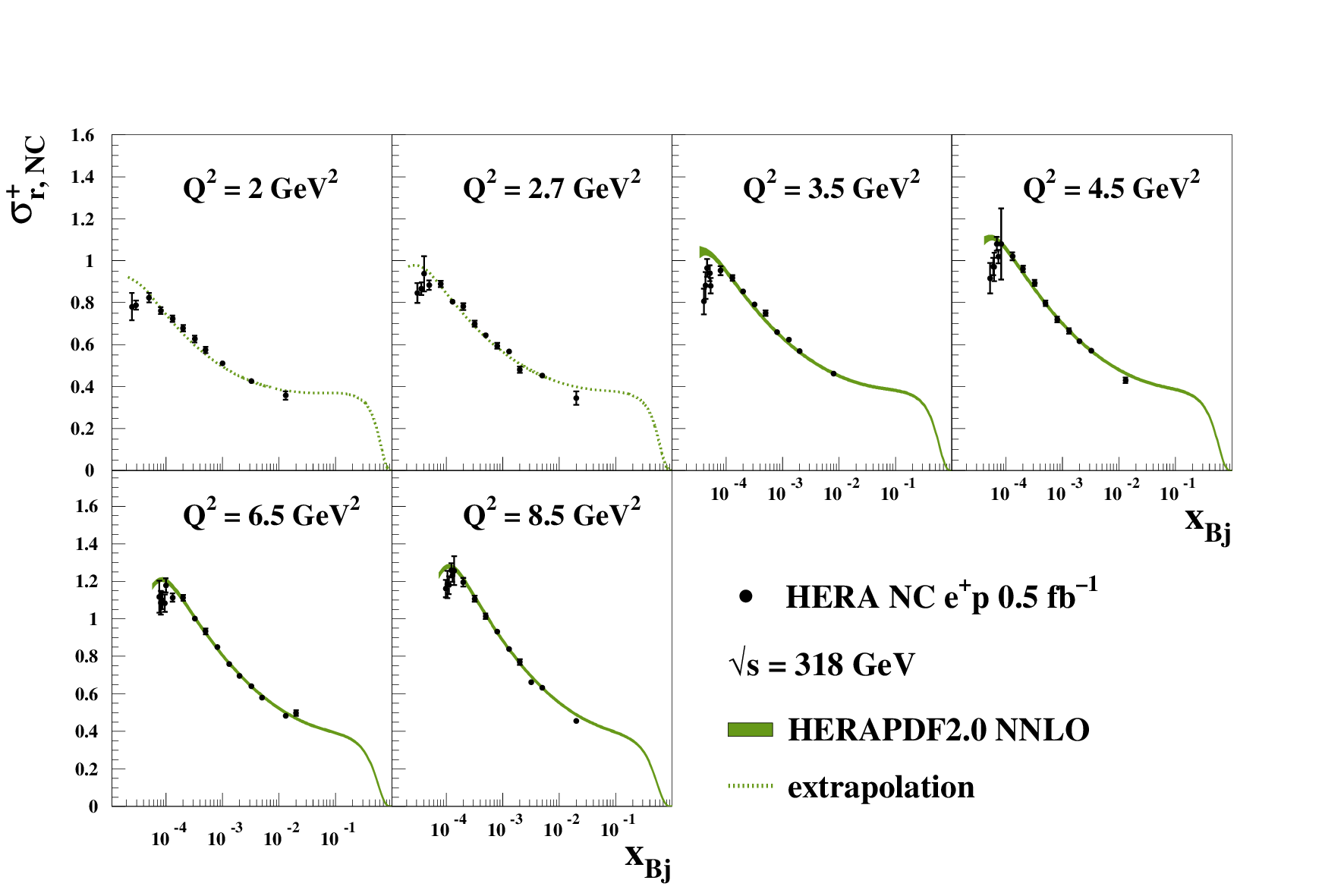,width=0.9\textwidth}}
\vspace{0.5cm}
\caption {The predictions of HHT NNLO (top) and HERAPDF2.0 NNLO 
          (bottom),
          both with $Q^2_{\rm min} = 3.5$\,GeV$^2$, 
          compared to the HERA measurements of $\sigma_{r}$. 
          The bands represent the experimental, i.e.\ fit, uncertainties.
          Extrapolations are indicated as dotted lines. From~\cite{Abt:2016vjh}
}
\label{fig:NNLO:HHT-HPDF}
\end{figure}

\begin{figure}[tbp]
\vspace{-0.3cm} 
\centerline{
\epsfig{figure=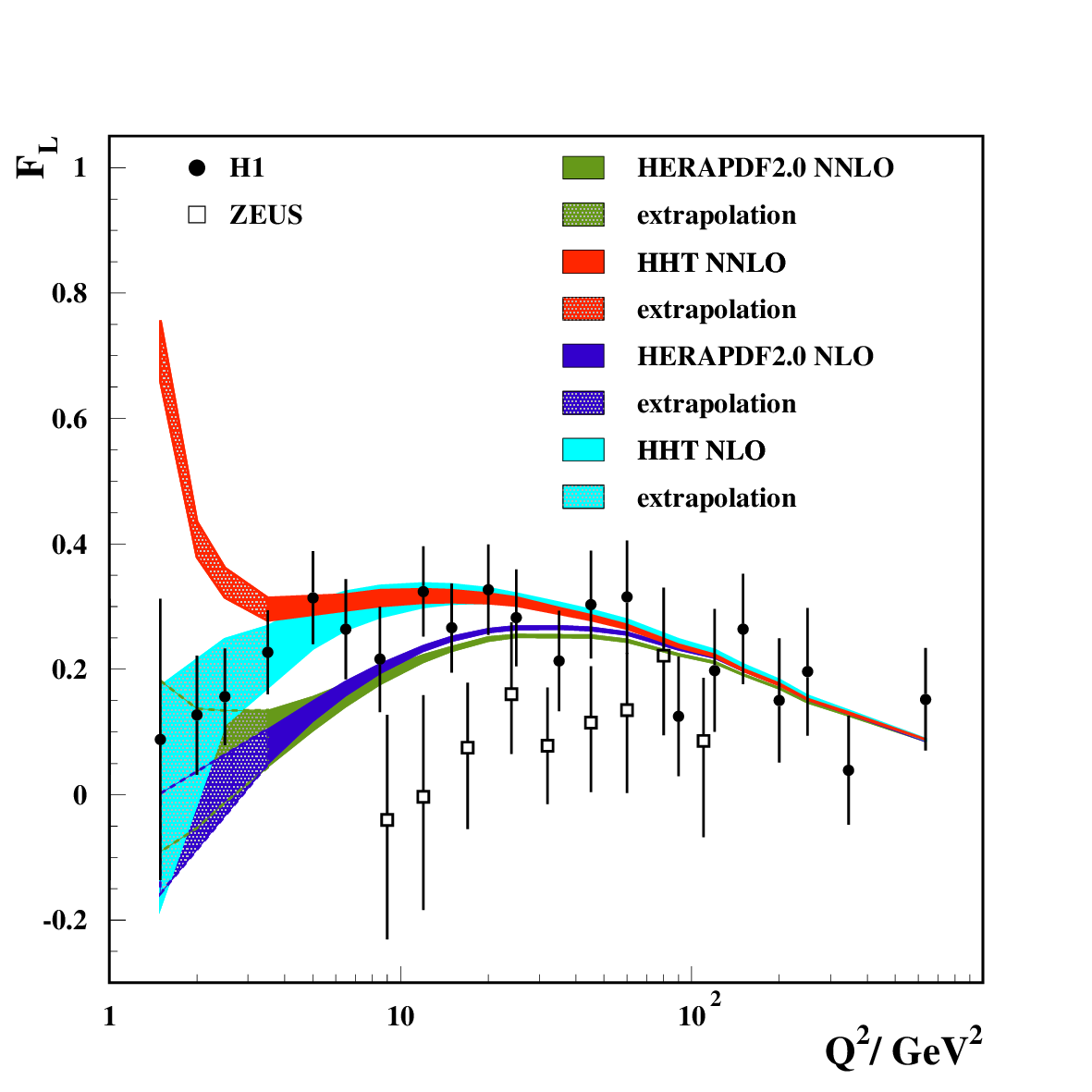,width=0.9\textwidth}}
\vspace{0.5cm}
\caption {The predictions for $F_{\rm L}$ from HHT and HERAPDF2.0,
          both with $Q^2_{\rm min} = 3.5$\,GeV$^2$, 
          compared to the separate direct measurements published by the 
          H1 and ZEUS collaborations.
          The bands represent the experimental, i.e.\ fit, uncertainties 
          on the predictions. Hatched bands represent extrapolations. From \cite{Abt:2016vjh}
}
\label{fig:flht}
\end{figure}

Fig~\ref{fig:NNLO:HHT-HPDF} shows the corresponding NNLO fits to the reduced cross section, with and without the higher twist term included in $F_L$.
The turn over of the reduced cross section at low-$x$ is much better described by the higher twist fit (HHT) because $F_L$ has
been increased by the higher twist contribution. Fig.~\ref{fig:flht} shows the improved fit to $F_L$ data for both NLO and NNLO fits. It also shows that this picture is not successful for very low $Q^2$, $Q^2 \leqsim 2$GeV$^2$.
Furthermore, the PDFs of HHT are very similar to the PDFs of HERAPDF2.0,
even though
the values of $A_{\rm L}^{\rm HT}$ extracted are quite high: e.g.  
$A_{\rm L}^{\rm HT}=5.5\pm0.6$~GeV$^2$ from the NNLO fit.
Thus the gluon PDF in this picture remains suppressed, which is somewhat contradictory to the idea that it is a high gluon
density reponsible for the gluon-gluon interactions which generate higher twist terms. 

\clearpage
\subsection{Interpretation of $F_2^{c\bar{c}}$}
\label{sec:f2ccinterp}

\begin{figure}[h]
\center
\epsfig{file=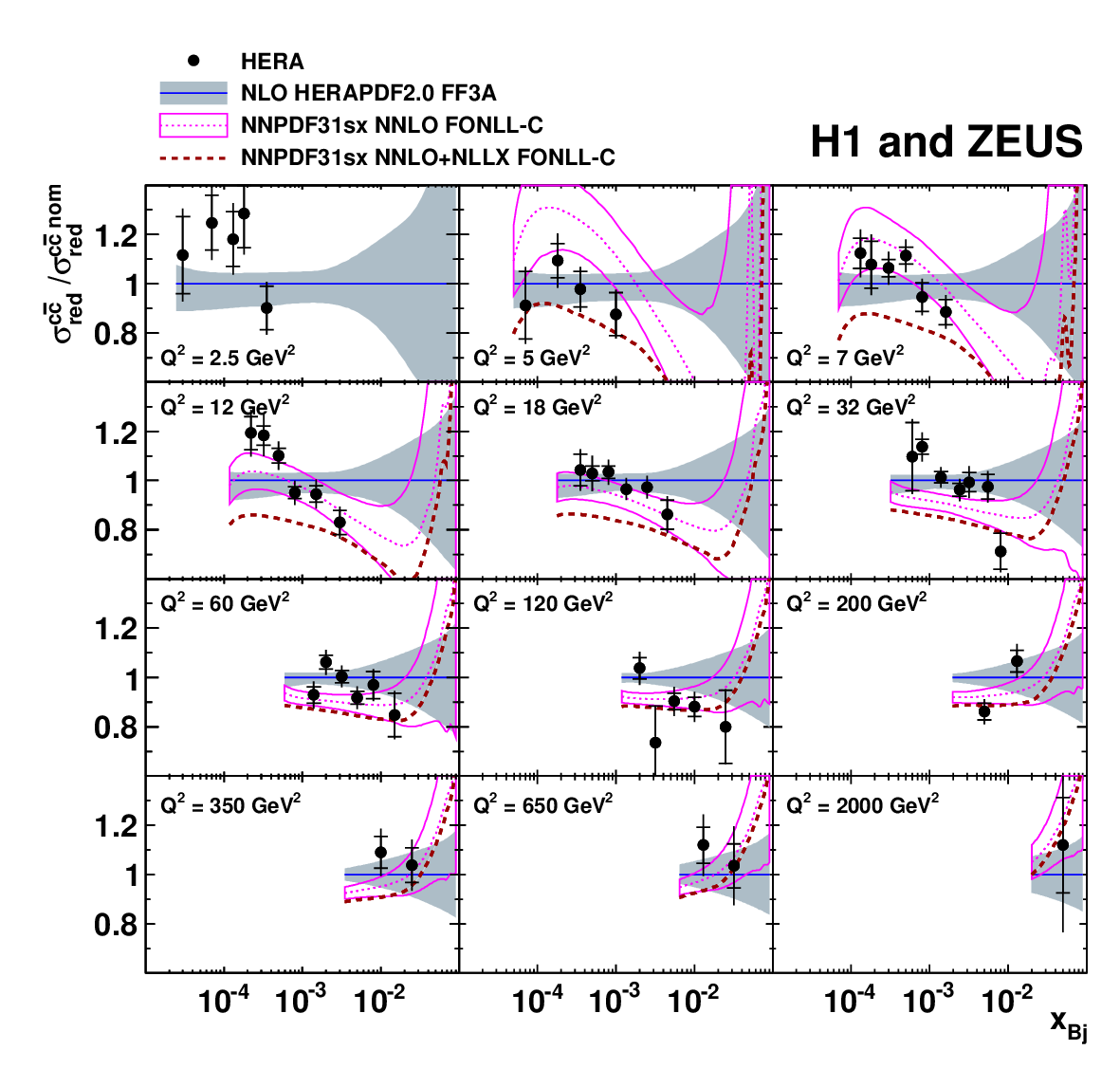 ,width=1.0\textwidth}
\caption{
Ratio of reduced charm cross sections as a function of $x$ for given values of $Q^2$ for the combined data~(full circles) and the NNLO VFNS predictions of the NNPDF group with respect to the FFNS predictions, $\sigma^{ c\overline{c}\ {\rm nom}}_{\rm red}$, obtained using HERAPDF2.0 FF3A (solid lines with shaded uncertainty bands).
Results from two different GMVFN calculations are shown: without (FONLL-C, dotted lines with uncertainty bands) and with low-$x$ resummation (FONLL-C+NLLsx, dashed lines). For the calculations the NNPDF3.1sx PDF set is used.
For better clarity of the presentation the uncertainties of the FONLL+NLLsx calculations are not shown. These are in size similar to those shown for the FONLL calculations. From ~\cite{H1:2018flt}.
 }  
\label{fig:data-charm-theory-nnpdf-ratio} 
\end{figure}
In Fig.~\ref{fig:data-charm-theory-nnpdf-ratio} we show the data on the charm reduced cross section, as a function of $x$ for
various values of $Q^2$,  in ratio to the predictions of the HERAPDF FFN analysis (as the central line) but also to two
different GMVFN calculations (FONLL) of the NNPDF collaboration both with and without $\ln(1/x)$ resummation.
In both cases the PDFs
used are the NNPDF3.1sx set, which included $\ln(1/x)$ resummation calculations for the inclusive DIS HERA data to which
it was fitted. Although neither of these calculations is fully successful they do desribe the shape of the data better than the standard FFN calculation.

This could be regarded as additonal evidence for $\ln(1/x)$ resummation in the HERA data, but one should beware that the path 
from the observation of of a $D$ or $D^*$ to the reduced charm cross section involves much more uncertainty than is involved in inclusive cross section measurements.

\subsection{Searching for BFKL in forward jets}
\label{sec:forwardjets}
Further evidence for BFKL type resummation from semi-inclusive processes comes from the observation of forward jets.
Considering the gluon emission diagrams of Fig.~\ref{fig:ladder} we can see that for a $p_t$ ordered gluon ladder such as
requited by the DGLAP framework (at leading order) there can be no possibility of a hard (ie high-$p_t$) jet in the forward direction
(ie collinear with the proton), because the gluons are assumed to have zero $p_t$ at the start of the ladder gaining $p_t$ at each splitting in an ordered manner.
However for a disordered BFKL ladder forward jet production is  possible.
Hence to look for hard jets emitted in the forward direction
the experiments making cuts on jets as follows: $E_{T,jet} > 3.5, 5$ GeV, $0.5 < E_{T,jet}^{2} / Q^{2} < 2$,
to ensure hard jets of
comparable scale to $Q^2$; $x_{jet} > 0.035$, $1.7 < \eta_{jet} < 2.8$ to ensure that they are forward jets.

\begin{figure*}
 \vspace{3cm}
 \begin{picture}(120,270)
 \put(-5,-50){\epsfig{file=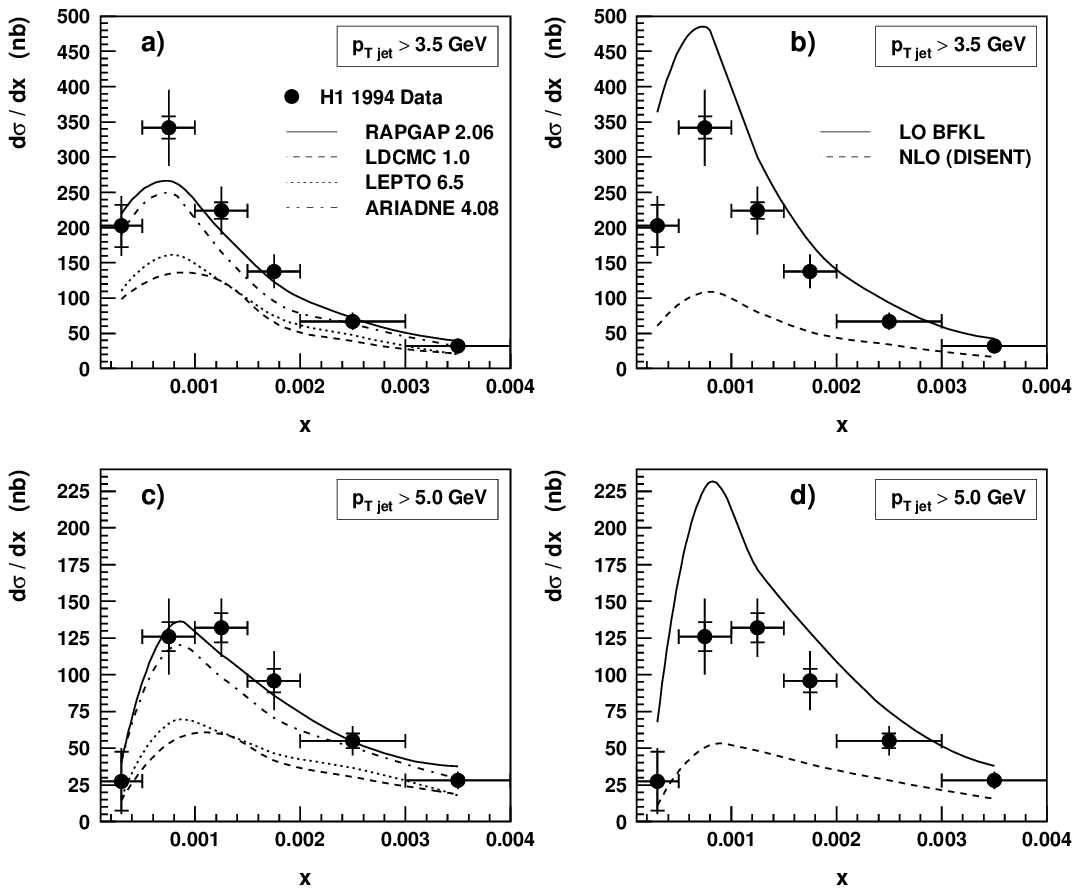,width=14cm,height=20cm}
        }
 \end{picture}
\vspace{-3cm}
\caption{Forward jet data of H1, \cite{H1:1998cuj}, as a function of Bjorken-$x$
 for two $E_{T_{jet}}$ cuts of $3.5$~GeV and $5.0$~GeV. (a) and (c)
 contain Monte-Carlo model predictions, whereas (b) and (d) show the 
 results of a LO BFKL (full) and a fixed NLO (dashed) calculation.
 \label{HERAf}} 
\end{figure*}

In Fig.~\ref{HERAf} the data from H1 in
the forward region is shown for two cuts $E_{T,jet} > 3.5$~GeV and 5~GeV, together with predictions from various
Monte-Carlo models, on the left hand-side, and from an LO BFKL calculation and an NLO (DGLAP) calculation (DISENT), on the
right-hand-side.

The Monte-Carlo models ARIADNE,
\cite{Lonnblad:1992tz}, LEPTO, \cite{Ingelman:1996mq}, RAPGAP,~\cite{Jung:1993gf},  and 
LDC, \cite{Kharraziha:1997dn}, have been used for predictions. ARIADNE includes one
of the main features present in the BFKL approach, which is the
absence of the strong $k_T$ ordering. LDC is based on the CCFM approach,~\cite{Ciafaloni:1987ur,Catani:1989sg,Catani:1989yc,Marchesini:1994wr},
(see Sec.~\ref{sec:lowq2interp}).  LEPTO is based on conventional
leading log DGLAP evolution and RAPGAP is also based on DGLAP evolution but
contains an additional resolved virtual photon component.
LDC and LEPTO lie below the data by a factor of 2, whereas
ARIADNE gives a reasonable good description. As an interesting result,
RAPGAP also gives a good description of the H1 data.

It is clear that the conventional DGLAP calculation of DISENT undershoots the data. However, is also
clear that the BFKL calculations at LO overshoot the data.
These early BFKL 
calculations suffer, however, from several deficiencies. They are
asymptotic and do not contain the correct kinematic constraints of the
produced jets. Furthermore they do not allow the implementation of a
jet algorithm as used in the experimental analysis. Also NLO
$\ln(1/x)$ terms in the BFKL kernel predict large negative
corrections which are expected to reduce the forward cross section as
well. Recently the BFKL calculations have been improved by taking into
account higher order consistency conditions as a way of including
sub-leading corrections to the BFKL equation, \cite{Kwiecinski:1999hv}. The
consistency constraint requires that the virtuality of the emitted
gluons along the chain should arise predominantly from the transverse
components of momentum.

\begin{figure*}
\vspace{-5cm}
 \begin{picture}(120,390)
  \put(30,-3){\psfig{file=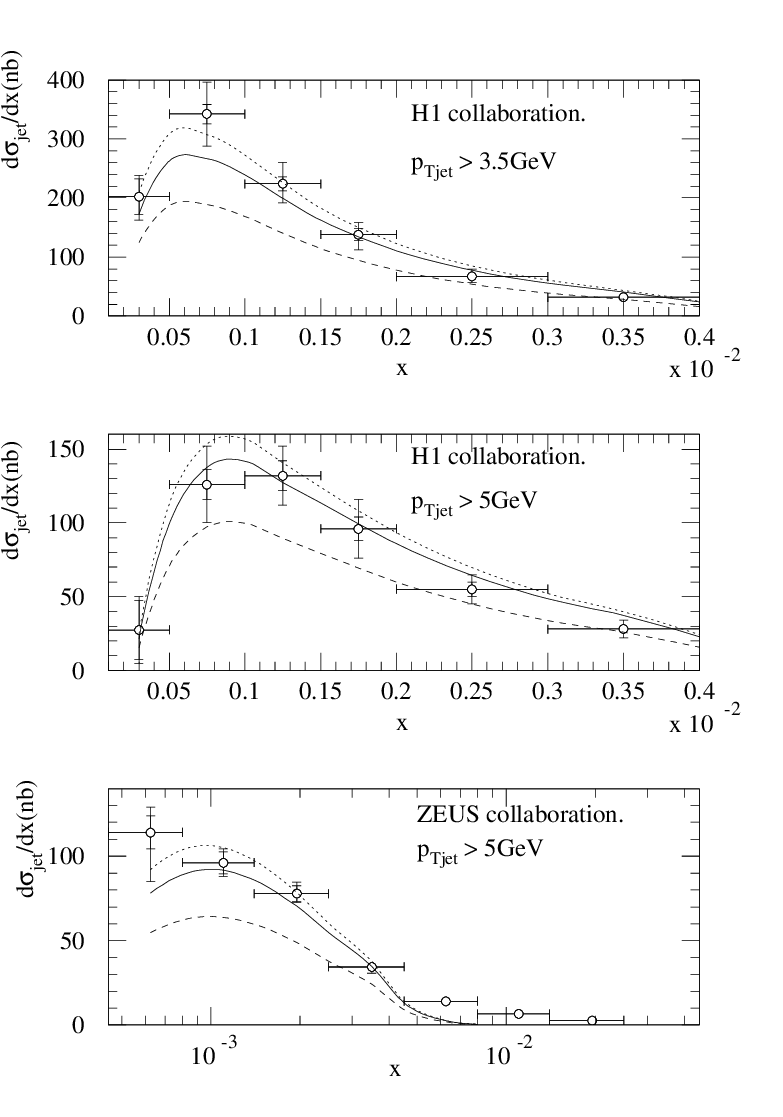,height=10cm}} 
 \end{picture}
\caption{Forward jet data of H1, \cite{H1:1998cuj}, and ZEUS, \cite{ZEUS:1998uhq}, compared
  to prediction based on the BFKL formalism including sub-leading
  corrections, \cite{Kwiecinski:1999hv}. The three curves correspond to variation of model choices, such as the choice of scale
  and of an infra-red cut off (perturbative calculations cannot be done at low scale when the string coupling parameter
  becomes large). \label{outh}}
\end{figure*}

Including the consistency conditions, good agreement between the predictions
and the forward jet data is found, for both the H1 and the ZEUS data,
as shown in Fig.~\ref{outh}. 

Thus the BFKL picture appears successful, but this is not defnitive because the conventional picture also describes the
data IF a resolved photon component is included. This involves the idea that the virtual photon exchanged in the DIS
scattering can have a 'resolved' as well as a 'point-like' component. The resolved comonent arises from the idea that the
virtual photon can split into a $q-\bar{q}$ pair and thence into more partons, i.e. that it has a structure. This phenomenon
is established in the lower $Q^2$, $Q^2\leqsim 5$GeV$^2$ data at HERA. Unfotunately, since BFKL effects are expected at low $x$, the kinematics of HERA implies low $x$ is only probed at relatively low $Q^2$ and thus it is not possible to distinguish
between these explanations cleanly. We simply note that the idea of the photon haveing a $q\bar{q}$ component has something in common with the dipole picture which we will discuss later in the context of saturation, see Sec.~\ref{sec:lowq2interp}.

\clearpage

\subsection{Very low $Q^2$ interpretation}
\label{sec:lowq2interp}

\subsubsection{$\sigma(\gamma^*p)$}

We have seen a change of behaviour of $F_2$ at low $Q^2$ such that it stops rising steeply.
To understand this it is helpful to look at the HERA data from another point of view. 
 
At small $x$ $F_2$ is
related to the $\gamma^*p$ total cross-section
\begin{equation}
\sigma_{tot}^{\gamma^*p}(W^2,Q^2)\equiv\sigma_T+\sigma_L=
{4\pi^2\alpha\over Q^2}F_2(x,Q^2),
\label{eqn:gamma*p}
\end{equation}
where $W^2\approx Q^2/x$ is the centre of mass energy for the $\gamma^*p$ subprocess . The above equation shows that
$F_2(x,Q^2)\to {\rm const.}Q^2$ as $Q^2\to 0$, since the real photon proton scattering cross section,
$\sigma_{tot}^{\gamma p}$ is finite. The surprising features of the HERA data are that perturbative QCD
appears to describe the data to really quite low values of $Q^2$
and that the strong rise of $F_2$ at small persists to even smaller
$Q^2$ values. How can we describe the data for $Q^2 \leqsim 1$GeV$^2$?

The relation~\ref{eqn:gamma*p} implies
that a rise of $F_2$ at small-$x$ corresponds to a rising $\gamma^* p$ 
cross-section with $W^2$. This suggests
another approach to understanding the behaviour of $F_2$ at low $x$ and
that is the framework of Regge theory that was used to explain the high energy
behaviour of hadron-hadron total cross-sections. Regge theory predicts that 
the high energy behaviour of hadronic scattering amplitudes is 
$\ds {\rm Im} A(ab \to cd) \sim \sum_i \beta_is^{\alpha_i}$, where
$\alpha_i$ is the intercept of a Regge trajectory which has the 
right quantum numbers for an exchange in the crossed channel 
$a\bar{c} \to \bar{b}d$. Using the optical theorem, the corresponding total cross-section
will vary as $\ds \sigma^{tot}(ab) \sim \sum_i \beta_is^{\alpha_i -1}$. 
For the corresponding forward elastic scattering amplitude, the leading  Regge 
trajectory has the quantum numbers of the vacuum and is known as the Pomeron,
for which the value $\alpha_P \approx 1.08$ has been determined from 
hadron-hadron data. This prediction also describes the high energy behaviour
of real-photon hadron cross-section measurements successfully. 
The model was extended to describe virtual-photon proton scattering by ~\cite{Donnachie:1993it} (DL) who assumed that any $Q^2$ 
dependence would reside in the functions $\beta_i(Q^2)$ only
and that the intercepts, $\alpha_i$ would be independent of $Q^2$.
Using Eq.~\ref{eqn:gamma*p} this approach then predicts a flattish input 
for the gluon and sea PDFs, since $\sigma(\gamma^* p) \sim s^{\alpha_i-1}$ 
implies $F_2 \sim x^{1-\alpha_i} = x^{-0.08}$. This prediction had been so successful in hadronic physics
that it led to our expectation that the $x$ dependence of the low $Q^2$ input, for the momemntum distributions of
sea quarks and gluons, to the DGLAP evolution should be flat. Indeed this was successful in the early DIS data which did not
probe lower than $x\sim 0.01$~\footnote{The Regge 
picture also led to the prediction $xF_3 \sim x^{+0.5}$, which was a successful description of the $xF_3$ structure funtion
measured in early (anti-)neutrino scattering}.

While the DL approach successfully described the pre-HERA low $x$ data for 
$Q^2$ values up to about $10\,$GeV$^2$, it cannot describe the steeply 
rising $F_2$ data measured at HERA. This has already been shown
in Fig.~\ref{fig:lambdaeff} which shows the slope
$\ds \lambda (Q^2) = {\partial\ln F_2\over \partial\ln (1/x)}$ or  
$\ds F_2 \sim x^{-\lambda(Q^2)}$, at low $x$. 
For $Q^2$ values less
than $1\,$GeV$^2$ or so, the value of $\lambda$ is consistent with that
from hadronic Regge theory, whereas for $Q^2>1\,$GeV$^2$ the slope
rises steadily to reach a value greater than $0.3$ by 
$Q^2\approx 100\,$GeV$^2$. This larger value of $\lambda$ is close to that expected from
using BFKL ideas. Indeed the
BFKL equation could be viewed as a method for calculating the slope of the Pomeron trajectory
using perturbative QCD. However, the result does not correspond to the measured Pomeron slope
in low energy hadron-hadron collisions. Thus the idea of two Pomerons arose: a hard or
perturbative Pomeron trajectory with intercept $\sim 1.3-1.5$ and the soft or 
non-perturbative Pomeron of hadronic physics with intercept
$\sim 1.08$. DL, \cite{Donnachie:2001he}, extended their Regge model 
by the addition of a hard Pomeron with intercept $1.44$, which allowed them
to describe the low $x$ HERA data up to $Q^2$ values of a few hundred
GeV$^2$. Theirs is one of many attempts, using ideas from Regge theory,
to model the `transition region' from real photoproduction through to deep 
inelastic scattering with $Q^2>>1\,$GeV$^2$.

\begin{figure}[tbp]
\vspace{-0.5cm} 
\centerline{
    \includegraphics[width=0.9\textwidth]{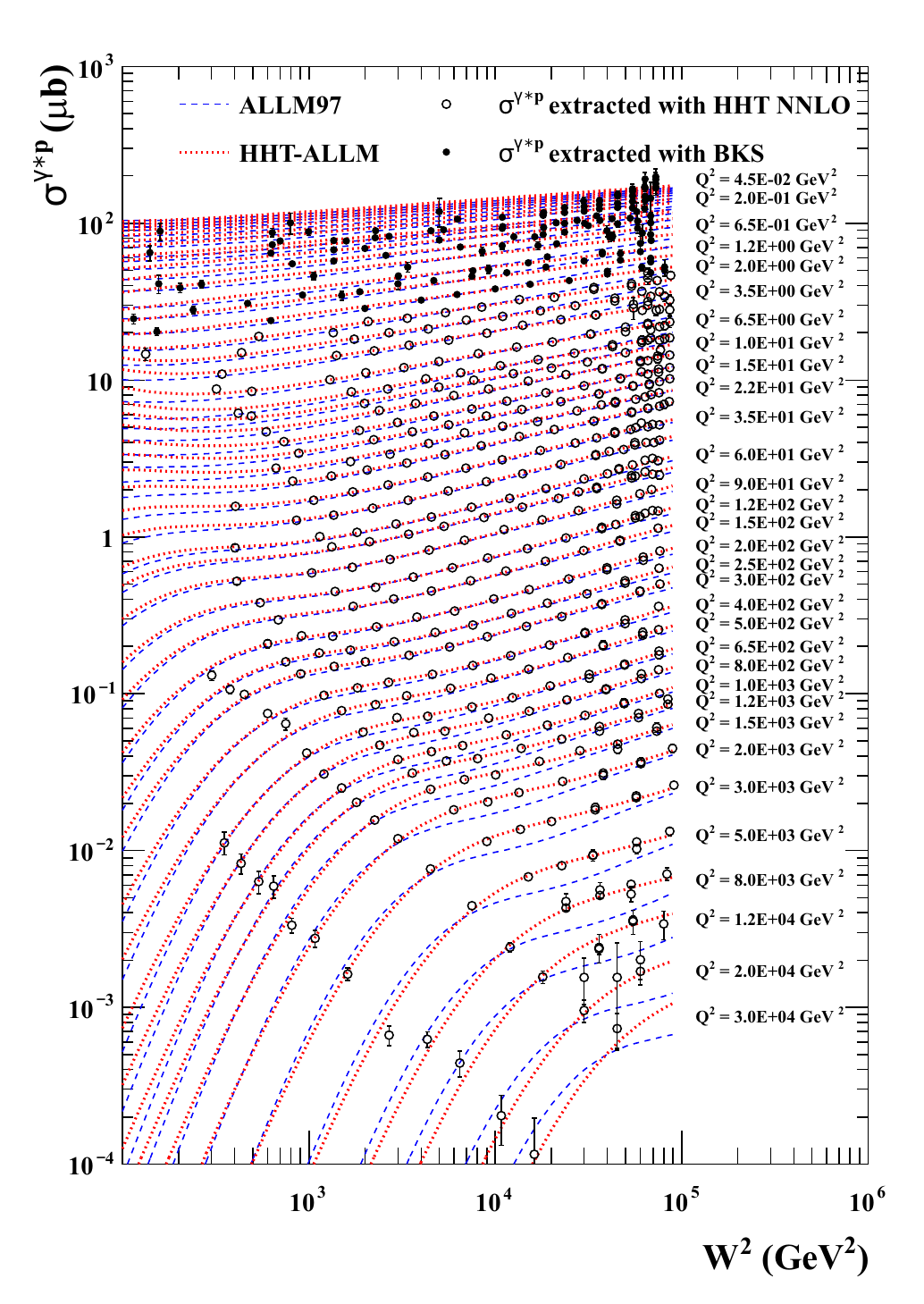}}
\caption{The virtual-photon proton cross section, 
         as extracted from the HERA
         $e^+p$ NC data at $\sqrt{s}=318$ and 300\,GeV, plotted against its centre of mass energy $W^2$.
          From~\cite{Abt:2017nkc}.}
\label{fig:sigma-W}
\end{figure}
\begin{figure}[tbp]
\vspace{-0.5cm} 
\centerline{
    \includegraphics[width=0.9\textwidth]{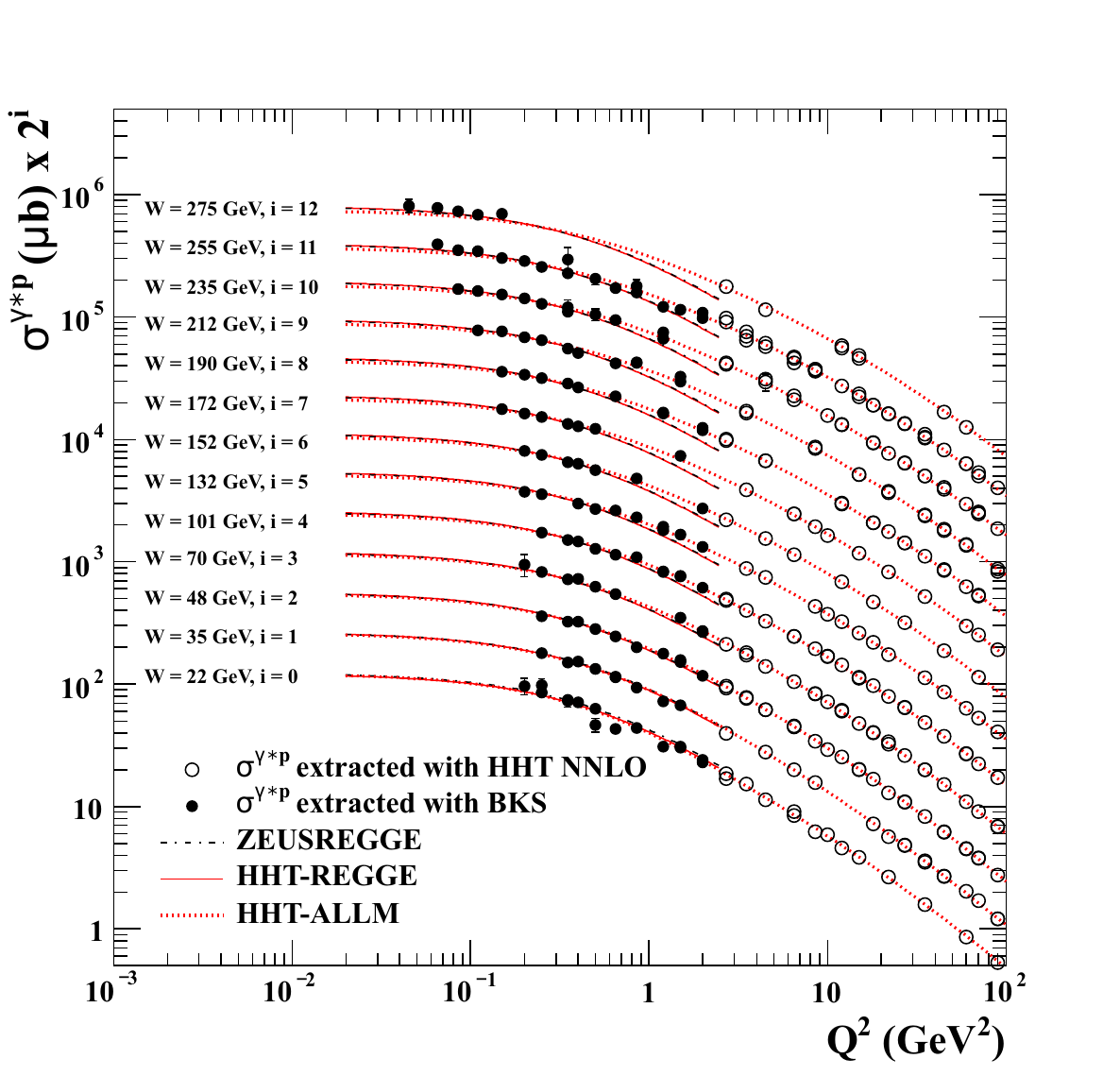}}
\caption{$\sigma^{\gamma^*p}(Q^2)$ in the low-$Q^2$ regime
         as extracted from the HERA
         $e^+p$ NC data at $\sqrt{s}=318$ and 300\,GeV for
         selected values of $W$, together with the predictions
         of the ALLM parametrisations and the Regge predictions.
         The values of $\sigma^{\gamma^*p}(Q^2)$ are multiplied
         by $2^{\rm i}$ for different values of $W$  to make
         the curves individually visible. From~\cite{Abt:2017nkc}.}
\label{fig:sigma-low-Q2}
\end{figure}
Figs.~\ref{fig:sigma-W} and~\ref{fig:sigma-low-Q2} show the data in terms of $\sigma^{\gamma^*p}$ as a function of $W^2$ and $Q^2$ respectively. Note that this cross section is derived from $F_2$ and thus the contribution of $F_L$ to the DIS reduced
cross section must be accounted for in its extraction. This is done in two ways: i) using a parametrisation of $F_L$ data
(BKS),~\cite{Badelek:1996ap}, and
ii) using a pQCD fit with higher twist (HHT),~\cite{Abt:2016vjh}.

Fig~\ref{fig:sigma-W} shows the rise of the $\sigma^{\gamma^*p}$ with $W^2$, for $Q^2$ values from $0.045$GeV$^2$ to
$30,000$GeV$^2$. At the lowest $Q^2$, the data show a slow rise compatible with the original Regge prediction of $(W^2)^{0.08}$.
As $Q^2$ rises, the slope also rises towards the hard Pomeron value $\sim 0.3$, as already summarise in terms of the slope of $F_2$ with $\ln(1/x)$ in Fig.~\ref{fig:lambdaeff}. Note that beyond $Q^2\sim 200$GeV$^2$ the data can no longer be described in terms of a simple slope. This is because these data are no longer at low $x$, $x \leqsim 0.01$, and Eqn.~\ref{eqn:gamma*p} does not apply.
The data for all $Q^2$ values are parametrised by the ALLM parametrisation,~\cite{Abramowicz:1991xz,Abramowicz:1997ms}, which used early HERAdata, and the HHT-ALLM,~\cite{Abt:2017nkc}, which used the final combined HERA I+II data.

Fig.~\ref{fig:sigma-low-Q2} shows the data as a function of $Q^2$ for data which are all in the low $x$ region, for different $W$ values. For all $W$, the data is compatible with a
smooth transition in the $Q^2$ range $0.1-1\,$GeV$^2$ from an 
approximately constant behaviour to $\sim 1/Q^2$ at higher $Q^2$. From Eqn~\ref{eqn:gamma*p} this implies approximate Bjorken
scaling. In fact, as we have already seen, these higher $Q^2$ data are fully compatible with the slow logarithmic scale breaking of pQCD. This figure also makes clear the limitations of the Regge descriptions which only give a good fit at low $Q^2$

The advantages of looking at the data in terms of $\sigma^{\gamma^*p}$ will become evident in the next subsection.

\clearpage
\subsubsection{Non-linear evolution equations, saturation}

Either considering the rise of $F_2$ as $x\to 0$ or the steep energy
dependence of $\sigma(\gamma^*p)$ at large $Q^2$, such behaviour will 
eventually violate the Froissart bound. If the origin of the rise arises
from a high gluon density then this problem may be avoided because the 
gluons can shadow each other from the $\gamma^*$. At even higher
densities the gluons will `recombine' via the process $gg\to g$ -- the inverse 
of gluon splitting -- and the gluon density, and hence $F_2$, will saturate.
These ideas were formalized in the simplest terms by Gribov, Levin \& Ryskin by the
addition of a non-linear term to the equation for gluon evolution
\begin{equation}
\frac{d^2 xg(x,Q^2)}{d lnQ^2 dln(1/x)} = \frac{3\alpha_s}{\pi} xg(x,Q^2) -
\frac{81\alpha_s^2}{16 Q^2 R^2}\left[xg(x,Q^2)\right]^2 
\end{equation}
where $R$ is the proton radius.
When $xg(x,Q^2) \sim \pi Q^2 R^2 /\alpha_s(Q^2)$ the non-linear
term cancels the linear term and evolution stops, this is saturation.
For a nucleon radious of 1fm = 5GeV$^{-1}$ and scale $Q^2= 1$GeV$^2$, taking $\alpha_s=0.5$, this is a very high gluon density $\sim 150$, but it has been suggested that there could be hot-spots wihin the nucleon of R=2GeV$^2$, yielding a gluon density of
$\sim 24$.

\begin{figure}[tbp]
\vspace*{13pt}
\centerline{\psfig{figure=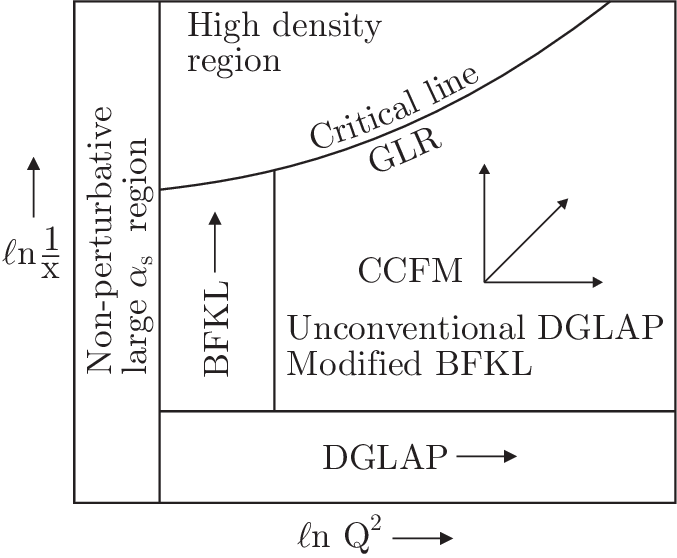,height=.35\textheight}} 
\caption{Approaches to physics at low-$x$. Courtesy of A D Martin}
\label{fig:admmap}
\end{figure}

The various evolution equations applicable across the $x,Q^2$ 
plane are summarised schematically 
in Fig.~\ref{fig:admmap}.
At small $Q^2$ non-perturbative effects must be accounted for no matter what the value of $x$ because the strong coupling is large.
For higher $Q^2$, but not too low $x$, the DGLAP $\ln(Q^2)$ resummation applies. At low $x$, for $Q^2$ above the non
perturbative region but not very high, the BFKL $\ln(1/x)$ resummation applies. At low-$x$ and higher $Q^2$ some combination of these resummations must be made. Such a 'modified' DGLAP approach was discussed in Section~\ref{sec:lnxresum}. A somewhat different picture, considering the $p_t$ of the gluon emissions, is the CCFM equation. 
Note the appearance of the `critical line' above which the gluon density
is in principle high enough for non-linear effects to be important. In this region,
for large enough $Q^2$, one has the possibility that $\alpha_s$ will be 
small enough for weak-coupling but non-perturbative methods to be applicable, 
of which the GLR equation would be just a first approximation.
No scales are given on the diagram of Fig~\ref{fig:admmap} because it is a matter of debate where each region begins. Some argue that the non-perturbative region should extend to higher $Q^2$ as x decreases
subsuming the high density region, others that the high density region starts lower in $x$ and subsumes the BFKL region.
One of the many hotly argued questions raised by the HERA data is does HERA have the reach to see high density effects and have they been
seen? We will first consider this in terms of inclusive variables.

The most successful such approach to the high density regime is that of the Colour Glass Condensate (CGC), which is an effective
field theory for saturated gluons. In this theory the usual parton model description is replaced by classical strong gluon fields.
However, quantum fluctuations around the classical
solution are enhanced by terms proportional to $\alpha_s ln(1/x)$ and these require resummation by non-linear
evolution equations. The Balitsky-Kovchegov (BK),~\cite{Balitsky:1995ub,Kovchegov:1999yj,Kovchegov:1999ua}, equation is such an equation, valid at high $N_c$, which reduces to the BFKL equation at
low gluon density. The Jalilian-Marian, Iancu, Mclerran, Weigert, Leonidov, Kovner (JIMWLK) equation is a more exact alternative.
The phenomenology of these approaches can be described in terms of the dipole models, to which we now turn.
The CGC is discussed in detail in the contribution of E. Iancu to this encyclopedia.

Dipole models have proved to be a very fruitful approach in exploring low-$x$ physics
by modelling the behaviour of $F_2$ or $\sigma(\gamma^*p)$
through the transition region towards $Q^2=0$. The idea is that the 
virtual photon splits into a $q \bar{q}$ pair\footnote{As for example shown
by the quark box at the top of Fig.~\ref{fig:ladder}.} 
(a colour dipole) with a transverse size $r\sim 1/Q$, the splitting 
occurring a `long' time $\sim 1/(mx)$ before the dipole interacts with the 
proton. The dipole then scatters coherently from the proton in a time 
which is short in comparison to $\sim 1/(mx)$. 
The $\gamma^*\to q \bar{q}$ 
process is described by QED and the strong interaction physics 
is then contained in the cross-section, $\hat{\sigma}$,  for the 
dipole-proton interaction.

Many authors have worked on these models in
the context of DIS processes both inclusive and diffractive (see Sec.~\ref{sec:diffraction} for a discussuion of diffractive proceses). Here
 we outline the idea using the simplest possible dipole model of Golec-Biernat and W\"usthoff (GBW), 
\cite{Golec-Biernat:1999qor}. The original GBW model provides a simple
parameterization of $\hat{\sigma}$ with explicit saturation, in the sense
that $\hat{\sigma}\to \sigma_0$ (a constant) as $r$ becomes large, but 
in a such a way that the approach to saturation is $x$ dependent, controlled
by the `saturation radius', $R_0$. Explicitly the dipole cross-section is
given by
\begin{equation}
\hat{\sigma}(x,r^2)=\sigma_0 (1-\exp(-\hat{r}^2)),~~~\hat{r}={r\over 2R_0(x)}
\label{eq:gbw-g}
\end{equation}
where $\sigma_0$ is a constant, and
\begin{equation}
R_0(x)^2={1\over Q_0}\left({x\over x_0}\right)^{\lambda},
\end{equation}
$R_0(x)^2$ is understood to be inversely proportional to the 
gluon density ($\lambda \sim \lambda_g$), so that $R_0$ is a measure of the
transverse separation of the gluons in the target. Thus when the dipole 
separation is large compared to the gluon separation  (small $Q^2$ and 
small $x$) the dipole cross-section saturates, $\hat{\sigma} \sim \sigma_0$, 
and $\sigma(\gamma^*p)$ also tends to a constant, giving 
$F_2 \propto 1/Q^2$ from Eq.~\ref{eqn:gamma*p}. When the dipole separation is 
small compared to the gluon separation (high $Q^2$ and large $x$),
$\hat{\sigma} \sim \sigma_0/(Q^2 R_0^2)$ 
and $\sigma(\gamma^*p)$ varies as $1/Q^2$, so that $F_2$ exhibits Bjorken 
scaling. Because of the $x$ dependence of $R_0$, $\hat{\sigma}$ saturates
for smaller dipole sizes as $x$ decreases. 

\begin{figure}[tbp]
\vspace*{13pt}
\centerline{\psfig{figure=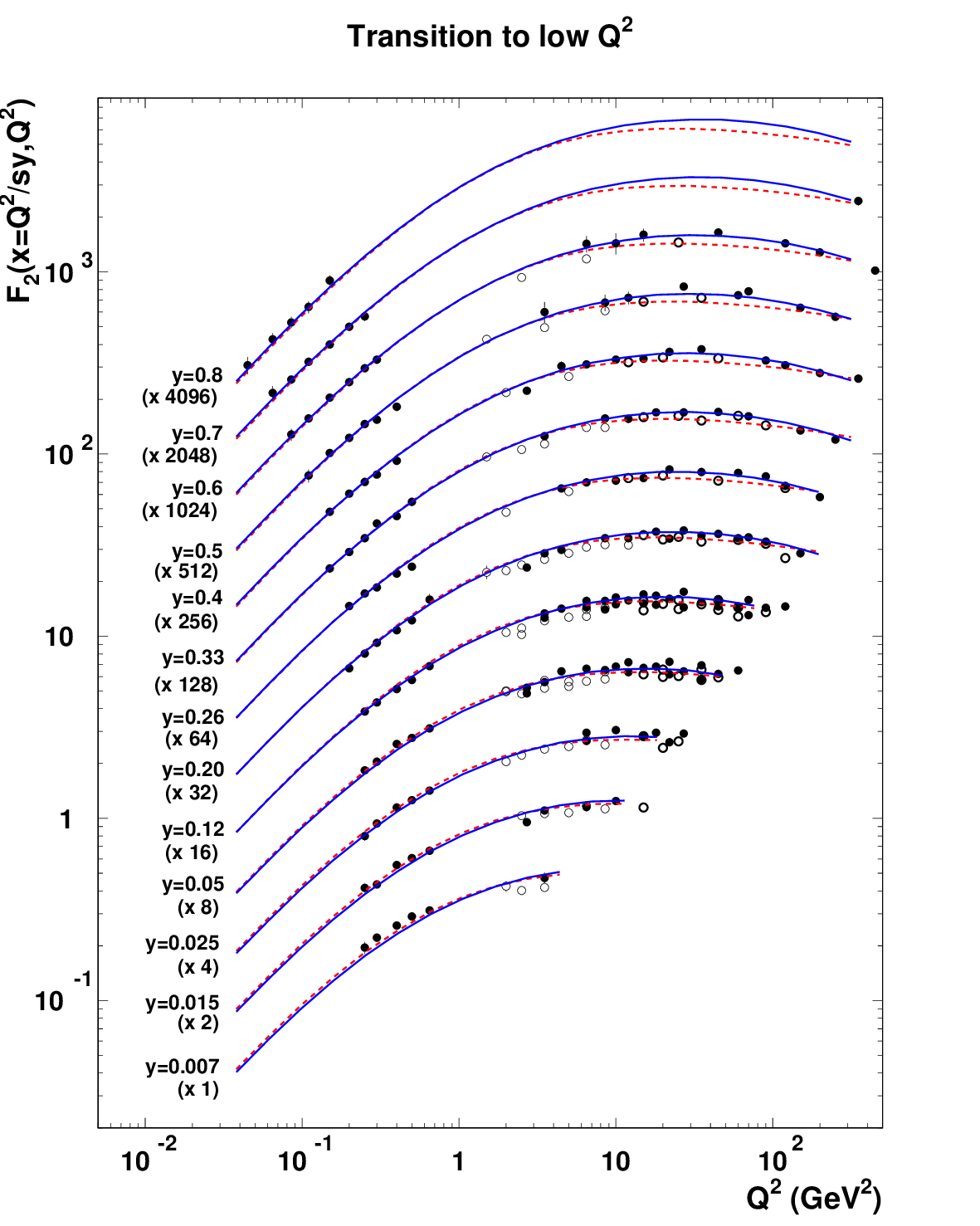,height=.45\textheight}} 
\caption{The GBW dipole model fit to $F_2$ data through the transition
region to photoproduction. The data are plotted versus $Q^2$ in bins
of $y$. The dashed curves show the original model, the full curve the
improved model with DGLAP evolution. From \cite{Bartels:2002cj}.}
\label{fig:gbw-f2}
\end{figure}

We have seen  this behaviour already in Fig.~\ref{fig:sigma-low-Q2}. Let us look at it again in Fig.~\ref{fig:gbw-f2} in terms
of the behaviour of $F_2$ as a function of $Q^2$ at fixed $y=Q^2/sx$.
In this figure the data are compared to the GBW dipole model.  We can see that 
the data do exhibit the features $F_2 \propto 1/Q^2$ at low $Q^2$ and $F_2\sim$ constant at high $Q^2$, as described above.
There is a clear change in behaviour 
around $Q^2\approx 1\,$GeV$^2$, which might then be taken as a rough estimate 
of the saturation scale for HERA data.
However, one needs to be somewhat careful before jumping to the conclusion that
the GBW dipole model shows that saturation effects have
been seen at HERA. Given Eq.~\ref{eqn:gamma*p},
$F_2 \sim 1/Q^2$ at low $Q^2$ is required in any model since
the photoproduction cross-section is finite, and at high $Q^2$ slow,
logarithmic scale breaking is required by pQCD. Thus the real 
challenge is not in describing the general
behaviour of $F_2$ and/or $\sigma(\gamma^*p)$  but in describing the exact shape of the data across the transition,
from low to high $Q^2$,
now that high precision data are available. 

The original
GBW model did not include QCD evolution in $Q^2$ and this limited its
success at larger $Q^2$. This deficiency was corrected in a second version,
\cite{Bartels:2002cj}, in which $\hat{\sigma}$ was related to the gluon density explictly, such that
gluon evolution with the scale $\mu^2 = C/r^2 + \mu_0^2$ is included.
\begin{equation}
\hat{\sigma}(x,r^2)=\sigma_0 (1-\exp(-\frac{\pi^2 r^2 \alpha_s(\mu^2)xg(x,\mu^2)}{3\sigma_0}) )
\label{eq:gbbk}
\end{equation}
Results
from both versions are shown in Fig.~\ref{fig:gbw-f2}.

\begin{figure}[tbp]
\vspace*{13pt}
\centerline{\psfig{figure=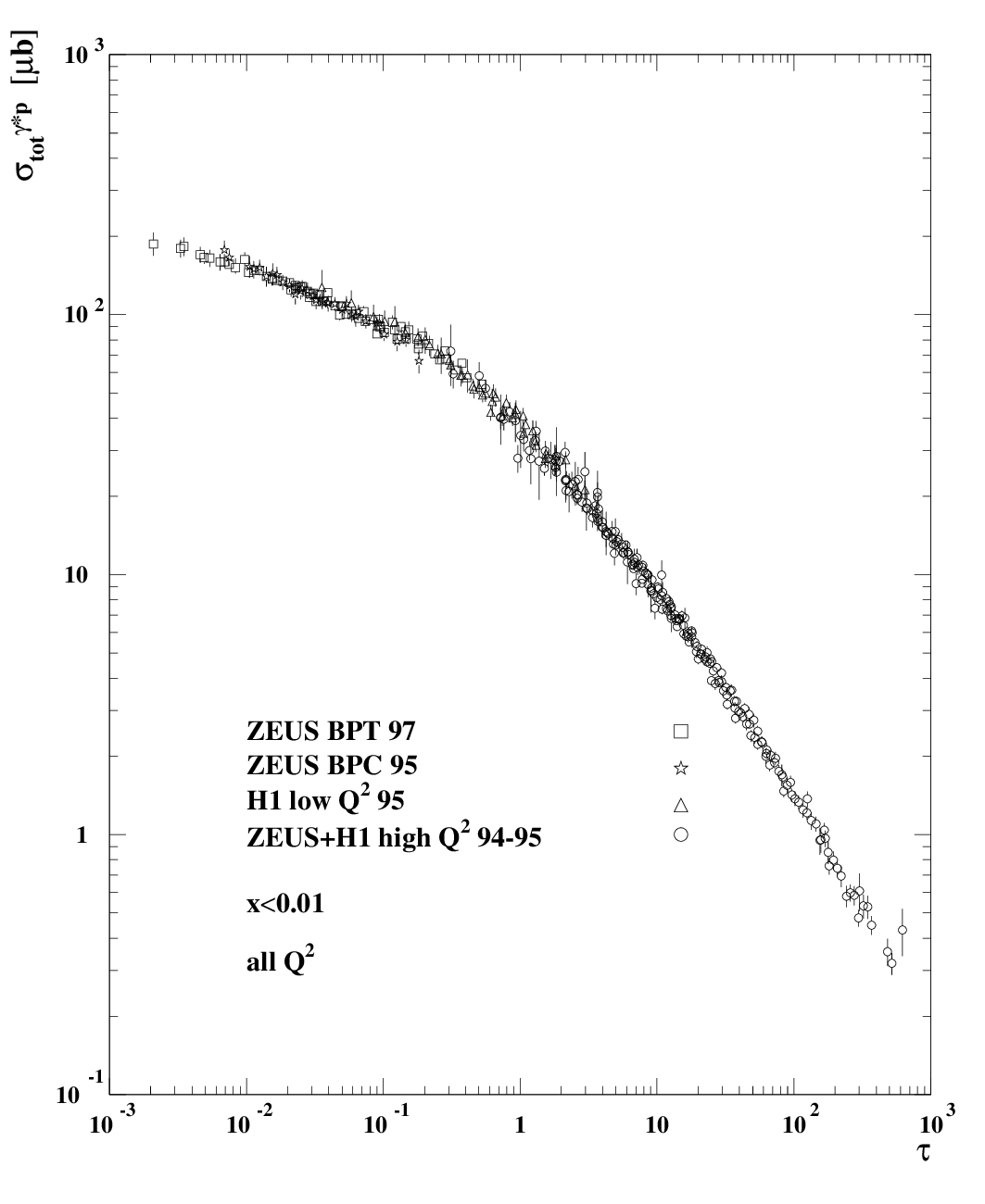,height=.45\textheight}} 
\caption{Geometrical scaling. Data on $\sigma(\gamma^*p)$ with
$x<0.01$ plotted versus the scaling variable $\tau=Q^2R^2_0(x)$.
From \cite{Stasto:2000er}.}
\label{fig:gscal}
\end{figure}
Another feature of saturation models which is illustrated in
the GBW model is the dependence of the dipole cross-section on a
scaled variable, here $r/R_0(x)$. Stasto, Golec-Biernat and Kwiecinski, 
\cite{Stasto:2000er}, have made the interesting observation that this leads to 
a new scaling 
property of $\sigma(\gamma^* p)$. At low $x$, $\sigma(\gamma^* p)$ should 
depend only on the dimensionless variable $\tau=Q^2R^2_0$. What is more, 
such a scaling property seems to be satisfied by HERA data. One expects
\[
\sigma(\gamma^* p)\sim \sigma_0~~(\tau~{\rm small}) \to
\sigma(\gamma^* p)\sim \sigma_0/\tau~~(\tau~{\rm large}).
\]
Fig.~\ref{fig:gscal} shows HERA data with $x<0.01$ as a function of
$\tau$. Not only do the data show this scaling very clearly, but also 
the expected change in behaviour from small to large $\tau$ values seems to occur around 
$\tau\sim 1$. This new scaling -- {\it 
geometrical scaling} -- holds for $Q^2$ values up to about $400\,$GeV$^2$
but only for $x<0.01$. Again, though it is not a proof, the remarkable
demonstration of geometrical scaling by the low $x$ HERA data does show that
they have many of the attributes of a saturated system.

\begin{figure}[t]       
\includegraphics[width=0.49\textwidth,clip]{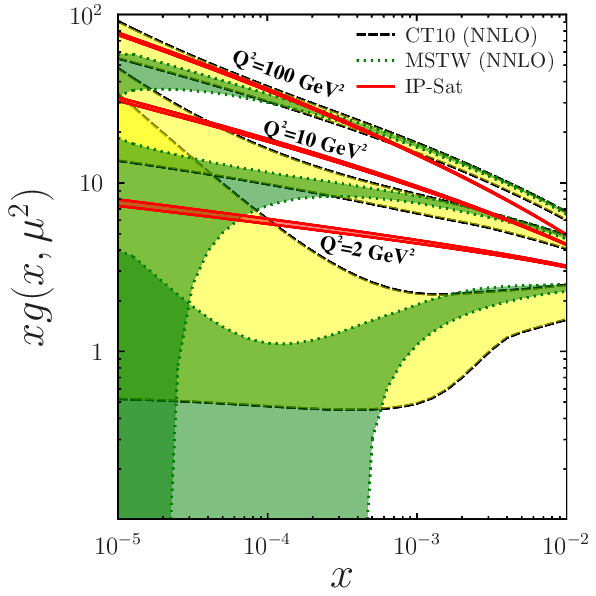}
\includegraphics[width=0.49\textwidth,clip]{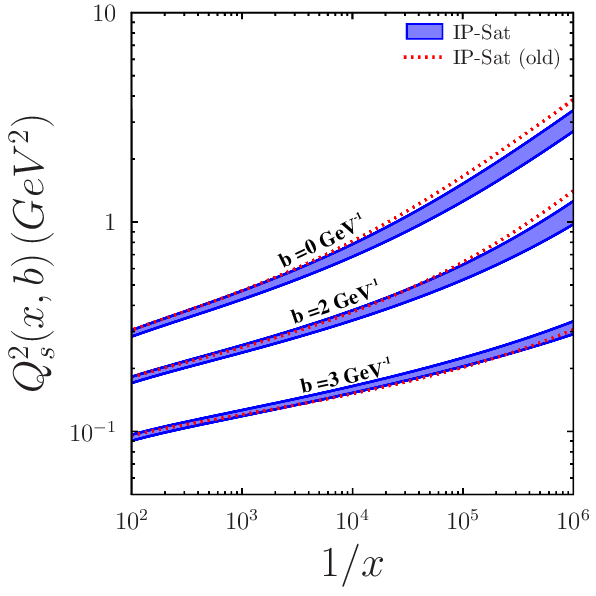} 
\caption{ Left: The gluon structure function as a function of $x$ for various fixed values of virtuality $Q^2$ extracted in the dipole saturation model (IP-Sat),  or the PDFs CT10 (NNLO),~\cite{Lai:2010vv}, and MSTW 2008 (NNLO),~\cite{Martin:2009iq}.  
  The corresponding theoretical uncertainties are represented with bands between solid, dashed and dotted lines for IP-Sat, CT10 and MSTW, respectively. Right: The saturation scale extracted from the IP-Sat model and from an older
  parameterization set from~\cite{Watt:2007nr}, labeled by IP-Sat (old), as a function of $1/x$ at various impact-parameter values. From~\cite{Rezaeian:2012ji}.}
\label{fig:glQs}           
\end{figure}

More recent successful dipole models include the Impact-Parameter dependent Saturation (IPSat) model,~\cite{Rezaeian:2012ji}, in which a Gaussian impact parameter, $b$,  dependence, $T_G(b)$ is given to the dipole
amplitude
\begin{equation}
\hat{\sigma}(x,r^2)=\sigma_0 \left[1-\exp(-\frac{\pi^2 r^2}{2N_c} \alpha_s(\mu^2)xg(x,\mu^2) T_G(b)) \right], T_G(b)=\frac{1}{2\pi B_G} \exp (-b^2/2B_G).
\label{eq:ipsat}
\end{equation}
This model is very successful in describing $F_2$, $F_2^{c\bar{c}}$ and $F_L$ at low-$x$, and
it also predicts a gluon which is not decreasing at low-$x$ and $Q^2$ in contrast to the predictions of DGLAP evolution, see Fig.~\ref{fig:glQs}. 
This figure also shows the value of the saturation scale, $Q_s$ as a function of $x$,
for various values of the impact parameter, $b$. We notice that the values of of the saturation scale are rather
low, $Q^2_S \leqsim 2$GeV$^2$, even for the lowest $x$ values reached at HERA $\sim 10^{-5}$. This makes it hard to
claim definitive evidence for saturation at HERA, since these low $Q^2$ values require the consideration of further
non-perturbative effects such as higher twist contributions, and most of the HERA data are not in the saturation
region. However, more compelling evidence can be found in semi-inclusive data sets on vector meson production and
Deeply Virtual Compton Scattering (DVCS), see Section~\ref{sec:diffraction}.

\begin{figure}[tbp]
    \centering
    \includegraphics[scale=0.5]{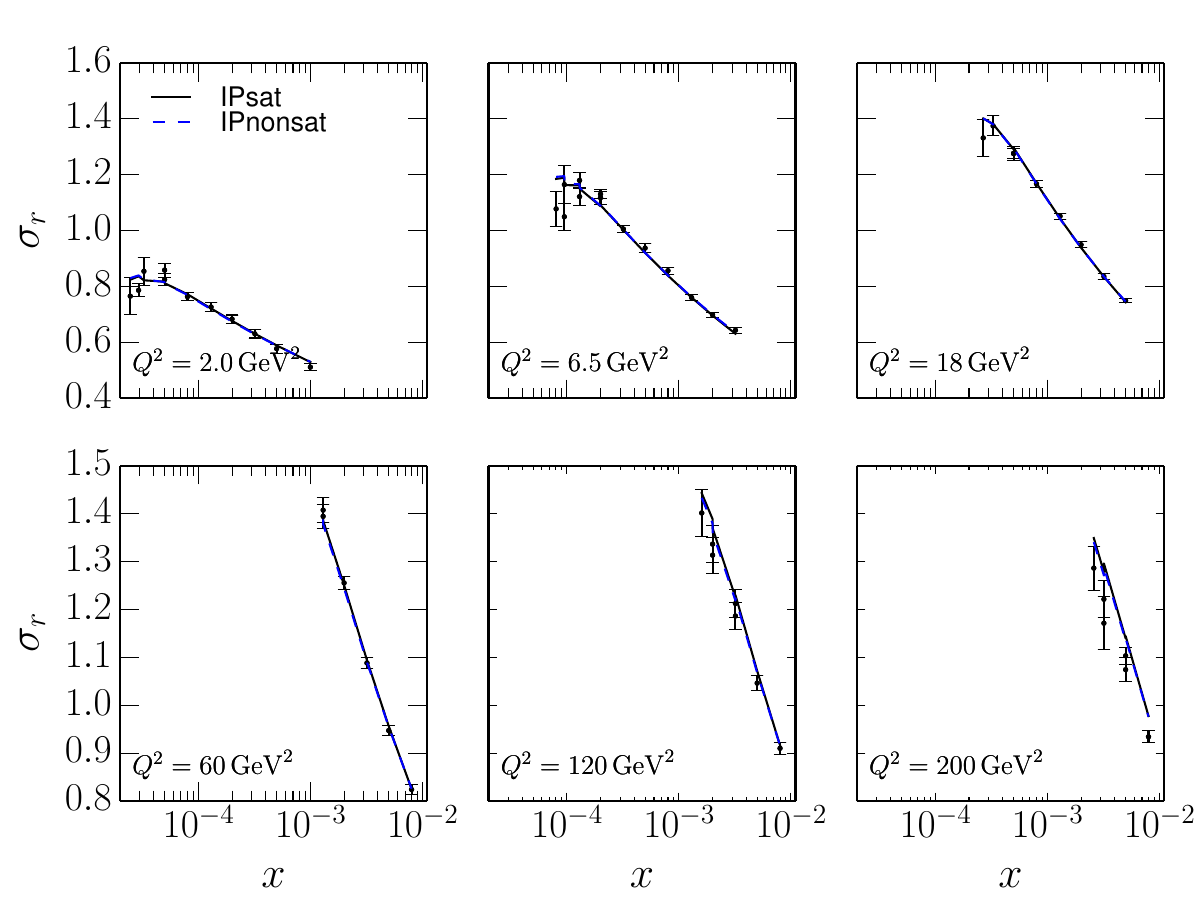}
    \caption{IPsat and IPnonsat (linearized IPsat) independent fits to inclusive reduced cross-section HERA data. Figure from \cite{Mantysaari:2018nng}.}
    \label{fig:satvsnonsat}
\end{figure}

Two further points can be made.
Firstly, it should be noted that
there are other dipole models that fit the HERA reduced cross section data as successfully without
an explicit $x$ dependent saturation scale. Fig.~\ref{fig:satvsnonsat} compares the IPsat model with a linearised
version of the same model, without saturation, IPnonsat. The fits are almost indistinguishable such that the good fit of the IPsat model cannot be claimed as defnitive evidence for saturation.

Secondly, saturation can also be achieved without the dipole picture.
The operator product expansion beyond leading twist has diagrams in which two, three or four gluons may be exchanged
in the t-channel such that these gluons may be viewed as recombining. This recombination could lead to gluon saturation.

\section{Diffraction}
\label{sec:diffraction}

\subsection{Introduction}
\label{diff:intro}

We now turn to diffractive processes, which many models predict may 
be a productive hunting ground for novel low $x$ dynamics such as saturation. 
Diffractive processes in proton-(anti)proton scattering, and to a lesser
extent, other projectiles scattering diffractively on proton targets,
have been the subject of sustained and intense study since well 
before HERA,~\cite{Goulianos:1982vk}. 
High energy data now also exist from the
LHC,~\cite{ATLAS:2012djz,CMS:2015inp,ATLAS:2019asg}.
Although it was not widely expected prior to data taking, the study of 
quasi-elastic and
diffractive processes has been one of the most successful areas of
study at HERA, and certainly one of the most prolific in terms of 
publications.
The narrative here closely follows that of~\cite{Newman:2013ada}.

The kinematics of HERA, with a strong forward boost of the hadronic
centre of mass relative to the lab frame for non-extreme $y$ values,
were particularly favourable for the study of diffractive excitations
of the real and virtual photon. 
Two different diffractive processes are of interest: quasi elastic scattering and diffractive dissociation.
In the (quasi-)elastic case, 
$\gamma^{(*)} p \rightarrow V p$,
the photon coupling to the beam lepton
either remains intact (Deeply 
Virtual Compton Scattering, DVCS) or converts to a vector 
meson $V$.
In the diffractive dissociation case, 
$\gamma^{(*)} p \rightarrow X p$
the system $X$ produced at the
proton vertex is a multi-particle state covering a continuum of 
invariant masses, $M_X$. 
Most of the interest has centred around the case where the 
proton remains intact and is scattered through a small angle (implying
a small absolute value of the 
squared four-momentum transfer at the proton vertex, $|t|$
(see Figs.~\ref{VM:feynman}(a) and~\ref{feynman}(a)). 


A QCD-based treatment of exclusive vector
meson production,~\cite{Ryskin:1992ui,Brodsky:1994kf,Frankfurt:1995jw}, by quasi-elastic scattering
predicts that cross sections should be $Q^2$-suppressed relative to 
inclusive DIS and relatable 
to the parton densities of the proton. The lowest 
order partonic exchange would be a pair of gluons with compensating colour 
charges, \cite{Low:1975sv,Nussinov:1975mw}. 

QCD models of single diffractive
photon dissociation need to be broken down into two categories. First
`hard' diffractive processes, similar to exclusive vector
meson production, are expected. In these the colour singlet exchange 
couples fully o the hard interaction with the photon.
Such an exchange
may again be perturbatively calculable starting from a 
knowledge of the proton structure.   
Second, `soft' diffraction, in which the 
colour singlet exchange is a composite virtual object, 
similar to the pomeron
of peripheral hadronic scattering,
in which a single parton participates in the
hard scattering, the remainder forming a low transverse momentum remnant 
system, cleanly separated in rapidity from the oppositely-travelling
intact proton.
This second type of configuration, \cite{Ingelman:1984ns,Donnachie:1987xh},
produces a leading twist contribution
with similar $Q^2$ dependence to the total cross section, 
limited only by kinematic 
constraints at large $Q^2$. 
In the DIS regime, it can be related
to a concept of diffractive parton densities.

A major issue in inclusive diffraction at HERA has been determining
in which kinematic regions each of these `soft' and `hard' processes are dominant and 
understanding the transition between the two. 

The detailed dependence on centre-of-mass energy of elastic and total 
hadron--hadron cross sections (via the optical theorem) has historically been remarkably well
described in a large kinematic domain by Regge
phenomenology,~\cite{Regge:1959mz,Regge:1960zc}. In this framework, 
interactions take place via the $t$-channel exchange of
reggeons, related to mesons,~\cite{Chew:1961yz}, and the pomeron ($\pom$),
which is the colourless, flavourless leading vacuum singularity,~\cite{Chew:1961ev}.
The pomeron is the mediator of diffractive scattering. 
Although the pomeron is essentially a phenomenological object,
associated mainly with soft processes, and now often called the 'soft' pomeron, it has also be calculated 
perturbatively. For example, it has been calculated by connecting two exchanged gluons in the t-channel
(the lowest order partonic process that can describe a flavourless, colourless excgange)  by adding
further gluon `rungs'. This leading-logarithmic $1/x$ approach 
leads to the `BFKL pomeron' or 'hard' pomeron discussed elsewhere in this contribution.

At asymptotically large energies, 
pomeron exchange dominates the elastic channel, such that both elastic and 
total cross sections display a slow increase with centre-of-mass
energy.  Interactions in which one
or both of the hadrons dissociates to higher 
mass states,~\cite{Fienberg:1956,Good:1960ba}, also
occur naturally in this approach. 
Such processes are characterised by the presence of
large regions of rapidity space in which no hadrons are produced.
They are dominated by diffractive exchange at large
centre-of-mass energy $\sqrt{s}$ and small dissociation 
masses $M_X$. The inclusive dissociation mass distribution may be treated via
Mueller's generalisation of the optical theorem,~\cite{Mueller:1970fa}, 
such that an appropriate Regge description involves diagrams that contain 
three-reggeon couplings.

Pomeron language has been adopted to a large extent at HERA, 
as a matter of convenience. This should not be taken to imply
the existence of a universal
$t$-channel pomeron exchange, since the detailed properties of the
exchange and its Regge trajectory 
clearly vary between processes and with kinematic variables at HERA,
as discussed in detail below.

\begin{figure}[htb]
\begin{center}
~\epsfig{file=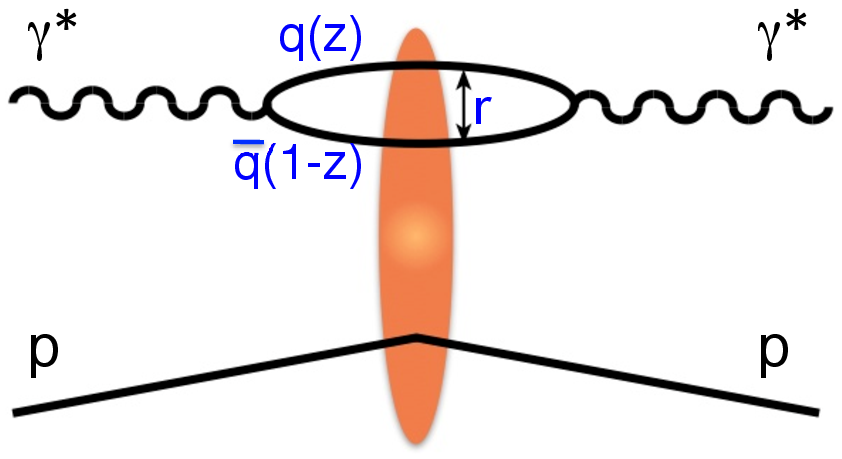,width=0.3\textwidth,angle=270}
\caption{Schematic illustration of the dipole model concept,
in which a virtual photon fluctuates to a $q \bar{q}$ pair with 
relative momentum fractions $z$ and $1-z$, forming a colour dipole
of transverse size $r \sim 1 / \surd Q^2$. The dipole then scatters elastically from the proton
according to a dipole cross section. The example shown corresponds
to the amplitude for the total inclusive $\gamma^* p$ cross section.
However, only small modifications are required to adapt this to the cases of
exclusive or inclusive diffraction.}
\label{Dipolefig}
\end{center}
\end{figure}

As an alternative to Regge theory, an $s$-channel
picture, \cite{Miettinen:1978jb,Dederichs:1989gk}, can be developed
 around the original Good \& Walker
approach, \cite{Good:1960ba}. In this approach 
different projectile Foch states are
absorbed differently on the target. 
The application of this $s$-channel approach to 
HERA data involves the elastic and quasi-elastic scattering
of colour dipoles, corresponding to $q \bar{q}$ (or 
higher multiplicity)
fluctuations of real and virtual photons on the 
proton, \cite{Mueller:1989st,Nikolaev:1990ja}. 
An attractive feature of this approach is the possible presence of `saturation' phenomena,
whereby the low $x$ growth of parton densities is tamed as ultimately required to satisfy unitarity.
This effect is easily incorporated at the parametric
level, \cite{GolecBiernat:1998js}, while more sophisticated treatments
are based for example on the Color Glass Condensate model, \cite{Gelis:2010nm}. Whilst such models
were originally applied at HERA to describe the total $\gamma^* p$ cross
section (see Fig.~\ref{Dipolefig}),
vector meson production and other exclusive processes are easily incorporated
with no further free parameters except those that describe the 
$t$ dependence,~\cite{Kowalski:2003hm,Kowalski:2006hc}.

Diffractive dissociation can also be 
included, \cite{Nikolaev:1991et},  in this approach 
although it has proved
more difficult to describe due to the need to incorporate higher
terms, corresponding for example to a dipole formed from a
$q \bar{q}$ pair and a gluon, \cite{Bartels:1998ea,GolecBiernat:1999qd}.
The dipole approach to diffraction at HERA is discussed further in 
Sections~\ref{vm:dipole} and~\ref{sec:dipoles}.

We now discuss experimental results on quasi-elastic scattering  and diffractive dissociation in in Section~\ref{sec:vmdvcs}
and Section~\ref{sec:diffdis}, respectively.

\subsection{Exclusive Production of Vector Mesons and Photons}
\label{sec:vmdvcs}
\subsubsection{Kinematics and Experimental Selection}
\label{vm:intro}

\begin{figure}[htb] \unitlength 1mm
  \begin{center}
    \begin{picture}(60,36)
      \put(-16,2){\epsfig{file=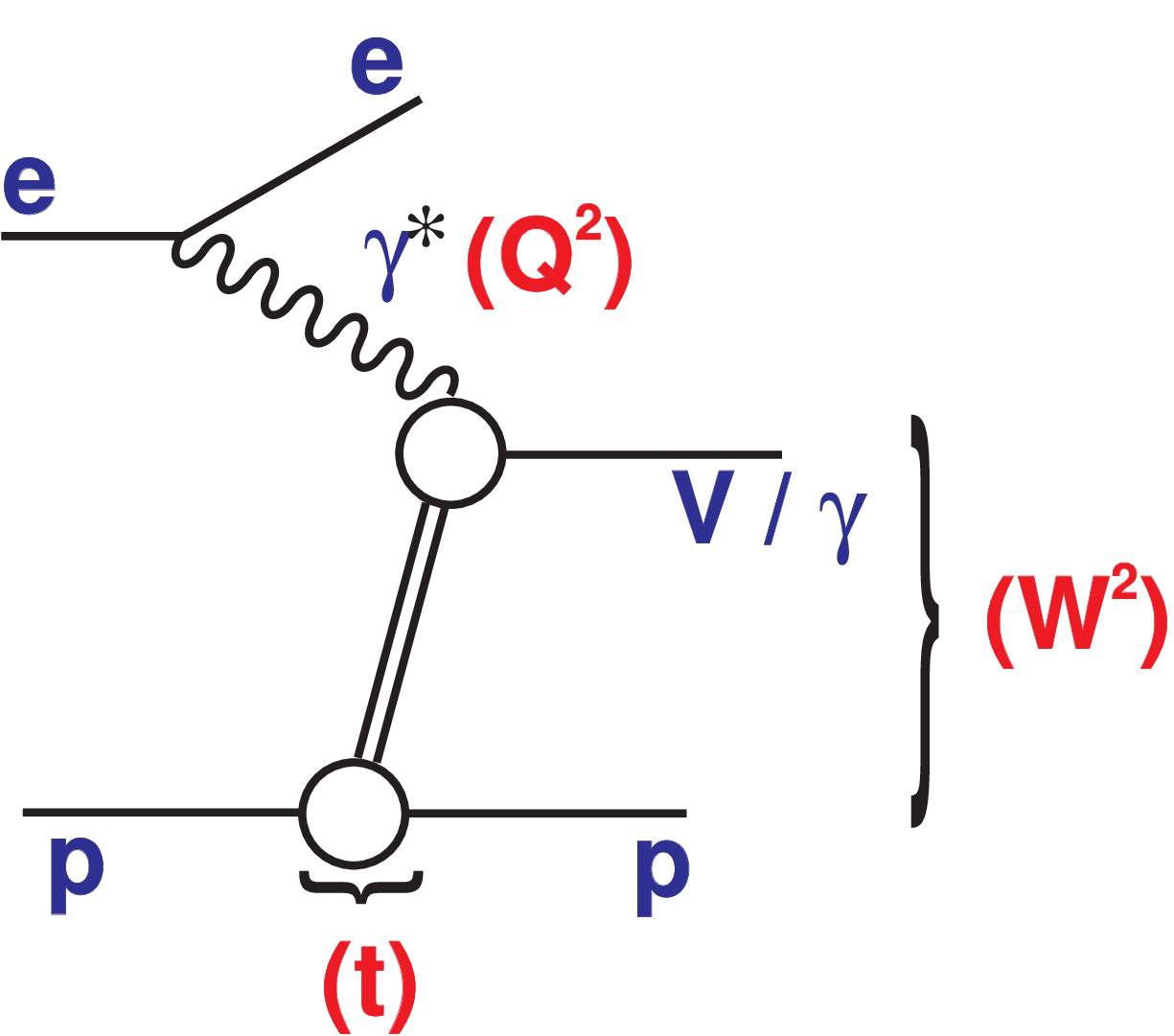,height=0.19\textwidth,clip}}
      \put(43,8){\epsfig{file=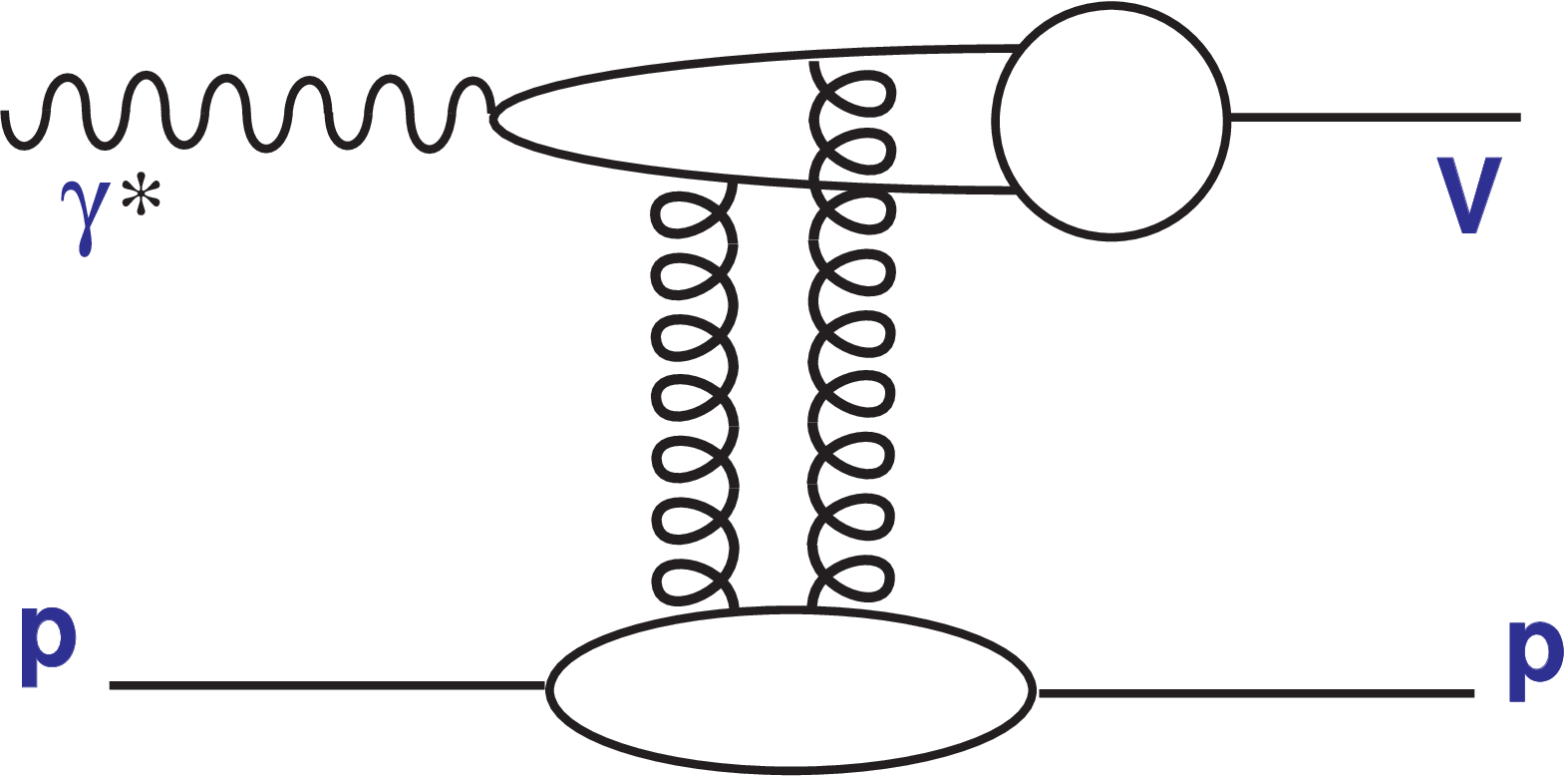,height=0.11\textwidth,clip}}
      \put(-2,-3){{\footnotesize{{\bf (a)}}}}
      \put(58,-3){{\footnotesize{{\bf (b)}}}}
    \end{picture}
  \end{center}
  \caption{(a) Generic representation of
exclusive vector meson production via the quasi-elastic
scattering of a real or virtual photon. The
commonly used kinematic variables
discussed in the text are indicated.
(b) Lowest order perturbative diagram for hard vector
meson production, 
involving the exchange of a
pair of gluons between the proton and the real or virtual photon.}
\label{VM:feynman}
\end{figure}

Quasi-elastic vector meson production and DVCS (Fig.~\ref{VM:feynman}(a))
are the simplest diffractive processes that have been studied at HERA.
For a fixed final state vector meson or photon, 
they are usually described in terms of the kinematic variables,
$Q^2$, $W$ and $t$. Distributions in all three of these variables have been 
measured in analyses covering the vector meson species,
$\rho$, $\omega$, $\phi$, $\rho^\prime$, $J/\psi$, 
$\psi^\prime$ and $\Upsilon$ as well as in DVCS. 
No evidence has been found for the exclusive production at the photon vertex
of particles with non-$1^{--}$ quantum numbers such as the $\pi^0$, \cite{Adloff:2002dw}, 
as would be expected for the exchange of the postulated negative $C$-parity partner 
of the pomeron, the odderon, \cite{Lukaszuk:1973nt}.

The most precise vector meson data are obtained
by reconstructing two-prong decays via charged decay products 
(notably $\rho^0 \rightarrow \pi^+ \pi^-$, $\phi \rightarrow K^+ K^-$
and $J/\psi \rightarrow e^+ e^-$ or $\mu^+ \mu^-$) and requiring no
further activity beyond the noise levels in the detector, except that
associated with the scattered beam electron. DVCS selections require a
similar lack of activity in the detector beyond the final state electron
and photon, the main 
experimental complication being separating the signal from
the competing purely electromagnetic Bethe--Heitler process, which generates
identical final states.   
In contrast to inclusive diffractive studies, the simplicity of these 
final states and the high precision of the tracking
detectors have allowed precision $t$ measurements  
without the need to tag the intact outgoing proton. 
This reconstruction method uses 
$|t - t_{\rm min}| = p_T^2 (V)$, where the vector meson transverse
momentum $p_T (V)$ is obtained from its decay products and $|t_{\rm min}|$,
the minimum kinematically accessible value of $|t|$, is usually negligible.

\subsubsection{General Characteristics of Vector Meson Production}
\label{vm:data}

Vector meson production has emerged as a sensitive probe of the transition
from the soft diffractive dynamics which are familiar from hadronic scattering
experiments to a harder regime which may be calculated perturbatively. 
The former regime is encountered wherever no hard scale is present, the 
classic example being $\rho^0$ 
photoproduction, \cite{Aid:1996bs,Breitweg:1999jy,Breitweg:1997ed}. Under
such circumstances, the energy dependence of the photon--proton process
is in good agreement with the form predicted by Regge asymptotics: 
\begin{equation}
\sigma^{ \gamma p \rightarrow Vp} \propto 
(W^2)^{2 \alphapom(t) - 2} \ ,
\label{vm:energy}
\end{equation}
where the pomeron trajectory,
$\alphapom(t) = \alphapom(0) + \alphapom^\prime t$, is
assumed to be linear and its intercept 
$\alphapom(0) \simeq 1.085$, \cite{Donnachie:1992ny,Cudell:1996sh}. This is the 'soft' pomeron.
This has been 
found to work well for $\rho^0$ photoproduction data, though 
interestingly, the slope
of the pomeron trajectory has been found, \cite{Breitweg:1999jy},
to be significantly smaller than the value of
$\alphapom^\prime \sim 0.25$, \cite{Donnachie:1984xq,Abe:1993xx}, 
obtained from soft $pp$ and $p \bar{p}$ 
scattering. 
A possible explanation for this
may be found in process-dependent absorptive
corrections, 
which are absent in DIS, present to some extent in photoproduction and to 
a larger extent in fully hadronic scattering. Detailed models of these effects 
can be found for example 
in \cite{Kaidalov:2003xf,Kaidalov:2009fp,Gotsman:2007pn}.
Further characteristics of this soft regime 
are, \cite{Aid:1996bs,Breitweg:1997ed}, a skewed
lineshape for the $\rho$ meson due to its interference with non-resonant
$\pi^+ \pi^-$ production and a large value,
$B \sim 10 \ {\rm GeV^{-2}}$,
of the slope parameter describing the $t$ dependence according to
\begin{equation}
\frac{{\rm d} \sigma^{\gamma p \rightarrow Vp}}{{\rm d} t} = 
\left( \frac{{\rm d} \sigma^{\gamma p \rightarrow Vp}}{{\rm d} t} \right)_{t=0} \
e^{Bt} \ .
\label{eq:tslope}
\end{equation}
These $B$ values suggest that the interaction 
takes place over
a transverse distance of typically $1 - 1.5 \ {\rm fm}$, when interpreted in a simple optical model 
as a measurement of the mean impact parameter between the incoming
photon and proton,  

\begin{figure}[htp]
\begin{center}
~\epsfig{file=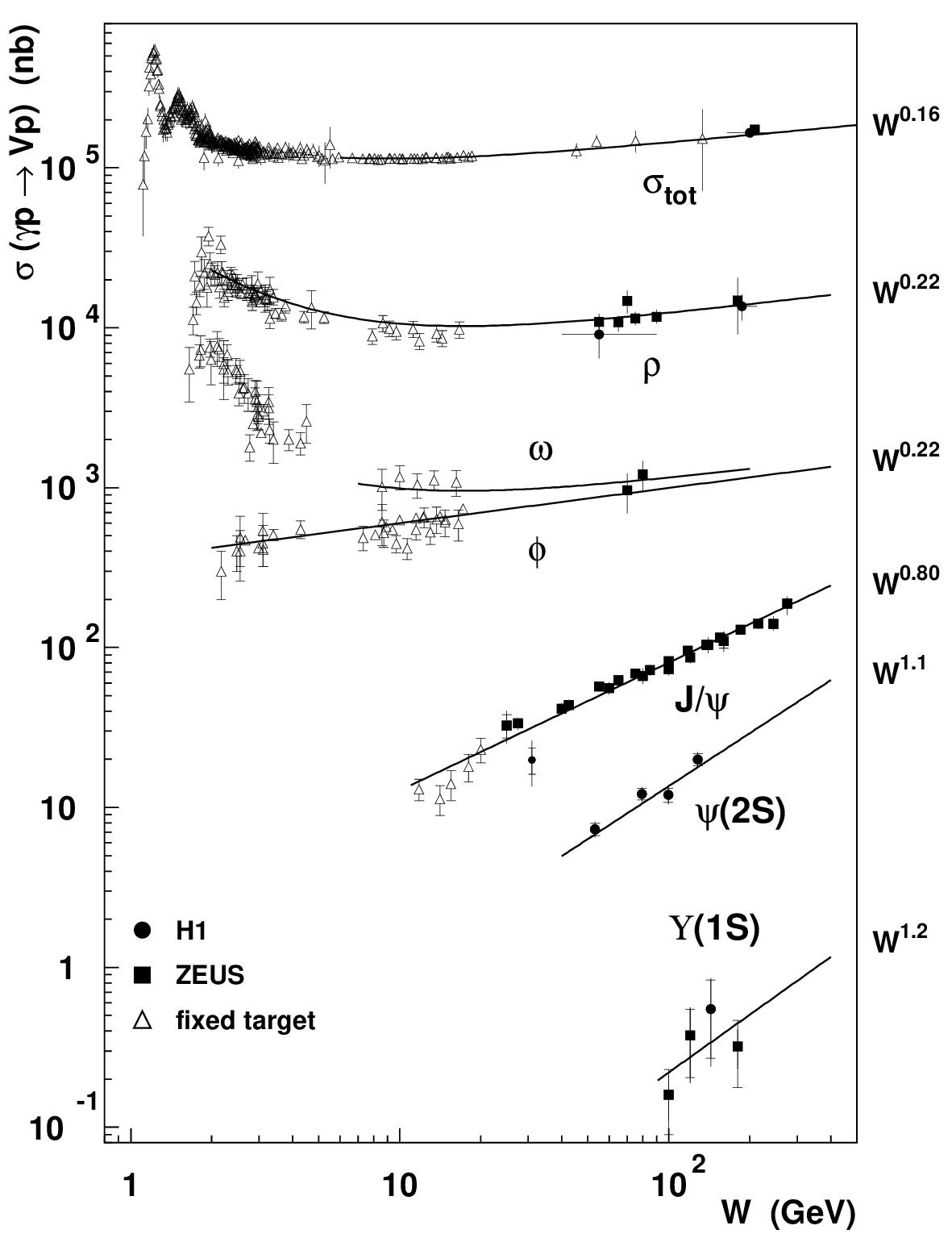,width=8.5cm}
\caption{Compilation of photoproduction cross section measurements as a function
of the $\gamma p$ centre-of-mass energy, $W$. The total cross section and 
various vector meson production cross sections are 
included, with the approximate power law dependences 
$\sigma \propto W^\delta$ indicated for each process. From~\cite{Levy:2009gy}.}
\label{vm:wdependence}
\end{center}
\end{figure}

Wherever hard scales are present, usually provided either
by heavy quarks in the vector meson or by large $Q^2$, but sometimes also
by large $|t|$, the qualitative picture changes. The energy dependence 
becomes progressively steeper, such that the $W$ dependence,
parametrised similarly to Eq.~\ref{vm:energy}, yields an increased
effective value of $\alphapom(0)$. This effect is illustrated in 
Fig.~\ref{vm:wdependence}, 
where a compilation of photoproduction data on the total
cross section and different vector meson production cross sections is 
shown as a function of $W$. 
The steepening of the dependence on $W$ for the heaviest vector mesons
can be interpreted in terms of the presence of a 'hard' pomeron exchange or alternatively in terms
of the scale dependence of the proton
gluon density at low $x$, as discussed further below. 

\begin{figure}[htp]
\begin{center}
~\epsfig{file=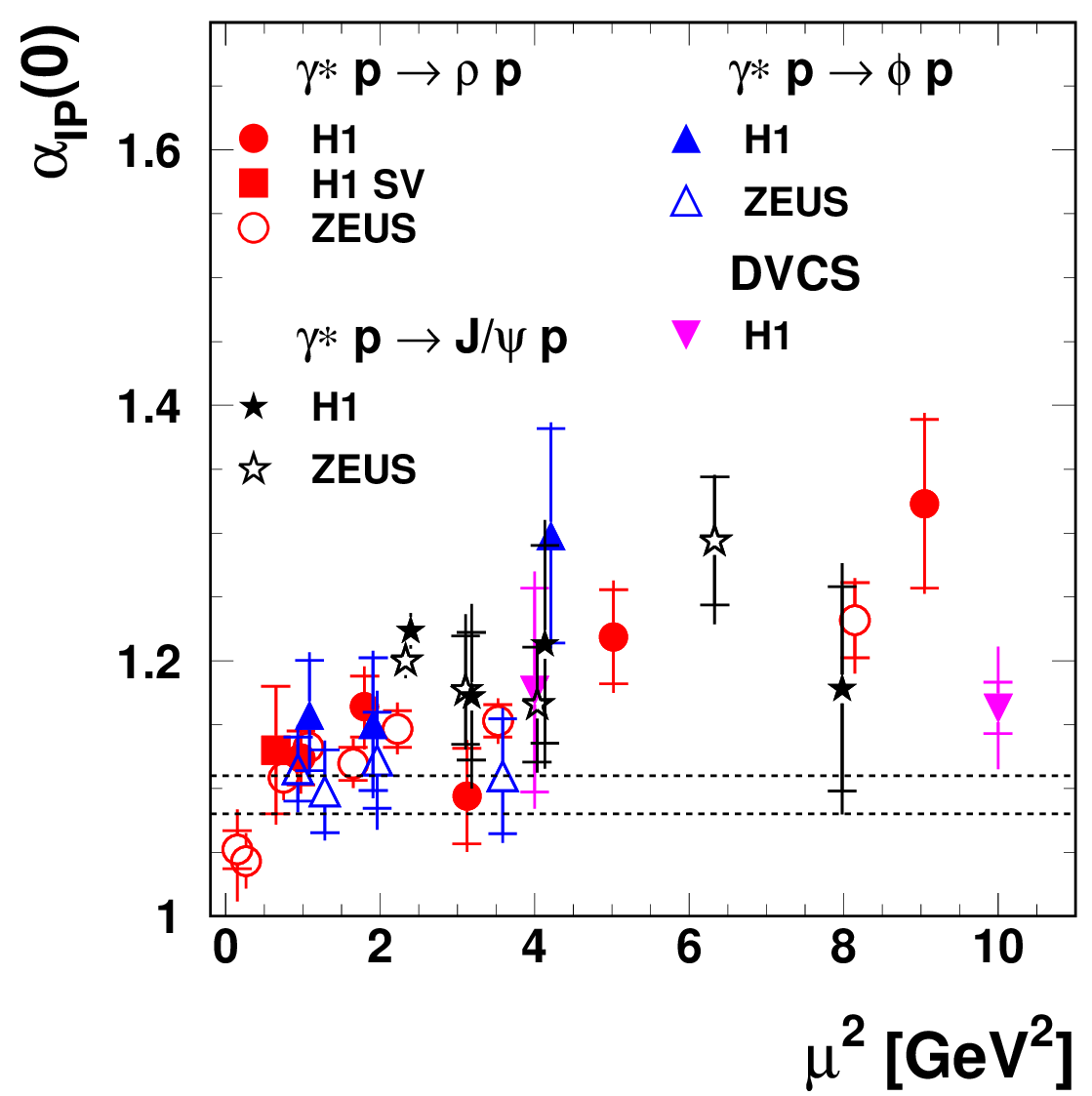,width=8cm}
\caption{Development of the effective 
pomeron intercept (see Eq.~\ref{vm:energy} and surrounding text)
with a characteristic scale, $\mu^2 = (Q^2 + M_V^2)/4$,  
derived from fits to the $W$ dependences 
of various vector meson production data,
as well as a DVCS measurement 
(shown at $\mu^2 = Q^2$). The dashed lines indicate
the range of values which are typical in soft diffraction.
From~\cite{Aaron:2009xp}.}
\label{vm:delta}
\end{center}
\end{figure}

The transition from soft hadronic to perturbative behaviour is 
neatly mapped out in a single process in $\rho^0$ electroproduction
data. As $Q^2$ increases, the $t$ slope parameter $B$ decreases, the 
$W$ dependence becomes steeper and the lineshape skewing disappears,
all in a manner which has been measured with good 
precision, \cite{Adloff:1999kg,Aaron:2009xp,Chekanov:2007zr,Breitweg:1998nh}.
An example, also including comparisons with other 
exclusive vector meson production
processes, is shown in Fig.~\ref{vm:delta}. Here, the effective 
pomeron intercept (Eq.~\ref{vm:energy})
is shown as a function of a scale, which is chosen to be
$(Q^2 + M_V^2)/4$ (see Section~\ref{vm:dipole}).
As the scale increases, the effective pomeron
intercept shifts from values typical
of soft hadronic scattering to values which are compatible with results
for the equivalent quantity 
$\alphapom(0) = 1 + \lambda$ in fits of inclusive low-$x$ 
HERA data to the form 
$F_2(x,Q^2) \propto x^{- \lambda(Q^2)}$, \cite{Adloff:2001rw,Chekanov:2008cw},
as also discussed here in Sec.~\ref{sec:lowq2} (Fig.~\ref{fig:lambdaeff}a). 

\begin{figure}[htp]
\begin{center}
~\epsfig{file=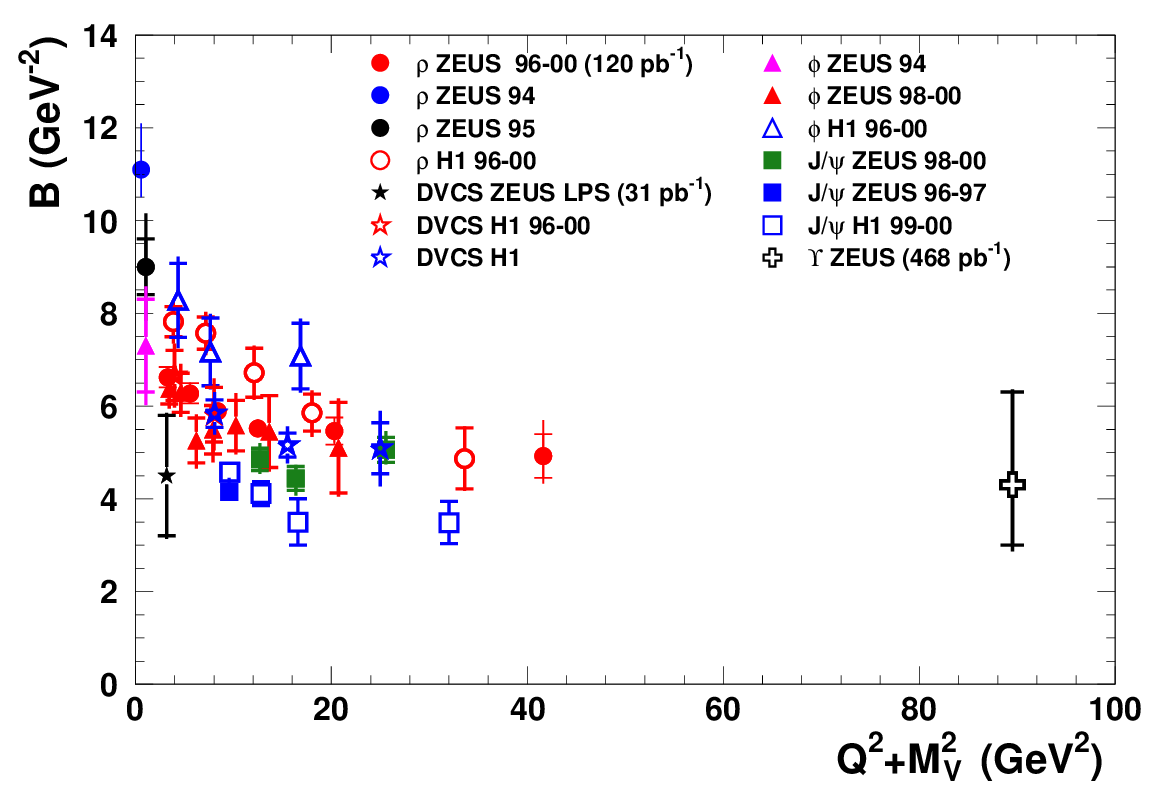,width=8.7cm}
\caption{Exponential $t$ slopes for vector meson electroproduction and 
photoproduction as a function of the characteristic scale
$Q^2 + M_V^2$. From~\cite{Abramowicz:2011fa}.}
\label{vm:tslopes}
\end{center}
\end{figure}

The exponential $t$ slopes of vector meson production processes are also 
found to vary systematically with scale 
and are approximately invariant in
$Q^2 + M_V^2$. A compendium of results is shown in 
Fig.~\ref{vm:tslopes}. Although the uncertainties are often large and
there is some scatter, the data suggest a convergence towards an 
asymptotic value of $B \sim 5 \ {\rm GeV^{-2}}$. 
In optical models, this can be interpreted as
the point at which the physics is entirely short-distance in nature, 
the size of the probe becomes negligible and the slope parameter measures the size of the proton. 
Quantitatively, this indicates an effective
proton size of around $0.6 \ {\rm fm}$, which is interestingly
smaller than the value of $\sim 0.8 \ {\rm fm}$ which is well measured
using electromagnetic probes. 
Interpreting vector meson production in terms of 
gluon exchange, this suggests that the
gluon radius of the proton may be smaller than its quark radius. 

\subsubsection{Vector Meson Production in QCD}
\label{vm:qcd}

The observed relationship between the $W$ dependences 
of inclusive DIS and vector meson production in the presence of a 
sufficiently large 
scale encourages a perturbative approach.
This has evolved considerably in the HERA era, to the point
where most observables can be successfully predicted. Although the 
basic quark charge counting SU(4) prediction for the ratio of cross sections
$\rho : \omega : \phi : J/\psi = 9:1:2:8$ holds approximately,
when viewed as a function of the 
scale $(Q^2 + M_V^2)/4$, \cite{Aaron:2009xp}, 
vector meson wavefunction effects remain a significant source of
uncertainty. 

A QCD collinear factorisation theorem, \cite{Collins:1996fb}, valid for the leading power
of $Q^2$ where $|t| \ll \Lambda_{\rm QCD}^2$, relates cross sections for
heavy vector meson production from longitudinally polarised photons
to the generalised parton densities (GPDs) of the proton, \cite{Diehl:2003ny},
(see Section~\ref{sec:dvcs}). 
Neglecting skewing effects, in which
the two exchanged partons carry different fractions of the proton
longitudinal momentum, the GPDs reduce to the 
square of the gluon density of the proton. The 
process is then driven to first approximation by the 
exchange of a pair of gluons from the proton structure, as illustrated in 
Fig.~\ref{VM:feynman}(b). 
Later approaches, \cite{Martin:1996bp},
incorporated transversely
polarised photon cross sections at sufficiently large $Q^2$
and light vector mesons in a similar framework, with some degree of
success.

\begin{figure}[htp]
\begin{center}
~\epsfig{file=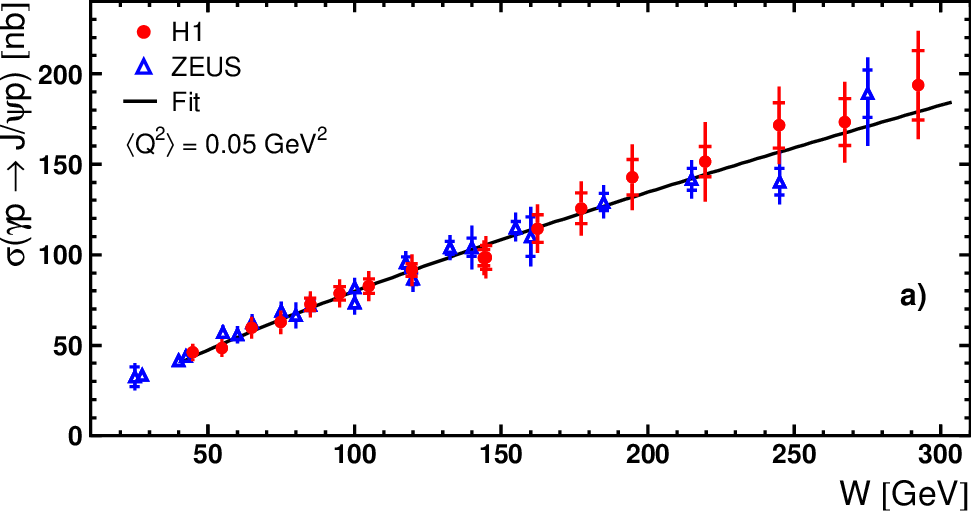,width=0.475\textwidth}
~\epsfig{file=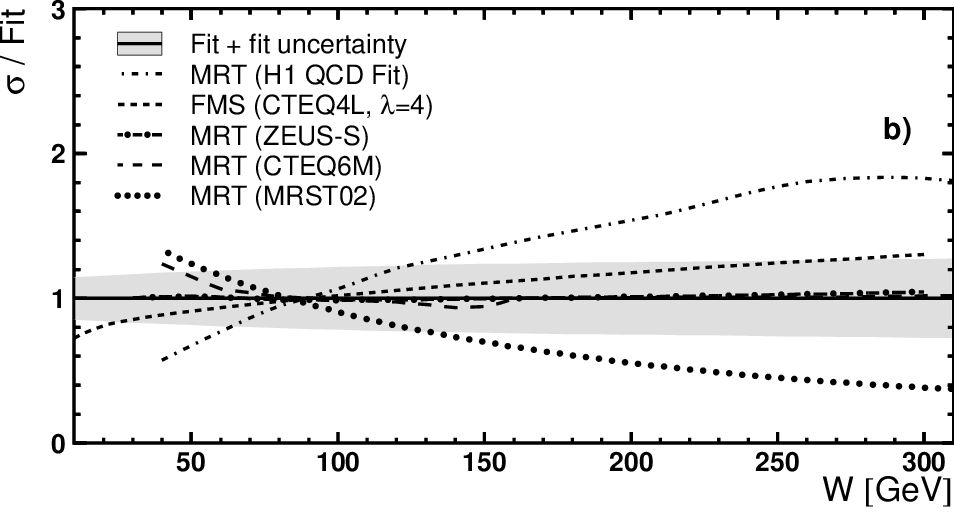,width=0.475\textwidth}
\caption{$J/\psi$ photoproduction cross section as a function of $W$,
compared with a QCD calculation, MRT \cite{Martin:1999wb}, based on 
four different parametrisations of the gluon density of the 
proton and a dipole model, FMS \cite{Frankfurt:2000ez}. 
The predictions are normalised to the data.
From~\cite{Aktas:2005xu}.}
\label{jpsi:wdep}
\end{center}
\end{figure}

Since it is relatively uncomplicated theoretically, has a reasonably large scale, 
probes small $x$ values and is experimentally clean, 
$J/\psi$ photoproduction is an ideal testing ground for these ideas and,
 a potentially competitive means of extracting the gluon density of the proton. 
The many measurements at 
HERA, \cite{Aktas:2005xu,Chekanov:2002xi,Alexa:2013xxa}, have now
been supplemented by measurements from ultra-peripheral
processes at the LHC, extending the $W$ range to 
beyond $1 \ {\rm TeV}$, \cite{LHCb:2013nqs,ALICE:2023mfc}. 
The $W$ dependence of the 
cross section for $J/\psi$ photoproduction has been calculated 
using the proton gluon density by a number of 
groups, \cite{Martin:1999wb,Martin:2007sb,Frankfurt:2000ez}. An example 
comparison with data is shown in Fig.~\ref{jpsi:wdep}.
From comparisons between the
predictions with different gluon densities, it is clear that there is
high sensitivity. However, 
the normalisation is not well predicted and 
there remains considerable model dependence.

More detailed comparisons with QCD models require an understanding of the 
helicity structure of vector meson production. The five independent
helicity amplitudes, describing transitions from either longitudinally
or transversely polarised photons to either 
longitudinally or transversely polarised vector mesons, can be extracted
from the production and decay angular distributions via spin density
matrix elements as parametrised in \cite{Schilling:1973ag}. 
Whilst helicity amplitudes for $J/\psi$ production 
at modest $|t|$ are consistent with 
$s$-channel helicity 
conservation, \cite{Aktas:2005xu}, significant `helicity flip'
amplitudes, for which the vector meson helicity differs from that of the 
photon, are observed for the case of 
$\rho$ 
production, \cite{Adloff:2002tb,Breitweg:1999fm,Aaron:2009xp,Chekanov:2007zr},
and for $J/\psi$ production at large 
$|t|$, \cite{Aktas:2003zi,Chekanov:2009ab}. 
This reflects the often unequal 
sharing of the longitudinal momentum between the two quarks,
corresponding to values of $z$ close to 0 or 1 in Fig.~\ref{Dipolefig}, 
particularly for the transverse polarisation case with $\rho$ mesons, 
This allows the helicity to flip at sufficiently 
large $t$, that perturbative QCD-based models give good predictions. 

Probably the most important feature of angular distribution analyses is that
they allow an unfolding of the 
cross sections $\sigma_L$ and $\sigma_T$,
corresponding to vector meson production from 
longitudinally and transversely 
polarised vector mesons, respectively.
The $Q^2$ dependences of $\sigma_L$ and $\sigma_T$ are separately
predicted by QCD models. The basic leading order dependences
at $t = 0$ are  $\sigma_L \propto 1/Q^6$ and 
$\sigma_T \propto 1/Q^8$, \cite{Brodsky:1994kf}. These are 
expected to be strongly
violated, in particular due to the $Q^2$ dependence of the
proton gluon density, but also due to the quark virtuality and the running of
$\alpha_s$, \cite{Frankfurt:1995jw}. The HERA data for
all vector meson species exhibit approximate scaling with 
power law dependences of $\sigma_L \propto (Q^2 + M_V^2)^{2.1}$ 
and $\sigma_T \propto (Q^2 + M_V^2)^{2.9}$, which are reasonably
well reproduced by both collinear factorisation and dipole-based 
models, \cite{Martin:1996bp,Kowalski:2006hc}. 

As discussed in Section~\ref{vm:data}
the variation of the slope parameter $B$ with centre-of-mass 
energy $W$ is usually parametrised in terms of the slope of
the pomeron trajectory $\alphapom^\prime$. Values close to zero
are measured with small uncertainties in some
hard exclusive processes, notably $J/\psi$ production at
large $|t|$, \cite{Aktas:2005xu,Chekanov:2002xi,Chekanov:2004mw}. 
This is consistent
with expectations in the BFKL 
approach, \cite{Nikolaev:1995ed,Brodsky:1998kn}, 
where $\alphapom^\prime$
is related to the average transverse momentum of partons along the
gluon ladder. However, it is hard to draw quantitative conclusions in light
of the smaller-than-expected $\alphapom^\prime$
values for soft processes such as $\rho^0$ photoproduction 
(Section~\ref{vm:data}). 

\subsubsection{Deeply Virtual Compton Scattering and Generalised
Parton Densities}
\label{sec:dvcs}

Due to the underlying exchange of a pair of gluons, which in general
differ in both longitudinal and transverse momentum, 
hard exclusive DIS processes have emerged in recent years as
candidates with which to understand correlations between partons
in the proton and thus the transverse spatial, momentum 
and spin distributions
of partons. This information
is encoded in GPDs, \cite{Diehl:2003ny}, which were introduced in
Section~\ref{vm:qcd}.
HERA data offer unique sensitivity to GPDs at low $x$. 
This topic is most commonly associated with the DVCS process, though
it is also relevant to vector meson production.

The dominant underlying parton level process for DVCS at low $x$
is similar to that shown in
Fig.~\ref{VM:feynman}(b), though there is also a 
`handbag' contribution in which the incoming (virtual) and 
outgoing (real) photons couple directly to quarks from the
GPDs. DVCS has the advantage of
avoiding the complication of the vector meson wavefunction, but 
cross sections are suppressed 
due to the final state photon coupling. 
Indeed, DVCS studies emerged relatively late in the lifetime
of HERA, due to the smaller cross sections compared with their vector meson
counterparts. For the
HERA kinematic region, integrated over azimuthal degrees of freedom, 
interference between DVCS and the competing Bethe--Heitler process
is small and 
DVCS cross sections can be extracted by statistically subtracting the
Bethe--Heitler component using Monte Carlo models tuned in regions in
which the DVCS contribution can be neglected. 

\begin{figure}[htp]
\begin{center}
~\epsfig{file=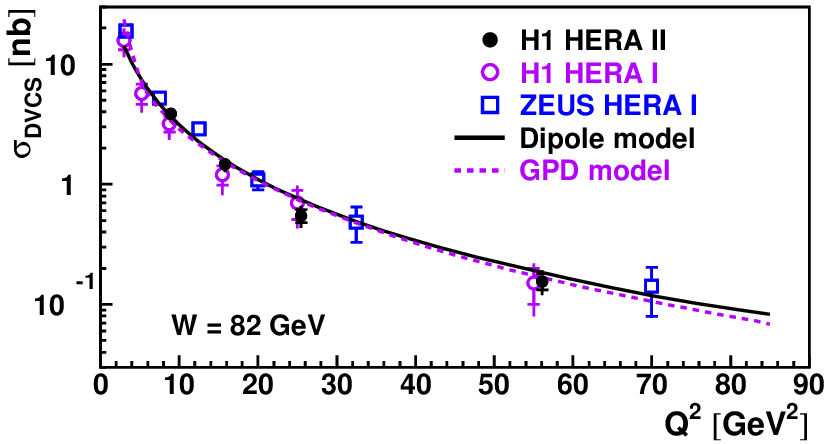,width=8.7cm}
\caption{Dependence of HERA DVCS cross section 
measurements on $Q^2$. 
The data are compared with a GPD-based model, \cite{Kumericki:2009uq},
and also with a dipole-based 
model, \cite{Marquet:2007qa}. See Section~\ref{vm:qcd} for 
an explanation of the latter. 
From~\cite{H1:2009wnw}.}
\label{dvcs:fig}
\end{center}
\end{figure}

By the end of HERA operation, $Q^2$, $W$ and $t$ 
distributions for DVCS had all been measured with good 
precision, \cite{Chekanov:2008vy,H1:2009wnw,H1:2007vrx}.
As shown in Figs.~\ref{vm:delta} and~\ref{vm:tslopes},  
the exponential $t$ slope is $B \simeq 5 \ {\rm GeV^{-2}}$ and 
parametrisations of the cross section in the usual $W^\delta$
form yield $\delta \approx 0.5 - 0.6$. 
Example $Q^2$ dependence data are shown in Fig.~\ref{dvcs:fig}.
The DVCS cross section falls more 
slowly (roughly as $(Q^2)^{-1.5}$)
with photon virtuality than is the case for vector meson
production.  
This observation is in line with theoretical 
predictions, \cite{Frankfurt:1997at}, and is partially explained
by the presence of the quark-driven handbag contribution, which
is not relevant to vector meson production.  
Together, these observations 
indicate that DVCS can be considered as a hard process, even at 
relatively small $Q^2$.

Asymmetries in the azimuthal degree of freedom between the lepton and
the (virtual-to-real) photon scattering
planes in DVCS give access to the real part of the amplitude, and 
are most sensitive to the non-diagonal GPDs. 
The beam charge asymmetry at HERA is clearly non-zero, \cite{H1:2009wnw},
and its azimuthal dependence 
has been 
used to confirm that the DVCS amplitude is dominantly imaginary for
HERA kinematics. 
It has not yet been possible to determine 
beam polarisation asymmetries in the manner successfully pursued
in the fixed target configuration by 
HERMES, \cite{Airapetian:2012pg,Airapetian:2012mq}.
Nonetheless, the H1 and ZEUS 
data provide important low $x$ ingredients in models of GPDs, the principal
sensitivity being to the GPD, $H$, \cite{Kumericki:2009uq}.

\subsubsection{Exclusive Processes in Dipole Models and 
Low {\boldmath $x$} Saturation}
\label{vm:dipole}

Viewed in the proton rest frame, vector meson production 
naturally
factorises into three separate processes, occurring on 
distinct time-scales (\emph{c.f.}\ Figs.~\ref{Dipolefig} and~\ref{VM:feynman}(b)). 
The
incoming real or virtual photon first fluctuates into a $q \bar{q}$ pair,
which creates a colour dipole of transverse size 
$r_\perp \sim 1 / \surd Q^2$, \cite{Wusthoff:1997fz}. 
The dipole then scatters elastically from the proton, 
a process which is generically described in terms of a dipole cross section. 
Finally, the outgoing colour dipole combines back to a photon in the
DVCS case, or else 
hadronises to a final state vector meson
in a manner which depends on the non-perturbative vector meson wavefunction. 
The scale choice of $(Q^2 + M_V^2) / 4$, already used in the previous
sections, corresponds to the inverse of the 
scanning radius of the dipole--proton interaction, for the case where the 
incoming photon longitudinal momentum 
is shared equally between the quark and antiquark
forming the dipole, \cite{Frankfurt:1995jw,Ivanov:2004ax}.
It is appropriate for heavy vector mesons and for light vector mesons
produced by longitudinally polarised photons. 

At small enough $r_\perp$, the dipole cross section
can be described perturbatively in the manner discussed 
in Section~\ref{vm:qcd}. At larger $r_\perp$, 
the dipole cross section is 
not known \emph{a priori}, but is
sensitive to any low $x$ 
saturation of the 
proton parton densities.
There are various parametrisations of the dipole cross section
which include saturation phenomena, 
some of which are
based on particular approximations to 
QCD with non-linear evolution such as the Colour Glass 
Condensate, \cite{Iancu:2003ge},
and some of which are purely 
phenomenological, \cite{GolecBiernat:1998js,GolecBiernat:1999qd,Forshaw:1999uf,Forshaw:2003ki}. 

In the most sophisticated dipole models of vector meson production,
the target matter density, and hence any saturation effects, 
depend on the distance from the centre of the proton, 
such that the dipole cross section varies with impact 
parameter, \cite{Kowalski:2003hm,Kowalski:2006hc,Rezaeian:2012ji}, 
which in turn is 
closely related to $1/t$ (see also \cite{Marquet:2007qa} for a 
complementary approach). 
When viewed differentially in $t$, the influence of the saturation 
effects thus becomes stronger as $t$ 
increases, \cite{Kowalski:2006hc,AbelleiraFernandez:2012cc}.
Such models adequately describe almost all
aspects of vector meson production and DVCS at HERA, though the energy frontier for 
such processes 
has now moved on to the Ultraperipheral Collision programme at the LHC.


\subsection{Inclusive Diffractive Dissociation}
\label{sec:diffdis}
\subsubsection{Kinematics and Experimental Selection}

\begin{figure*}[tb]
\centerline{\hspace*{0.1cm}
            {\Large{\bf{(a)}}}
            \includegraphics[height=0.3\textwidth]{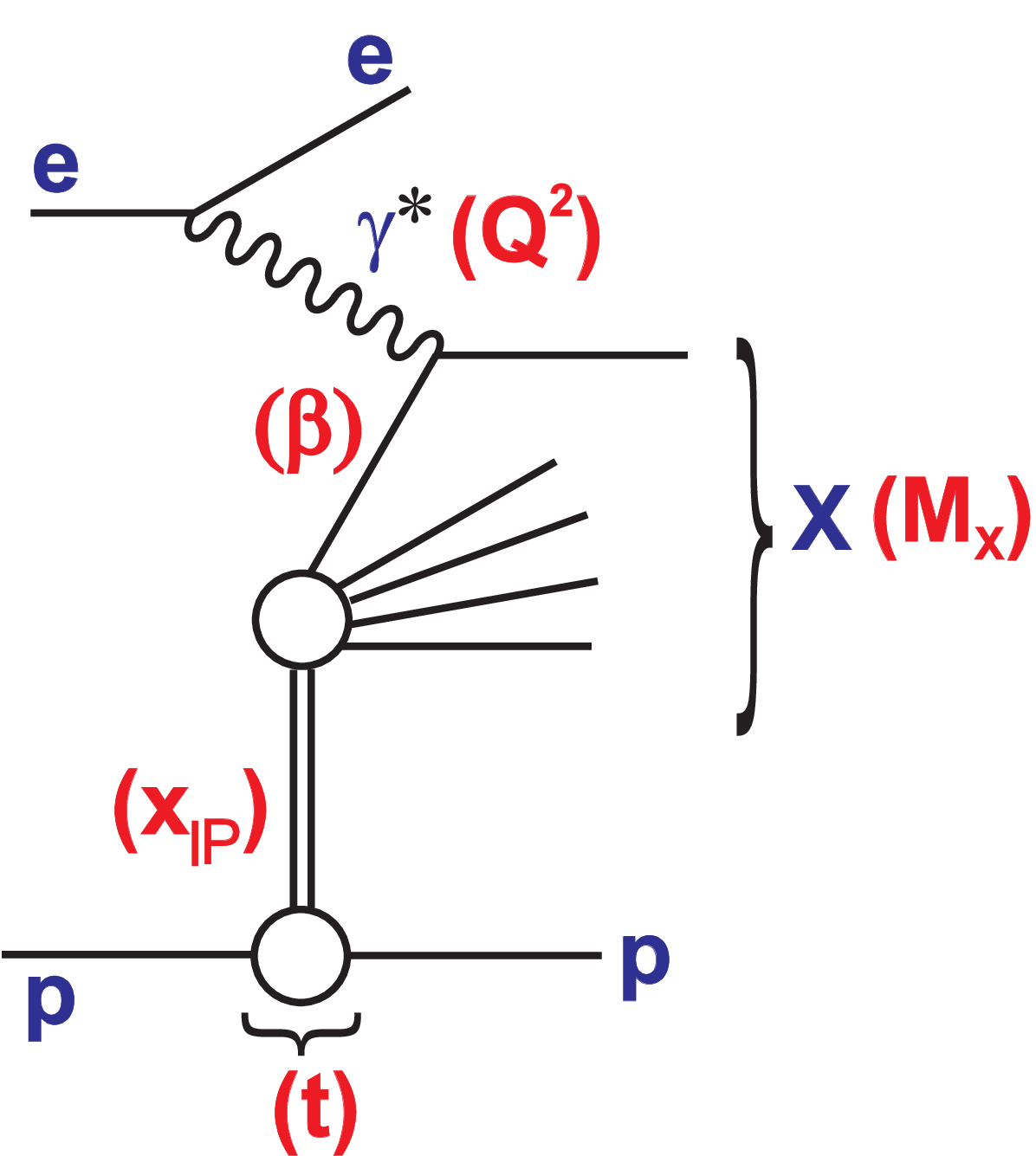}
            \hspace*{0.1cm}
            {\Large{\bf{(b)}}}
            \includegraphics[height=0.3\textwidth]{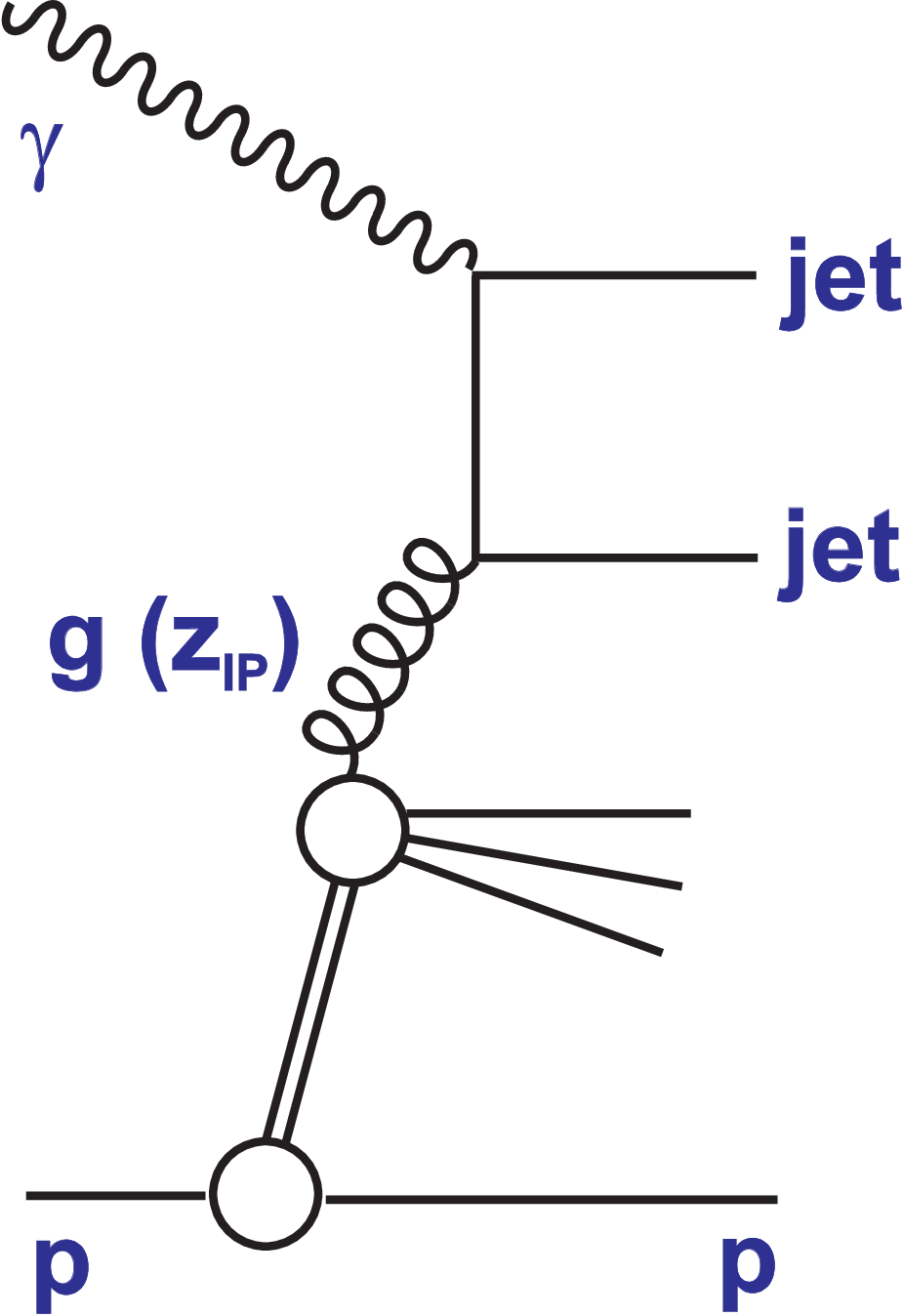}
            \hspace*{0.6cm}
            {\Large{\bf{(c)}}}
            \includegraphics[height=0.3\textwidth]{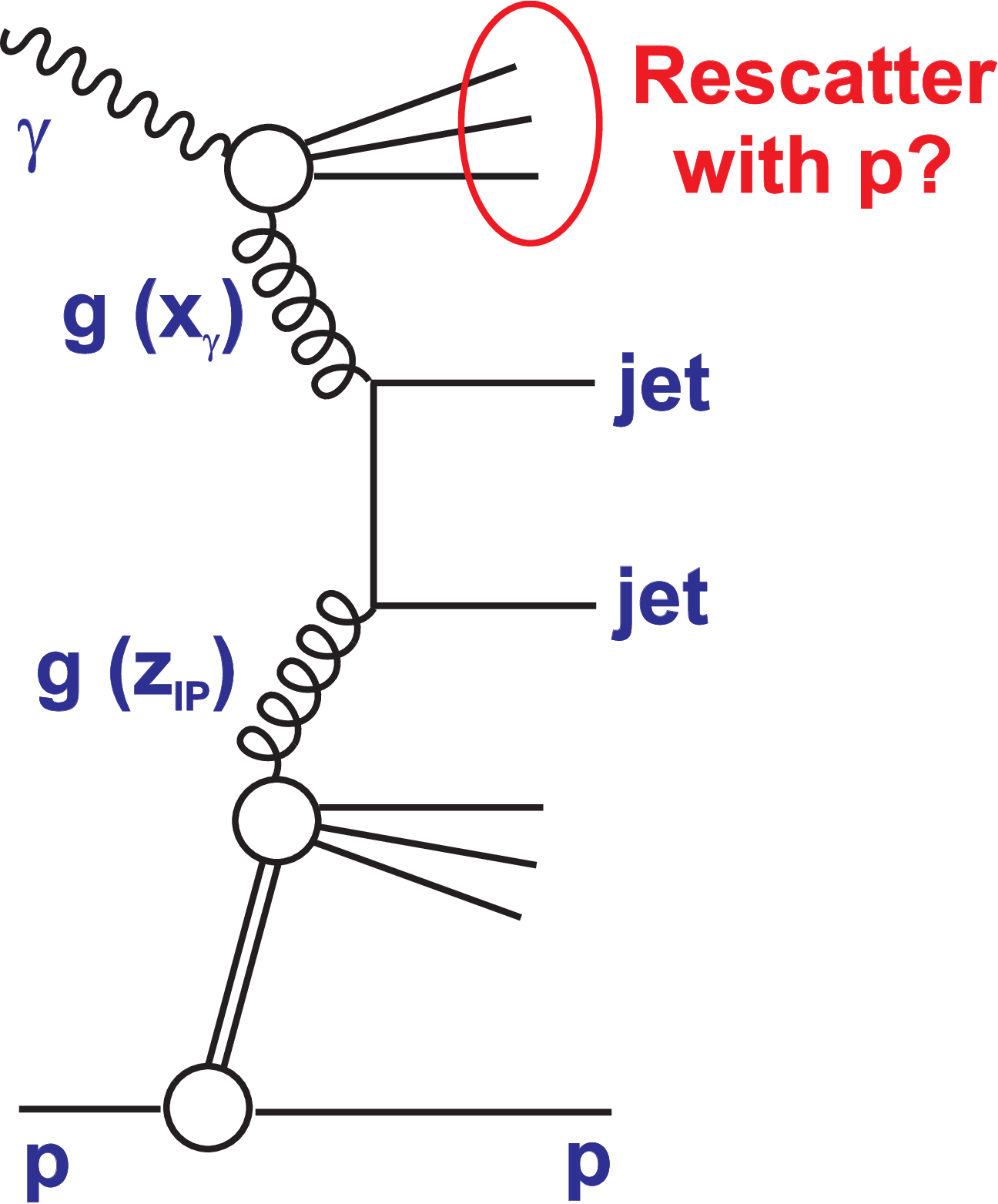}}
\caption{Sketches of diffractive $ep$ processes.
(a) Inclusive DDIS at the level of the quark parton model,
illustrating the kinematic variables 
discussed in the text. (b) Dominant leading order QCD diagram for 
hard scattering in DDIS or direct photoproduction, in
which a parton of momentum fraction $z_{I\!\!P}$ from the DPDFs
enters the hard scattering. (c) A leading
order process in resolved photoproduction involving a parton 
of momentum fraction $x_\gamma$ relative to the photon.}
\label{feynman}
\end{figure*}

The kinematic variables describing the inclusive
diffractive DIS (DDIS) process $ep \rightarrow eXp$ are 
illustrated in Fig.~\ref{feynman}(a).
In addition to $x$, $Q^2$ and $t \, (<0)$, there is one further 
non-trivial invariant, 
corresponding to the mass $M_X$ 
of the diffractive final state. 
In practice, the variable $M_X$ is usually expressed in terms of
\begin{equation}
\beta \;  = \; \frac{Q^2}{Q^2+M_X^2-t}\ .
\end{equation}
Small values of $\beta$ 
thus refer to events large diffractive masses, larger than $Q^2$,
whereas exclusive processes have values of $\beta$ close to unity.
In models based on a factorisable 
pomeron (Section~\ref{factorisation}), $\beta$ may be interpreted as
the fraction of the pomeron longitudinal momentum 
which is carried by the struck parton.
The longitudinal momentum fraction
of the colourless
exchange with respect to the incoming
proton is denoted $x_{_{I\!\!P}}$, such that $\beta \, x_{_{I\!\!P}} = x$.

Experimentally, diffractive $ep$ scattering 
is characterised by the presence of a leading proton in the 
final state, retaining most of the initial state proton energy, and
by a lack of any other hadronic activity in the 
forward (outgoing proton) direction, such that the
system $X$ is cleanly separated and $M_X$ may be measured in the central
detector components.  
These signatures have been widely exploited at HERA to select 
diffractive events by tagging the outgoing proton
in the H1 Forward Proton Spectrometer or 
the ZEUS Leading Proton Spectrometer
(proton-tagging method) or
by requiring the presence of a large gap in the rapidity 
distribution of hadronic final state particles 
in the forward region (LRG method). 
In a third approach 
($M_X$ method,~\cite{Breitweg:1998gc,Chekanov:2005vv,Chekanov:2008cw}), the inclusive
DIS sample is
decomposed into diffractive and non-diffractive contributions based
on their characteristic dependences on $M_X$.
Whilst the LRG and $M_X$-based techniques yield better
statistics than the LPS method, they suffer from systematic uncertainties 
associated with an admixture 
of proton dissociation to baryon states with small masses, 
typically $M_Y\, \leqsim\, 1.6 \ {\rm GeV}$, which 
is irreducible due to the limited forward detector acceptance.
Detailed comparisons between 
cross sections obtained by the LRG and proton 
tagging methods 
can be found in \cite{Chekanov:2008fh,Aaron:2010aa}.
Neither collaboration observes 
any evidence for deviations from a constant 
ratio of cross sections measured by the two methods, 
indicating that the ratio of probabilities for
the proton to scatter elastically and to undergo a low mass excitation 
is independent of the 
inclusive kinematic quantities. 

\subsubsection{Inclusive Diffraction Data}
\label{f2d}

Observations of diffractive
dissociation in DIS, \cite{Derrick:1993xh,Ahmed:1994nw}, 
and of jet production in diffractive 
photoproduction, \cite{Ahmed:1994ui,Derrick:1994yg}, and 
DIS, \cite{Derrick:1994ze}, 
were among the earliest HERA publications.
These were based on rapidity gaps which were much larger than could conceivably
occur at significant rates in standard fragmentation 
models, \cite{Andersson:1983ia}.
As increasingly larger datasets became available, 
the precision on the inclusive diffraction
cross sections improved correspondingly, until the final measurements covered
three-fold differential cross sections with a precision of a few 
percent, \cite{Aaron:2012ad,Chekanov:2008fh}, over large phase space regions. 

\begin{figure}[htp]
\begin{center}
~\epsfig{file=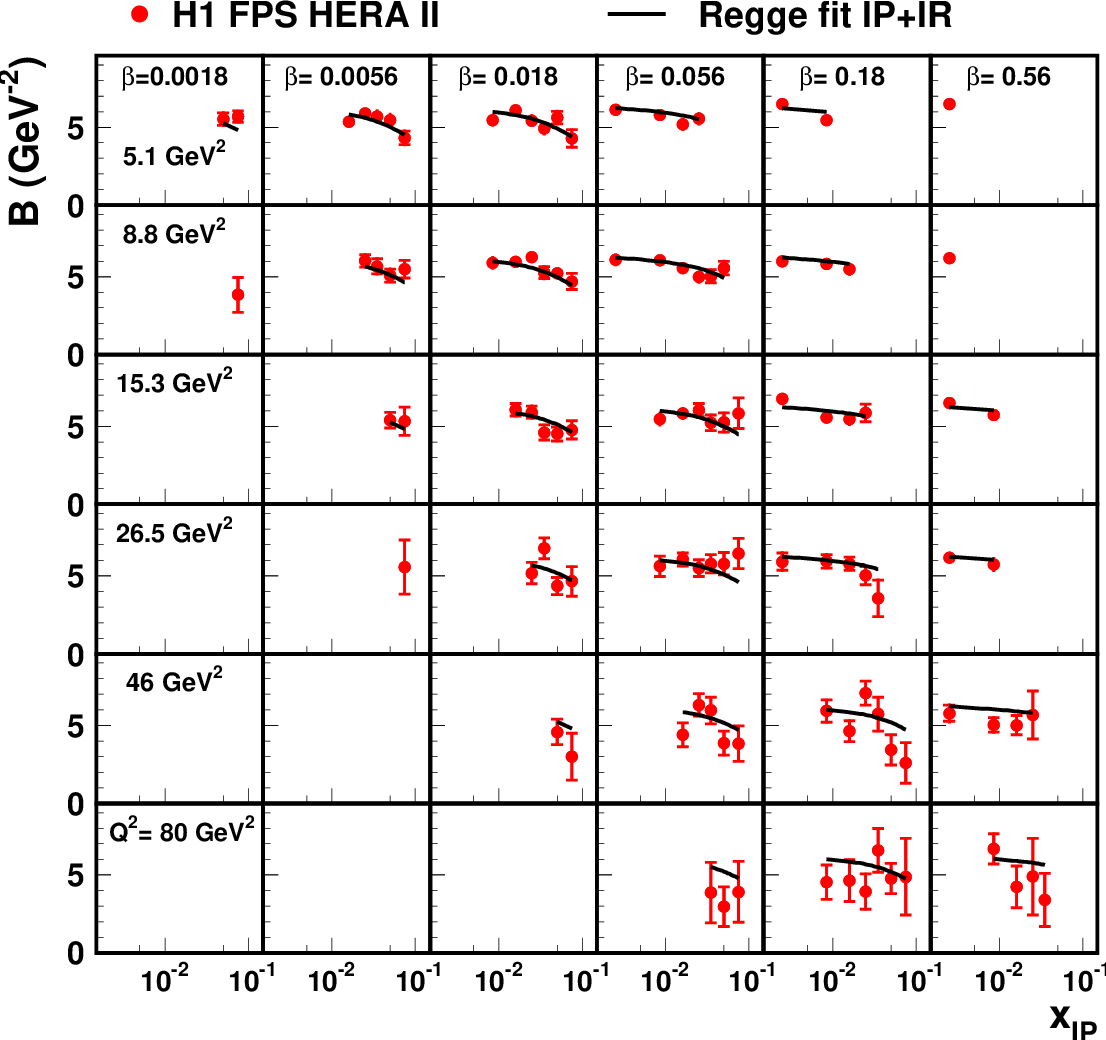,width=8.5cm}
\caption{Example data on the dynamics of the slope parameter, $B$,
describing the $t$ dependence of DDIS,
as measured using the H1 FPS.  From~\cite{Aaron:2010aa}.}
\label{fig:ddistslope}
\end{center}
\end{figure}

In analyses of data with protons tagged in 
the LPS and FPS, the $t$ dependence of DDIS has been measured and,
as in the case of exclusive vector meson production, found to
be compatible with an exponential parametrisation of the form of
Eq~\ref{eq:tslope}.
Example H1 data on the 
slope parameter $B$
are shown in Fig.~\ref{fig:ddistslope}, \cite{Aaron:2010aa}.
Similar results, with globally slightly larger $B$ values, are
obtained by ZEUS, \cite{Chekanov:2008fh}.
Beyond the low $M_X$ resonance region,
$B$ is typically around $6-7 \ {\rm GeV^{-2}}$,
independently of $\beta$ and $Q^2$, but with some dependence 
on $x_{_{I\!\!P}}$, which is further discussed in Section~\ref{sec:regge}. This is 
indicative of an 
approximate factorisation of the proton vertex from the virtual photon
vertex in DDIS (Section~\ref{factorisation}). 

The semi-inclusive DDIS cross section is usually presented in 
the form of a diffractive reduced cross section $\sigma_r^{D(3)}$,
which is related to the experimentally measured
differential cross section analogously to the inclusive case (Sec.~\ref{chap:exp-dat}), such that
\begin{equation}
\frac{{\rm d}^3\sigma^{ep \rightarrow e X p}}{\mathrm{d} x_{_{I\!\!P}} \ \mathrm{d} x \ \mathrm{d} Q^2} = \frac{2\pi
  \alpha^2}{x Q^4} \cdot Y_+ \cdot \sigma_{r}^{D(3)}(x_{_{I\!\!P}},x,Q^2) \ ,
\label{sigmar}
\end{equation}
where $Y_+ = 1 + (1-y)^2$ and $y$ is the usual Bjorken variable.
The reduced cross section depends 
at moderate scales, $Q^2$, on two
diffractive structure functions
$F_2^{D(3)}$ and $F_L^{D(3)}$ according to
\begin{equation}
\sigma_r^{D(3)} =
F_2^{D(3)} - \frac{y^2}{Y_+} F_L^{D(3)}.
\label{sfdef}
\end{equation}
For $y$ not too close to unity,
$\sigma_r^{D(3)} = F_2^{D(3)}$ holds to very good approximation.
Measurements have also been made of the 
full four dimensional reduced cross section, dependent also 
on $t$, \cite{Chekanov:2008fh,Aaron:2010aa}. 
However, since the 
$t$ dependence factorises to good approximation, this degree of 
freedom is usually integrated out.
A more detailed exposition of the formalism 
of diffractive DIS can be found for example in \cite{Aktas:2006hy}.

\begin{figure*}[tb]
\begin{center}
\epsfig{file=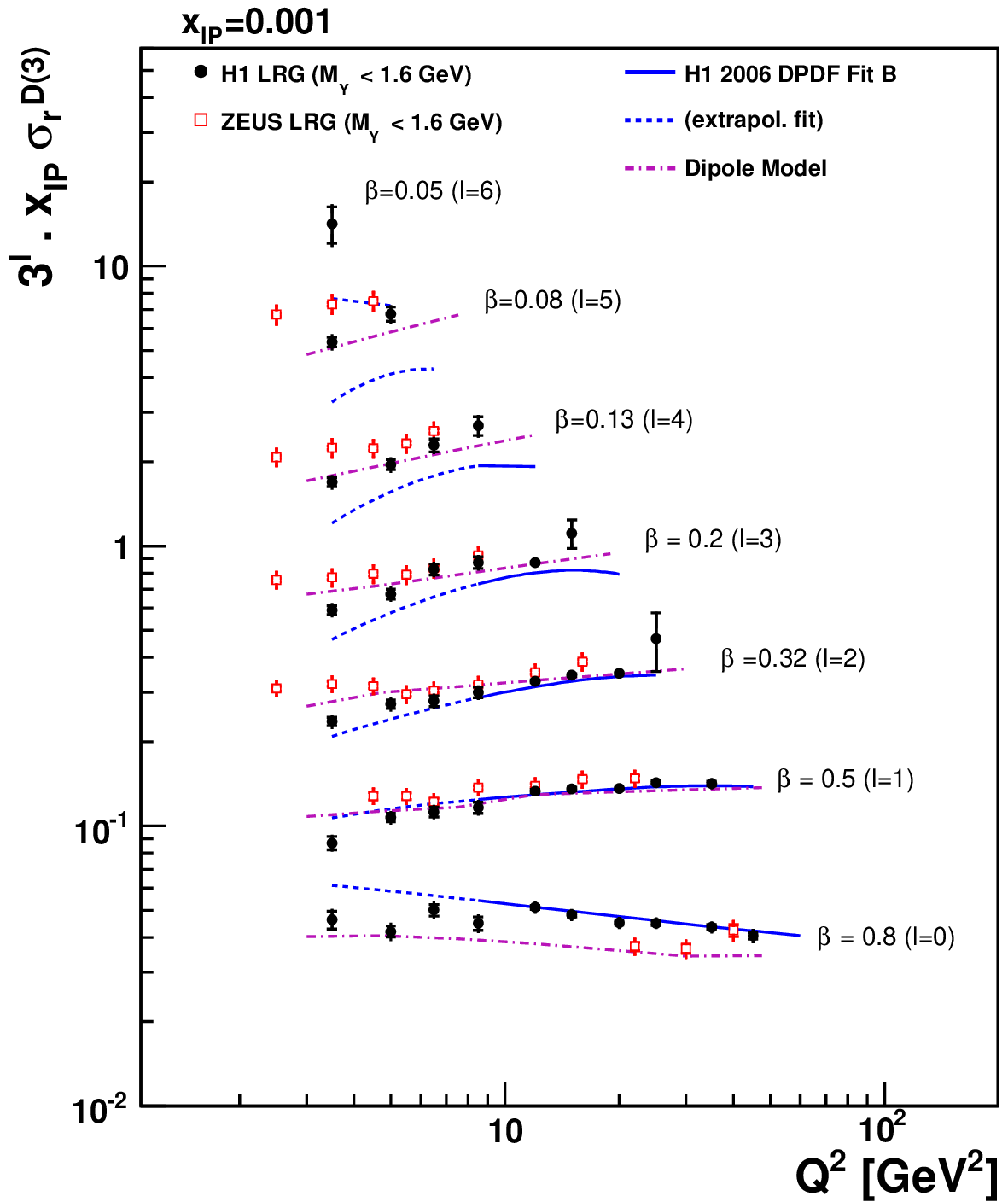,height=6.8cm}
\hspace*{2cm}
\epsfig{file=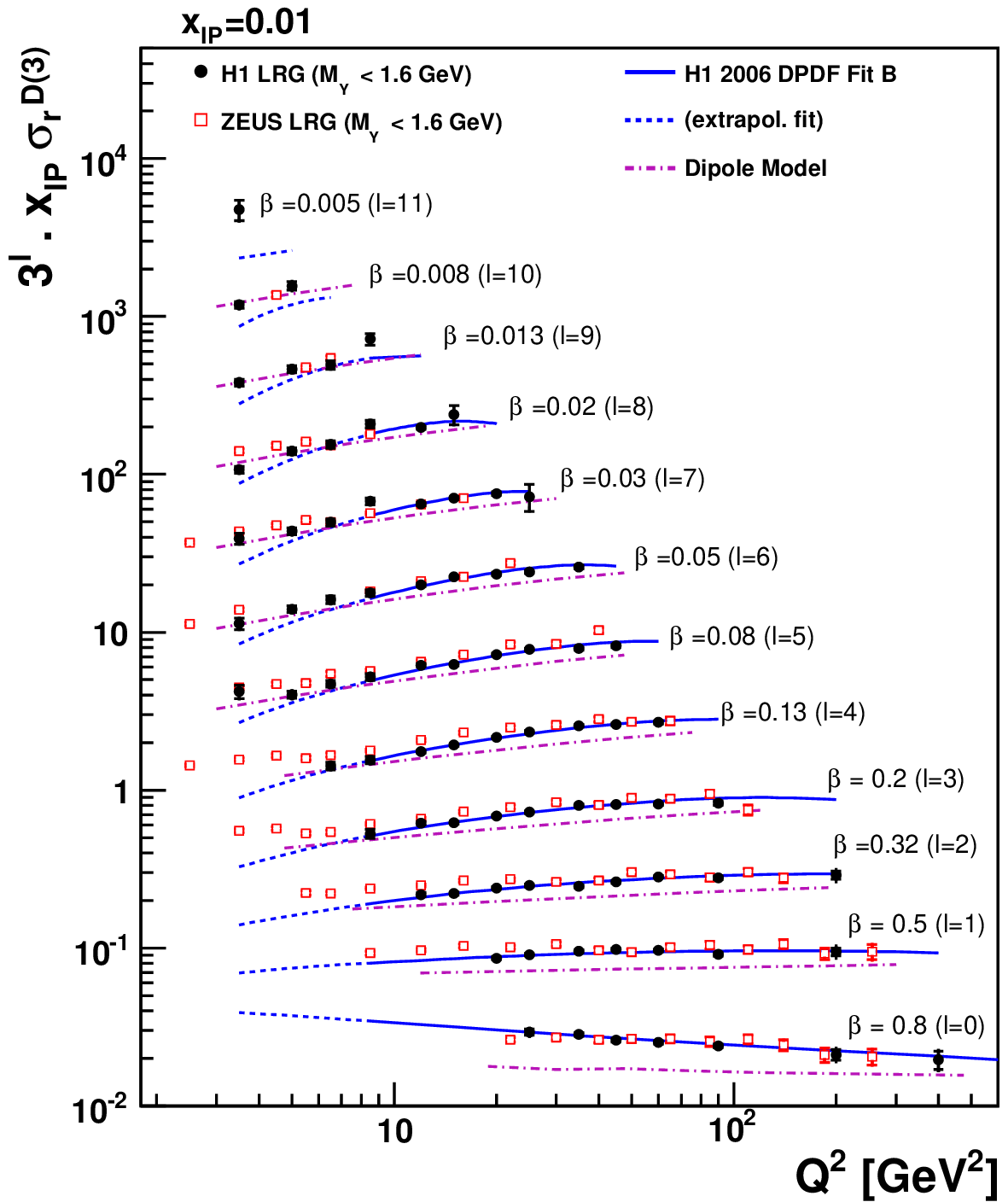,height=6.8cm}
\caption{Comparisons between H1 and ZEUS DDIS data measured using the
LRG method at example values of
$\xpom = 0.001$ and $\xpom = 0.01$. Data for these and other $\xpom$ values are presented in~\cite{Aaron:2012ad}.}
\label{LRGcomp}
\end{center}
\end{figure*}

A combination of data from the H1 and ZEUS experiments using the proton tagging method
is published in \cite{:2012vx}. 
The cross-calibration of the systematic 
uncertainties between the two experiments implicit in the averaging
procedure leads to an improvement in precision well beyond that
expected from statistical considerations alone.
Full combinations of data from the two collaborations for the LRG method
have not been performed, mainly due to some non-trivial discrepancies,
usually at the edges of the accessible phase space.
An investigation of the residual differences and crude combination may be 
found in \cite{Newman:2009km}. The comparison of final data in the
most precise region 
is illustrated in Fig.~\ref{LRGcomp}. 
Whilst there is good agreement in the bulk of the phase space, 
residual disagreements are apparent at small $\beta$ and
small $x_{_{I\!\!P}}$.
These discrepancies 
have only a small influence on the
interpretation of the data discussed in the following. 

Various models have been used to interpret DDIS data. Since it has become 
standard, Sections~\ref{factorisation}\,--\ref{gap:survival} 
focus mainly on the `proton vertex
factorisation' approach, in which the diffractive exchange (`pomeron') has
a resolved structure. The 
relevance of DDIS to dipole models is
discussed separately in Section~\ref{sec:dipoles}.
It is worth noting that a third approach, 
in which rapidity
gaps are randomly produced from final state colour interactions in otherwise
standard DIS events, has been at least partially successful. This model
is attractive in its simplicity. Indeed an early 
incarnation, \cite{Buchmuller:1995qa},
in which diffractive final states simply emerge randomly 
due to colour rotations in one ninth of all low $x$ DIS events works remarkably
well for inclusive data. Closely related ideas are encoded in 
the `Semi-classical model', \cite{Buchmuller:1997eb,Buchmuller:1998jv}.
The `Soft Colour Interaction' model, \cite{Edin:1995gi},
offers a rather different, but again related, approach, 
which later developed into the `Generalised Area Law', \cite{Rathsman:1998tp}
model. All of these approaches have enjoyed success,
particularly in comparisons with $F_2^D$ and $\sigma_r^D$ data,
but none provides a comprehensive description, 
breaking down for example when confronted with more exclusive 
processes such as diffractive
dijet production, \cite{Adloff:2000qi}.

\subsubsection{Factorisation Studies}
\label{factorisation}

Cleanly interpreting a three (or even four)-fold differential cross section is 
far from easy. It is simplified considerably if the 
dependences on the relevant kinematic variables can be 
factorised. QCD hard scattering collinear factorisation, when applied to
diffractive DIS, \cite{Collins:1997sr}, implies that the cross section for the process 
$ep \rightarrow eXY$ can be written in terms of convolutions of 
partonic cross sections $\hat{\sigma}^{e i} (x, Q^2)$ with 
diffractive parton density functions (DPDFs) $f_i^D$ as
\begin{align}
{\rm d} \sigma^{ep \rightarrow eXY} (x, Q^2, \xpom, t) = 
\sum_i \ 
& f_i^D(x, Q^2, \xpom, t) \nonumber \\
 \otimes \
& {\rm d} \hat{\sigma}^{ei}(x,Q^2) \ ,
\label{equ:diffpdf}
\end{align}
with additional implicit dependences on $\mu_F$, $\mu_R$ and $\alpha_S$
as in the analogous inclusive case.
The partonic cross sections
are the same as those for inclusive DIS.
This factorisation formula is valid for 
sufficiently large $Q^2$ and fixed
$\xpom$ and $t$.

The experimental data over most of the
accessible phase space are compatible with `proton vertex' factorisation, \cite{Ingelman:1984ns},
in which the DPDFs may be further factorised into a term containing only 
variables associated with the virtual photon vertex ($\beta$, $Q^2$) and a term containing 
variables associated with the proton vertex ($\xpom$, $t$):
\begin{equation}
f_i^D(x,Q^2,\xpom,t) = f_{\pom/p}(\xpom,t) \cdot
f_i (\beta=x/\xpom,Q^2) \, .
\label{reggefac}
\end{equation}
This is equivalent to considering
the diffractive exchange as a `pomeron'
with a partonic structure given by the parton
distributions $f_i (\beta,Q^2)$, the variable $\beta$ 
corresponding to the fraction of the pomeron longitudinal momentum
carried by the struck quark. The term
 $f_{\pom/p}(\xpom,t)$ represents the
probability that a pomeron with particular values of $\xpom$ and $t$
couples to the proton and is called the `pomeron flux factor'. 
Whilst this approach cannot be valid to ultimate precision (e.g.\ it
is clearly violated in the high $\beta$ region where higher twist processes
such as exclusive vector meson production play an important
role), it is justified theoretically for the leading
twist case, \cite{Blumlein:2001xf}.

As first shown in \cite{Adloff:1997sc}, a good description of the data
at relatively large $\xpom$ cannot be obtained by considering pomeron exchange alone. 
As in the case of total hadronic cross sections, \cite{Donnachie:1992ny}, 
sub-leading
exchanges become important away from the asymptotic limit ($\xpom \rightarrow 0$ in this case). 
An additional sub-leading
exchange ($\reg$) is therefore usually considered, contributing at low $\xpom$ and $\beta$ 
and exhibiting a similar factorisation to the pomeron term, such that 
Eq.~(\ref{reggefac}) is modified to
\begin{align}
f_i^D(x,Q^2,\xpom,t) & =  f_{\pom/p}(\xpom,t) \cdot f_i (\beta,Q^2) \nonumber \\
& + n_\reg \cdot f_{\reg/p}(\xpom,t) \cdot f_i^\reg (\beta,Q^2) \, ,
\label{reggefac2}
\end{align} 
where $n_\reg$ sets the relative normalisation of the sub-leading term.
In practice, the sub-leading exchange is not well constrained in the HERA kinematic regime.

With the above ansatz, a rather complete description is obtained of
all HERA inclusive diffractive data. 
The energy dependence of DDIS is encoded in the
pomeron flux factor, which is parametrised based on Regge phenomenology,
as discussed further in Section~\ref{sec:regge}. The DPDFs are treated in
a similar way to the case of inclusive DIS, as described in 
Section~\ref{dpdfs}. 
Full details of the standard fitting scheme adopted for the most recent data
by both collaborations can be found 
in \cite{Aktas:2006hy,Aktas:2006hx}. An improved heavy flavour
treatment is described in \cite{Chekanov:2009aa}.

\subsubsection{Energy Dependences and Soft Phenomenology}
\label{sec:regge}

In the fits to DDIS data described in Section~\ref{factorisation}, 
the $\xpom$ and $t$ dependences are
parametrised using a flux factor 
motivated by Regge theory,
\begin{eqnarray}
f_{\pom/p}(\xpom, t) = A_\pom \cdot 
\frac{e^{B_\pom t}}{\xpom^{2\alphapom (t)-1}} \ .
\label{eq:fluxfac}
\end{eqnarray}
The parameters
$B_\pom$ and $\alphapom^\prime$ and their uncertainties are obtained from
fits to FPS or LPS data, \cite{Aktas:2006hx}, which also take sub-leading contributions
into account. The $\xpom$ dependence of the data principally determines the
effective pomeron intercept $\alphapom (0)$, appropriate to the DDIS data. The 
most precise determinations are obtained from LRG data, with remarkable 
consistency
between statistically independent data sets: 
\begin{align}
\alphapom(0) = 1.118 \ \pm 0.008  \ {\rm{(exp.)}}  
& ^{+ 0.029}_{- 0.010} \ {\rm{(model)}}
& \text{\cite{Aktas:2006hy}} \, ; \nonumber  \\
\alphapom(0) = 1.117 \ \pm 0.006 \ {\rm{(exp.)}}   
& ^{+ 0.022}_{- 0.007} \ {\rm{(model)}} 
& \text{\cite{Chekanov:2008fh}} \, ; \nonumber \\
\alphapom(0) = 1.113 \ \pm 0.002 \ {\rm{(exp.)}}   
& ^{+ 0.029}_{- 0.015} \ {\rm{(model)}} 
& \text{\cite{Aaron:2012ad}} \, , \nonumber 
\end{align}
the model dependence uncertainties being highly correlated between
the different data sets. 

These $\alphapom(0)$ 
values are only slightly larger than their counterparts obtained
from fits to 
total, \cite{Donnachie:1992ny,Cudell:1996sh},
and diffractive, \cite{Goulianos:1982vk,ATLAS:2012djz},
cross sections in $pp$ and $p \bar{p}$ scattering 
and are compatible with results from soft photoproduction at 
HERA, \cite{Adloff:1997mi,Breitweg:1997za}. They are 
significantly smaller than the results obtained from the
energy dependences of hard exclusive processes at HERA 
(Section~\ref{vm:data}) and are 
much smaller than values predicted
based on leading-logarithmic 
BFKL, \cite{Balitsky:1978ic,Kuraev:1976ge,Kuraev:1977fs}, 
or other `hard pomeron', \cite{Donnachie:1998gm}, approaches. 
This supports the picture of the dominating process
in DDIS being the application of a hard virtual photon probe 
to an approximately factorisable
exchange object, closely related to the 'soft pomeron' of 
soft hadronic scattering.

The slope of the effective DDIS
pomeron trajectory can be determined from the $\xpom$ dependence of
the slope parameter $B$, as obtained using the LPS and FPS detectors. In this case, the
H1 and ZEUS results are:
\begin{align}
\alphapom^\prime  = -0.01 \ \pm 0.06 \ {\rm{(stat.)}}  
& ^{+ 0.04}_{- 0.08} \ {\rm{(syst.)}} \ 
\pm 0.04 \ {\rm{(model)}} 
& \text{\cite{Chekanov:2008fh}} \, ; \nonumber \\
\alphapom^\prime  = \hspace{0.27cm} 0.04 \ \pm 0.02 \ {\rm{(exp.)}}   
& ^{+ 0.08}_{- 0.06} \ {\rm{(model)}} 
& \text{\cite{Aaron:2010aa}} \nonumber .
\end{align}
Whilst the precision on these results is limited, the trajectory slope is
clearly smaller than the canonical value from 
soft $pp$ and $p \bar{p}$ scattering of
$\alphapom^\prime \sim 0.25$, \cite{Donnachie:1984xq,Abe:1993xx}.
As discussed in Section~\ref{vm:data},
soft vector meson photoproduction data also show lower
$\alphapom^\prime$ values than those usually taken from $pp$ and $p \bar{p}$ 
scattering, \cite{Breitweg:1999jy},
suggesting that this is a heavily process dependent parameter,
for example due to different absorptive corrections. 

\begin{figure}[htp]
\begin{center}
~\epsfig{file=DESY-08-175_21.eps,width=9cm}
\caption{Summary 
of ZEUS data on tests of proton vertex factorisation in the form 
of possible $Q^2$ variations of the pomeron intercept in 
fits to LRG and LPS data, 
as well as `FPC' data obtained by the $M_X$ decomposition method. From~\cite{Chekanov:2008fh}.}
\label{alphapom:fac}
\end{center}
\end{figure}

The validity of the factorisation assumption implicit in the fits to 
inclusive diffractive
data is tested by adding further free parameters, which allow the 
pomeron intercept and the
slope parameter $B$ to vary freely between different 
$Q^2$ or $\beta$ bins. A summary of ZEUS
results on possible variations of $\alphapom(0)$ with $Q^2$ is shown in 
Fig.~\ref{alphapom:fac}. Both here and in corresponding H1 studies 
(see e.g.~\cite{Aaron:2012ad}), there is no evidence for any variation of the 
effective pomeron intercept with $Q^2$ or $\beta$ when considering 
data taken using either the LRG or LPS/FPS methods. 
Similar searches for a $Q^2$ or $\beta$ dependence of $\alphapom^\prime$
or $B$ have also yielded null results, \cite{Aaron:2010aa}, 
albeit with lesser precision.

\begin{figure}[htp]
\begin{center}
~\epsfig{file=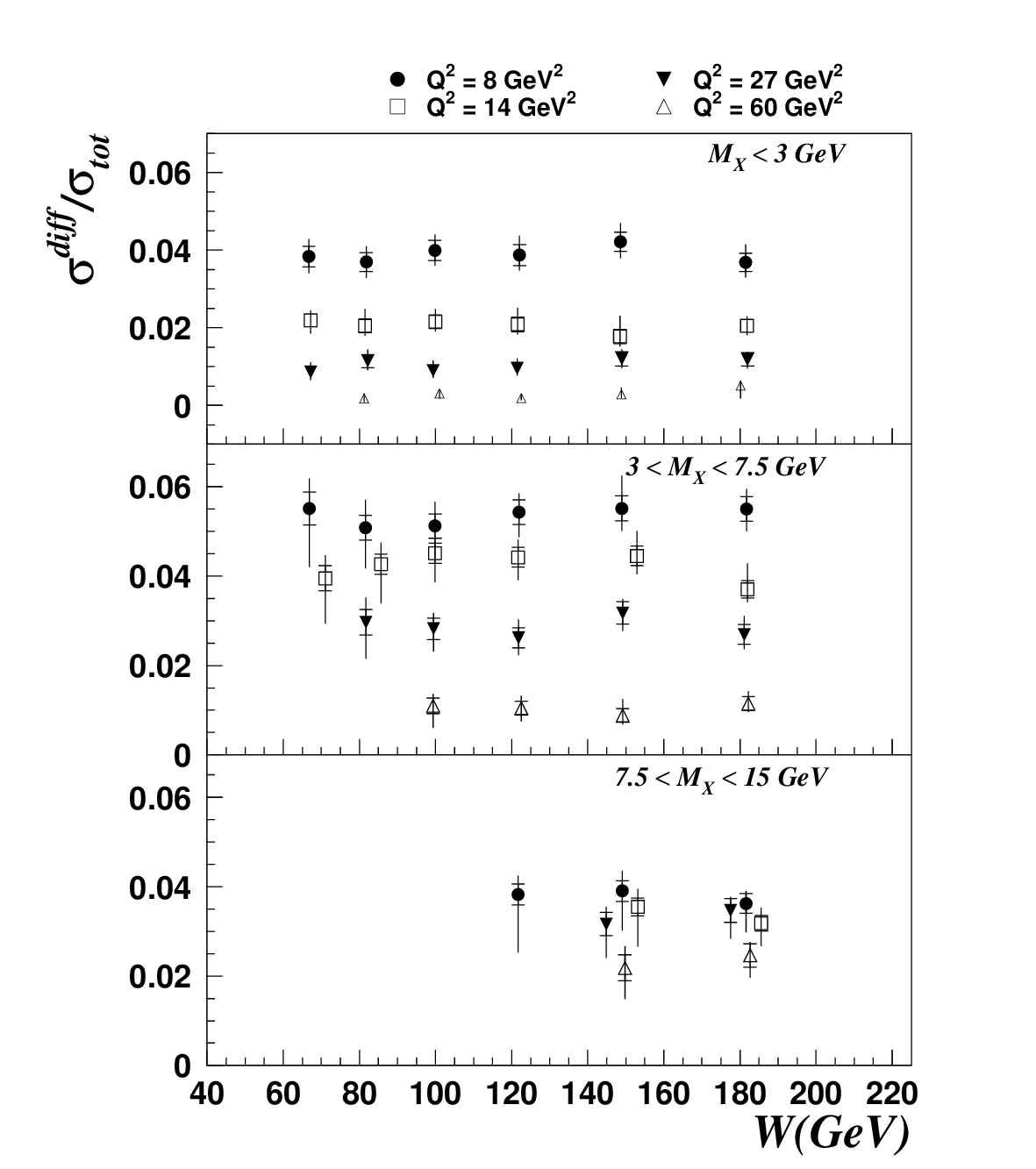,width=9cm}
\caption{The ratio of diffractive to inclusive cross sections 
as a function 
of $W$, integrated over different $M_X$ regions at 
fixed $Q^2$. Obtained using ZEUS 1994 data.  From~\cite{Breitweg:1998gc}.}
\label{zeus:difratio}
\end{center}
\end{figure}

It is informative to compare the photon--proton centre-of-mass energy 
($W$ or equivalently
$1/x$ at fixed $Q^2$) dependences of the diffractive and the inclusive cross
sections. Basic Regge pole phenomenology predicts that the growth 
with centre-of-mass energy of the
diffractive cross section
($\sim (W^2)^{2 \alphapom(t) - 2}$ at fixed $Q^2$ and $\beta$, with 
$\langle t \rangle \sim 1/B \sim 0.2\,{\rm GeV^2}$) 
should be 
faster than that of the total cross section ($\sim (W^2)^{\alphapom(0) - 1}$). 
However, in numerous
studies of HERA data, the ratio 
of diffractive to inclusive cross sections
has been found
to be relatively flat as a function of $W$. 
The first example, \cite{Breitweg:1998gc}, is shown in Fig.~\ref{zeus:difratio}.  See also
\cite{Aktas:2006hy,Aaron:2012ad}. 
This flatness represents a clear breakdown of the 
simple Regge approach, showing 
conclusively that there is no universal pomeron 
in virtual photon--proton scattering, as already
discussed in Section~\ref{vm:data}. The flatness of the ratio has been 
interpreted, \cite{GolecBiernat:1998js,GolecBiernat:1999qd},
as evidence for saturation of low $x$ parton densities, which 
is expected to be visible in diffraction at
higher $x$ than in the inclusive case, due to the exchange of two, 
rather than one, gluon in the simplest interpretation. 
Saturation effects would tend to reduce the diffractive cross section
relative to the inclusive cross section in a manner which 
becomes more important as $x$ falls or $W$ grows.
Whilst the flatness of the ratio of these cross sections is thus a highly suggestive observation, 
it has also been explained without invoking saturation, 
for example in models which explain rapidity gap formation as
a random process, due to  
soft colour rearrangements, \cite{Buchmuller:1998jv,Edin:1995gi}.

\subsubsection{QCD Phenomenology and Diffractive Parton Densities}
\label{dpdfs}

Fig.~\ref{LRGcomp} immediately
leads to the conclusion that the scaling violations in diffraction remain of
positive sign ($\partial \sigma_r^D / \partial Q^2 > 0$) up to large values of  
$\beta\, \sim\, 0.5$, which may be compared with $x \sim 0.1$
for the
inclusive cross section.\footnote{See \cite{Aktas:2006hy}
for an alternative comparison between scaling violations in 
diffractive and inclusive DIS considered at the same $x$ values in
each case.} 
This is indicative of a 
large role of gluons in the diffractive exchange, \cite{Ahmed:1995ns}. 
This qualitative conclusion
is formalised through the extraction of DPDFs, as described in more detail
in this section. 

The DPDFs in Eq.~\ref{equ:diffpdf}
represent probability distributions for partons $i$ in the proton
under the constraint that the proton is scattered to a system $Y$ with a specified four-momentum. 
They are essentially equivalent to the fracture functions developed
in an earlier approach, \cite{Trentadue:1993ka}.
They are not known from first principles, but 
can be determined from fits to the data using the DGLAP evolution
equations,~\cite{Blumlein:2001xf}.
Due to kinematic constraints,
it is not possible to access the full range of $\beta$ and $Q^2$
using data from only one value of $\xpom$ at HERA.
The parametrisations of the $\xpom$ 
dependence of the DPDFs described in Section~\ref{sec:regge}, together with the 
proton vertex factorisation assumption introduced in 
Section~\ref{factorisation},
is therefore adopted, such that data from multiple $\xpom$ values can 
be used simultaneously in extracting DPDFs which then depend on
$\beta$ and $Q^2$ only.

DPDFs have been extracted using standard NLO QCD procedures, similar to those 
employed to extract inclusive proton PDFs (Sec.~\ref{sec:lowxinterp}), but with smaller numbers of free parameters
for the momentum fraction dependence of the parton densities at the starting
scale for evolution. This momentum fraction is usually denoted $z$ or 
$z_\pom$ (it is equal to $\beta$ when used to describe the quark coupling to the
exchange boson). Only two distinct parton densities
are generally considered: a gluon density and a singlet quark density, which is assumed
to be flavour symmetric 
between up, down and strange quarks and their antiquarks. 
This latter assumption, particularly in regard to the strange quark density, 
is rather ad hoc and remains to be tested in detail.

\begin{figure}[htp]
\begin{center}
~\epsfig{file=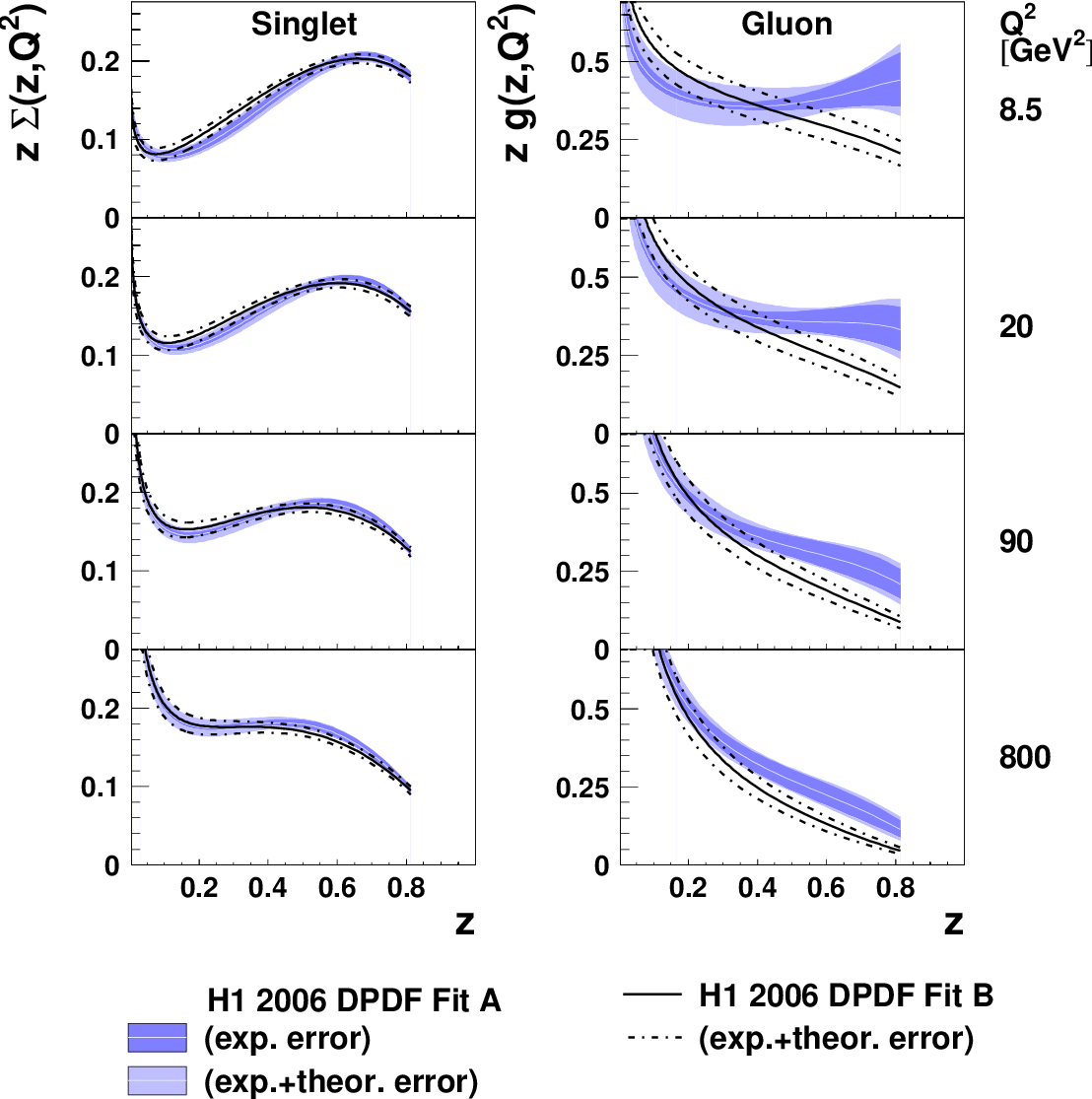,width=9cm}
\caption{DPDFs extracted from an H1 fit to inclusive LRG data 
only. The two different sets of DPDFs shown 
(Fit A and Fit B) have very similar fit qualities, illustrating the
lack of sensitivity to the gluon density at large momentum fractions. 
From~\cite{Aktas:2006hy}.}
\label{H1:DPDFs}
\end{center}
\end{figure}

\begin{figure}[htb] \unitlength 1mm
  \begin{center}
    \begin{picture}(60,30)
      \put(-15,-3){\epsfig{file=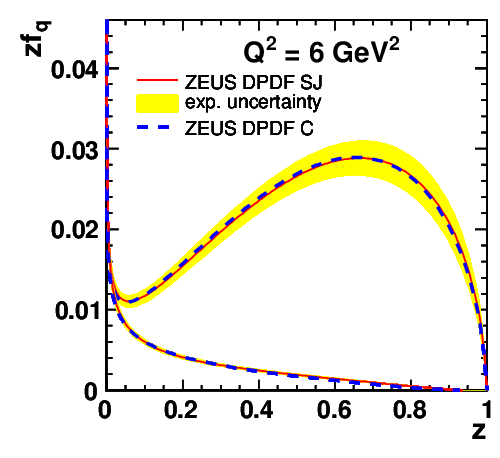,height=0.22\textwidth,clip}}
      \put(30,-3){\epsfig{file=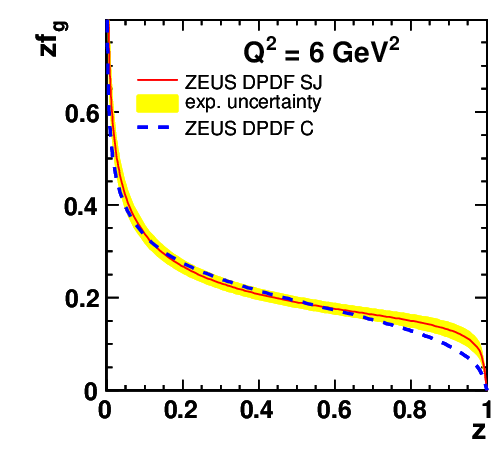,height=0.22\textwidth,clip}}
      \put(8,6){{\scriptsize{charm}}}
      \put(14,18){{\scriptsize{light}}}
    \end{picture}
  \end{center}
  \caption{Diffractive quark (left) and 
gluon (right) distributions 
extracted from a ZEUS fit to inclusive LRG and LPS 
data, with diffractive dijet data also included.  From~\cite{Chekanov:2009aa}.}
\label{ZEUS:DPDFs}
\end{figure}

Example
results obtained when fitting inclusive $\sigma_r^D$ data alone can be found
in \cite{Aktas:2006hy,Chekanov:2009aa}, see Fig.~\ref{H1:DPDFs}. 
Since $\sigma_r^D$ measures essentially the
charge-squared-weighted sum of quarks, these fits result in tight constraints
(to around 5\%) on the singlet quark density. From the scaling violations, the
gluon distribution is also constrained (to around $10 - 15 \%$) at moderate $z$ values.
However, the sensitivity to the gluon density with $z \, \geqsim \, 0.5$ is poor, due to
the inevitable dominance of the evolution by the
$q \rightarrow qg$ splitting as $z \rightarrow 1$. 
An additional constraint can come from diffractive jet production (see Section~\ref{apply:dpdf}).
such that the high $z$ gluon precision is
improved considerably. However, this comes at the expense of additional theoretical
assumptions on the large $\beta$ dynamics. Results of fits of this type can be
found in \cite{Aktas:2007bv,Chekanov:2009aa}, see Fig.~\ref{ZEUS:DPDFs}.

In all fits performed, the DPDFs are dominated by the gluon density, 
which extends
to large values of $z$ and accounts for typically 
60 \% (ZEUS \cite{Chekanov:2009aa}) to
70\% (H1 \cite{Aktas:2006hy}) of the total 
longitudinal momentum of the diffractive 
exchange. 
It is interesting to note that the
DPDFs, particularly the quark 
distribution, resemble the parton densities of the 
photon, \cite{Nisius:1999cv}. 
The hadronic structure of the photon is generated from an initial 
$\gamma \rightarrow q \bar{q}$ splitting, with lower $x$ structure emerging
from further splittings and evolution. This superficial similarity is
as expected if diffraction emerges from a single gluon exchange,
dressed in a manner which neutralises the colour. The structure then develops
from $g \rightarrow q \bar{q}$ similarly to the photon case, but with an
admixture of $g \rightarrow gg$ initial splittings.

More sophisticated approaches to DPDFs exist, notably those
in which a direct, hard pomeron contribution is also
included, \cite{Martin:2006td}, or those which include a higher
twist contribution at large $\beta$ 
(see Section~\ref{sec:dipoles}), \cite{Royon:2000wg,GolecBiernat:2007kv}.
Whilst these contributions arise naturally
in QCD and must be present at some level, the evidence both from these 
fits and from hadronic final state comparisons 
(Section~\ref{apply:dpdf}) is that they
are numerically small, compared with the standard 
resolved structure contribution. 
The DPDF approach does, however, appear to undergo an infrared
breakdown at larger $Q^2$ 
scales than is the case in inclusive QCD 
fits, \cite{Aktas:2006hy,Chekanov:2009aa}. 
Whilst this may provide further evidence for saturation 
effects in DDIS, \cite{Frankfurt:2001av}, 
it may also be a consequence of enhanced higher twist
contributions, or of a less quickly convergent QCD order expansion in the 
diffractive than the inclusive case.

\subsubsection{Applications of Diffractive Parton Densities}
\label{apply:dpdf}

The DPDFs described in Section~\ref{dpdfs} have been 
used to predict a wide variety of observables in DIS and photoproduction
at HERA, as well as at the Tevatron and the LHC. At one level, this is
done through implementations 
(usually of the H1 Fit B DPDFs, \cite{Aktas:2006hy}) in Monte Carlo 
generators, \cite{Jung:1993gf,Cox:2000jt,Navin:2010kk}. 
In particular,
the {\sc Rapgap} model has been used extensively as an experimental tool for modelling
both inclusive and hadronic final state diffractive data at HERA. 
The DPDFs have also been 
interfaced to various NLO QCD
calculations in order to compare their predictions with measurements. 


Early measurements of
inclusive final state observables in DDIS
were sufficient to rule out a diffractive exchange
with a quark-dominated structure, by showing that the basic event
topology is consistent with the boson--gluon fusion process, 
$\gamma^* g \rightarrow q \bar{q}$ (Fig.~\ref{feynman}(b)), 
yielding copious high $p_T$ particle production.
The full list of inclusive final state observables measured in DDIS comprises
charged particle spectra, multiplicities and their rapidity 
correlations, \cite{Adloff:1998ds,Adloff:1998dw,Chekanov:2001ph},
energy flows, \cite{Adloff:1998ds,Chekanov:2001ph}, and event
shapes (thrust and sphericity), \cite{Breitweg:1997ug,Adloff:1997nn,Chekanov:2001ph}. 
These measurements are universally in agreement with predictions based on DPDFs
extracted from $\sigma_r^D$ data, reinforcing the 
picture of a diffractive exchange dominated by gluons extending to large momentum fractions. 

The most precise tests of DPDFs in diffractive DIS have been obtained from exclusive final
state observables formed from relatively high transverse momentum jet
production, the rate for which is
closely related to the diffractive gluon density. Dijet measurements have been 
made as a function of many different 
variables, \cite{Aktas:2007bv,Aktas:2007hn,Chekanov:2007aa,Aaron:2011mp},
and have been extensively used in comparisons with DPDF-based NLO 
predictions, \cite{Aktas:2007bv,Chekanov:2009aa}. 
An example is shown in Fig.~\ref{diffdis:jets}.
This sort of comparison has again been universally successful.
The strong experimental evidence for proton vertex factorisation and the applicability
of DPDFs to diffractive dijets in DIS, together with the relatively high precision 
with which the momentum fraction $z$ can be reconstructed
from the kinematics of the 
diffractive final state and the jet pair, led to 
the jet data being 
used as an input to DPDF extractions, \cite{Aktas:2007bv,Chekanov:2009aa},
as discussed in Section~\ref{dpdfs}.

\begin{figure}[htb]
\begin{center}
~\epsfig{file=DESY-09-191_7.eps, width=0.52\textwidth}
\caption{Comparison of diffractive DIS dijet cross sections, measured
double differentially in $z_\pom$ and $Q^2$,  
with NLO QCD predictions based on DPDFs. In this case, the data shown were used together 
with inclusive diffraction measurements in the fit, to improve the high $\beta$
sensitivity. 
From~\cite{Chekanov:2009aa}.}
\label{diffdis:jets}
\end{center}
\end{figure}

\begin{figure*}[tb]
\begin{center}
~\epsfig{file=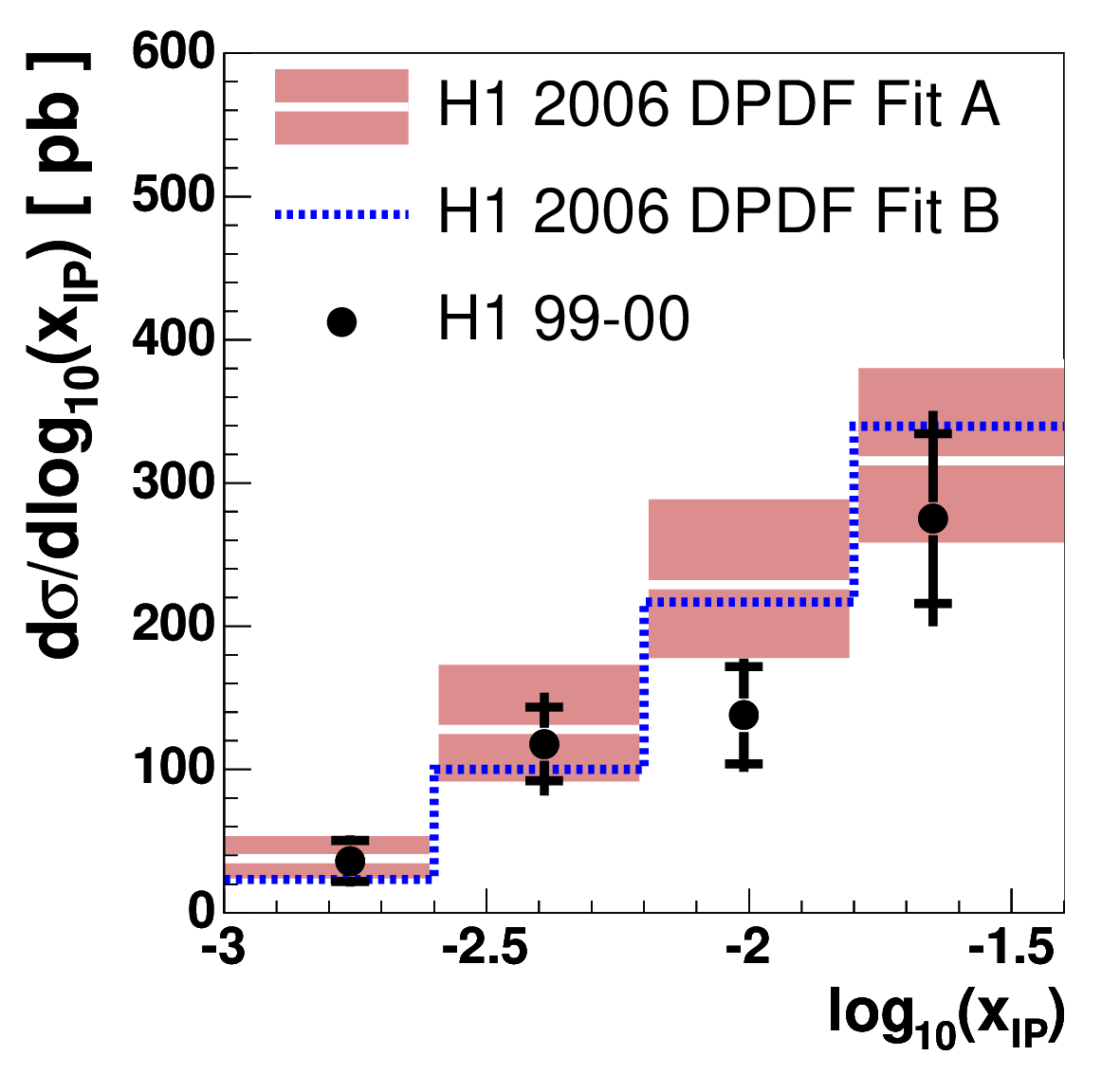,height=4.5cm}
~\epsfig{file=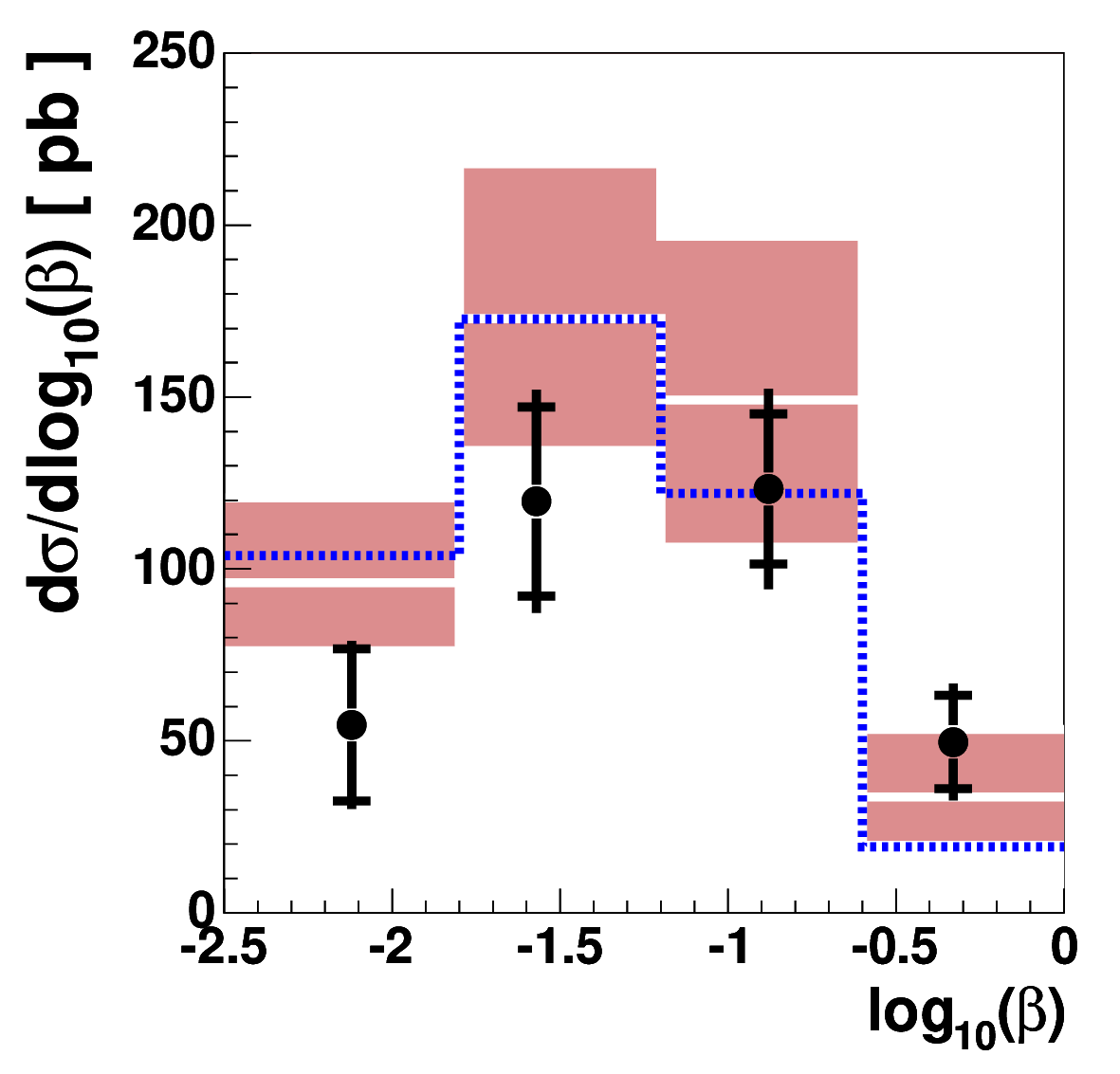,height=4.5cm}
~\epsfig{file=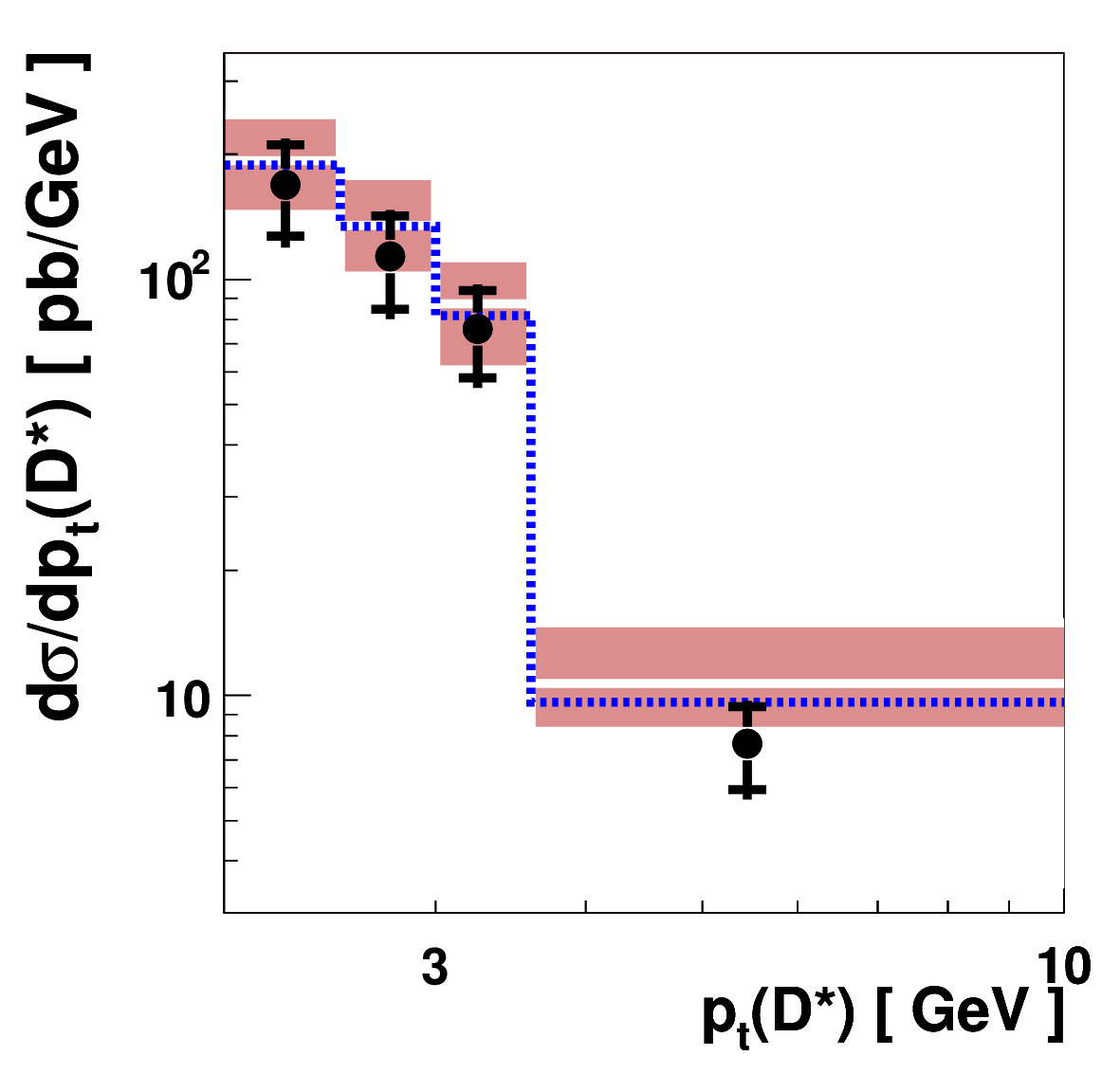,height=4.5cm}
\caption{Comparisons between diffractive open charm ($D^*$) production measurements
and NLO QCD predictions based on DPDFs.
From~\cite{Aktas:2006up}.}
\label{diffdis:charm}
\end{center}
\end{figure*}

Diffractive open charm production in DIS, 
\cite{Aktas:2006up,Chekanov:2003gt},
provides a further exclusive final
state with high sensitivity to the diffractive gluon density via the lowest
order contributing process, $\gamma^* g \rightarrow c \bar{c}$. 
The charm quarks are usually tagged through the reconstruction of a $D^*$
meson using the $D^* - D^0$ mass difference 
technique.
However, due to larger backgrounds
and smaller statistics, charm observables have not yielded the precision 
tests of diffractive factorisation and DPDFs that have been achieved with jets. In
principle, due to the weaker kinematic constraints on $M_X$, 
it should be possible to probe smaller $z$ values by using charm as a
tag. However, the problems of forward-going track reconstruction experienced by
both collaborations has limited progress in this direction. Nonetheless, as can be seen
from the example comparisons in Fig.~\ref{diffdis:charm}, open charm production is
well described by DPDF-based models, providing complementary support for this approach
and a constraint on the heavy flavour treatment in diffractive QCD fits.

\begin{figure}[htb]
\begin{center}
~\epsfig{file=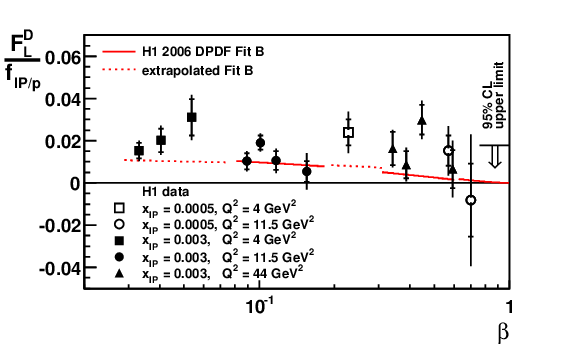}
\caption{Summary of H1 measurements of the diffractive longitudinal structure 
function $F_L^D$, exploiting data taken with variations in the proton beam energy. 
To allow comparisons between measurements at different $\xpom$ values,
the data are normalised by a diffractive flux factor $f_{\pom/p}$ 
(see Section~\ref{sec:regge}).  The data are compared with an NLO QCD prediction 
based on DPDFs.  
From~\cite{Aaron:2012zz}.}
\label{FLDfig}
\end{center}
\end{figure}

A final class 
tests of the consistency of the DPDF model in DIS arises from the inclusive diffractive
cross sections. 
An H1 measurement of the diffractive charged current cross
section is shown to be consistent
with DPDF-based predictions in \cite{Aktas:2006hy}. The rather ad hoc flavour-democratic assumptions 
made in the standard QCD fitting procedures (Section~\ref{dpdfs}) are
consistent with the data within the large statistical uncertainties, but the diffractive
charged current cross section is the only measurement made to date which is sensitive to the 
light quark flavour decomposition of the diffractive quark density.
The diffractive
longitudinal structure function, $F_L^D$, \cite{Newman:2005mm}, is a further independent observable
with sensitivity to both the quark and the gluon densities in novel ways. 
In particular, it provides a
unique test of the low-$x$ gluon.
Significantly non-zero results have been 
obtained, \cite{Aaron:2012zz},
using the Rosenbluth technique, comparing diffractive reduced cross section data at
fixed $\xpom$, $\beta$ and $Q^2$, but different $ep$ centre-of-mass energy, 
exploiting the
data taken at $\sqrt{s} = 460 \ {\rm GeV}$ and $\sqrt{s} = 525 \ {\rm GeV}$ at the
very end of the HERA running. The sensitivity to $F_L^D$ requires highly challenging
measurements at large $y$ (small scattered electron energy). Nonetheless, this structure
function has been measured over a fairly large kinematic region. As summarised in 
Fig.~\ref{FLDfig}, the data are again supportive of the DPDF approach. 

\subsubsection{Limitations of Diffractive Parton Densities}
\label{gap:survival}

Whilst models based on proton vertex factorisation and DPDFs work well to
describe all diffractive processes beyond the lowest $M_X$ resonance region
in DIS, they fail spectacularly when DPDFs extracted from HERA $\sigma_r^D$
data are applied to diffractive $p \bar{p}$ scattering at the Tevatron or LHC. 
For example, predictions for diffractive dijet production at the 
Tevatron exceed the data by  
a factor of around 10, \cite{Affolder:2000vb,Klasen:2009bi}. This 
limitation is expected as part of the QCD hard scattering factorisation
theorem for diffraction, \cite{Collins:1997sr}, and is
usually interpreted in terms of
multiple scattering, or `absorptive' effects, which occur in the 
presence of beam remnants.
These effects can be parametrised in terms of a `rapidity gap survival 
probability', \cite{Dokshitzer:1991he,Bjorken:1992er}. 
Similar effects are emerging at the LHC, \cite{Chatrchyan:2012vc}. 

\begin{figure}[htb]
\begin{center}
~\epsfig{file=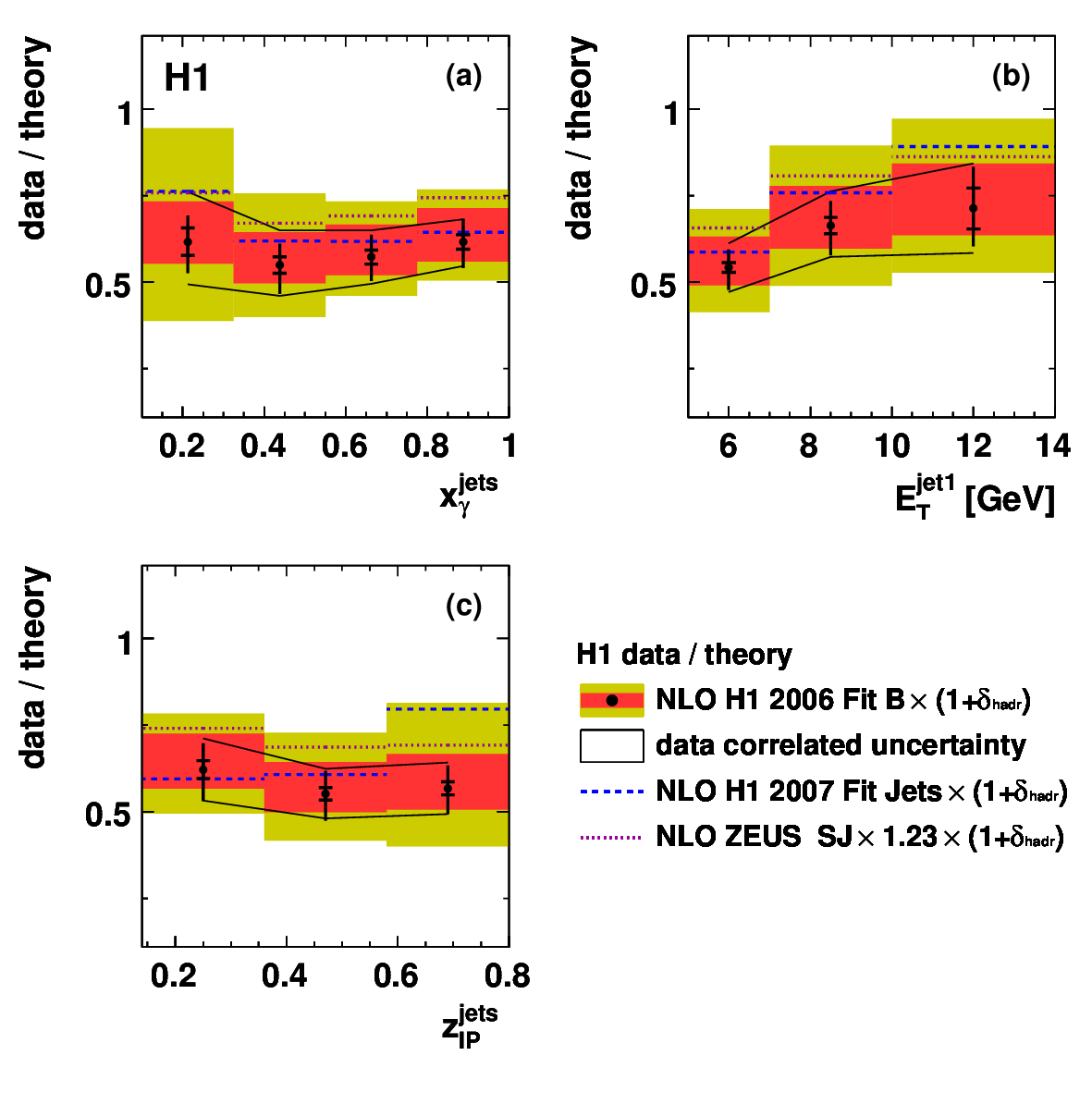,width=0.47\textwidth}
~\epsfig{file=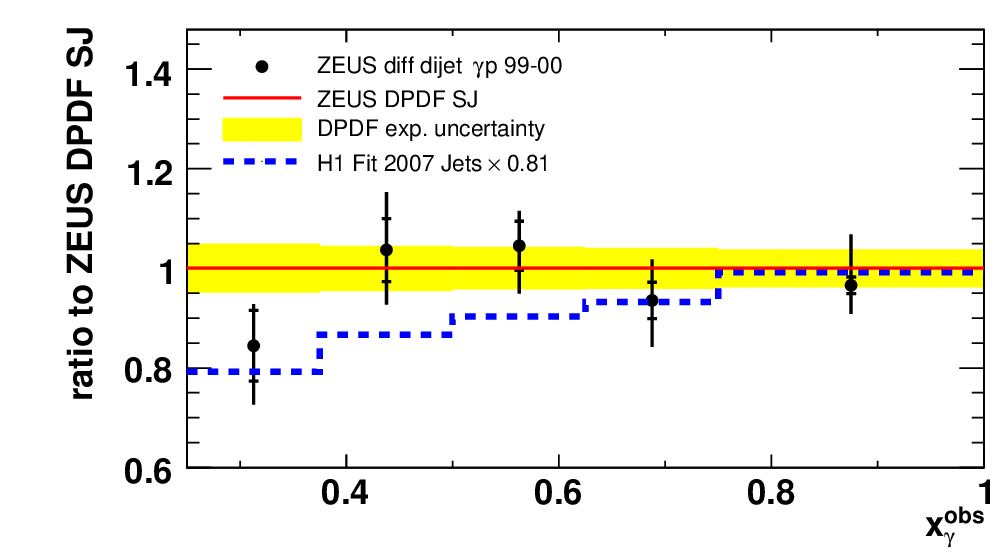,width=0.47\textwidth}
\caption{Fraction of Diffractive dijet photoproduction data shown as ratios of measured
cross sections to DPDF-based NLO predictions (i.e. as measurements of the
rapidity gap survival probability). Left: H1 data, from \cite{Aaron:2010su} 
Right: ZEUS data, from \cite{Chekanov:2009aa}.}
\label{gammap:difjet}
\end{center}
\end{figure}

Measurements of diffractive dijet 
photoproduction in DIS have been pursued as a control experiment
for gap destruction models.
In a lowest order interpretation, direct photon processes 
should be unaffected by such effects,
whereas they are expected in
resolved photon processes, where the photon interacts through its hadronic
structure (Fig.~\ref{feynman}(c)). 
For example, a theoretical model which successfully describes the Tevatron result
predicts a suppression of the resolved contribution by a factor 
of 0.34, \cite{Kaidalov:2003xf}.
Multiple measurements of diffractive dijet photoproduction have been made by H1 and 
ZEUS, and have been compared with NLO 
calculations based on DPDFs extracted in DDIS.
Whilst the experimental data are just about compatible
between the two collaborations, the conclusions from
H1 and ZEUS differ. 
Notably, neither collaboration sees any evidence for $x_\gamma$
dependence of the survival probability. 
H1 data, \cite{Aktas:2007hn,Aaron:2010su}, suggest a suppression 
of the data by a factor of around 0.6 relative to NLO QCD predictions, 
independently of $x_\gamma$. ZEUS results, which correspond to larger jet transverse 
energies, \cite{Chekanov:2007rh,Chekanov:2009aa},
suggest a smaller level of suppression and are, in fact, 
consistent with no suppression
whatsoever. Selected data from both collaborations are shown in 
Fig.~\ref{gammap:difjet}.
These apparent inconsistencies are at least partially resolved by a
model in which a more careful treatment of point-like, as distinct from fully resolved, photon 
structure is introduced, \cite{Kaidalov:2009fp}, leading to a much larger, and $E_T$ dependent,
survival probability for $0.1 \, \leqsim \, x_\gamma < 1$.


\subsubsection{Diffractive DIS in Dipole Models}
\label{sec:dipoles}

\begin{figure}[htb]
\begin{center}
~\epsfig{file=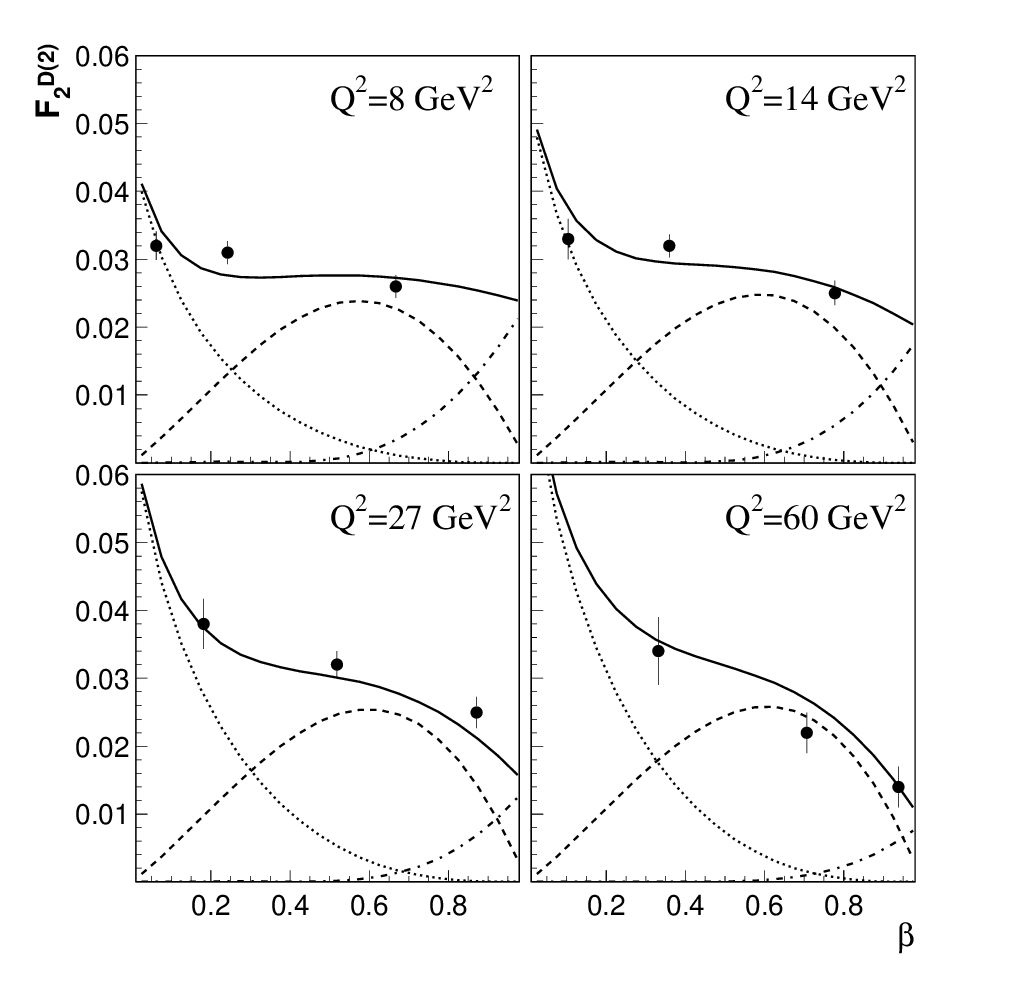,width=0.45\textwidth}
\caption{An example decomposition of the $\beta$ dependence of ZEUS
DDIS data, \cite{Breitweg:1998gc}, into different dipole terms, 
according to the parametrisation in \cite{GolecBiernat:1999qd}. 
The dashed, dotted and dot-dashed
curves correspond to leading twist transverse $q \bar{q}$, 
leading twist transverse $q \bar{q} g$ and
higher twist longitudinal $q \bar{q}$ contributions, respectively.}
\label{kgbfig}
\end{center}
\end{figure}

The application of the dipole picture to inclusive DDIS is challenging 
due to the need for higher multiplicity fluctuations of the virtual photon
($q \bar{q} g$ and perhaps others) in order to describe the large $M_X$,
small $\beta$ region. This need was first shown in a quantified
manner through the `BEKW' fits to 
$F_2^D$ data in \cite{Bartels:1998ea}.
In this parametrisation, the data at low and moderate $\beta$ are 
described in terms of $q \bar{q}$ and $q \bar{q} g$
dipole fluctuations of transversely polarised photons,
whilst the high $\beta$ region contains a $Q^2$-suppressed higher twist 
contribution from $q \bar{q}$ fluctuations of longitudinally polarised
photons. This approach was further developed for comparisons with
inclusive DDIS data, for example 
in the saturation model, \cite{GolecBiernat:1999qd,GolecBiernat:2007kv}.
An example decomposition of the $\beta$ dependence of DDIS data is shown
in Fig.~\ref{kgbfig}. 
Predictions have also been made for 
hadronic final state observables, based on a two-gluon exchange model
of exclusive $q \bar{q}$, \cite{Bartels:1996ne,Lotter:1996zu}, and 
$q \bar{q} g$, \cite{Bartels:1999tn,Bartels:2002ri}, production. 
However, progress has been limited by significant theoretical uncertainties 
associated with the low $\beta$ $q \bar{q} g$ contribution and a lack of 
direct evidence for the high $\beta$ higher twist $q \bar{q}$ term.

\subsection{Very Hard Diffraction and the BFKL Pomeron}
\label{bfkl}

For the diffractive processes discussed in the previous 
sections, $|t|$ is generally smaller than typical hadronic mass
scales. However, diffractive processes have also been studied at
HERA at $|t|$ values that are large enough to provide a hard scale
in their own right. Under such circumstances, the $t$ channel pomeron
exchange ought to be genuinely hard in the sense that it 
couples as a whole to individual partons.
Cases where the hard sub-process satisfies
$\hat{s} \gg -\hat{t} \gg \Lambda_{\rm QCD}$ correspond to a perturbatively
calculable limit of Regge theory. There is no particular ordering
in transverse momentum within the colour singlet exchange, such that the 
dynamics may be driven primarily by BFKL 
evolution. Diffractive processes
at large $|t|$ therefore represent a promising area in which to search
for low $x$ parton dynamics driven by 
the BFKL pomeron. 

\begin{figure}[htb]
\begin{center}
~\epsfig{file=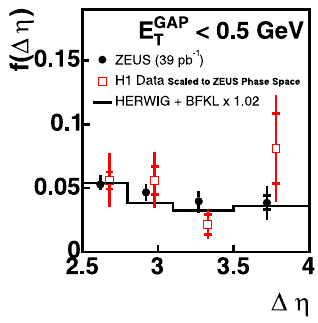,width=0.4\textwidth}
\hspace*{1cm}
~\epsfig{file=DESY-09-137_6.eps,width=0.45\textwidth}
\caption{Example data at very large $|t|$. 
Left: Fraction of dijet photoproduction events in which the two
jets with the highest transverse energy are separated by a pseudorapidity
'gap' of size $\Delta \eta$ in which the total transvere energy is less
than $0.5 \ {\rm GeV}$,  \cite{Chekanov:2006pw}. ZEUS and H1 data are
compared with a prediction that includes a BFKL pomeron contribution.
From~\cite{Chekanov:2006pw}. 
Right: Dependence on $|t|$ of proton dissociative $J/\psi$ 
production integrated over proton dissociation masses satisfying
$M_Y^2 / W^2 < 0.05$. The data are compared with a prediction 
(`GLMN LL', \cite{Gotsman:2001ne}) which
is based on DGLAP evolution and is expected to be valid
for $|t| < m_\psi^2$. They are also compared
with a BFKL-based prediction (`EMP LL', \cite{Enberg:2002zy})
and a more phenomenological approach (`FSZ', \cite{Frankfurt:2008er}.
From~\cite{Chekanov:2009ab}.}
\label{Hight:Jpsi}
\end{center}
\end{figure}

One striking signature for perturbative colour singlet exchange
is the photoproduction of pairs of jets separated by a large rapidity
gap. For such `gap between jets'
configurations, $|t| \approx E_{\rm T, jet}^2$ is very large,
typically $|t| > 25 \ {\rm GeV^2}$ in HERA studies. Calculations using
the leading-logarithmic BFKL approximation, \cite{Mueller:1992pe} 
(with appropriate modifications, \cite{Forshaw:1997wn,Enberg:2001ev}),
can be applied to the
scattering between a parton from the proton and a parton from the 
resolved structure of the photon.
However, the situation is complicated by the possibility of secondary
scattering and low rapidity gap survival.
The HERA data, \cite{Derrick:1995pb,Adloff:2002em,Chekanov:2006pw},
have shown clear evidence for events with little 
energy flow between jets, which  occur more frequently than 
expected from hadronisation fluctuations in standard Monte Carlo models of jet photoproduction. 
The order of magnitude of the signal is also in agreement with that
expected from the BFKL calculations. 
Example data from both ZEUS and H1 are shown in Fig.~\ref{Hight:Jpsi} (left).
The fraction of dijet events $f(\Delta\eta)$ ($E_{\rm T, jet}^1 > 6 \ {\rm GeV}$, $E_{\rm T, jet}^2 > 5 \ {\rm GeV}$)
for which the pseudorapidity region between the jets contains 
transverse energy of less than $0.5 \ {\rm GeV}$ is shown as a function of the pseudorapidity 
separation between the jets $\Delta\eta$. The data are
successfully described by a HERWIG Monte Carlo prediction that incorporates a BFKL pomeron contribution, whereas
models without such a contribution are unable to describe the data. 
Whilst this is highly 
suggestive, and is in fact perhaps the best evidence for BFKL dynamics
at HERA, more quantitative conclusions
have been precluded by the uncertainties in the
perturbative calculations, the gap survival probability and the 
residual non-diffractive contributions.

Another candidate process in which to observe perturbative 
colour singlet exchange is quasi-elastic vector meson production at large
$|t|$. Although $|t|$ values are smaller here than in the gaps-between-jets
case and the proton essentially always dissociates,
and the exclusive production mechanism at the photon vertex
(Fig.~\ref{VM:feynman}(b)) implies significantly reduced complications 
from gap destruction effects, even in photoproduction. 
The possibility of perturbative colour singlet
exchange has long been considered for this 
process, \cite{Ginzburg:1985tp}, and detailed calculations have been performed 
in the leading-logarithmic BFKL 
framework,~\cite{Forshaw:1995ax,Bartels:1996fs,Forshaw:2001pf,Enberg:2003jw,Poludniowski:2003yk}.

BFKL evolution is 
expected to be relevant for $|t| > m_V^2$, where
$m_V$ is the vector meson mass. 
For the case of heavy vector meson production, there is an 
interim region where
$|t| < m_V^2$, where
$|t|$ remains large enough to apply perturbative techniques. In this case, 
the transverse momenta along the gluon ladder remain ordered and DGLAP
dynamics are expected to apply, \cite{Gotsman:2001ne,Blok:2010ds}.
Studies have been made at HERA for 
light vector mesons, \cite{Aktas:2006qs,Chekanov:2002rm}, and
$J/\psi$ mesons, \cite{Aktas:2003zi,Chekanov:2009ab}, 
as well as exclusively produced photons, \cite{Aaron:2008ab}. 
The $J/\psi$ channel offers adequate statistics as well as a 
relatively clean theoretical interpretation, and thus it
 has yielded the most precise tests. An example
analysis is shown in Fig.~\ref{Hight:Jpsi} (right).
As expected in a BFKL based prediction, the $t$ dependence generally follows the expected
approximate power law behaviour at large $|t|$ and 
the effective trajectory slope $\alphapom^\prime$ is 
much smaller than that describing soft diffraction.
However, the specific BFKL model predictions compared with the 
data leave room for improvement.

\section{Summary}

In summary, the low $x$ data from HERA have produced many surprising and interesting
results that can be understood in terms of the perturbative dynamics of a gluon-rich system. 
Inclusive DIS at low $x$ reveals a progressively growing gluon distribution driven by the 
$g \rightarrow gg$ (and ultimately $g \rightarrow q \bar{q}$) splittings.  
The dominant feature of diffractive DIS data 
is scattering from
a composite net-colour-singlet exchange with a structure dominated by 
gluons at high fractional momenta, while exclusive processes in the perturbative domain are 
produced by the exchange of gluon pairs. 

To fully understand the inclusive data, 
it seems to be necessary to go beyond the conventional DGLAP $\ln(Q^2)$ resummation to
the include $\ln(1/x)$ resummation of BFKL. There are similar hints in charm production data and
forward-jet production, while in diffraction, describing data on rapidity gaps between
jets and very high $|t|$ $J/\psi$ production also requires BFKL contributions. 

Whether saturation effects have been
demonstrated at HERA is controversial, but one needs the $\ln(1/x)$ resummations in order to preserve the high
density gluon system that is a pre-requisite of saturation. 
Particularly significant are the
attempts to describe high density gluon dynamics in terms of semi-classical gluon fileds
-- the so-called `colour glass condensate', and the dipole models which have been developed to confront these
ideas with data. 



\seealso{other contributions to the encylopedia}




\bibliographystyle{Harvard}
\bibliography{reference}

@article{Gribov:1972ri,
    author = "Gribov, V. N. and Lipatov, L. N.",
    title = "{Deep inelastic e p scattering in perturbation theory}",
    reportNumber = "IPTI-381-71",
    journal = "Sov. J. Nucl. Phys.",
    volume = "15",
    pages = "438--450",
    year = "1972"
}

@article{Gribov:1972rt,
    author = "Gribov, V. N. and Lipatov, L. N.",
    title = "{e+ e- pair annihilation and deep inelastic e p scattering in perturbation theory}",
    journal = "Sov. J. Nucl. Phys.",
    volume = "15",
    pages = "675--684",
    year = "1972"
}

@article{Lipatov:1974qm,
    author = "Lipatov, L. N.",
    title = "{The parton model and perturbation theory}",
    journal = "Yad. Fiz.",
    volume = "20",
    pages = "181--198",
    year = "1974"
}

@article{Dokshitzer:1977sg,
    author = "Dokshitzer, Yuri L.",
    title = "{Calculation of the Structure Functions for Deep Inelastic Scattering and e+ e- Annihilation by Perturbation Theory in Quantum Chromodynamics.}",
    journal = "Sov. Phys. JETP",
    volume = "46",
    pages = "641--653",
    year = "1977"
}

@article{Altarelli:1977zs,
    author = "Altarelli, Guido and Parisi, G.",
    title = "{Asymptotic Freedom in Parton Language}",
    journal = "Nucl. Phys. B",
    volume = "126",
    pages = "298--318",
    year = "1977"
}

@article{Bailey:2020ooq,
    author = "Bailey, S. and Cridge, T. and Harland-Lang, L. A. and Martin, A. D. and Thorne, R. S.",
    title = "{Parton distributions from LHC, HERA, Tevatron and fixed target data: MSHT20 PDFs}",
    journal = "Eur. Phys. J. C",
    volume = "81",
    number = "4",
    pages = "341",
    year = "2021"
}

@article{Hou:2019efy,
    author = "Hou, Tie-Jiun and others",
    title = "{New CTEQ global analysis of quantum chromodynamics with high-precision data from the LHC}",
    journal = "Phys. Rev. D",
    volume = "103",
    number = "1",
    pages = "014013",
    year = "2021"
}

@article{NNPDF:2021njg,
    author = "Ball, Richard D. and others",
    collaboration = "NNPDF",
    title = "{The path to proton structure at 1{\%} accuracy}",
    journal = "Eur. Phys. J. C",
    volume = "82",
    number = "5",
    pages = "428",
    year = "2022"
}

@article{H1:2015ubc,
    author = "Abramowicz, H. and others",
    collaboration = "H1, ZEUS",
    title = "{Combination of measurements of inclusive deep inelastic ${e^{\pm }p}$ scattering cross sections and QCD analysis of HERA data}",
    journal = "Eur. Phys. J. C",
    volume = "75",
    number = "12",
    pages = "580",
    year = "2015"
}

@article{BCDMS:1989qop,
    author = "Benvenuti, A. C. and others",
    collaboration = "BCDMS",
    title = "{A High Statistics Measurement of the Proton Structure Functions F(2) (x, Q**2) and R from Deep Inelastic Muon Scattering at High Q**2}",
    journal = "Phys. Lett. B",
    volume = "223",
    pages = "485--489",
    year = "1989"
}

@article{NewMuon:1996fwh,
    author = "Arneodo, M. and others",
    collaboration = "New Muon",
    title = "{Measurement of the proton and deuteron structure functions, F2(p) and F2(d), and of the ratio sigma-L / sigma-T}",
    journal = "Nucl. Phys. B",
    volume = "483",
    pages = "3--43",
    year = "1997"
}

@article{ZEUS:1993ppj,
    author = "Derrick, M. and others",
    collaboration = "ZEUS",
    title = "{Measurement of the proton structure function F2 in e p scattering at HERA}",
    journal = "Phys. Lett. B",
    volume = "316",
    pages = "412--426",
    year = "1993"
}

@article{ZEUS:2002xjx,
    author = "Chekanov, S. and others",
    collaboration = "ZEUS",
    title = "{A ZEUS next-to-leading-order QCD analysis of data on deep inelastic scattering}",
    journal = "Phys. Rev. D",
    volume = "67",
    pages = "012007",
    year = "2003"
}

@article{H1:2013ktq,
    author = "Andreev, V. and others",
    collaboration = "H1",
    title = "{Measurement of inclusive $e p$ cross sections at high $Q^2$ at $\sqrt s =$ 225 and 252 GeV and of the longitudinal proton structure function $F_L$ at HERA}",
    journal = "Eur. Phys. J. C",
    volume = "74",
    number = "4",
    pages = "2814",
    year = "2014"
}

@article{Bartels:2000hv,
    author = "Bartels, Jochen and Golec-Biernat, Krzysztof J. and Peters, Krisztian",
    title = "{An Estimate of higher twist at small x(B) and low Q**2 based upon a saturation model}",
    journal = "Eur. Phys. J. C",
    volume = "17",
    pages = "121--128",
    year = "2000"
}

@article{H1:2009pze,
    author = "Aaron, F. D. and others",
    collaboration = "H1, ZEUS",
    title = "{Combined Measurement and QCD Analysis of the Inclusive e+- p Scattering Cross Sections at HERA}",
    journal = "JHEP",
    volume = "01",
    pages = "109",
    year = "2010"
}

@article{Lonnblad:1992tz,
    author = "Lonnblad, Leif",
    title = "{ARIADNE version 4: A Program for simulation of QCD cascades implementing the color dipole model}",
    journal = "Comput. Phys. Commun.",
    volume = "71",
    pages = "15--31",
    year = "1992"
}

@article{Ingelman:1996mq,
    author = "Ingelman, G. and Edin, A. and Rathsman, J.",
    title = "{LEPTO 6.5: A Monte Carlo generator for deep inelastic lepton - nucleon scattering}",
    journal = "Comput. Phys. Commun.",
    volume = "101",
    pages = "108--134",
    year = "1997"
}

@article{Kharraziha:1997dn,
    author = "Kharraziha, Hamid and Lonnblad, Leif",
    title = "{The Linked dipole chain Monte Carlo}",
    journal = "JHEP",
    volume = "03",
    pages = "006",
    year = "1998"
}

@article{Jung:1993gf,
    author = "Jung, Hannes",
    title = "{Hard diffractive scattering in high-energy e p collisions and the Monte Carlo generator RAPGAP}",
    journal = "Comput. Phys. Commun.",
    volume = "86",
    pages = "147--161",
    year = "1995"
}

@article{ZEUS:1998agx,
    author = "Breitweg, J. and others",
    collaboration = "ZEUS",
    title = "{ZEUS results on the measurement and phenomenology of F(2) at low x and low Q**2}",
    eprint = "hep-ex/9809005",
    archivePrefix = "arXiv",
    journal = "Eur. Phys. J. C",
    volume = "7",
    pages = "609--630",
    year = "1999"
}

@article{H1:1998cuj,
    author = "Adloff, C. and others",
    collaboration = "H1",
    title = "{Forward jet and particle production at HERA}",
    journal = "Nucl. Phys. B",
    volume = "538",
    pages = "3--22",
    year = "1999"
}

@article{ZEUS:1998uhq,
    author = "Breitweg, J. and others",
    collaboration = "ZEUS",
    title = "{Forward jet production in deep inelastic scattering at HERA}",
    journal = "Eur. Phys. J. C",
    volume = "6",
    pages = "239--252",
    year = "1999"
}

@article{Kwiecinski:1999hv,
    author = "Kwiecinski, J. and Martin, Alan D. and Outhwaite, J. J.",
    title = "{Small x QCD effects in DIS with a forward jet or a forward pi0}",
    journal = "Eur. Phys. J. C",
    volume = "9",
    pages = "611--622",
    year = "1999"
}

@article{Fadin:1975cb,
    author = "Fadin, Victor S. and Kuraev, E. A. and Lipatov, L. N.",
    title = "{On the Pomeranchuk Singularity in Asymptotically Free Theories}",
    journal = "Phys. Lett. B",
    volume = "60",
    pages = "50--52",
    year = "1975"
}

@article{Kuraev:1976ge,
    author = "Kuraev, E. A. and Lipatov, L. N. and Fadin, Victor S.",
    title = "{Multiregge processes in the Yang-Mills theory}",
    journal = "Sov. Phys. JETP",
    volume = "44",
    number = "3",
    pages = "443--451",
    year = "1976"
}

@article{Kuraev:1977fs,
    author = "Kuraev, E. A. and Lipatov, L. N. and Fadin, Victor S.",
    title = "{The Pomeranchuk singularity in nonabelian gauge theories}",
    journal = "Sov. Phys. JETP",
    volume = "45",
    pages = "199--204",
    year = "1977"
}

@article{Balitsky:1978ic,
    author = "Balitsky, I. I. and Lipatov, L. N.",
    title = "{The Pomeranchuk Singularity in Quantum Chromodynamics}",
    journal = "Sov. J. Nucl. Phys.",
    volume = "28",
    pages = "822--829",
    year = "1978"
}

@article{xFitterDevelopersTeam:2018hym,
    author = "Abdolmaleki, Hamed and others",
    collaboration = "xFitter Developers' Team",
    title = "{Impact of low-$x$ resummation on QCD analysis of HERA data}",
    journal = "Eur. Phys. J. C",
    volume = "78",
    number = "8",
    pages = "621",
    year = "2018"
}

@article{Abt:2016vjh,
    author = "Abt, I. and Cooper-Sarkar, A. M. and Foster, B. and Myronenko, V. and Wichmann, K. and Wing, M.",
    title = "{Study of HERA ep data at low Q$^2$ and low $x_{Bj}$ and the need for higher-twist corrections to standard perturbative QCD fits}",
    journal = "Phys. Rev. D",
    volume = "94",
    number = "3",
    pages = "034032",
    year = "2016"
}

@article{H1:2018flt,
    author = "Abramowicz, H. and others",
    collaboration = "H1, ZEUS",
    title = "{Combination and QCD analysis of charm and beauty production cross-section measurements in deep inelastic $ep$ scattering at HERA}",
    journal = "Eur. Phys. J. C",
    volume = "78",
    number = "6",
    pages = "473",
    year = "2018"
}

@article{Badelek:1996ap,
    author = "Badelek, B. and Kwiecinski, J. and Stasto, A.",
    title = "{A Model for F(L) and R = F(L) / F(T) at low x and low Q**2}",
    journal = "Z. Phys. C",
    volume = "74",
    pages = "297--306",
    year = "1997"
}

@article{Abt:2017nkc,
    author = "Abt, I. and Cooper-Sarkar, A. M. and Foster, B. and Myronenko, V. and Wichmann, K. and Wing, M.",
    title = "{Investigation into the limits of perturbation theory at low $Q^2$ using HERA deep inelastic scattering data}",
    journal = "Phys. Rev. D",
    volume = "96",
    number = "1",
    pages = "014001",
    year = "2017"
}

@article{Abramowicz:1991xz,
    author = "Abramowicz, H. and Levin, E. M. and Levy, A. and Maor, U.",
    title = "{A Parametrization of sigma-T (gamma* p) above the resonance region Q**2 {\ensuremath{>}}= 0}",
    journal = "Phys. Lett. B",
    volume = "269",
    pages = "465--476",
    year = "1991"
}

@article{Abramowicz:1997ms,
    author = "Abramowicz, H. and Levy, A.",
    title = "{The ALLM parameterization of sigma(tot)(gamma* p): An Update}",
    reportNumber = "DESY-97-251",
    month = "12",
    year = "1997"
}

@article{Ball:2017otu,
    author = "Ball, Richard D. and Bertone, Valerio and Bonvini, Marco and Marzani, Simone and Rojo, Juan and Rottoli, Luca",
    title = "{Parton distributions with small-x resummation: evidence for BFKL dynamics in HERA data}",
    reportNumber = "EDINBURGH-2017-15, NIKHEF-2017-027, OUTP-17-12P",
    journal = "Eur. Phys. J. C",
    volume = "78",
    number = "4",
    pages = "321",
    year = "2018"
}

@article{Bonvini:2017ogt,
    author = "Bonvini, Marco and Marzani, Simone and Muselli, Claudio",
    title = "{Towards parton distribution functions with small-$x$ resummation: HELL 2.0}",
    journal = "JHEP",
    volume = "12",
    pages = "117",
    year = "2017"
}

@article{Ball:1994du,
    author = "Ball, Richard D. and Forte, Stefano",
    title = "{Double asymptotic scaling at HERA}",
    journal = "Phys. Lett. B",
    volume = "335",
    pages = "77--86",
    year = "1994"
}

@article{DeRujula:1974mnv,
    author = "De Rujula, A. and Glashow, S. L. and Politzer, H. David and Treiman, S. B. and Wilczek, Frank and Zee, A.",
    title = "{Possible NonRegge Behavior of Electroproduction Structure Functions}",
    journal = "Phys. Rev. D",
    volume = "10",
    pages = "1649",
    year = "1974"
}

@inproceedings{Gayler:2002ah,
    author = "Gayler, Joerg",
    title = "{Experimental approaches to low x at HERA}",
    booktitle = "{32nd International Symposium on Multiparticle Dynamics}",
    eprint = "hep-ex/0211072",
    pages = "371--374",
    month = "11",
    year = "2002"
}

@article{Ciafaloni:1987ur,
    author = "Ciafaloni, Marcello",
    title = "{Coherence Effects in Initial Jets at Small q**2 / s}",
    journal = "Nucl. Phys. B",
    volume = "296",
    pages = "49--74",
    year = "1988"
}

@article{Catani:1989sg,
    author = "Catani, S. and Fiorani, F. and Marchesini, G.",
    title = "{Small x Behavior of Initial State Radiation in Perturbative QCD}",
    journal = "Nucl. Phys. B",
    volume = "336",
    pages = "18--85",
    year = "1990"
}

@article{Catani:1989yc,
    author = "Catani, S. and Fiorani, F. and Marchesini, G.",
    title = "{QCD Coherence in Initial State Radiation}",
    journal = "Phys. Lett. B",
    volume = "234",
    pages = "339--345",
    year = "1990"
}

@article{Marchesini:1994wr,
    author = "Marchesini, Giuseppe",
    title = "{QCD coherence in the structure function and associated distributions at small x}",
    journal = "Nucl. Phys. B",
    volume = "445",
    pages = "49--80",
    year = "1995"
}

@article{Donnachie:1993it,
    author = "Donnachie, A. and Landshoff, P. V.",
    title = "{Proton structure function at small q**2}",
    journal = "Z. Phys. C",
    volume = "61",
    pages = "139--146",
    year = "1994"
}

@article{Donnachie:2001he,
    author = "Donnachie, A. and Landshoff, P. V.",
    title = "{New data and the hard Pomeron}",
    journal = "Phys. Lett. B",
    volume = "518",
    pages = "63--71",
    year = "2001"
}

@article{Golec-Biernat:1999qor,
    author = "Golec-Biernat, Krzysztof J. and Wusthoff, M.",
    title = "{Saturation in diffractive deep inelastic scattering}",
    journal = "Phys. Rev. D",
    volume = "60",
    pages = "114023",
    year = "1999"
}

@article{Bartels:2002cj,
    author = "Bartels, J. and Golec-Biernat, Krzysztof J. and Kowalski, H.",
    title = "{A modification of the saturation model: DGLAP evolution}",
    journal = "Phys. Rev. D",
    volume = "66",
    pages = "014001",
    year = "2002"
}

@article{Stasto:2000er,
    author = "Stasto, A. M. and Golec-Biernat, Krzysztof J. and Kwiecinski, J.",
    title = "{Geometric scaling for the total gamma* p cross-section in the low x region}",
    journal = "Phys. Rev. Lett.",
    volume = "86",
    pages = "596--599",
    year = "2001"
}

@article{Balitsky:1995ub,
    author = "Balitsky, I.",
    title = "{Operator expansion for high-energy scattering}",
    journal = "Nucl. Phys. B",
    volume = "463",
    pages = "99--160",
    year = "1996"
}

@article{Kovchegov:1999yj,
    author = "Kovchegov, Yuri V.",
    title = "{Small x F(2) structure function of a nucleus including multiple pomeron exchanges}",
    journal = "Phys. Rev. D",
    volume = "60",
    pages = "034008",
    year = "1999"
}

@article{Kovchegov:1999ua,
    author = "Kovchegov, Yuri V.",
    title = "{Unitarization of the BFKL pomeron on a nucleus}",
    journal = "Phys. Rev. D",
    volume = "61",
    pages = "074018",
    year = "2000"
}

@article{Rezaeian:2012ji,
    author = "Rezaeian, Amir H. and Siddikov, Marat and Van de Klundert, Merijn and Venugopalan, Raju",
    title = "{Analysis of combined HERA data in the Impact-Parameter dependent Saturation model}",
    journal = "Phys. Rev. D",
    volume = "87",
    number = "3",
    pages = "034002",
    year = "2013"
}

@article{Lai:2010vv,
    author = "Lai, Hung-Liang and Guzzi, Marco and Huston, Joey and Li, Zhao and Nadolsky, Pavel M. and Pumplin, Jon and Yuan, C. -P.",
    title = "{New parton distributions for collider physics}",
    journal = "Phys. Rev. D",
    volume = "82",
    pages = "074024",
    year = "2010"
}

@article{Martin:2009iq,
    author = "Martin, A. D. and Stirling, W. J. and Thorne, R. S. and Watt, G.",
    title = "{Parton distributions for the LHC}",
    journal = "Eur. Phys. J. C",
    volume = "63",
    pages = "189--285",
    year = "2009"
}

@article{Watt:2007nr,
    author = "Watt, G. and Kowalski, H.",
    title = "{Impact parameter dependent colour glass condensate dipole model}",
    journal = "Phys. Rev. D",
    volume = "78",
    pages = "014016",
    year = "2008"
}

@article{Mantysaari:2018nng,
    author = {M{\"a}ntysaari, Heikki and Zurita, Pia},
    title = "{In depth analysis of the combined HERA data in the dipole models with and without saturation}",
    journal = "Phys. Rev. D",
    volume = "98",
    pages = "036002",
    year = "2018"
}

@Article{Breitweg:1997ug,
     author    = "Breitweg, J. and others",
 collaboration = "ZEUS",
     title     = "{Event shape analysis of deep inelastic scattering events
                  with a large  rapidity gap at HERA}",
     journal   = "Phys. Lett.",
     volume    = "B421",
     year      = "1998",
     pages     = "368-384",
     eprint    = "hep-ex/9710027",
     archivePrefix = "arXiv",
     SLACcitation  = "%%CITATION = HEP-EX/9710027;%%"
}

@Article{Adloff:2002dw,
      author         = "Adloff, C. and others",
      title          = "{Search for odderon induced contributions to exclusive
                        pi0 photoproduction at HERA}",
      collaboration  = "H1",
      journal        = "Phys.Lett.",
      volume         = "B544",
      pages          = "35-43",
      year           = "2002",
      eprint         = "hep-ex/0206073",
      archivePrefix  = "arXiv",
      primaryClass   = "hep-ex",
      reportNumber   = "DESY-02-087",
      SLACcitation   = "%%CITATION = HEP-EX/0206073;%%",
}

@article{Devenish:2004pb,
      author         = "Devenish, R. and Cooper-Sarkar, A.",
      title          = "{Deep inelastic scattering}",
      journal        = "\emph{Deep inelastic scattering}",
      year           = "2004",
      volume          = "OUP",
      pages           = "403\,p",
      SLACcitation   = "%%CITATION = INSPIRE-647004;%%",
}

@Article{Goulianos:1982vk,
     author    = "Goulianos, Konstantin A.",
     title     = "{Diffractive Interactions of Hadrons at High-Energies}",
     journal   = "Phys. Rept.",
     volume    = "101",
     year      = "1983",
     pages     = "169",
     SLACcitation  = "%%CITATION = PRPLC,101,169;%%"
}

@Article{Good:1960ba,
     author    = "Good, M. L. and Walker, W. D.",
     title     = "{Diffraction disssociation of beam particles}",
     journal   = "Phys. Rev.",
     volume    = "120",
     year      = "1960",
     pages     = "1857-1860",
     SLACcitation  = "%%CITATION = PHRVA,120,1857;%%"
}

@Article{Regge:1959mz,
     author    = "Regge, T.",
     title     = "{Introduction to complex orbital momenta}",
     journal   = "Nuovo Cim.",
     volume    = "14",
     year      = "1959",
     pages     = "951",
     SLACcitation  = "%%CITATION = NUCIA,14,951;%%"
}

@Article{Regge:1960zc,
     author    = "Regge, T.",
     title     = "{Bound states, shadow states and Mandelstam
                  representation}",
     journal   = "Nuovo Cim.",
     volume    = "18",
     year      = "1960",
     pages     = "947-956",
     SLACcitation  = "%%CITATION = NUCIA,18,947;%%"
}

@Article{Chew:1961yz,
     author    = "Chew, Geoffrey F. and Frautschi, Steven C. and Mandelstam,
                  Stanley",
     title     = "{Regge poles in pi pi scattering}",
     journal   = "Phys. Rev.",
     volume    = "126",
     year      = "1961",
     pages     = "1202-1208",
     SLACcitation  = "%%CITATION = PHRVA,126,1202;%%"
}

@Article{Chew:1961ev,
     author    = "Chew, G. F. and Frautschi, Steven C.",
     title     = "{PRINCIPLE OF EQUIVALENCE FOR ALL STRONGLY INTERACTING
                  PARTICLES WITHIN THE S MATRIX FRAMEWORK}",
     journal   = "Phys. Rev. Lett.",
     volume    = "7",
     year      = "1961",
     pages     = "394-397",
     SLACcitation  = "%%CITATION = PRLTA,7,394;%%"
}

@Article{Fienberg:1956,
     author    = "Feinberg, E. and Pomeran\u{c}uk, I.",
     title     = "{Fill me in}",
     journal   = "Suppl. Nuovo Cimento",
     volume    = "3",
     year      = "1956",
     pages     = "652"
}

@Article{Miettinen:1978jb,
     author    = "Miettinen, Hannu I. and Pumplin, Jon",
     title     = "{Diffraction Scattering and the Parton Structure of
                  Hadrons}",
     journal   = "Phys. Rev.",
     volume    = "D18",
     year      = "1978",
     pages     = "1696",
     SLACcitation  = "%%CITATION = PHRVA,D18,1696;%%"
}

@Article{Dederichs:1989gk,
     author    = "Dederichs, K. H. and Faessler, M. A.",
     title     = "{BREMSSTRAHLUNG AND DIFFRACTIVE DISSOCIATION}",
     journal   = "Phys. Lett.",
     volume    = "B232",
     year      = "1989",
     pages     = "405-412",
     SLACcitation  = "%%CITATION = PHLTA,B232,405;%%"
}

@Article{Mueller:1970fa,
     author    = "Mueller, Alfred H.",
     title     = "{O(2,1) ANALYSIS OF SINGLE PARTICLE SPECTRA AT HIGH-
                  ENERGY}",
     journal   = "Phys. Rev.",
     volume    = "D2",
     year      = "1970",
     pages     = "2963-2968",
     SLACcitation  = "%%CITATION = PHRVA,D2,2963;%%"
}

@Article{Ingelman:1984ns,
     author    = "Ingelman, G. and Schlein, P. E.",
     title     = "{Jet Structure in High Mass Diffractive Scattering}",
     journal   = "Phys. Lett.",
     volume    = "B152",
     year      = "1985",
     pages     = "256",
     SLACcitation  = "%%CITATION = PHLTA,B152,256;%%"
}

@Article{Donnachie:1987xh,
     author    = "Donnachie, A. and Landshoff, P. V.",
     title     = "{Diffractive Deep Inelastic Lepton Scattering}",
     journal   = "Phys. Lett.",
     volume    = "B191",
     year      = "1987",
     pages     = "309",
     note = "Erratum-ibid.\ 1987, {\bf B198}, 590",
     SLACcitation  = "%%CITATION = PHLTA,B191,309;%%"
}

@Article{Collins:1997sr,
     author    = "Collins, John C.",
     title     = "{Proof of factorization for diffractive hard scattering}",
     journal   = "Phys. Rev.",
     volume    = "D57",
     year      = "1998",
     pages     = "3051-3056",
     eprint    = "hep-ph/9709499",
     archivePrefix = "arXiv",
     note = "Erratum-ibid.\ 2000, {\bf D61}, 019902",
     SLACcitation  = "%%CITATION = HEP-PH/9709499;%%"
}

@Article{Derrick:1993xh,
     author    = "Derrick, M. and others",
 collaboration = "ZEUS",
     title     = "{Observation of events with a large rapidity gap in deep
                  inelastic scattering at HERA}",
     journal   = "Phys. Lett.",
     volume    = "B315",
     year      = "1993",
     pages     = "481-493",
     SLACcitation  = "%%CITATION = PHLTA,B315,481;%%"
}

@Article{Ahmed:1994nw,
     author    = "Ahmed, T. and others",
 collaboration = "H1",
     title     = "{Deep inelastic scattering events with a large rapidity gap
                  at HERA}",
     journal   = "Nucl. Phys.",
     volume    = "B429",
     year      = "1994",
     pages     = "477-502",
     SLACcitation  = "%%CITATION = NUPHA,B429,477;%%"
}

@Article{Ahmed:1994ui,
     author    = "Ahmed, T. and others",
 collaboration = "H1",
     title     = "{Observation of hard processes in rapidity gap events in
                  gamma p interactions at HERA}",
     journal   = "Nucl. Phys.",
     volume    = "B435",
     year      = "1995",
     pages     = "3-22",
     SLACcitation  = "%%CITATION = NUPHA,B435,3;%%"
}

@Article{Derrick:1994yg,
     author    = "Derrick, M. and others",
 collaboration = "ZEUS",
     title     = "{Observation of hard scattering in photoproduction events
                  with a large rapidity gap at HERA}",
     journal   = "Phys. Lett.",
     volume    = "B346",
     year      = "1995",
     pages     = "399-414",
     eprint    = "hep-ex/9501011",
     archivePrefix = "arXiv",
     SLACcitation  = "%%CITATION = HEP-EX/9501011;%%"
}

@Article{Derrick:1994ze,
     author    = "Derrick, M. and others",
 collaboration = "ZEUS",
     title     = "{Observation of jet production in deep inelastic scattering
                  with a large rapidity gap at HERA}",
     journal   = "Phys. Lett.",
     volume    = "B332",
     year      = "1994",
     pages     = "228-243",
     SLACcitation  = "%%CITATION = PHLTA,B332,228;%%"
}

@Article{Breitweg:1997za,
     author    = "Breitweg, J. and others",
 collaboration = "ZEUS",
     title     = "{Study of photon dissociation in diffractive
                  photoproduction at HERA}",
     journal   = "Z. Phys.",
     volume    = "C75",
     year      = "1997",
     pages     = "421-435",
     eprint    = "hep-ex/9704008",
     archivePrefix = "arXiv",
     SLACcitation  = "%%CITATION = HEP-EX/9704008;%%"
}

@Article{Breitweg:1998gc,
     author    = "Breitweg, J. and others",
 collaboration = "ZEUS",
     title     = "{Measurement of the diffractive cross-section in deep
                  inelastic scattering using ZEUS 1994 data}",
     journal   = "Eur. Phys. J.",
     volume    = "C6",
     year      = "1999",
     pages     = "43-66",
     eprint    = "hep-ex/9807010",
     archivePrefix = "arXiv",
     SLACcitation  = "%%CITATION = HEP-EX/9807010;%%"
}

@Article{Chekanov:2001ph,
     author    = "Chekanov, S. and others",
 collaboration = "ZEUS",
     title     = "{Properties of hadronic final states in diffractive deep
                  inelastic e p  scattering at HERA}",
     journal   = "Phys. Rev.",
     volume    = "D65",
     year      = "2002",
     pages     = "052001",
     eprint    = "hep-ex/0107052",
     archivePrefix = "arXiv",
     SLACcitation  = "%%CITATION = HEP-EX/0107052;%%"
}

@Article{Derrick:1995pb,
     author    = "Derrick, M. and others",
 collaboration = "ZEUS",
     title     = "{Rapidity Gaps between Jets in Photoproduction at HERA}",
     journal   = "Phys. Lett.",
     volume    = "B369",
     year      = "1996",
     pages     = "55-68",
     eprint    = "hep-ex/9510012",
     archivePrefix = "arXiv",
     SLACcitation  = "%%CITATION = HEP-EX/9510012;%%"
}

@article{Adloff:1997mi,
      author         = "Adloff, C. and others",
      title          = "{Diffraction dissociation in photoproduction at HERA}",
      collaboration  = "H1",
      journal        = "Z.Phys.",
      volume         = "C74",
      pages          = "221-236",
      year           = "1997",
      eprint         = "hep-ex/9702003",
      archivePrefix  = "arXiv",
      primaryClass   = "hep-ex",
      reportNumber   = "DESY-97-009",
      SLACcitation   = "%%CITATION = HEP-EX/9702003;%%",
}

@article{ATLAS:2012djz,
    author = "Aad, Georges and others",
    collaboration = "ATLAS",
    title = "{Rapidity gap cross sections measured with the ATLAS detector in $pp$ collisions at $\sqrt{s}=7$ TeV}",
    eprint = "1201.2808",
    archivePrefix = "arXiv",
    primaryClass = "hep-ex",
    reportNumber = "CERN-PH-EP-2011-220",
    journal = "Eur. Phys. J. C",
    volume = "72",
    pages = "1926",
    year = "2012"
}

@article{Affolder:2000vb,
      author         = "Affolder, T. and others",
      title          = "{Diffractive dijets with a leading antiproton in
                        $\bar{p}p$ collisions at $\sqrt{s} = 1800$ GeV}",
      collaboration  = "CDF",
      journal        = "Phys.Rev.Lett.",
      volume         = "84",
      pages          = "5043-5048",
      year           = "2000",
      reportNumber   = "FERMILAB-PUB-00-055-E",
      SLACcitation   = "%%CITATION = PRLTA,84,5043;%%",
}

@article{Mueller:1989st,
      author         = "Mueller, Alfred H.",
      title          = "{Small x Behavior and Parton Saturation: A QCD Model}",
      journal        = "Nucl.Phys.",
      volume         = "B335",
      pages          = "115",
      year           = "1990",
      reportNumber   = "CU-TP-441a",
      SLACcitation   = "%%CITATION = NUPHA,B335,115;%%",
}

@article{Nikolaev:1990ja,
      author         = "Nikolaev, Nikolai N. and Zakharov, B.G.",
      title          = "{Color transparency and scaling properties of nuclear
                        shadowing in deep inelastic scattering}",
      journal        = "Z.Phys.",
      volume         = "C49",
      pages          = "607-618",
      year           = "1991",
      reportNumber   = "OUTP-90-23-P",
      SLACcitation   = "%%CITATION = ZEPYA,C49,607;%%",
}

@article{Kowalski:2003hm,
      author         = "Kowalski, Henri and Teaney, Derek",
      title          = "{An Impact parameter dipole saturation model}",
      journal        = "Phys.Rev.",
      volume         = "D68",
      pages          = "114005",
      year           = "2003",
      eprint         = "hep-ph/0304189",
      archivePrefix  = "arXiv",
      primaryClass   = "hep-ph",
      SLACcitation   = "%%CITATION = HEP-PH/0304189;%%",
}

@article{Kowalski:2006hc,
      author         = "Kowalski, H. and Motyka, L. and Watt, G.",
      title          = "{Exclusive diffractive processes at HERA within the
                        dipole picture}",
      journal        = "Phys.Rev.",
      volume         = "D74",
      pages          = "074016",
      year           = "2006",
      eprint         = "hep-ph/0606272",
      archivePrefix  = "arXiv",
      primaryClass   = "hep-ph",
      reportNumber   = "DESY-06-095",
      SLACcitation   = "%%CITATION = HEP-PH/0606272;%%",
}

@article{GolecBiernat:1998js,
      author         = "Golec-Biernat, Krzysztof J. and Wusthoff, M.",
      title          = "{Saturation effects in deep inelastic scattering at low
                        Q**2 and its implications on diffraction}",
      journal        = "Phys.Rev.",
      volume         = "D59",
      pages          = "014017",
      year           = "1998",
      eprint         = "hep-ph/9807513",
      archivePrefix  = "arXiv",
      primaryClass   = "hep-ph",
      reportNumber   = "DTP-98-50",
      SLACcitation   = "%%CITATION = HEP-PH/9807513;%%",
}

@article{GolecBiernat:1999qd,
      author         = "Golec-Biernat, Krzysztof J. and Wusthoff, M.",
      title          = "{Saturation in diffractive deep inelastic scattering}",
      journal        = "Phys.Rev.",
      volume         = "D60",
      pages          = "114023",
      year           = "1999",
      eprint         = "hep-ph/9903358",
      archivePrefix  = "arXiv",
      primaryClass   = "hep-ph",
      reportNumber   = "DTP-99-20",
      SLACcitation   = "%%CITATION = HEP-PH/9903358;%%",
}

@article{Bartels:1998ea,
      author         = "Bartels, Jochen and Ellis, John R. and Kowalski, H. and
                        Wusthoff, M.",
      title          = "{An Analysis of diffraction in deep inelastic
                        scattering}",
      journal        = "Eur.Phys.J.",
      volume         = "C7",
      pages          = "443-458",
      year           = "1999",
      eprint         = "hep-ph/9803497",
      archivePrefix  = "arXiv",
      primaryClass   = "hep-ph",
      reportNumber   = "CERN-TH-98-67, DESY-98-034, DTP-98-02",
      SLACcitation   = "%%CITATION = HEP-PH/9803497;%%",
}

@article{Nikolaev:1991et,
      author         = "Nikolaev, Nikolai and Zakharov, Bronislav G.",
      title          = "{Pomeron structure function and diffraction dissociation
                        of virtual photons in perturbative QCD}",
      journal        = "Z.Phys.",
      volume         = "C53",
      pages          = "331-346",
      year           = "1992",
      reportNumber   = "DFTT-5-91",
      SLACcitation   = "%%CITATION = ZEPYA,C53,331;%%",
}

@article{Chekanov:2008cw,
      author         = "Chekanov, S. and others",
      title          = "{Deep inelastic inclusive and diffractive scattering at
                        $Q^2$ values from 25 to 320 GeV$^2$ with the ZEUS forward
                        plug calorimeter}",
      collaboration  = "ZEUS",
      journal        = "Nucl.Phys.",
      volume         = "B800",
      pages          = "1-76",
      year           = "2008",
      eprint         = "0802.3017",
      archivePrefix  = "arXiv",
      primaryClass   = "hep-ex",
      reportNumber   = "DESY-08-011",
      SLACcitation   = "%%CITATION = ARXIV:0802.3017;%%",
}

@article{Chekanov:2005vv,
      author         = "Chekanov, S. and others",
      title          = "{Study of deep inelastic inclusive and diffractive
                        scattering with the ZEUS forward plug calorimeter}",
      collaboration  = "ZEUS",
      journal        = "Nucl.Phys.",
      volume         = "B713",
      pages          = "3-80",
      year           = "2005",
      eprint         = "hep-ex/0501060",
      archivePrefix  = "arXiv",
      primaryClass   = "hep-ex",
      reportNumber   = "DESY-05-011",
      SLACcitation   = "%%CITATION = HEP-EX/0501060;%%",
}

@article{Newman:2009km,
      author         = "Newman, Paul and Ruspa, Marta",
      title          = "{Towards a Combined HERA Diffractive Deep Inelastic
                        Scattering Measurement}",
      pages          = "401-411",
      year           = "2009",
      eprint         = "0903.2957",
      archivePrefix  = "arXiv",
      primaryClass   = "hep-ex",
      SLACcitation   = "%%CITATION = ARXIV:0903.2957;%%",
}

@article{:2012vx,
      author         = "Aaron, F.D. and others",
      title          = "{Combined inclusive diffractive cross sections measured
                        with forward proton spectrometers in deep inelastic $ep$
                        scattering at HERA}",
      collaboration  = "H1 and ZEUS",
      journal        = "Eur.Phys.J.",
      volume         = "C72",
      pages          = "2175",
      year           = "2012",
      eprint         = "1207.4864",
      archivePrefix  = "arXiv",
      primaryClass   = "hep-ex",
      reportNumber   = "DESY-12-100",
      SLACcitation   = "%%CITATION = ARXIV:1207.4864;%%",
}

@article{Aaron:2012ad,
      author         = "Aaron, F.D. and others",
      title          = "{Inclusive Measurement of Diffractive Deep-Inelastic
                        Scattering at HERA}",
      collaboration  = "H1",
      journal        = "Eur.Phys.J.",
      volume         = "C72",
      pages          = "2074",
      year           = "2012",
      eprint         = "1203.4495",
      archivePrefix  = "arXiv",
      primaryClass   = "hep-ex",
      reportNumber   = "DESY-12-041",
      SLACcitation   = "%%CITATION = ARXIV:1203.4495;%%",
}

@article{Chekanov:2008fh,
      author         = "Chekanov, S. and others",
      title          = "{Deep inelastic scattering with leading protons or large
                        rapidity gaps at HERA}",
      collaboration  = "ZEUS",
      journal        = "Nucl.Phys.",
      volume         = "B816",
      pages          = "1-61",
      year           = "2009",
      eprint         = "0812.2003",
      archivePrefix  = "arXiv",
      primaryClass   = "hep-ex",
      reportNumber   = "DESY-08-175",
      SLACcitation   = "%%CITATION = ARXIV:0812.2003;%%",
}

@article{Aktas:2006hy,
      author         = "Aktas, A. and others",
      title          = "{Measurement and QCD analysis of the diffractive
                        deep-inelastic scattering cross-section at HERA}",
      collaboration  = "H1",
      journal        = "Eur.Phys.J.",
      volume         = "C48",
      pages          = "715-748",
      year           = "2006",
      eprint         = "hep-ex/0606004",
      archivePrefix  = "arXiv",
      primaryClass   = "hep-ex",
      reportNumber   = "DESY-06-049",
      SLACcitation   = "%%CITATION = HEP-EX/0606004;%%",
}

@article{Aaron:2010aa,
      author         = "Aaron, F.D. and Alexa, C. and Andreev, V. and Backovic,
                        S. and Baghdasaryan, A. and others",
      title          = "{Measurement of the cross section for diffractive
                        deep-inelastic scattering with a leading proton at HERA}",
      journal        = "Eur.Phys.J.",
      volume         = "C71",
      pages          = "1578",
      year           = "2011",
      eprint         = "1010.1476",
      archivePrefix  = "arXiv",
      primaryClass   = "hep-ex",
      reportNumber   = "DESY-10-095",
      SLACcitation   = "%%CITATION = ARXIV:1010.1476;%%",
}

@article{Aktas:2006hx,
      author         = "Aktas, A. and others",
      title          = "{Diffractive deep-inelastic scattering with a leading
                        proton at HERA}",
      collaboration  = "H1",
      journal        = "Eur.Phys.J.",
      volume         = "C48",
      pages          = "749-766",
      year           = "2006",
      eprint         = "hep-ex/0606003",
      archivePrefix  = "arXiv",
      primaryClass   = "hep-ex",
      reportNumber   = "DESY-06-048",
      SLACcitation   = "%%CITATION = HEP-EX/0606003;%%",
}

@article{Adloff:1997sc,
      author         = "Adloff, C. and others",
      title          = "{Inclusive measurement of diffractive deep inelastic ep
                        scattering}",
      collaboration  = "H1",
      journal        = "Z.Phys.",
      volume         = "C76",
      pages          = "613-629",
      year           = "1997",
      eprint         = "hep-ex/9708016",
      archivePrefix  = "arXiv",
      primaryClass   = "hep-ex",
      reportNumber   = "DESY-97-158",
      SLACcitation   = "%%CITATION = HEP-EX/9708016;%%",
}

@article{Donnachie:1992ny,
      author         = "Donnachie, A. and Landshoff, P.V.",
      title          = "{Total cross-sections}",
      journal        = "Phys.Lett.",
      volume         = "B296",
      pages          = "227-232",
      year           = "1992",
      eprint         = "hep-ph/9209205",
      archivePrefix  = "arXiv",
      primaryClass   = "hep-ph",
      reportNumber   = "CERN-TH-6635-92",
      SLACcitation   = "%%CITATION = HEP-PH/9209205;%%",
}

@article{Chekanov:2009aa,
      author         = "Chekanov, S. and others",
      title          = "{A QCD analysis of ZEUS diffractive data}",
      collaboration  = "ZEUS",
      journal        = "Nucl.Phys.",
      volume         = "B831",
      pages          = "1-25",
      year           = "2010",
      eprint         = "0911.4119",
      archivePrefix  = "arXiv",
      primaryClass   = "hep-ex",
      reportNumber   = "DESY-09-191",
      SLACcitation   = "%%CITATION = ARXIV:0911.4119;%%",
}

@article{Cudell:1996sh,
      author         = "Cudell, Jean Rene and Kang, Kyungsik and Kim, Sung Ku",
      title          = "{Simple pole fits to p p and anti-p p total
                        cross-sections and real parts}",
      journal        = "Phys.Lett.",
      volume         = "B395",
      pages          = "311-331",
      year           = "1997",
      eprint         = "hep-ph/9601336",
      archivePrefix  = "arXiv",
      primaryClass   = "hep-ph",
      reportNumber   = "SNUTP-95-103, BROWN-HET-1016, ULG-PNT-95-1-JRC",
      SLACcitation   = "%%CITATION = HEP-PH/9601336;%%",
}

@article{Low:1975sv,
      author         = "Low, F.E.",
      title          = "{A Model of the Bare Pomeron}",
      journal        = "Phys.Rev.",
      volume         = "D12",
      pages          = "163-173",
      year           = "1975",
      reportNumber   = "MIT-CTP-438",
      SLACcitation   = "%%CITATION = PHRVA,D12,163;%%",
}

@article{Nussinov:1975mw,
      author         = "Nussinov, S.",
      title          = "{Colored Quark Version of Some Hadronic Puzzles}",
      journal        = "Phys.Rev.Lett.",
      volume         = "34",
      pages          = "1286-1289",
      year           = "1975",
      reportNumber   = "TAUP-462-75",
      SLACcitation   = "%%CITATION = PRLTA,34,1286;%%",
}

@article{Blumlein:2001xf,
      author         = "Blumlein, Johannes and Robaschik, Dieter",
      title          = "{On the scaling violations of diffractive structure
                        functions: Operator approach}",
      journal        = "Phys.Lett.",
      volume         = "B517",
      pages          = "222-232",
      year           = "2001",
      eprint         = "hep-ph/0106037",
      archivePrefix  = "arXiv",
      primaryClass   = "hep-ph",
      reportNumber   = "DESY-01-072",
      SLACcitation   = "%%CITATION = HEP-PH/0106037;%%",
}

@article{Trentadue:1993ka,
      author         = "Trentadue, L. and Veneziano, G.",
      title          = "{Fracture functions: An Improved description of inclusive
                        hard processes in QCD}",
      journal        = "Phys.Lett.",
      volume         = "B323",
      pages          = "201-211",
      year           = "1994",
      SLACcitation   = "%%CITATION = PHLTA,B323,201;%%",
}

@article{Aktas:2007bv,
      author         = "Aktas, A. and others",
      title          = "{Dijet Cross Sections and Parton Densities in Diffractive
                        DIS at HERA}",
      collaboration  = "H1",
      journal        = "JHEP",
      volume         = "0710",
      pages          = "042",
      year           = "2007",
      eprint         = "0708.3217",
      archivePrefix  = "arXiv",
      primaryClass   = "hep-ex",
      reportNumber   = "DESY-07-115",
      SLACcitation   = "%%CITATION = ARXIV:0708.3217;%%",
}

@article{Adloff:1997nn,
      author         = "Adloff, C. and others",
      title          = "{Thrust jet analysis of deep inelastic large rapidity gap
                        events}",
      collaboration  = "H1",
      journal        = "Eur.Phys.J.",
      volume         = "C1",
      pages          = "495-507",
      year           = "1998",
      eprint         = "hep-ex/9711006",
      archivePrefix  = "arXiv",
      primaryClass   = "hep-ex",
      reportNumber   = "DESY-97-210",
      SLACcitation   = "%%CITATION = HEP-EX/9711006;%%",
}

@article{Adloff:1998ds,
      author         = "Adloff, C. and others",
      title          = "{Hadron production in diffractive deep inelastic
                        scattering}",
      collaboration  = "H1",
      journal        = "Phys.Lett.",
      volume         = "B428",
      pages          = "206-220",
      year           = "1998",
      eprint         = "hep-ex/9803032",
      archivePrefix  = "arXiv",
      primaryClass   = "hep-ex",
      reportNumber   = "DESY-98-029",
      SLACcitation   = "%%CITATION = HEP-EX/9803032;%%",
}

@article{Adloff:1998dw,
      author         = "Adloff, C. and others",
      title          = "{Multiplicity structure of the hadronic final state in
                        diffractive deep inelastic scattering at HERA}",
      collaboration  = "H1",
      journal        = "Eur.Phys.J.",
      volume         = "C5",
      pages          = "439-452",
      year           = "1998",
      eprint         = "hep-ex/9804012",
      archivePrefix  = "arXiv",
      primaryClass   = "hep-ex",
      reportNumber   = "DESY-98-044",
      SLACcitation   = "%%CITATION = HEP-EX/9804012;%%",
}

@article{Aaron:2011mp,
      author         = "Aaron, F.D. and others",
      title          = "{Measurement of Dijet Production in Diffractive
                        Deep-Inelastic Scattering with a Leading Proton at HERA}",
      collaboration  = "H1",
      journal        = "Eur.Phys.J.",
      volume         = "C72",
      pages          = "1970",
      year           = "2012",
      eprint         = "1111.0584",
      archivePrefix  = "arXiv",
      primaryClass   = "hep-ex",
      reportNumber   = "DESY-11-166",
      SLACcitation   = "%%CITATION = ARXIV:1111.0584;%%",
}

@article{Adloff:2000qi,
      author         = "Adloff, C. and others",
      title          = "{Diffractive jet production in deep inelastic $e^{+} p$
                        collisions at HERA}",
      collaboration  = "H1",
      journal        = "Eur.Phys.J.",
      volume         = "C20",
      pages          = "29-49",
      year           = "2001",
      eprint         = "hep-ex/0012051",
      archivePrefix  = "arXiv",
      primaryClass   = "hep-ex",
      reportNumber   = "DESY-00-174",
      SLACcitation   = "%%CITATION = HEP-EX/0012051;%%",
}

@article{Aktas:2007hn,
      author         = "Aktas, A. and others",
      title          = "{Tests of QCD factorisation in the diffractive production
                        of dijets in deep-inelastic scattering and photoproduction
                        at HERA}",
      collaboration  = "H1",
      journal        = "Eur.Phys.J.",
      volume         = "C51",
      pages          = "549-568",
      year           = "2007",
      eprint         = "hep-ex/0703022",
      archivePrefix  = "arXiv",
      primaryClass   = "hep-ex",
      reportNumber   = "DESY-07-018",
      SLACcitation   = "%%CITATION = HEP-EX/0703022;%%",
}

@article{Martin:2006td,
      author         = "Martin, A.D. and Ryskin, M.G. and Watt, G.",
      title          = "{Diffractive parton distributions from H1 data}",
      journal        = "Phys.Lett.",
      volume         = "B644",
      pages          = "131-135",
      year           = "2007",
      eprint         = "hep-ph/0609273",
      archivePrefix  = "arXiv",
      primaryClass   = "hep-ph",
      reportNumber   = "IPPP-06-66, DCPT-06-132",
      SLACcitation   = "%%CITATION = HEP-PH/0609273;%%",
}

@article{Aktas:2006up,
      author         = "Aktas, A. and others",
      title          = "{Diffractive open charm production in deep-inelastic
                        scattering and photoproduction at HERA}",
      collaboration  = "H1",
      journal        = "Eur.Phys.J.",
      volume         = "C50",
      pages          = "1-20",
      year           = "2007",
      eprint         = "hep-ex/0610076",
      archivePrefix  = "arXiv",
      primaryClass   = "hep-ex",
      reportNumber   = "DESY-06-164",
      SLACcitation   = "%%CITATION = HEP-EX/0610076;%%",
}

@article{Chekanov:2007aa,
      author         = "Chekanov, S. and others",
      title          = "{Dijet production in diffractive deep inelastic
                        scattering at HERA}",
      collaboration  = "ZEUS",
      journal        = "Eur.Phys.J.",
      volume         = "C52",
      pages          = "813-832",
      year           = "2007",
      eprint         = "0708.1415",
      archivePrefix  = "arXiv",
      primaryClass   = "hep-ex",
      reportNumber   = "DESY-07-126",
      SLACcitation   = "%%CITATION = ARXIV:0708.1415;%%",
}

@article{Chekanov:2003gt,
      author         = "Chekanov, S. and others",
      title          = "{Measurement of the open charm contribution to the
                        diffractive proton structure function}",
      collaboration  = "ZEUS",
      journal        = "Nucl.Phys.",
      volume         = "B672",
      pages          = "3-35",
      year           = "2003",
      eprint         = "hep-ex/0307068",
      archivePrefix  = "arXiv",
      primaryClass   = "hep-ex",
      reportNumber   = "DESY-03-094",
      SLACcitation   = "%%CITATION = HEP-EX/0307068;%%",
}

@article{Newman:2005mm,
      author         = "Newman, Paul R.",
      title          = "{Prospects for F(L)**D measurements at HERA-II}",
      year           = "2005",
      eprint         = "hep-ex/0511047",
      archivePrefix  = "arXiv",
      primaryClass   = "hep-ex",
      SLACcitation   = "%%CITATION = HEP-EX/0511047;%%",
}

@article{Aaron:2012zz,
      author         = "Aaron, F.D. and others",
      title          = "{Measurement of the Diffractive Longitudinal Structure
                        Function $F_L^D$ at HERA}",
      collaboration  = "H1",
      journal        = "Eur.Phys.J.",
      volume         = "C72",
      pages          = "1836",
      year           = "2012",
      eprint         = "1107.3420",
      archivePrefix  = "arXiv",
      primaryClass   = "hep-ex",
      reportNumber   = "DESY-11-084",
      SLACcitation   = "%%CITATION = ARXIV:1107.3420;%%",
}

@article{Aaron:2010su,
      author         = "Aaron, F.D. and others",
      title          = "{Diffractive Dijet Photoproduction in ep Collisions at
                        HERA}",
      collaboration  = "H1",
      journal        = "Eur.Phys.J.",
      volume         = "C70",
      pages          = "15-37",
      year           = "2010",
      eprint         = "1006.0946",
      archivePrefix  = "arXiv",
      primaryClass   = "hep-ex",
      reportNumber   = "DESY-10-043, DESY10-043",
      SLACcitation   = "%%CITATION = ARXIV:1006.0946;%%",
}

@article{Chekanov:2007rh,
      author         = "Chekanov, Sergei and others",
      title          = "{Diffractive photoproduction of dijets in ep collisions
                        at HERA}",
      collaboration  = "ZEUS",
      journal        = "Eur.Phys.J.",
      volume         = "C55",
      pages          = "177-191",
      year           = "2008",
      eprint         = "0710.1498",
      archivePrefix  = "arXiv",
      primaryClass   = "hep-ex",
      reportNumber   = "DESY-07-161",
      SLACcitation   = "%%CITATION = ARXIV:0710.1498;%%",
}

@article{Kaidalov:2009fp,
      author         = "Kaidalov, A.B. and Khoze, V.A. and Martin, A.D. and
                        Ryskin, M.G.",
      title          = "{Factorization breaking in diffractive dijet
                        photoproduction at HERA}",
      journal        = "Eur.Phys.J.",
      volume         = "C66",
      pages          = "373-376",
      year           = "2010",
      eprint         = "0911.3716",
      archivePrefix  = "arXiv",
      primaryClass   = "hep-ph",
      reportNumber   = "IPPP-09-89, DCPT-09-178",
      SLACcitation   = "%%CITATION = ARXIV:0911.3716;%%",
}

@article{Kaidalov:2003xf,
      author         = "Kaidalov, A.B. and Khoze, V.A. and Martin, A.D. and
                        Ryskin, M.G.",
      title          = "{Unitarity effects in hard diffraction at HERA}",
      journal        = "Phys.Lett.",
      volume         = "B567",
      pages          = "61-68",
      year           = "2003",
      eprint         = "hep-ph/0306134",
      archivePrefix  = "arXiv",
      primaryClass   = "hep-ph",
      reportNumber   = "IPPP-03-32, DCPT-03-64",
      SLACcitation   = "%%CITATION = HEP-PH/0306134;%%",
}

@article{Dokshitzer:1991he,
      author         = "Dokshitzer, Yuri L. and Khoze, Valery A. and Sjostrand,
                        T.",
      title          = "{Rapidity gaps in Higgs production}",
      journal        = "Phys.Lett.",
      volume         = "B274",
      pages          = "116-121",
      year           = "1992",
      reportNumber   = "DTP-91-66, CERN-TH-6269-91, LU-TP-91-24",
      SLACcitation   = "%%CITATION = PHLTA,B274,116;%%",
}

@article{Bjorken:1992er,
      author         = "Bjorken, J.D.",
      title          = "{Rapidity gaps and jets as a new physics signature in
                        very high-energy hadron hadron collisions}",
      journal        = "Phys.Rev.",
      volume         = "D47",
      pages          = "101-113",
      year           = "1993",
      reportNumber   = "SLAC-PUB-5616",
      SLACcitation   = "%%CITATION = PHRVA,D47,101;%%",
}

@article{Chatrchyan:2012vc,
      author         = "Chatrchyan, Serguei and others",
      title          = "{Observation of a diffractive contribution to dijet
                        production in proton-proton collisions at $\sqrt{s}=7$
                        TeV}",
      collaboration  = "CMS",
      journal        = "Phys.Rev.",
      volume         = "D87",
      pages          = "012006",
      year           = "2013",
      eprint         = "1209.1805",
      archivePrefix  = "arXiv",
      primaryClass   = "hep-ex",
      reportNumber   = "CMS-FWD-10-004, CERN-PH-EP-2012-248",
      SLACcitation   = "%%CITATION = ARXIV:1209.1805;%%",
}

@article{Buchmuller:1995qa,
      author         = "Buchmuller, W. and Hebecker, Arthur",
      title          = "{A Parton model for diffractive processes in deep
                        inelastic scattering}",
      journal        = "Phys.Lett.",
      volume         = "B355",
      pages          = "573-578",
      year           = "1995",
      eprint         = "hep-ph/9504374",
      archivePrefix  = "arXiv",
      primaryClass   = "hep-ph",
      reportNumber   = "DESY-95-077",
      SLACcitation   = "%%CITATION = HEP-PH/9504374;%%",
}

@article{Edin:1995gi,
      author         = "Edin, A. and Ingelman, G. and Rathsman, J.",
      title          = "{Soft color interactions as the origin of rapidity gaps
                        in DIS}",
      journal        = "Phys.Lett.",
      volume         = "B366",
      pages          = "371-378",
      year           = "1996",
      eprint         = "hep-ph/9508386",
      archivePrefix  = "arXiv",
      primaryClass   = "hep-ph",
      reportNumber   = "TSL-ISV-95-0125, DESY-95-163",
      SLACcitation   = "%%CITATION = HEP-PH/9508386;%%",
}

@article{Rathsman:1998tp,
      author         = "Rathsman, Johan",
      title          = "{A Generalized area law for hadronic string
                        re-interactions}",
      journal        = "Phys.Lett.",
      volume         = "B452",
      pages          = "364-371",
      year           = "1999",
      eprint         = "hep-ph/9812423",
      archivePrefix  = "arXiv",
      primaryClass   = "hep-ph",
      reportNumber   = "SLAC-PUB-8034",
      SLACcitation   = "%%CITATION = HEP-PH/9812423;%%",
}

@article{Donnachie:1984xq,
      author         = "Donnachie, A. and Landshoff, P.V.",
      title          = "{Elastic Scattering and Diffraction Dissociation}",
      journal        = "Nucl.Phys.",
      volume         = "B244",
      pages          = "322",
      year           = "1984",
      reportNumber   = "DAMTP 84/6, M/C TH-84/8",
      SLACcitation   = "%%CITATION = NUPHA,B244,322;%%",
}

@article{Abe:1993xx,
      author         = "Abe, F. and others",
      title          = "{Measurement of small angle $\bar{p}p$ elastic scattering
                        at $\sqrt{s} = 546$ GeV and 1800 GeV}",
      collaboration  = "CDF",
      journal        = "Phys.Rev.",
      volume         = "D50",
      pages          = "5518-5534",
      year           = "1994",
      reportNumber   = "FERMILAB-PUB-93-232-E, CDF-PUB-MIN-BIAS-PUBLIC-2049,
                        ANL-HEP-PR-95-13",
      SLACcitation   = "%%CITATION = PHRVA,D50,5518;%%",
}

@article{Breitweg:1999jy,
      author         = "Breitweg, J. and others",
      title          = "{Measurement of diffractive photoproduction of vector
                        mesons at large momentum transfer at HERA}",
      collaboration  = "ZEUS",
      journal        = "Eur.Phys.J.",
      volume         = "C14",
      pages          = "213-238",
      year           = "2000",
      eprint         = "hep-ex/9910038",
      archivePrefix  = "arXiv",
      primaryClass   = "hep-ex",
      reportNumber   = "DESY-99-160, ANL-HEP-PR-99-119",
      SLACcitation   = "%%CITATION = HEP-EX/9910038;%%",
}

@article{Ahmed:1995ns,
      author         = "Ahmed, T. and others",
      title          = "{First measurement of the deep inelastic structure of
                        proton diffraction}",
      collaboration  = "H1",
      journal        = "Phys.Lett.",
      volume         = "B348",
      pages          = "681-696",
      year           = "1995",
      eprint         = "hep-ex/9503005",
      archivePrefix  = "arXiv",
      primaryClass   = "hep-ex",
      reportNumber   = "DESY-95-036",
      SLACcitation   = "%%CITATION = HEP-EX/9503005;%%",
}

@article{Nisius:1999cv,
      author         = "Nisius, Richard",
      title          = "{The Photon structure from deep inelastic electron photon
                        scattering}",
      journal        = "Phys.Rept.",
      volume         = "332",
      pages          = "165-317",
      year           = "2000",
      eprint         = "hep-ex/9912049",
      archivePrefix  = "arXiv",
      primaryClass   = "hep-ex",
      SLACcitation   = "%%CITATION = HEP-EX/9912049;%%",
}

@article{Klasen:2009bi,
      author         = "Klasen, Michael and Kramer, Gustav",
      title          = "{Survival probability for diffractive dijet production in
                        p anti-p collisions from next-to-leading order
                        calculations}",
      journal        = "Phys.Rev.",
      volume         = "D80",
      pages          = "074006",
      year           = "2009",
      eprint         = "0908.2531",
      archivePrefix  = "arXiv",
      primaryClass   = "hep-ph",
      reportNumber   = "DESY-09-130, LPSC-09-112",
      SLACcitation   = "%%CITATION = ARXIV:0908.2531;%%",
}

@article{Abramowicz:2011fa,
     author         = "Abramowicz, H. and others",
     title          = "{Measurement of the t dependence in exclusive
                       photoproduction of Upsilon(1S) mesons at HERA}",
     collaboration  = "ZEUS",
     journal        = "Phys.Lett.",
     volume         = "B708",
     pages          = "14-20",
     year           = "2012",
     eprint         = "1111.2133",
     archivePrefix  = "arXiv",
     primaryClass   = "hep-ex",
     reportNumber   = "DESY-11-186",
     SLACcitation   = "%%CITATION = ARXIV:1111.2133;%%",
}

@article{Aid:1996bs,
      author         = "Aid, S. and others",
      title          = "{Elastic photoproduction of rho0 mesons at HERA}",
      collaboration  = "H1",
      journal        = "Nucl.Phys.",
      volume         = "B463",
      pages          = "3-32",
      year           = "1996",
      eprint         = "hep-ex/9601004",
      archivePrefix  = "arXiv",
      primaryClass   = "hep-ex",
      reportNumber   = "DESY-95-251",
      SLACcitation   = "%%CITATION = HEP-EX/9601004;%%",
}

@article{Breitweg:1997ed,
      author         = "Breitweg, J. and others",
      title          = "{Elastic and proton dissociative $\rho^0$ photoproduction
                        at HERA}",
      collaboration  = "ZEUS",
      journal        = "Eur.Phys.J.",
      volume         = "C2",
      pages          = "247-267",
      year           = "1998",
      eprint         = "hep-ex/9712020",
      archivePrefix  = "arXiv",
      primaryClass   = "hep-ex",
      reportNumber   = "DESY-97-237",
      SLACcitation   = "%%CITATION = HEP-EX/9712020;%%",
}

@article{Aaron:2009xp,
      author         = "Aaron, F.D. and others",
      title          = "{Diffractive Electroproduction of rho and phi Mesons at
                        HERA}",
      collaboration  = "H1",
      journal        = "JHEP",
      volume         = "1005",
      pages          = "032",
      year           = "2010",
      eprint         = "0910.5831",
      archivePrefix  = "arXiv",
      primaryClass   = "hep-ex",
      reportNumber   = "DESY-09-093",
      SLACcitation   = "%%CITATION = ARXIV:0910.5831;%%",
}

@article{Adloff:1999kg,
      author         = "Adloff, C. and others",
      title          = "{Elastic electroproduction of rho mesons at HERA}",
      collaboration  = "H1",
      journal        = "Eur.Phys.J.",
      volume         = "C13",
      pages          = "371-396",
      year           = "2000",
      eprint         = "hep-ex/9902019",
      archivePrefix  = "arXiv",
      primaryClass   = "hep-ex",
      reportNumber   = "DESY-99-010",
      SLACcitation   = "%%CITATION = HEP-EX/9902019;%%",
}

@article{Chekanov:2007zr,
      author         = "Chekanov, S. and others",
      title          = "{Exclusive rho0 production in deep inelastic scattering
                        at HERA}",
      collaboration  = "ZEUS",
      journal        = "PMC Phys.",
      volume         = "A1",
      pages          = "6",
      year           = "2007",
      eprint         = "0708.1478",
      archivePrefix  = "arXiv",
      primaryClass   = "hep-ex",
      reportNumber   = "DESY-07-118",
      SLACcitation   = "%%CITATION = ARXIV:0708.1478;%%",
}

@article{Breitweg:1998nh,
      author         = "Breitweg, J. and others",
      title          = "{Exclusive electroproduction of $\rho^0$ and $J/\psi$
                        mesons at HERA}",
      collaboration  = "ZEUS",
      journal        = "Eur.Phys.J.",
      volume         = "C6",
      pages          = "603-627",
      year           = "1999",
      eprint         = "hep-ex/9808020",
      archivePrefix  = "arXiv",
      primaryClass   = "hep-ex",
      reportNumber   = "DESY-98-107, ANL-HEP-PR-98-95",
      SLACcitation   = "%%CITATION = HEP-EX/9808020;%%",
}

@article{Adloff:2001rw,
      author         = "Adloff, C. and others",
      title          = "{On the rise of the proton structure function F(2)
                        towards low x}",
      collaboration  = "H1",
      journal        = "Phys.Lett.",
      volume         = "B520",
      pages          = "183-190",
      year           = "2001",
      eprint         = "hep-ex/0108035",
      archivePrefix  = "arXiv",
      primaryClass   = "hep-ex",
      reportNumber   = "DESY-01-104",
      SLACcitation   = "%%CITATION = HEP-EX/0108035;%%",
}

@article{Adloff:2002tb,
      author         = "Adloff, C. and others",
      title          = "{A Measurement of the t dependence of the helicity
                        structure of diffractive rho meson electroproduction at
                        HERA}",
      collaboration  = "H1",
      journal        = "Phys.Lett.",
      volume         = "B539",
      pages          = "25-39",
      year           = "2002",
      eprint         = "hep-ex/0203022",
      archivePrefix  = "arXiv",
      primaryClass   = "hep-ex",
      reportNumber   = "DESY-02-027",
      SLACcitation   = "%%CITATION = HEP-EX/0203022;%%",
}

@article{Breitweg:1999fm,
      author         = "Breitweg, J. and others",
      title          = "{Measurement of the spin density matrix elements in
                        exclusive electroproduction of rho0 mesons at HERA}",
      collaboration  = "ZEUS",
      journal        = "Eur.Phys.J.",
      volume         = "C12",
      pages          = "393-410",
      year           = "2000",
      eprint         = "hep-ex/9908026",
      archivePrefix  = "arXiv",
      primaryClass   = "hep-ex",
      reportNumber   = "DESY-99-102, ANL-HEP-PR-99-93",
      SLACcitation   = "%%CITATION = HEP-EX/9908026;%%",
}

@article{Brodsky:1994kf,
      author         = "Brodsky, Stanley J. and Frankfurt, L. and Gunion, J.F.
                        and Mueller, Alfred H. and Strikman, M.",
      title          = "{Diffractive leptoproduction of vector mesons in QCD}",
      journal        = "Phys.Rev.",
      volume         = "D50",
      pages          = "3134-3144",
      year           = "1994",
      eprint         = "hep-ph/9402283",
      archivePrefix  = "arXiv",
      primaryClass   = "hep-ph",
      reportNumber   = "SLAC-PUB-6412, CU-TP-617, UCD-93-36",
      SLACcitation   = "%%CITATION = HEP-PH/9402283;%%",
}

@article{Collins:1996fb,
      author         = "Collins, John C. and Frankfurt, Leonid and Strikman,
                        Mark",
      title          = "{Factorization for hard exclusive electroproduction of
                        mesons in QCD}",
      journal        = "Phys.Rev.",
      volume         = "D56",
      pages          = "2982-3006",
      year           = "1997",
      eprint         = "hep-ph/9611433",
      archivePrefix  = "arXiv",
      primaryClass   = "hep-ph",
      reportNumber   = "CERN-TH-96-314, PSU-TH-168",
      SLACcitation   = "%%CITATION = HEP-PH/9611433;%%",
}

@article{Ryskin:1992ui,
      author         = "Ryskin, M.G.",
      title          = "{Diffractive J / psi electroproduction in LLA QCD}",
      journal        = "Z.Phys.",
      volume         = "C57",
      pages          = "89-92",
      year           = "1993",
      reportNumber   = "LU-TP-92-12",
      SLACcitation   = "%%CITATION = ZEPYA,C57,89;%%",
}

@article{Martin:1999wb,
      author         = "Martin, Alan D. and Ryskin, M.G. and Teubner, T.",
      title          = "{Q**2 dependence of diffractive vector meson
                        electroproduction}",
      journal        = "Phys.Rev.",
      volume         = "D62",
      pages          = "014022",
      year           = "2000",
      eprint         = "hep-ph/9912551",
      archivePrefix  = "arXiv",
      primaryClass   = "hep-ph",
      reportNumber   = "DTP-99-98",
      SLACcitation   = "%%CITATION = HEP-PH/9912551;%%",
}

@article{Martin:2007sb,
      author         = "Martin, A.D. and Nockles, C. and Ryskin, Mikhail G. and
                        Teubner, Thomas",
      title          = "{Small x gluon from exclusive J/psi production}",
      journal        = "Phys.Lett.",
      volume         = "B662",
      pages          = "252-258",
      year           = "2008",
      eprint         = "0709.4406",
      archivePrefix  = "arXiv",
      primaryClass   = "hep-ph",
      reportNumber   = "IPPP-07-50, DCPT-07-100, LTH-755",
      SLACcitation   = "%%CITATION = ARXIV:0709.4406;%%",
}

@article{Frankfurt:2000ez,
      author         = "Frankfurt, L. and McDermott, M. and Strikman, M.",
      title          = "{A Fresh look at diffractive J / psi photoproduction at
                        HERA, with predictions for THERA}",
      journal        = "JHEP",
      volume         = "0103",
      pages          = "045",
      year           = "2001",
      eprint         = "hep-ph/0009086",
      archivePrefix  = "arXiv",
      primaryClass   = "hep-ph",
      SLACcitation   = "%%CITATION = HEP-PH/0009086;%%",
}

@article{Aktas:2005xu,
      author         = "Aktas, A. and others",
      title          = "{Elastic J/psi production at HERA}",
      collaboration  = "H1",
      journal        = "Eur.Phys.J.",
      volume         = "C46",
      pages          = "585-603",
      year           = "2006",
      eprint         = "hep-ex/0510016",
      archivePrefix  = "arXiv",
      primaryClass   = "hep-ex",
      reportNumber   = "DESY-05-161",
      SLACcitation   = "%%CITATION = HEP-EX/0510016;%%",
}

@article{Chekanov:2002xi,
      author         = "Chekanov, S. and others",
      title          = "{Exclusive photoproduction of J / psi mesons at HERA}",
      collaboration  = "ZEUS",
      journal        = "Eur.Phys.J.",
      volume         = "C24",
      pages          = "345-360",
      year           = "2002",
      eprint         = "hep-ex/0201043",
      archivePrefix  = "arXiv",
      primaryClass   = "hep-ex",
      reportNumber   = "DESY-02-008",
      SLACcitation   = "%%CITATION = HEP-EX/0201043;%%",
}

@article{Adloff:2002em,
      author         = "Adloff, C. and others",
      title          = "{Energy flow and rapidity gaps between jets in
                        photoproduction at HERA}",
      collaboration  = "H1",
      journal        = "Eur.Phys.J.",
      volume         = "C24",
      pages          = "517-527",
      year           = "2002",
      eprint         = "hep-ex/0203011",
      archivePrefix  = "arXiv",
      primaryClass   = "hep-ex",
      reportNumber   = "DESY-02-023",
      SLACcitation   = "%%CITATION = HEP-EX/0203011;%%",
}

@article{Chekanov:2006pw,
      author         = "Chekanov, S. and others",
      title          = "{Photoproduction of events with rapidity gaps between
                        jets at HERA}",
      collaboration  = "ZEUS",
      journal        = "Eur.Phys.J.",
      volume         = "C50",
      pages          = "283-297",
      year           = "2007",
      eprint         = "hep-ex/0612008",
      archivePrefix  = "arXiv",
      primaryClass   = "hep-ex",
      reportNumber   = "DESY-06-215",
      SLACcitation   = "%%CITATION = HEP-EX/0612008;%%",
}

@article{Mueller:1992pe,
      author         = "Mueller, Alfred H. and Tang, Wai-Keung",
      title          = "{High-energy parton-parton elastic scattering in QCD}",
      journal        = "Phys.Lett.",
      volume         = "B284",
      pages          = "123-126",
      year           = "1992",
      SLACcitation   = "%%CITATION = PHLTA,B284,123;%%",
}

@article{Enberg:2001ev,
      author         = "Enberg, R. and Ingelman, G. and Motyka, L.",
      title          = "{Hard color singlet exchange and gaps between jets at the
                        Tevatron}",
      journal        = "Phys.Lett.",
      volume         = "B524",
      pages          = "273-282",
      year           = "2002",
      eprint         = "hep-ph/0111090",
      archivePrefix  = "arXiv",
      primaryClass   = "hep-ph",
      reportNumber   = "TSL-ISV-2001-0255",
      SLACcitation   = "%%CITATION = HEP-PH/0111090;%%",
}

@article{Forshaw:1997wn,
      author         = "Forshaw, Jeffrey R. and Sutton, P.J.",
      title          = "{Diffusion and the BFKL pomeron}",
      journal        = "Eur.Phys.J.",
      volume         = "C1",
      pages          = "285-291",
      year           = "1998",
      eprint         = "hep-ph/9703225",
      archivePrefix  = "arXiv",
      primaryClass   = "hep-ph",
      reportNumber   = "MC-TH-97-01, LU-TP-97-01",
      SLACcitation   = "%%CITATION = HEP-PH/9703225;%%",
}

@article{Aaron:2008ab,
      author         = "Aaron, F.D. and others",
      title          = "{Measurement of Diffractive Scattering of Photons with
                        Large Momentum Transfer at HERA}",
      collaboration  = "H1",
      journal        = "Phys.Lett.",
      volume         = "B672",
      pages          = "219-226",
      year           = "2009",
      eprint         = "0810.3096",
      archivePrefix  = "arXiv",
      primaryClass   = "hep-ex",
      reportNumber   = "DESY-08-077",
      SLACcitation   = "%%CITATION = ARXIV:0810.3096;%%",
}

@article{Aktas:2006qs,
      author         = "Aktas, A. and others",
      title          = "{Diffractive photoproduction of rho mesons with large
                        momentum transfer at HERA}",
      collaboration  = "H1",
      journal        = "Phys.Lett.",
      volume         = "B638",
      pages          = "422-431",
      year           = "2006",
      eprint         = "hep-ex/0603038",
      archivePrefix  = "arXiv",
      primaryClass   = "hep-ex",
      reportNumber   = "DESY-06-023",
      SLACcitation   = "%%CITATION = HEP-EX/0603038;%%",
}

@article{Aktas:2003zi,
      author         = "Aktas, A. and others",
      title          = "{Diffractive photoproduction of $J/\psi$ mesons with
                        large momentum transfer at HERA}",
      collaboration  = "H1",
      journal        = "Phys.Lett.",
      volume         = "B568",
      pages          = "205-218",
      year           = "2003",
      eprint         = "hep-ex/0306013",
      archivePrefix  = "arXiv",
      primaryClass   = "hep-ex",
      reportNumber   = "DESY-03-061",
      SLACcitation   = "%%CITATION = HEP-EX/0306013;%%",
}

@article{Chekanov:2002rm,
      author         = "Chekanov, S. and others",
      title          = "{Measurement of proton dissociative diffractive
                        photoproduction of vector mesons at large momentum
                        transfer at HERA}",
      collaboration  = "ZEUS",
      journal        = "Eur.Phys.J.",
      volume         = "C26",
      pages          = "389-409",
      year           = "2003",
      eprint         = "hep-ex/0205081",
      archivePrefix  = "arXiv",
      primaryClass   = "hep-ex",
      reportNumber   = "DESY-02-072",
      SLACcitation   = "%%CITATION = HEP-EX/0205081;%%",
}

@article{Chekanov:2009ab,
      author         = "Chekanov, S. and others",
      title          = "{Measurement of J/psi photoproduction at large momentum
                        transfer at HERA}",
      collaboration  = "ZEUS",
      journal        = "JHEP",
      volume         = "1005",
      pages          = "085",
      year           = "2010",
      eprint         = "0910.1235",
      archivePrefix  = "arXiv",
      primaryClass   = "hep-ex",
      reportNumber   = "DESY-09-137",
      SLACcitation   = "%%CITATION = ARXIV:0910.1235;%%",
}

@article{Ivanov:2004ax,
      author         = "Ivanov, I.P. and Nikolaev, N.N. and Savin, A.A.",
      title          = "{Diffractive vector meson production at HERA: From soft
                        to hard QCD}",
      journal        = "Phys.Part.Nucl.",
      volume         = "37",
      pages          = "1-85",
      year           = "2006",
      eprint         = "hep-ph/0501034",
      archivePrefix  = "arXiv",
      primaryClass   = "hep-ph",
      reportNumber   = "DESY-04-243",
      SLACcitation   = "%%CITATION = HEP-PH/0501034;%%",
}

@article{Forshaw:1995ax,
      author         = "Forshaw, Jeffrey R. and Ryskin, M.G.",
      title          = "{Diffractive vector meson production at large momentum
                        transfer}",
      journal        = "Z.Phys.",
      volume         = "C68",
      pages          = "137-148",
      year           = "1995",
      eprint         = "hep-ph/9501376",
      archivePrefix  = "arXiv",
      primaryClass   = "hep-ph",
      reportNumber   = "DESY-94-162-REV, RAL-94-058-REV, DESY-94-162, RAL-94-058",
      SLACcitation   = "%%CITATION = HEP-PH/9501376;%%",
}

@article{Bartels:1996fs,
      author         = "Bartels, Jochen and Forshaw, Jeffrey R. and Lotter, H.
                        and Wusthoff, M.",
      title          = "{Diffractive production of vector mesons at large t}",
      journal        = "Phys.Lett.",
      volume         = "B375",
      pages          = "301-309",
      year           = "1996",
      eprint         = "hep-ph/9601201",
      archivePrefix  = "arXiv",
      primaryClass   = "hep-ph",
      reportNumber   = "M-C-TH-95-23, ANL-HEP-PR-95-89, DESY-95-253",
      SLACcitation   = "%%CITATION = HEP-PH/9601201;%%",
}

@article{Forshaw:2001pf,
      author         = "Forshaw, Jeffrey R. and Poludniowski, G.",
      title          = "{Vector meson photoproduction at high t and comparison to
                        HERA data}",
      journal        = "Eur.Phys.J.",
      volume         = "C26",
      pages          = "411-415",
      year           = "2003",
      eprint         = "hep-ph/0107068",
      archivePrefix  = "arXiv",
      primaryClass   = "hep-ph",
      reportNumber   = "MC-TH-01-05",
      SLACcitation   = "%%CITATION = HEP-PH/0107068;%%",
}

@article{Gotsman:2001ne,
      author         = "Gotsman, E. and Levin, E. and Maor, U. and Naftali, E.",
      title          = "{Momentum transfer dependence of the differential
                        cross-section for J / psi production}",
      journal        = "Phys.Lett.",
      volume         = "B532",
      pages          = "37-47",
      year           = "2002",
      eprint         = "hep-ph/0110256",
      archivePrefix  = "arXiv",
      primaryClass   = "hep-ph",
      reportNumber   = "TAUP-2691-2001",
      SLACcitation   = "%%CITATION = HEP-PH/0110256;%%",
}

@article{Ginzburg:1985tp,
      author         = "Ginzburg, I.F. and Panfil, S.L. and Serbo, V.G.",
      title          = "{POSSIBILITY OF THE EXPERIMENTAL INVESTIGATION OF THE QCD
                        POMERON IN SEMIHARD PROCESSES AT THE gamma gamma
                        COLLISIONS}",
      journal        = "Nucl.Phys.",
      volume         = "B284",
      pages          = "685-705",
      year           = "1987",
      reportNumber   = "TF-26-139",
      SLACcitation   = "%%CITATION = NUPHA,B284,685;%%",
}

@article{Enberg:2002zy,
      author         = "Enberg, Rikard and Motyka, Leszek and Poludniowski,
                        Gavin",
      title          = "{Diffractive heavy vector meson production from the BFKL
                        equation}",
      journal        = "Eur.Phys.J.",
      volume         = "C26",
      pages          = "219-228",
      year           = "2002",
      eprint         = "hep-ph/0207027",
      archivePrefix  = "arXiv",
      primaryClass   = "hep-ph",
      SLACcitation   = "%%CITATION = HEP-PH/0207027;%%",
}

@article{Frankfurt:2008er,
      author         = "Frankfurt, Leonid and Strikman, Mark and Zhalov, Michael",
      title          = "{Large t diffractive J/psi photoproduction with proton
                        dissociation in ultraperipheral pA collisions at LHC}",
      journal        = "Phys.Lett.",
      volume         = "B670",
      pages          = "32-36",
      year           = "2008",
      eprint         = "0807.2208",
      archivePrefix  = "arXiv",
      primaryClass   = "hep-ph",
      SLACcitation   = "%%CITATION = ARXIV:0807.2208;%%",
}

@article{Chekanov:2008vy,
      author         = "Chekanov, S. and others",
      title          = "{A Measurement of the Q**2, W and t dependences of deeply
                        virtual Compton scattering at HERA}",
      collaboration  = "ZEUS",
      journal        = "JHEP",
      volume         = "0905",
      pages          = "108",
      year           = "2009",
      eprint         = "0812.2517",
      archivePrefix  = "arXiv",
      primaryClass   = "hep-ex",
      reportNumber   = "DESY-08-132",
      SLACcitation   = "%%CITATION = ARXIV:0812.2517;%%",
}

@article{H1:2009wnw,
    author = "Aaron, F. D. and others",
    collaboration = "H1",
    title = "{Deeply Virtual Compton Scattering and its Beam Charge Asymmetry in e+- Collisions at HERA}",
    eprint = "0907.5289",
    archivePrefix = "arXiv",
    primaryClass = "hep-ex",
    reportNumber = "DESY-09-109, DESY09-109",
    journal = "Phys. Lett. B",
    volume = "681",
    pages = "391--399",
    year = "2009"
}

@article{H1:2007vrx,
    author = "Aaron, F. D. and others",
    collaboration = "H1",
    title = "{Measurement of deeply virtual Compton scattering and its t-dependence at HERA}",
    eprint = "0709.4114",
    archivePrefix = "arXiv",
    primaryClass = "hep-ex",
    reportNumber = "DESY-07-142",
    journal = "Phys. Lett. B",
    volume = "659",
    pages = "796--806",
    year = "2008"
}

@article{Diehl:2003ny,
      author         = "Diehl, M.",
      title          = "{Generalized parton distributions}",
      journal        = "Phys.Rept.",
      volume         = "388",
      pages          = "41-277",
      year           = "2003",
      eprint         = "hep-ph/0307382",
      archivePrefix  = "arXiv",
      primaryClass   = "hep-ph",
      reportNumber   = "DESY-THESIS-2003-018",
      SLACcitation   = "%%CITATION = HEP-PH/0307382;%%",
}

@article{Kumericki:2009uq,
      author         = "Kumericki, Kresimir and Mueller, Dieter",
      title          = "{Deeply virtual Compton scattering at small x(B) and the
                        access to the GPD H}",
      journal        = "Nucl.Phys.",
      volume         = "B841",
      pages          = "1-58",
      year           = "2010",
      eprint         = "0904.0458",
      archivePrefix  = "arXiv",
      primaryClass   = "hep-ph",
      SLACcitation   = "%%CITATION = ARXIV:0904.0458;%%",
}

@article{Marquet:2007qa,
      author         = "Marquet, C. and Peschanski, Robert B. and Soyez, G.",
      title          = "{Exclusive vector meson production at HERA from QCD with
                        saturation}",
      journal        = "Phys.Rev.",
      volume         = "D76",
      pages          = "034011",
      year           = "2007",
      eprint         = "hep-ph/0702171",
      archivePrefix  = "arXiv",
      primaryClass   = "HEP-PH",
      SLACcitation   = "%%CITATION = HEP-PH/0702171;%%",
}

@article{Wusthoff:1997fz,
      author         = "Wusthoff, M.",
      title          = "{Large rapidity gap events in deep inelastic scattering}",
      journal        = "Phys.Rev.",
      volume         = "D56",
      pages          = "4311-4321",
      year           = "1997",
      eprint         = "hep-ph/9702201",
      archivePrefix  = "arXiv",
      primaryClass   = "hep-ph",
      reportNumber   = "ANL-HEP-PR-97-03",
      SLACcitation   = "%%CITATION = HEP-PH/9702201;%%",
}

@article{Airapetian:2012pg,
      author         = "Airapetian, A. and others",
      title          = "{Beam-helicity asymmetry arising from deeply virtual
                        Compton scattering measured with kinematically complete
                        event reconstruction}",
      collaboration  = "HERMES",
      journal        = "JHEP",
      volume         = "1210",
      pages          = "042",
      year           = "2012",
      eprint         = "1206.5683",
      archivePrefix  = "arXiv",
      primaryClass   = "hep-ex",
      reportNumber   = "DESY-12-095",
      SLACcitation   = "%%CITATION = ARXIV:1206.5683;%%",
}

@article{Airapetian:2012mq,
      author         = "Airapetian, A. and others",
      title          = "{Beam-helicity and beam-charge asymmetries associated
                        with deeply virtual Compton scattering on the unpolarised
                        proton}",
      collaboration  = "HERMES",
      journal        = "JHEP",
      volume         = "1207",
      pages          = "032",
      year           = "2012",
      eprint         = "1203.6287",
      archivePrefix  = "arXiv",
      primaryClass   = "hep-ex",
      reportNumber   = "DESY-12-040",
      SLACcitation   = "%%CITATION = ARXIV:1203.6287;%%",
}

@article{Gotsman:2007pn,
      author         = "Gotsman, E. and Kormilitzin, A. and Levin, E. and Maor,
                        U.",
      title          = "{Survival probability for high mass diffraction}",
      journal        = "Eur.Phys.J.",
      volume         = "C52",
      pages          = "295-304",
      year           = "2007",
      eprint         = "hep-ph/0702053",
      archivePrefix  = "arXiv",
      primaryClass   = "hep-ph",
      reportNumber   = "TAUP-2846-07",
      SLACcitation   = "%%CITATION = HEP-PH/0702053;%%",
}

@article{Frankfurt:1995jw,
      author         = "Frankfurt, Leonid and Koepf, Werner and Strikman, Mark",
      title          = "{Hard diffractive electroproduction of vector mesons in
                        QCD}",
      journal        = "Phys.Rev.",
      volume         = "D54",
      pages          = "3194-3215",
      year           = "1996",
      eprint         = "hep-ph/9509311",
      archivePrefix  = "arXiv",
      primaryClass   = "hep-ph",
      reportNumber   = "TAUP-2290-95",
      SLACcitation   = "%%CITATION = HEP-PH/9509311;%%",
}

@article{Martin:1996bp,
      author         = "Martin, Alan D. and Ryskin, M.G. and Teubner, T.",
      title          = "{The QCD description of diffractive rho meson
                        electroproduction}",
      journal        = "Phys.Rev.",
      volume         = "D55",
      pages          = "4329-4337",
      year           = "1997",
      eprint         = "hep-ph/9609448",
      archivePrefix  = "arXiv",
      primaryClass   = "hep-ph",
      reportNumber   = "DTP-96-64",
      SLACcitation   = "%%CITATION = HEP-PH/9609448;%%",
}

@article{Schilling:1973ag,
      author         = "Schilling, K. and Wolf, G.",
      title          = "{How to analyze vector meson production in inelastic
                        lepton scattering}",
      journal        = "Nucl.Phys.",
      volume         = "B61",
      pages          = "381-413",
      year           = "1973",
      SLACcitation   = "%%CITATION = NUPHA,B61,381;%%",
}

@article{Iancu:2003ge,
      author         = "Iancu, E. and Itakura, K. and Munier, S.",
      title          = "{Saturation and BFKL dynamics in the HERA data at small
                        x}",
      journal        = "Phys.Lett.",
      volume         = "B590",
      pages          = "199-208",
      year           = "2004",
      eprint         = "hep-ph/0310338",
      archivePrefix  = "arXiv",
      primaryClass   = "hep-ph",
      reportNumber   = "SACLAY-SPHT-T03-156, CPHT-RR-046-1003",
      SLACcitation   = "%%CITATION = HEP-PH/0310338;%%",
}

@article{Forshaw:1999uf,
      author         = "Forshaw, Jeffrey R. and Kerley, G. and Shaw, Graham",
      title          = "{Extracting the dipole cross-section from photoproduction
                        and electroproduction total cross-section data}",
      journal        = "Phys.Rev.",
      volume         = "D60",
      pages          = "074012",
      year           = "1999",
      eprint         = "hep-ph/9903341",
      archivePrefix  = "arXiv",
      primaryClass   = "hep-ph",
      reportNumber   = "CERN-TH-99-58, MC-TH-99-3",
      SLACcitation   = "%%CITATION = HEP-PH/9903341;%%",
}

@article{Forshaw:2003ki,
      author         = "Forshaw, Jeffrey R. and Sandapen, R. and Shaw, Graham",
      title          = "{Color dipoles and rho, phi electroproduction}",
      journal        = "Phys.Rev.",
      volume         = "D69",
      pages          = "094013",
      year           = "2004",
      eprint         = "hep-ph/0312172",
      archivePrefix  = "arXiv",
      primaryClass   = "hep-ph",
      SLACcitation   = "%%CITATION = HEP-PH/0312172;%%",
}

@article{AbelleiraFernandez:2012cc,
      author         = "Abelleira Fernandez, J.L. and others",
      title          = "{A Large Hadron Electron Collider at CERN: Report on the
                        Physics and Design Concepts for Machine and Detector}",
      collaboration  = "LHeC Study Group",
      journal        = "J.Phys.",
      volume         = "G39",
      pages          = "075001",
      year           = "2012",
      eprint         = "1206.2913",
      archivePrefix  = "arXiv",
      primaryClass   = "physics.acc-ph",
      reportNumber   = "SLAC-R-999, CERN-OPEN-2012-015, LHEC-NOTE-2012-001-GEN",
      SLACcitation   = "%%CITATION = ARXIV:1206.2913;%%",
}

@article{Frankfurt:1997at,
      author         = "Frankfurt, L.L. and Freund, A. and Strikman, M.",
      title          = "{Diffractive exclusive photoproduction in DIS at HERA}",
      journal        = "Phys.Rev.",
      volume         = "D58",
      pages          = "114001",
      year           = "1998",
      eprint         = "hep-ph/9710356",
      archivePrefix  = "arXiv",
      primaryClass   = "hep-ph",
      SLACcitation   = "%%CITATION = HEP-PH/9710356;%%",
}

@article{Andersson:1983ia,
      author         = "Andersson, Bo and Gustafson, G. and Ingelman, G. and
                        Sjostrand, T.",
      title          = "{Parton Fragmentation and String Dynamics}",
      journal        = "Phys.Rept.",
      volume         = "97",
      pages          = "31-145",
      year           = "1983",
      reportNumber   = "LU-TP-83-10",
      SLACcitation   = "%%CITATION = PRPLC,97,31;%%",
}

@article{Buchmuller:1998jv,
      author         = "Buchmuller, W. and Gehrmann, T. and Hebecker, Arthur",
      title          = "{Inclusive and diffractive structure functions at small
                        x}",
      journal        = "Nucl.Phys.",
      volume         = "B537",
      pages          = "477-500",
      year           = "1999",
      eprint         = "hep-ph/9808454",
      archivePrefix  = "arXiv",
      primaryClass   = "hep-ph",
      reportNumber   = "DESY-98-113, DAMTP-98-113",
      SLACcitation   = "%%CITATION = HEP-PH/9808454;%%",
}

@article{Buchmuller:1997eb,
      author         = "Buchmuller, W. and McDermott, M.F. and Hebecker, Arthur",
      title          = "{High p(T) jets in diffractive electroproduction}",
      journal        = "Phys.Lett.",
      volume         = "B410",
      pages          = "304-310",
      year           = "1997",
      eprint         = "hep-ph/9706354",
      archivePrefix  = "arXiv",
      primaryClass   = "hep-ph",
      reportNumber   = "DESY-97-099, DAMTP-97-59",
      SLACcitation   = "%%CITATION = HEP-PH/9706354;%%",
}

@article{Donnachie:1998gm,
      author         = "Donnachie, A. and Landshoff, P.V.",
      title          = "{Small x: Two pomerons!}",
      journal        = "Phys.Lett.",
      volume         = "B437",
      pages          = "408-416",
      year           = "1998",
      eprint         = "hep-ph/9806344",
      archivePrefix  = "arXiv",
      primaryClass   = "hep-ph",
      reportNumber   = "DAMTP-98-34, MC-TH-98-9",
      SLACcitation   = "%%CITATION = HEP-PH/9806344;%%",
}

@article{GolecBiernat:2007kv,
      author         = "Golec-Biernat, Krzysztof J. and Luszczak, A.",
      title          = "{Diffractive parton distributions from the analysis with
                        higher twist}",
      journal        = "Phys.Rev.",
      volume         = "D76",
      pages          = "114014",
      year           = "2007",
      eprint         = "0704.1608",
      archivePrefix  = "arXiv",
      primaryClass   = "hep-ph",
      SLACcitation   = "%%CITATION = ARXIV:0704.1608;%%",
}

@article{Royon:2000wg,
      author         = "Royon, C. and Schoeffel, L. and Bartels, Jochen and Jung,
                        H. and Peschanski, Robert B.",
      title          = "{QCD analysis of the diffractive structure function
                        F**D(3)(2)}",
      journal        = "Phys.Rev.",
      volume         = "D63",
      pages          = "074004",
      year           = "2001",
      eprint         = "hep-ph/0010015",
      archivePrefix  = "arXiv",
      primaryClass   = "hep-ph",
      reportNumber   = "DAPNIA-SPP-00-19",
      SLACcitation   = "%%CITATION = HEP-PH/0010015;%%",
}

@article{Frankfurt:2001av,
      author         = "Frankfurt, L. and Guzey, V. and McDermott, M. and
                        Strikman, M.",
      title          = "{Electron nucleus collisions at THERA}",
      year           = "2001",
      eprint         = "hep-ph/0104252",
      archivePrefix  = "arXiv",
      primaryClass   = "hep-ph",
      reportNumber   = "ADP-01-15-T450",
      SLACcitation   = "%%CITATION = HEP-PH/0104252;%%",
}

@article{Cox:2000jt,
      author         = "Cox, Brian E. and Forshaw, Jeffrey R.",
      title          = "{POMWIG: HERWIG for diffractive interactions}",
      journal        = "Comput.Phys.Commun.",
      volume         = "144",
      pages          = "104-110",
      year           = "2002",
      eprint         = "hep-ph/0010303",
      archivePrefix  = "arXiv",
      primaryClass   = "hep-ph",
      reportNumber   = "MAN-HEP-2000-3, MC-TH-00-09",
      SLACcitation   = "%%CITATION = HEP-PH/0010303;%%",
}

@article{Navin:2010kk,
      author         = "Navin, Sparsh",
      title          = "{Diffraction in Pythia}",
      year           = "2010",
      eprint         = "1005.3894",
      archivePrefix  = "arXiv",
      primaryClass   = "hep-ph",
      reportNumber   = "LUTP-09-23, MCNET-10-09",
      SLACcitation   = "%%CITATION = ARXIV:1005.3894;%%",
}

@article{Bartels:1996ne,
      author         = "Bartels, Jochen and Lotter, H. and Wusthoff, M.",
      title          = "{Quark - anti-quark production in DIS diffractive
                        dissociation}",
      journal        = "Phys.Lett.",
      volume         = "B379",
      pages          = "239-248",
      year           = "1996",
      eprint         = "hep-ph/9602363",
      archivePrefix  = "arXiv",
      primaryClass   = "hep-ph",
      reportNumber   = "DESY-96-026, ANL-HEP-PR-96-11",
      SLACcitation   = "%%CITATION = HEP-PH/9602363;%%",
}

@article{Lotter:1996zu,
      author         = "Lotter, H.",
      title          = "{Production of charm quark jets in DIS diffractive
                        dissociation}",
      journal        = "Phys.Lett.",
      volume         = "B406",
      pages          = "171-177",
      year           = "1997",
      eprint         = "hep-ph/9612415",
      archivePrefix  = "arXiv",
      primaryClass   = "hep-ph",
      reportNumber   = "DESY-96-260",
      SLACcitation   = "%%CITATION = HEP-PH/9612415;%%",
}

@article{Bartels:1999tn,
      author         = "Bartels, Jochen and Jung, H. and Wusthoff, M.",
      title          = "{Quark - anti-quark gluon jets in DIS diffractive
                        dissociation}",
      journal        = "Eur.Phys.J.",
      volume         = "C11",
      pages          = "111-125",
      year           = "1999",
      eprint         = "hep-ph/9903265",
      archivePrefix  = "arXiv",
      primaryClass   = "hep-ph",
      reportNumber   = "DESY-99-027, DTP-99-10, LUNFD6-NFFL-7166-1999",
      SLACcitation   = "%%CITATION = HEP-PH/9903265;%%",
}

@article{Bartels:2002ri,
      author         = "Bartels, J. and Jung, H. and Kyrieleis, A.",
      title          = "{Massive c anti-c g: Calculation in diffractive DIS and
                        diffractive D* production at HERA}",
      journal        = "Eur.Phys.J.",
      volume         = "C24",
      pages          = "555-560",
      year           = "2002",
      eprint         = "hep-ph/0204269",
      archivePrefix  = "arXiv",
      primaryClass   = "hep-ph",
      reportNumber   = "DESY-01-116, LUNFD6-NFFL-7204-2001",
      SLACcitation   = "%%CITATION = HEP-PH/0204269;%%",
}

@article{Alexa:2013xxa,
      author         = "Alexa, C. and others",
      title          = "{Elastic and Proton-Dissociative Photoproduction of J/psi
                        Mesons at HERA}",
      collaboration  = "H1",
      year           = "2013",
      eprint         = "1304.5162",
      archivePrefix  = "arXiv",
      primaryClass   = "hep-ex",
      reportNumber   = "DESY-13-058",
      SLACcitation   = "%%CITATION = ARXIV:1304.5162;%%",
}

@article{Chekanov:2004mw,
      author         = "Chekanov, S. and others",
      title          = "{Exclusive electroproduction of J/psi mesons at HERA}",
      collaboration  = "ZEUS",
      journal        = "Nucl.Phys.",
      volume         = "B695",
      pages          = "3-37",
      year           = "2004",
      eprint         = "hep-ex/0404008",
      archivePrefix  = "arXiv",
      primaryClass   = "hep-ex",
      reportNumber   = "DESY-04-052",
      SLACcitation   = "%%CITATION = HEP-EX/0404008;%%",
}

@article{Lukaszuk:1973nt,
      author         = "Lukaszuk, L. and Nicolescu, B.",
      title          = "{A Possible interpretation of p p rising total
                        cross-sections}",
      journal        = "Lett.Nuovo Cim.",
      volume         = "8",
      pages          = "405-413",
      year           = "1973",
      SLACcitation   = "%%CITATION = NCLTA,8,405;%%",
}

@article{LHCb:2013nqs,
    author = "Aaij, R and others",
    collaboration = "LHCb",
    title = "{Exclusive $J/\psi$ and $\psi$(2S) production in pp collisions at $ \sqrt{s} = 7$ TeV}",
    eprint = "1301.7084",
    archivePrefix = "arXiv",
    primaryClass = "hep-ex",
    reportNumber = "CERN-PH-EP-2013-005, LHCB-PAPER-2012-044",
    journal = "J. Phys. G",
    volume = "40",
    pages = "045001",
    year = "2013"
}

@article{ATLAS:2019asg,
    author = "Aad, Georges and others",
    collaboration = "ATLAS",
    title = "{Measurement of differential cross sections for single diffractive dissociation in $\sqrt{s} = 8$ TeV $pp$ collisions using the ATLAS ALFA spectrometer}",
    eprint = "1911.00453",
    archivePrefix = "arXiv",
    primaryClass = "hep-ex",
    reportNumber = "CERN-EP-2019-190",
    journal = "JHEP",
    volume = "02",
    pages = "042",
    year = "2020",
    note = "[Erratum: JHEP 10, 182 (2020)]"
}

@article{CMS:2015inp,
    author = "Khachatryan, Vardan and others",
    collaboration = "CMS",
    title = "{Measurement of diffraction dissociation cross sections in pp collisions at $\sqrt{s}$ = 7 TeV}",
    eprint = "1503.08689",
    archivePrefix = "arXiv",
    primaryClass = "hep-ex",
    reportNumber = "CMS-FSQ-12-005, CERN-PH-EP-2015-062",
    journal = "Phys. Rev. D",
    volume = "92",
    number = "1",
    pages = "012003",
    year = "2015"
}

@inproceedings{Levy:2009gy,
    author = "Levy, Aharon",
    title = "{Electroproduction of Vector Mesons}",
    booktitle = "{17th International Workshop on Deep-Inelastic Scattering and Related Subjects}",
    eprint = "0907.2178",
    archivePrefix = "arXiv",
    primaryClass = "hep-ex",
    pages = "177",
    month = "7",
    year = "2009"
}

@article{ALICE:2023mfc,
    author = "Acharya, Shreyasi and others",
    collaboration = "ALICE",
    title = "{Exclusive and dissociative J/{\ensuremath{\psi}} photoproduction, and exclusive dimuon production, in p-Pb collisions at sNN=8.16{\,}{\,}TeV}",
    eprint = "2304.12403",
    archivePrefix = "arXiv",
    primaryClass = "nucl-ex",
    reportNumber = "CERN-EP-2023-059",
    journal = "Phys. Rev. D",
    volume = "108",
    number = "11",
    pages = "112004",
    year = "2023"
}

@article{Brodsky:1998kn,
    author = "Brodsky, Stanley J. and Fadin, Victor S. and Kim, Victor T. and Lipatov, Lev N. and Pivovarov, Grigorii B.",
    title = "{The QCD pomeron with optimal renormalization}",
    eprint = "hep-ph/9901229",
    archivePrefix = "arXiv",
    reportNumber = "SLAC-PUB-8037, IITAP-98-010",
    journal = "JETP Lett.",
    volume = "70",
    pages = "155--160",
    year = "1999"
}

@article{Nikolaev:1995ed,
    author = "Nikolaev, Nikolai N. and Zakharov, B. G. and Zoller, V. R.",
    title = "{Exploratory study of shrinkage of the diffraction cone for the generalized BFKL pomeron}",
    eprint = "hep-ph/9506281",
    archivePrefix = "arXiv",
    reportNumber = "KFA-IKP-TH-1995-07",
    journal = "Phys. Lett. B",
    volume = "366",
    pages = "337--344",
    year = "1996"
}

@article{Gelis:2010nm,
    author = "Gelis, Francois and Iancu, Edmond and Jalilian-Marian, Jamal and Venugopalan, Raju",
    title = "{The Color Glass Condensate}",
    eprint = "1002.0333",
    archivePrefix = "arXiv",
    primaryClass = "hep-ph",
    journal = "Ann. Rev. Nucl. Part. Sci.",
    volume = "60",
    pages = "463--489",
    year = "2010"
}

@article{Poludniowski:2003yk,
    author = "Poludniowski, G. G. and Enberg, R. and Forshaw, Jeffrey R. and Motyka, L.",
    title = "{Vector meson photoproduction from the BFKL equation. 2. Phenomenology}",
    eprint = "hep-ph/0311017",
    archivePrefix = "arXiv",
    journal = "JHEP",
    volume = "12",
    pages = "002",
    year = "2003"
}

@article{Enberg:2003jw,
    author = "Enberg, R. and Forshaw, Jeffrey R. and Motyka, L. and Poludniowski, G.",
    title = "{Vector meson photoproduction from the BFKL equation. 1. Theory}",
    eprint = "hep-ph/0306232",
    archivePrefix = "arXiv",
    reportNumber = "TSL-ISV-2003-0269",
    journal = "JHEP",
    volume = "09",
    pages = "008",
    year = "2003"
}

@article{Blok:2010ds,
    author = "Blok, B. and Frankfurt, L. and Strikman, M.",
    title = "{DGLAP versus perturbative Pomeron in large momentum transfer hard diffractive processes at HERA and LHC}",
    eprint = "1002.3048",
    archivePrefix = "arXiv",
    primaryClass = "hep-ph",
    journal = "Phys. Lett. B",
    volume = "690",
    pages = "159--163",
    year = "2010"
}

@article{Newman:2013ada,
    author = "Newman, Paul and Wing, Matthew",
    title = "{The Hadronic Final State at HERA}",
    eprint = "1308.3368",
    archivePrefix = "arXiv",
    primaryClass = "hep-ex",
    journal = "Rev. Mod. Phys.",
    volume = "86",
    number = "3",
    pages = "1037",
    year = "2014"
}

\end{document}